\documentclass[10pt,preprint]{aastex}
\usepackage[utf8]{inputenc}
\usepackage{float}

\title{Eclipsing Binaries as Benchmarks for Trigonometric Parallaxes in the {\it Gaia} Era}
\author{Keivan G.\ Stassun\altaffilmark{1,2} and Guillermo Torres\altaffilmark{3}}

\altaffiltext{1}{Vanderbilt University, Department of Physics \& Astronomy, 6301 Stevenson Center Ln., Nashville, TN  37235, USA; {\tt keivan.stassun@vanderbilt.edu}}
\altaffiltext{2}{Fisk University, Department of Physics, 1000 17th Ave. N., Nashville, TN  37208, USA}
\altaffiltext{3}{Harvard-Smithsonian Center for Astrophysics, 60 Garden St., Cambridge, MA 02138, USA}

\newcommand{\lbol}{$L_{\rm bol}$}
\newcommand{\teff}{$T_{\rm eff}$}
\newcommand{\logg}{$\log g$}
\newcommand{\feh}{[Fe/H]}
\newcommand{\fbol}{$F_{\rm bol}$}
\newcommand{\rsun}{$R_\odot$}

\newcommand{\nebs}{{158}}
\newcommand{\ngoodseds}{{156}}
\newcommand{\nbadseds}{{2}}
\newcommand{\nhip}{{99}}
\newcommand{\ngoodhip}{{86}}
\newcommand{\nav}{{67}}

\newcommand{\parprecper}{{5\%}}
\newcommand{\parprec}{200}
\newcommand{\parprecwunit}{{\parprec}~$\mu$as}
\newcommand{\fbolprecper}{{3\%}}
\newcommand{\lbolprecper}{{10\%}}

\begin{document}

\begin{abstract}
We present fits to the broadband photometric spectral energy distributions (SEDs) of \nebs\ eclipsing binaries (EBs) in the {\it Tycho-2\/} catalog. These EBs were selected because they have highly precise stellar radii, effective temperatures, and in many cases metallicities previously determined in the literature, and thus have bolometric luminosities that are typically good to $\lesssim$\lbolprecper. In most cases the available broadband photometry spans a wavelength range 0.4--10~$\mu$m, and in many cases spans 0.15--22~$\mu$m. The resulting SED fits, which have only extinction
as a free parameter, provide a virtually model-independent measure of the bolometric flux at Earth. The SED fits are satisfactory for \ngoodseds\ of the EBs, for which we achieve typical precisions in the bolometric flux of $\approx$\fbolprecper. Combined with the accurately known bolometric luminosity, the result for each EB is a predicted parallax that is typically precise to $\lesssim$\parprecper. These predicted parallaxes---with typical uncertainties of \parprecwunit---are 4--5 times more precise than those determined by {\it Hipparcos\/} for \nhip\ of the EBs in our sample, with which we find excellent agreement. There is no evidence among this sample for significant systematics in the {\it Hipparcos\/} parallaxes of the sort that notoriously afflicted the Pleiades measurement.
The EBs are distributed over the entire sky, span more than 10 mag in brightness, reach distances of more than 5 kpc, and in many cases our predicted parallaxes should also be more precise than those expected from the {\it Gaia\/} first data release. The EBs studied here can thus serve as empirical, independent benchmarks for these upcoming fundamental parallax measurements.
\end{abstract}

\section{Introduction\label{sec:intro}}
There is arguably no astronomical measurement more fundamental than distance from trigonometric parallax. Such parallax measurements are foundational to the cosmic distance scale generally and to stellar astrophysics specifically, including our basic understanding of stellar evolution, stellar populations, and Galactic structure. 
Thus, the parallaxes provided by the {\it Hipparcos\/} mission \citep{Perryman:1997, vanLeeuwen:2007} have had important, unrivaled impact across many areas of study for the past 20 years. 

Yet the {\it Hipparcos\/} parallaxes were not without problems, perhaps the most significant of which was the aberrant distance to the Pleiades which the {\it Hipparcos\/} parallax placed at $120.2\pm1.9$ pc \citep{vanLeeuwen:2009} as compared to the broadly accepted distance of $\approx$135 pc from a variety of methods \citep[e.g.,][and others cited therein]{Munari:2004, Zwahlen:2004, Soderblom:2005, Groenewegen:2007, Melis:2014, Madler:2016}. Several of these recent determinations are formally highly precise ($\sigma = 1.2$--1.7 pc).
The discrepancy in the {\it Hipparcos\/} distance for the Pleiades was identified almost immediately \citep[e.g.,][]{Pinsonneault:1998, Soderblom:1998}, thanks to the unusual combination of proximity and cluster richness afforded by the Pleiades. In general, however, heretofore there have been very few {\it fundamental} benchmarks against which to test the {\it Hipparcos\/} parallaxes across the sky, across various stellar environments, and across stellar parameter space. One notable
example is the use of spatially resolved double-lined spectroscopic binaries, which can yield highly accurate and precise orbital parallaxes \citep[see, e.g.,][]{Tomkin:2005}.

Eclipsing binary (EB) stars have long served as fundamental benchmarks for stellar astrophysics. Through analysis of the light curve and radial velocities, EBs yield direct measurement of the component stellar masses and radii, and temperatures, with accuracies of $\sim$1\% for the best cases. As a result, the bolometric luminosity (\lbol) of an EB can be determined to very high precision and---importantly---without need of a distance measurement because it depends only on the measured radii and effective temperatures. 

EBs with such accurately determined \lbol\ can therefore serve as empirical, independent benchmarks for stellar distances obtained by other methods such as trigonometric parallax, if the bolometric flux at Earth (\fbol) can also be measured empirically and with sufficient precision. 
For example, in the recent analysis of newly discovered Pleiades EBs in the {\it K2\/} (successor to {\it Kepler\/}) mission data, \citet{David:2016} used the available broadband photometry from {\it GALEX\/} ultraviolet bands through {\it WISE\/} mid-infrared bands to determine an empirical distance of 132$\pm$5~pc to the Pleiades EB HCG~76, consistent with the consensus distance and again showing the {\it Hipparcos\/} distance to be biased. 
Impressively, this new empirical EB-based distance has a precision of better than 4\%, significantly more precise than the 12\% discrepancy in the {\it Hipparcos} distance, making it useful as a meaningful benchmark. 

Variants of this idea have been used to determine distances to EBs in the near field (relying on bolometric corrections) and even in external galaxies such as M31, the LMC or the SMC \citep[e.g.,][]{Ribas:2005, Pietrzynski:2009, Graczyk:2014}, sometimes based on application of surface brightness relations. These latter EB studies and others like them have been extremely important in establishing the lower rungs of the cosmological distance ladder, serving as calibrators for other methods reaching larger distances.

The upcoming {\it Gaia\/} mission holds great promise for many areas of stellar and Galactic astrophysics through the provision of fundamental trigonometric parallax measurements for $\sim 10^9$ stars. At the same time, there now exists a large sample of benchmark-grade EBs with accurate radii and temperatures, and in many cases metallicities, which provide accurate and distance-independent \lbol. In addition, there now exist all-sky, broadband photometric measurements for stars spanning a very broad range of wavelengths, from the {\it GALEX\/} far-UV at $\sim$0.1~$\mu$m to the {\it WISE} mid-IR at $\sim$22~$\mu$m. These measurements permit construction of spectral energy distributions (SEDs) that effectively sample the majority of the flux for all but the hottest stars. Consequently the bolometric fluxes, and in turn the distances to the EBs, can in principle be determined in a largely empirical manner that preserves the accuracy of the fundamental EB parameters. 

In this paper, we use available broadband photometry to construct empirical SEDs and to calculate \fbol\ for \nebs\ EBs in the {\it Tycho-2\/} catalog whose fundamental physical properties have previously been established with accuracies of better than 3\% \citep[e.g.,][]{Torres:2010}. The wavelength coverage of the SEDs for most of the EBs in our sample is sufficiently large that the resulting \fbol\ are typically precise to \fbolprecper, leading to predicted parallaxes that are in most cases precise to better than \parprecper. For \nhip\ of the EBs in our study sample, previous {\it Hipparcos\/} parallaxes are available for direct comparison to the parallaxes predicted from the EBs.

In Section~\ref{sec:data}, we present our study sample, the data that we use from the literature, and our SED fitting procedures. The main results of the work, including empirical \fbol\ and predicted parallaxes for the full sample, and a comparison to {\it Hipparcos\/} parallaxes where available, are presented in Section~\ref{sec:results}. In Section~\ref{sec:disc} we briefly discuss the upcoming applicability of this work to {\it Gaia}, which is expected to include all of these EBs as early as its first public data release. We summarize our conclusions in Section~\ref{sec:summary}.

\section{Data and Methods\label{sec:data}}

\subsection{Benchmark EB study sample}

For this work we have focused on detached EBs with well determined
radii and effective temperatures, to allow the derivation of
bolometric luminosities of the highest precision. We set a threshold
of 3\% for the uncertainties in the absolute radii, although a handful
of our objects were allowed to exceed this limit slightly, as
described below. More than half of the sample was drawn directly from
the compilation by \cite{Torres:2010}, featuring binaries that have been
carefully vetted with special attention paid to the number and quality of
the observations, the consistency of the light curve and radial-velocity
curve solutions and other details of the analysis, external checks, and
efforts to assess and control systematic errors.  Additional systems
were gathered from the more recent (or sometimes earlier) literature
including \citet{Stassun:2014} and the DEBCAT list\footnote{\url{http://www.astro.keele.ac.uk/jkt/debcat/}} maintained by \cite{Southworth:2015}, though some
have not necessarily received the same level of vetting beyond the requirement
of 3\% or better formal uncertainties in the radii. While we have attempted to
capture the most relevant systems for our study, our search was not
intended to be exhaustive so we make no claims of completeness. All EBs
were required to have entries in the {\it Tycho-2\/} catalog \citep{Hog:2000},
and many are contained also in the {\it Hipparcos\/} catalog.

Spectral types in our sample span a very wide range from late O to mid
M.  Most are main-sequence stars, but a few are giants. The challenges
involved in analyzing the spectroscopic and photometric observations 
of these binaries vary greatly depending on the nature of the system,
and in some cases unrecognized systematic errors or other problems may
not be fully reflected in the formal errors we have adopted.
It is also important to note that the temperature estimates are much
less fundamental in nature than the radius estimates. This is largely
because temperatures often rely on external calibrations, and have 
been derived in many different ways by different authors. While a  
light curve typically contains very precise information about the 
ratio between the primary and secondary temperatures, the value for
the primary that sets the absolute scale for the system must generally
be determined independently. This is sometimes done from an analysis
of disentangled spectra, from color indices, or even simply from its
spectral classification. The radii, on the other hand, are always 
based purely on geometry and dynamics. It is not surprising,
therefore, that the temperature uncertainties in our sample can be as
large as 10\% in some cases.

Small departures from spherical star shapes for most of the binaries in
our sample, or even moderate departures for the closest ones, along with
associated limb-darkening, reflection, tidal/rotational, and other effects,
are assumed to have been adequately accounted for in the light-curve
modeling, as suggested by the typically good agreement these systems show
when compared against stellar evolution models in the original publications.
In particular, the stellar sizes we have adopted are the volumetric radii,
as published.

The complete list of \nebs\ EBs for this study is given in
Table~\ref{tab:sample}, sorted by {\it Tycho\/} number.
We have made an effort to collect 
available estimates of the interstellar reddening for these systems,
as well as spectroscopic or photometric measures of the metallicity,
both of which can serve to constrain the SED fits described later. A
number of the binary components have been found to be metallic-line A 
or F stars and are so noted in the table, with estimates of the iron 
abundances given when available.  Four of our EBs belong to open
clusters for which the {\it Hipparcos\/} Mission has provided highly precise
average parallaxes based on individual measurements for half a dozen
or more members \citep{vanLeeuwen:2009}. They are V906~Sco (in   
NGC~6475), GV~Car (NGC~3532), V392~Car (NGC~2516), and TX~Cnc 
(Praesepe). The first also has an individual {\it Hipparcos\/} parallax.
GV~Car and TX~Cnc have somewhat poorer radius and/or temperature
determinations than the rest (and are only preliminary in the first
case), but they were included to enable a check on the {\it Hipparcos\/}
cluster distances. TX~Cnc is an over-contact (W~UMa) system rather than a
detached one, though this should not affect its usefulness for distance
determinations \citep[see, e.g.,][]{Wilson:2010}.


\subsection{Broadband photometric data from the literature} 
The \nebs\ EBs that comprise our study sample are, by virtue of having previously determined EB solutions and being in the {\it Tycho-2\/} catalog, relatively bright and well studied. Therefore, in most cases the EBs appear in many published photometric catalogs. In order to systematize and simplify our procedures, we opted to assemble for each EB the available broadband photometry from only the following large, all-sky catalogs (listed here in approximate order by wavelength coverage) via the {\tt VizieR}\footnote{\url{http://vizier.u-strasbg.fr/}} query service: 
\begin{itemize}
    \item {\it GALEX} All-sky Imaging Survey (AIS): FUV and NUV at $\approx$0.15 \micron\ and $\approx$0.22 \micron, respectively. 
    \item Catalog of Homogeneous Means in the $UBV$ System for bright stars from \citet{Mermilliod:2006}: Johnson $UBV$ bands ($\approx$0.35--0.55 \micron).
    \item {\it Tycho-2\/}: Tycho $B$ ($B_{\rm T}$) and Tycho $V$ ($V_{\rm T}$) bands ($\approx$0.42 \micron\ and $\approx$0.54 \micron, respectively). 
    \item Str\"omgren Photometric Catalog by \citet{Paunzen:2015}: Str\"omgren $uvby$ bands ($\approx$0.34--0.55 \micron).
    \item AAVSO Photometric All-Sky Survey (APASS) DR6 (obtained from the UCAC-4 catalog): Johnson $BV$ and SDSS $gri$ bands ($\approx$0.45--0.75 \micron). 
    \item Two-Micron All-Sky Survey (2MASS): $JHK_S$ bands ($\approx$1.2--2.2 \micron). 
    \item All-WISE: {\it WISE1--4} bands ($\approx$3.5--22 \micron). 
\end{itemize}

We found $B_{\rm T} V_{\rm T}$, $JHK_S$, and {\it WISE1--3} photometry---spanning a wavelength range $\approx$0.4--10 \micron---for nearly all of the EBs in our study sample. Most of the EBs also have {\it WISE4\/} photometry, and many of the EBs also have Str\"omgren and/or {\it GALEX\/} photometry, thus extending the wavelength coverage to $\approx$0.15--22~\micron. We adopted the reported measurement uncertainties unless they were less than 0.01 mag, in which case we assumed an uncertainty of 0.01 mag. In addition, to account for an artifact in the Kurucz atmospheres at 10 \micron, we artificially inflated the {\it WISE3\/} uncertainty to 0.1 mag unless the reported uncertainty was already larger than 0.1 mag.

Although the likelihood is small, it is always possible that some of these brightness measurements were obtained during an
eclipse, in which case they would underestimate the total flux of the system. However, most of the catalogs listed above
report averages of multi-epoch observations, from as many as 100 or more individual measurements taken over three years in the case of
{\it Tycho-2}, so they are less likely to be affected. Single-epoch measurements such as those in the 2MASS catalog, on
the other hand, are more susceptible to this problem. We found eleven systems in which the 2MASS measurements were
clearly obtained in eclipse, and in those cases we applied adjustments by referring back to the original optical light
curves at the exact phase of the observation, with small corrections from the optical to the near-infrared for binaries
with unequal component temperatures, or corrections for third light if significant. The adjustments in $JHK_S$ range from about 0.25~mag to 0.68~mag.
The assembled SEDs are presented in Appendix \ref{sec:sed_appendix}.

\subsection{Spectral energy distribution fitting\label{sec:fitting}} 
The observed SEDs were fitted with standard stellar atmosphere models. For the EBs in our sample with \teff\ $>$ 4000~K (all but two EBs), we adopted the atmospheres of \citet{Kurucz:2013}, whereas for the two EBs with \teff\ $<$ 4000~K (CU Cnc and YY Gem) we adopted the NextGen atmospheres of \citet{Hauschildt:1999}. The model atmosphere grids are parametrized by \teff, \logg, and \feh, in steps of approximately 100~K, 0.5 dex, and 0.1 dex, respectively. 

As summarized in Table~\ref{tab:sample}, for each star in each EB we have \teff\ and radius (with the masses listed in the original publications, the radius also gives \logg), and in many cases \feh\ as well (we assume \feh\ = $+0.2$ for the Am-type stars, solar metallicity otherwise). We interpolated in the model grid to obtain the appropriate model atmosphere for each star in units of emergent flux, and then summed the two model atmospheres scaled by the stars' surface areas to produce the total SED model for the EB. To redden the SED model, we adopted the interstellar extinction law of \citet{Cardelli:1989}. We then fitted the summed atmosphere model to the flux measurements to minimize $\chi^2$ by varying just two fit parameters: extinction ($A_{\rm V}$) and overall normalization. (The adopted stellar radii and \teff\ also have associated uncertainties, of course; these are handled in a later step via the propagation of errors through \lbol; see Section \ref{sec:eb_parallax}.) 
Where an $A_{\rm V}$ estimate was available from the literature, we adopted it as an initial guess but allowed the fit to vary $A_{\rm V}$ by as much as 3$\sigma$ or 20\%, whichever was larger. Where no prior $A_{\rm V}$ estimate was available, the $A_{\rm V}$ fit was unconstrained except that we limited the maximum value to that from the \citet{Schlegel:1998} dust maps for the given line of sight. 

The best-fit model SED with extinction is shown for each EB in Appendix \ref{sec:sed_appendix}, and the reduced $\chi^2$ values ($\chi_\nu^2$) are given in Table~\ref{tab:results}. 
The fits were satisfactory in \ngoodseds\ cases, leaving \nbadseds\ EBs with very large $\chi_\nu^2$ flagged in Table~\ref{tab:results}. We were not able to discern the cause of the very poor fits in these \nbadseds\ cases---we rechecked that the stellar radii and \teff\ from the original publications appear reliable and that the photometric measurements are not flagged as bad. Thus we simply discarded these \nbadseds\ cases for the remainder of our analyses. 

The primary quantity of interest for each EB is \fbol, which we obtained via direct summation of the fitted SED, {\it without} extinction, over all wavelengths. The formal uncertainty in \fbol\ was determined according to the standard criterion of $\Delta\chi^2 = 2.30$ for the case of two fitted parameters \citep[e.g.,][]{Press:1992}, where we first renormalized the $\chi^2$ of the fits such that $\chi_\nu^2 = 1$ for the best fit model. Because $\chi_\nu^2$ is in almost all cases greater than 1 (see Table~\ref{tab:results}), this $\chi^2$ renormalization is equivalent to inflating the photometric measurement errors by a constant factor and results in a more conservative final uncertainty in \fbol\ according to the $\Delta\chi^2$ criterion. 
While not strictly equivalent to 1$\sigma$ errors, we consider these uncertainties to be representative of our true errors, and evidence presented in Section \ref{sec:reliability} supports this. 

Finally, because the model atmosphere grids do not extend to wavelengths shorter than 0.1 \micron, we found it necessary to augment the model atmospheres at the blue end for hot stars with \teff\ $>$ 15,000 K, for which the emergent flux at $\lambda < 0.1$ \micron\ becomes comparable to the typical uncertainty in \fbol\ of \fbolprecper. Therefore, for these hot stars we appended to the model SED a simple blackbody representing the sum of two blackbodies corresponding to the two stars' temperatures scaled by their surface areas. To account for the non-blackbody nature of the SED at $\lambda > 0.1$ \micron, we adjusted the blackbody portion at $\lambda < 0.1$ \micron\ by the flux difference of the actual SED relative to a blackbody at $\lambda > 0.1$ \micron.

\subsection{How model-dependent are the bolometric fluxes?\label{sec:model_depend}}
For the purposes of the present work, the ultimate aim of the SED fitting is to obtain a measure of \fbol\ for each EB that is as model-independent as possible. It could be argued that the procedure is dependent on the model atmospheres used, which of course it is to some extent. This model dependence is mitigated, however, by the very large wavelength range covered by the actual flux measurements, which for most of the EBs includes a very large fraction of the emergent stellar flux. 

To quantify this, we have calculated the fraction of each EB's \fbol\ that is from beyond the span of the flux measurements, which for hot stars is most important at the blue end. 
For 50\% of the EBs this flux fraction is less than 4\%, and for only 25\% of the EBs is it greater than 25\%; for 5\% of the EBs, representing the very hottest stars, it is greater than 90\%. 
Given the large span of the flux measurements, in principle one could perform a simple linear interpolation between the measurements and, say, a simple polynomial extrapolation at the ends. 
The atmosphere model essentially serves as a more intelligent, more physically motivated way of performing the interpolation (and extrapolation, where needed), grounded in basic stellar astrophysics. Hot EBs, for which the extension of the SED model to the blue represents a relatively large contribution of the total \fbol, could be of concern. However, as we show in Sec.~\ref{sec:results}, the efficacy of the procedure appears to be independent of \teff. 

Other concerns may be that we have had to assume solar metallicity for some of the EBs. We therefore performed a check by varying the adopted \feh\ from $-0.5$ to $+0.3$---representing the range of metallicity for the vast majority of Milky Way stars---for several EBs in our sample over the full range of \teff. We find that the effect on the resulting \fbol\ is negligible for the hot stars and as much as $\sim$0.5\% for the cool stars, in all cases much smaller than the typical \fbol\ uncertainty of \fbolprecper\ (Table~\ref{tab:results}).

\begin{figure}[!ht]
    \centering
    \includegraphics[trim=0 0 0 75,clip,width=0.5\linewidth]{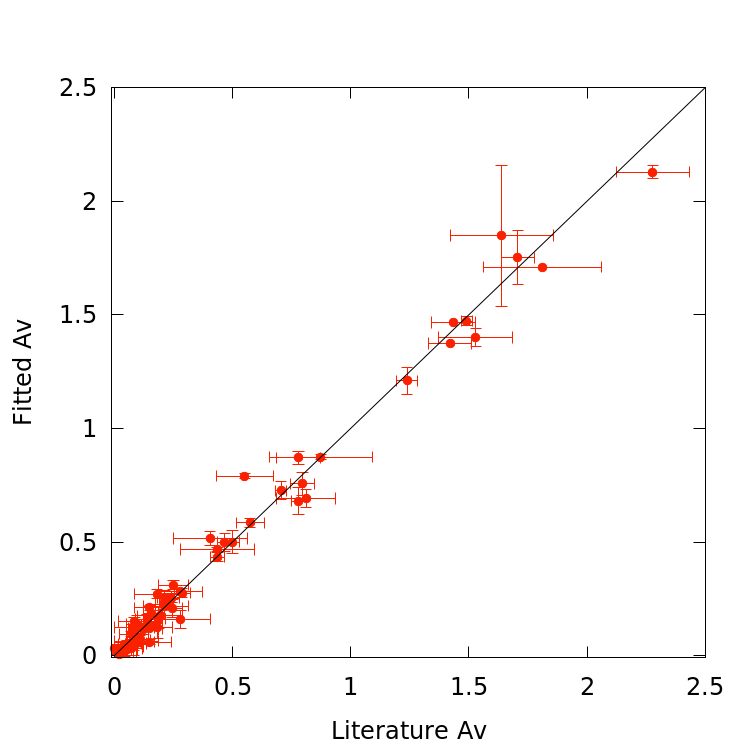}
    \caption{Comparison of fitted $A_{\rm V}$ versus literature $A_{\rm V}$ for the EBs possessing such literature measurements.}
    \label{fig:av_comp}
\end{figure}

Arguably the most important purpose of the fitting procedure is to determine $A_{\rm V}$, for determining what \fbol\ would be in the absence of extinction, thus permitting the distance to be calculated simply via 
\begin{equation}
\label{eq:dist}
    d = (L_{\rm bol}/4\pi F_{\rm bol})^{1/2}
\end{equation}
For \nav\ of the \ngoodseds\ EBs with good SED fits, $A_{\rm V}$ estimates can be derived from published reddening values previously reported in the literature via $A_{\rm V} = 3.1\,E(B-V)$. The comparison between our fitted $A_{\rm V}$ and the literature values is shown in Fig.~\ref{fig:av_comp}, where the agreement is very good. 
Indeed, we expect that the $A_{\rm V}$ values newly determined here should represent an improvement over the original values in many cases. 
We have adopted a single ratio of total-to-selective extinction, $R_{\rm V} = 3.1$ in our fits. $R_{\rm V}$ values in the literature span the range $\approx$2.5--4 for most Galactic sight lines, and thus in principle fitting for $R_{\rm V}$ could further improve the SED fits. However, we have opted for simplicity not to introduce a third free parameter to the SED fitting procedure. In any event, if any of the SED fits are poorer due to our choice of $R_{\rm V}$, the resulting increased $\chi_\nu^2$ will in turn result in more conservative uncertainties on \fbol.

\section{Results\label{sec:results}}
The results of the SED fitting procedure described in Section \ref{sec:fitting} are summarized in Table~\ref{tab:results}, and the full set of \nebs\ SED fits provided in Figure Set \ref{fig:seds} in Appendix \ref{sec:sed_appendix}. 
In this section we discuss several representative SED fits that demonstrate the range of cases, and then present the resulting \fbol\ for our sample of EBs. Next, we present the predicted parallaxes that result from combining \fbol\ together with the \lbol\ values from the literature. Finally, we compare our predicted parallaxes with the subset measured by {\it Hipparcos}, and we assess the reliability of the uncertainties in our predicted parallaxes.

\subsection{SED Fits and Bolometric Fluxes}
\begin{figure}[!ht]
    \centering
    \includegraphics[trim=0 0 0 75,clip,width=0.75\linewidth]{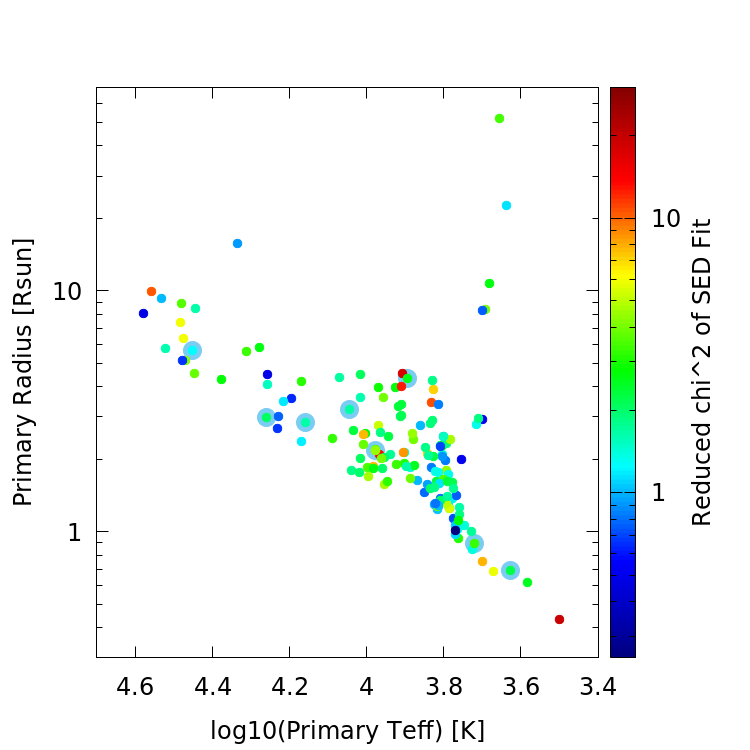}
    \caption{\teff-radius diagram of the primary stars in the \ngoodseds\ EBs with satisfactory SED fits, color coded by the $\chi_\nu^2$ of the fit. Points highlighted with blue halos represent EBs with otherwise satisfactory SED fits that are flagged by {\it Hipparcos\/} as having sub-arcsecond companions that could compromise the \fbol.}
    \label{fig:hrd}
\end{figure}

Satisfactory SED fits were achieved for \ngoodseds\ of the \nebs\ EBs in our study sample. For context, these \ngoodseds\ EBs are represented in the \teff-radius plane in Figure~\ref{fig:hrd}, color coded according to the $\chi_\nu^2$ of the SED fit. 

\begin{figure}[H]
    \centering
    \includegraphics[trim=50 50 50 50,clip,width=0.49\linewidth]{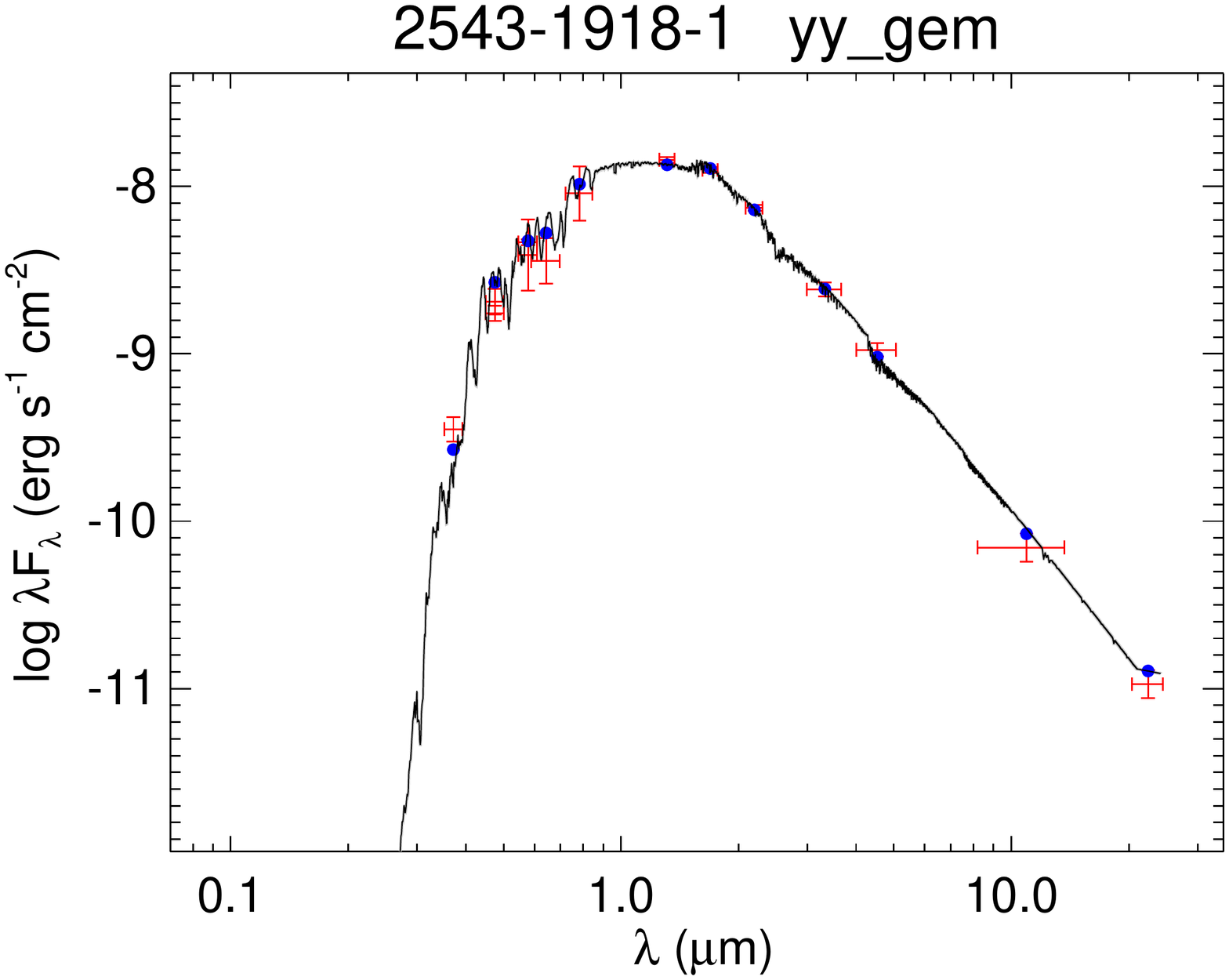}
    \includegraphics[trim=50 50 50 50,clip,width=0.49\linewidth]{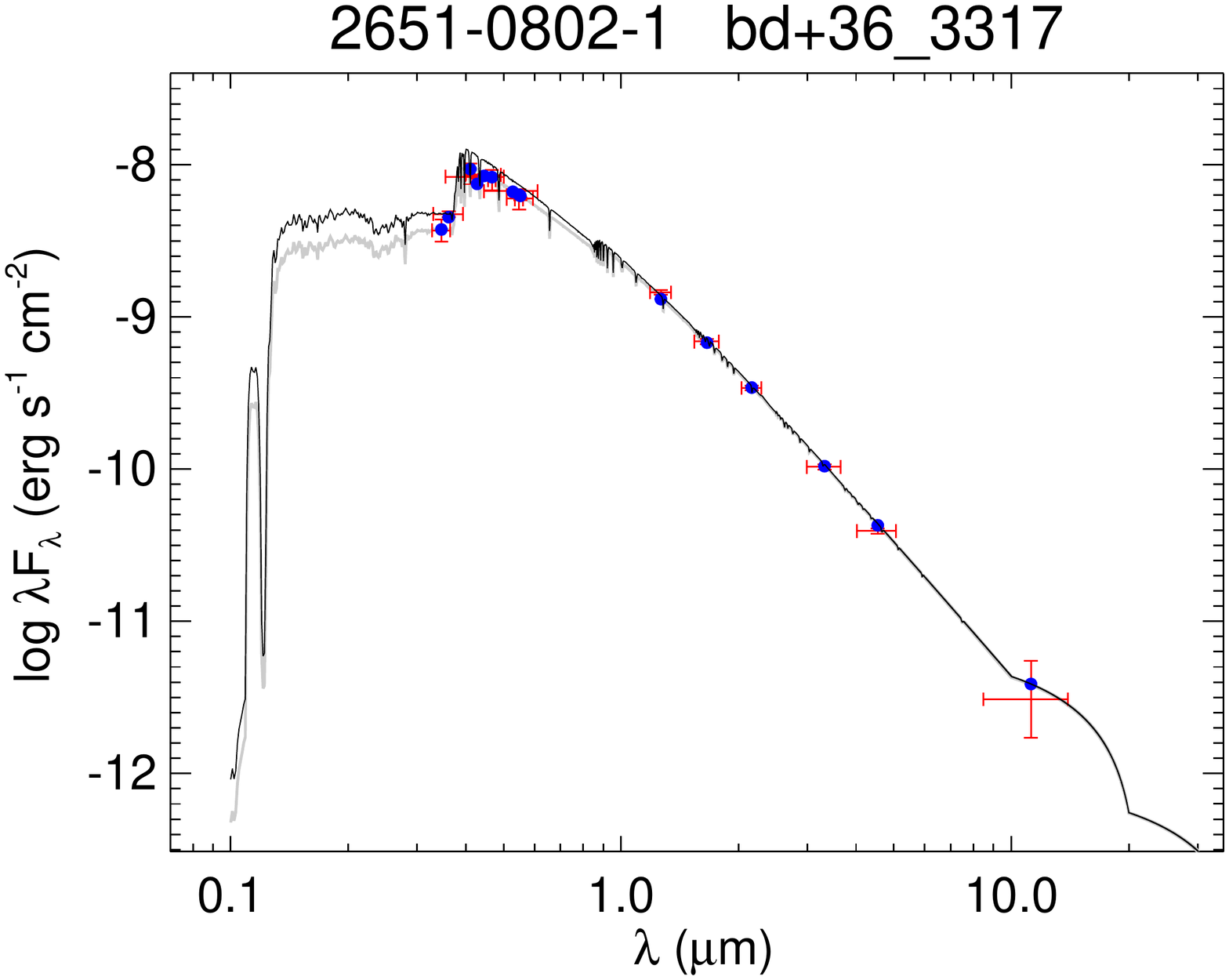}
    \includegraphics[trim=50 50 50 50,clip,width=0.49\linewidth]{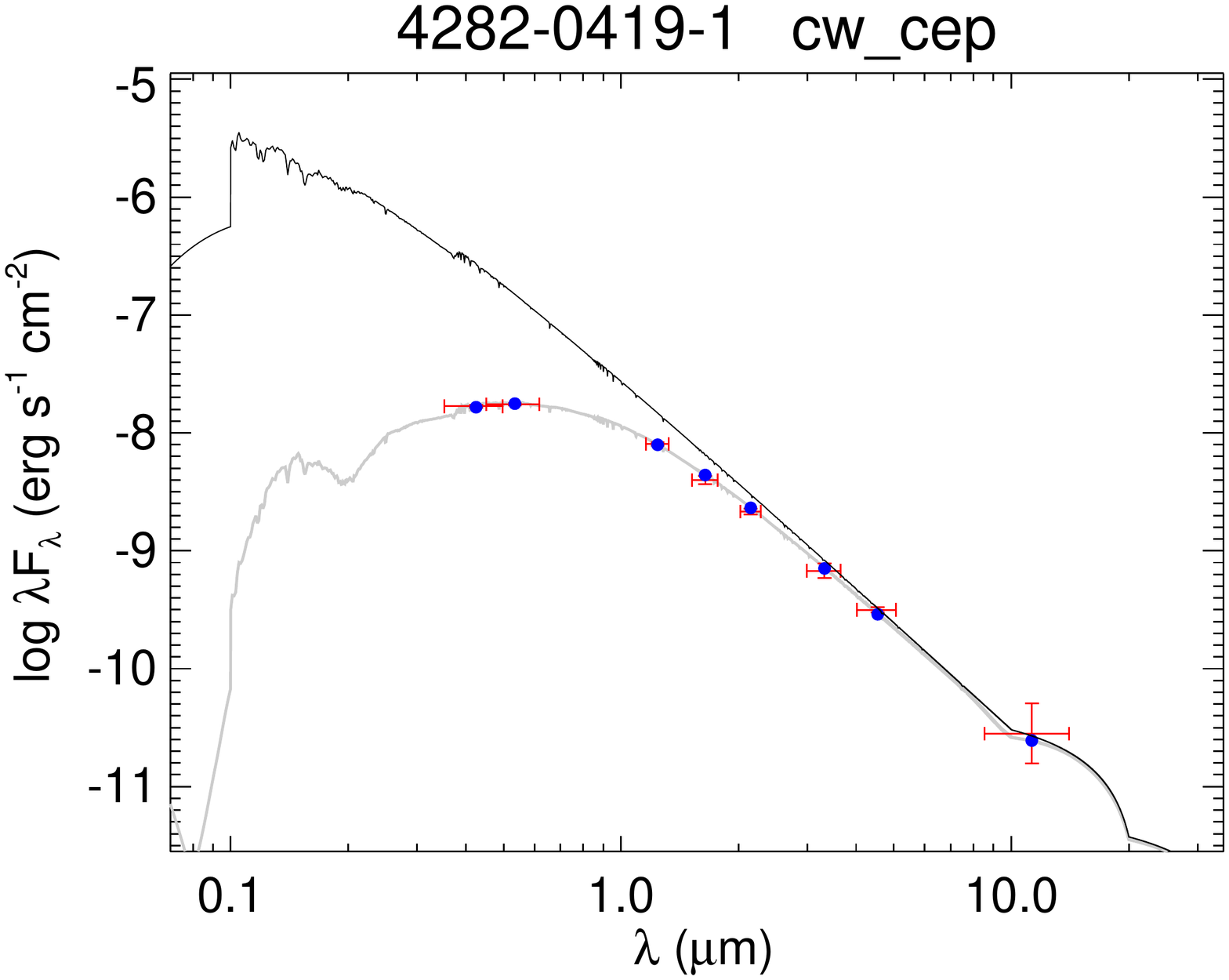}
    \includegraphics[trim=50 50 50 50,clip,width=0.49\linewidth]{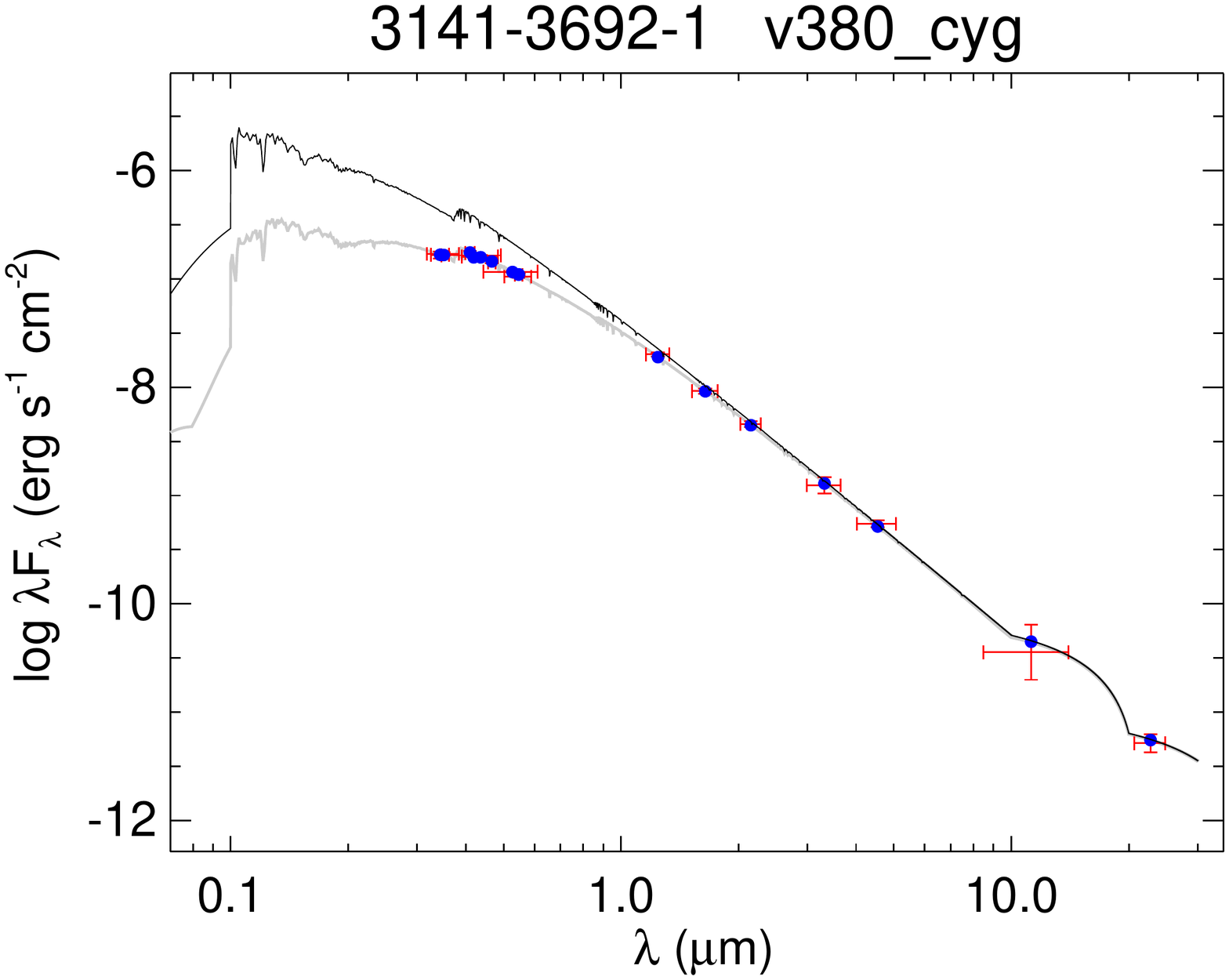}
    \includegraphics[trim=50 70 50 50,clip,width=0.49\linewidth]{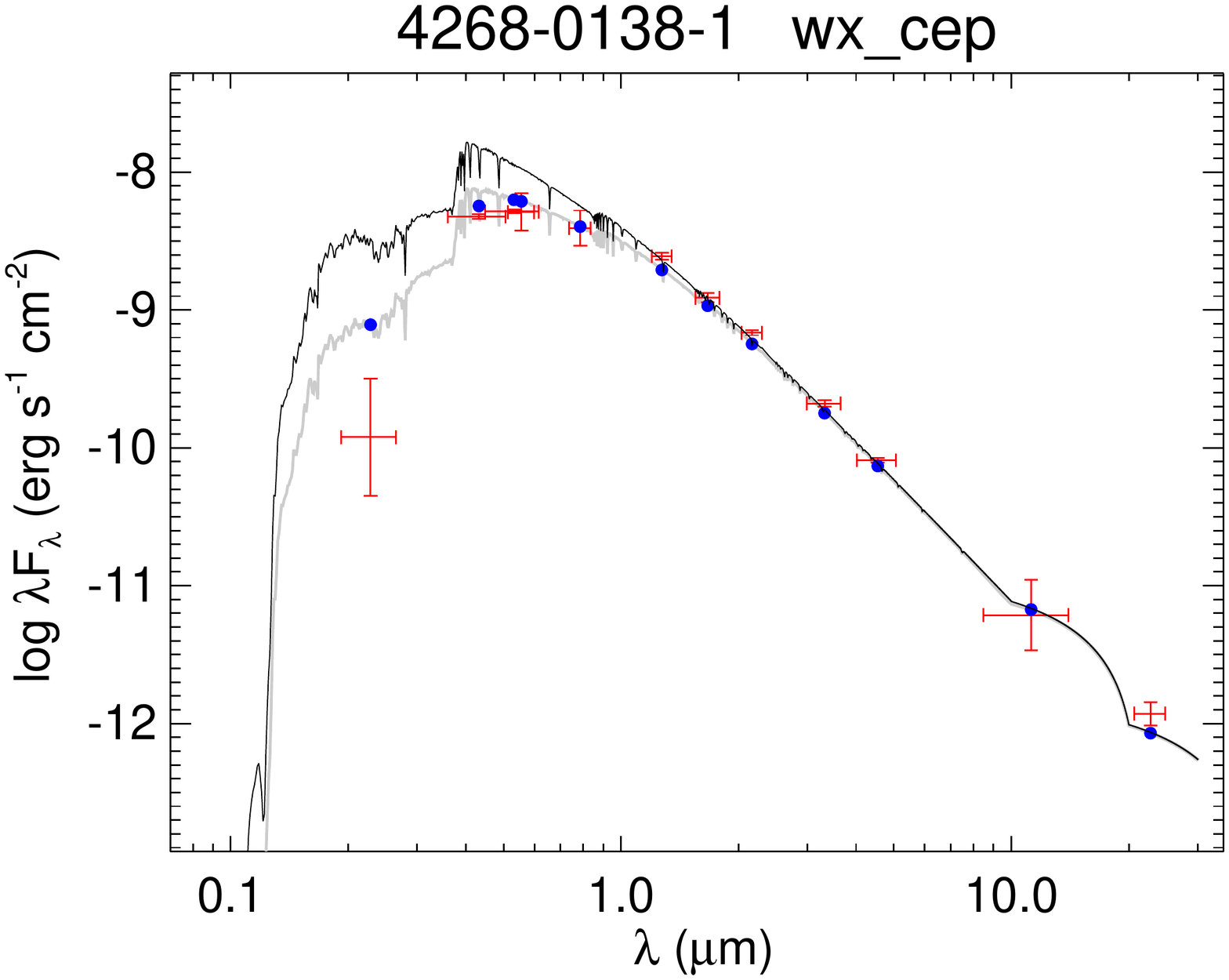}
    \includegraphics[trim=50 70 50 50,clip,width=0.49\linewidth]{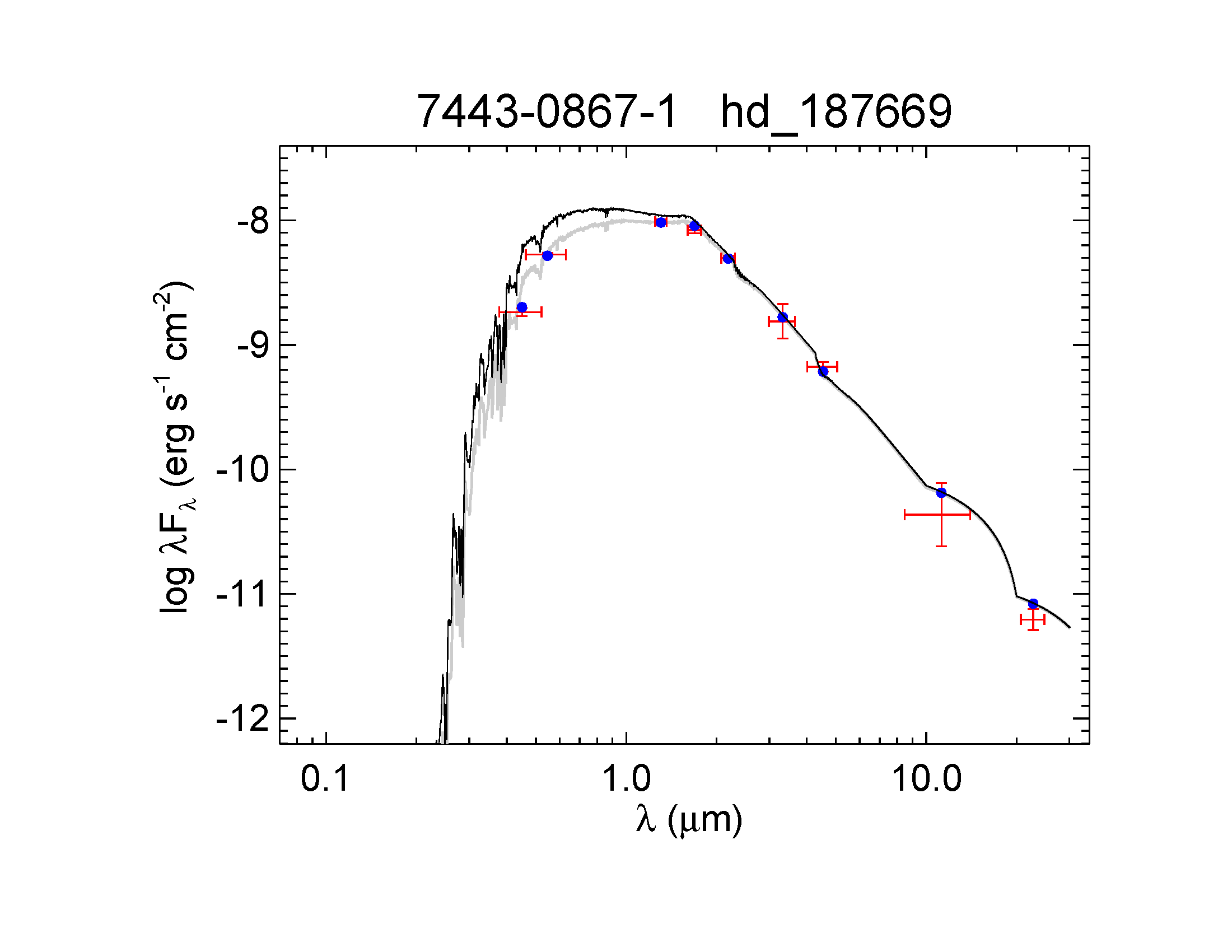}
    \caption{Representative SED fits for six EBs spanning a range of primary \teff, primary radius, and $\chi_\nu^2$ from Figure~\ref{fig:hrd}: YY~Gem (\teff\ = 3820 K, $R = 0.62$ \rsun, $\chi_\nu^2 = 2.49$), BD +36:3317 (\teff\ = 10450 K, $R = 1.76$ \rsun, $\chi_\nu^2 = 1.96$), CW~Cep (\teff\ = 28300 K, $R = 5.64$ \rsun, $\chi_\nu^2 = 1.22$), V380~Cyg (\teff\ = 21700 K, $R = 15.71$ \rsun, $\chi_\nu^2 = 0.90$), WX~Cep (\teff\ = 8150 K, $R = 4.00$ \rsun, $\chi_\nu^2 = 12.47$), and HD~187669 (\teff\ = 4330 K, $R = 22.62$ \rsun, $\chi_\nu^2 = 1.13$). Note that CW Cep is flagged by {\it Hipparcos\/} as possessing a sub-arcsecond companion and thus is excluded from our analyses (see the text). The discontinuity in the model SED at 0.1 \micron\ among the hot stars is due to the blackbody addition to the model for $\lambda < 0.1$ \micron\ for stars with \teff\ $>$ 15,000 K (see Section~\ref{sec:fitting}). All labels, symbols, lines, and colors are as in Figure Set \ref{fig:seds} in Appendix \ref{sec:sed_appendix}.}
    \label{fig:rep_seds}
\end{figure}

Six representative SED fits, covering a range of \teff, radius, and $\chi_\nu^2$ from Fig.~\ref{fig:hrd}, are presented in Fig.~\ref{fig:rep_seds}: YY~Gem, BD~$+$36:3317, CW~Cep, V380~Cyg, WX~Cep, and HD~187669. 
V380~Cyg, with $\chi_\nu^2 = 0.90$, is a good example of the best fits achieved by our procedures. WX~Cep, with $\chi_\nu^2 = 12.5$, is an example of one of the worst fits that we nonetheless deem acceptable. In this case, the quality of the fit has been affected by the 2MASS data, which appear systematically offset upward relative to the rest of the fit. BD~$+$36:3317 is an example of a case in which the original 2MASS measurements have been corrected for having been observed during eclipse (see Sec.~\ref{sec:data}), with a very good resulting $\chi_\nu^2 = 1.96$. 

\begin{figure}[H]
    \centering
    \includegraphics[trim=50 50 50 50,clip,width=0.49\linewidth]{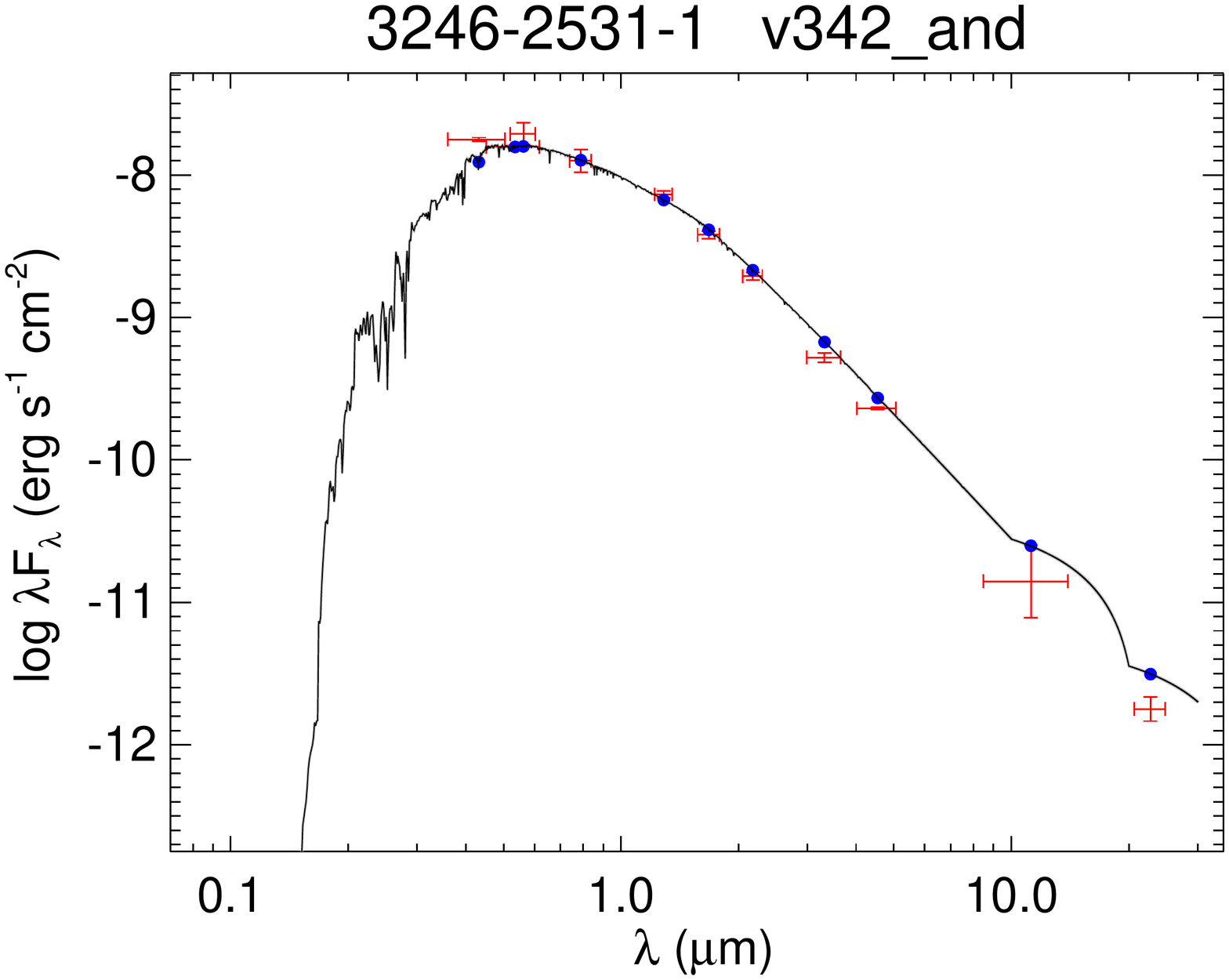}
    \includegraphics[trim=50 50 50 50,clip,width=0.49\linewidth]{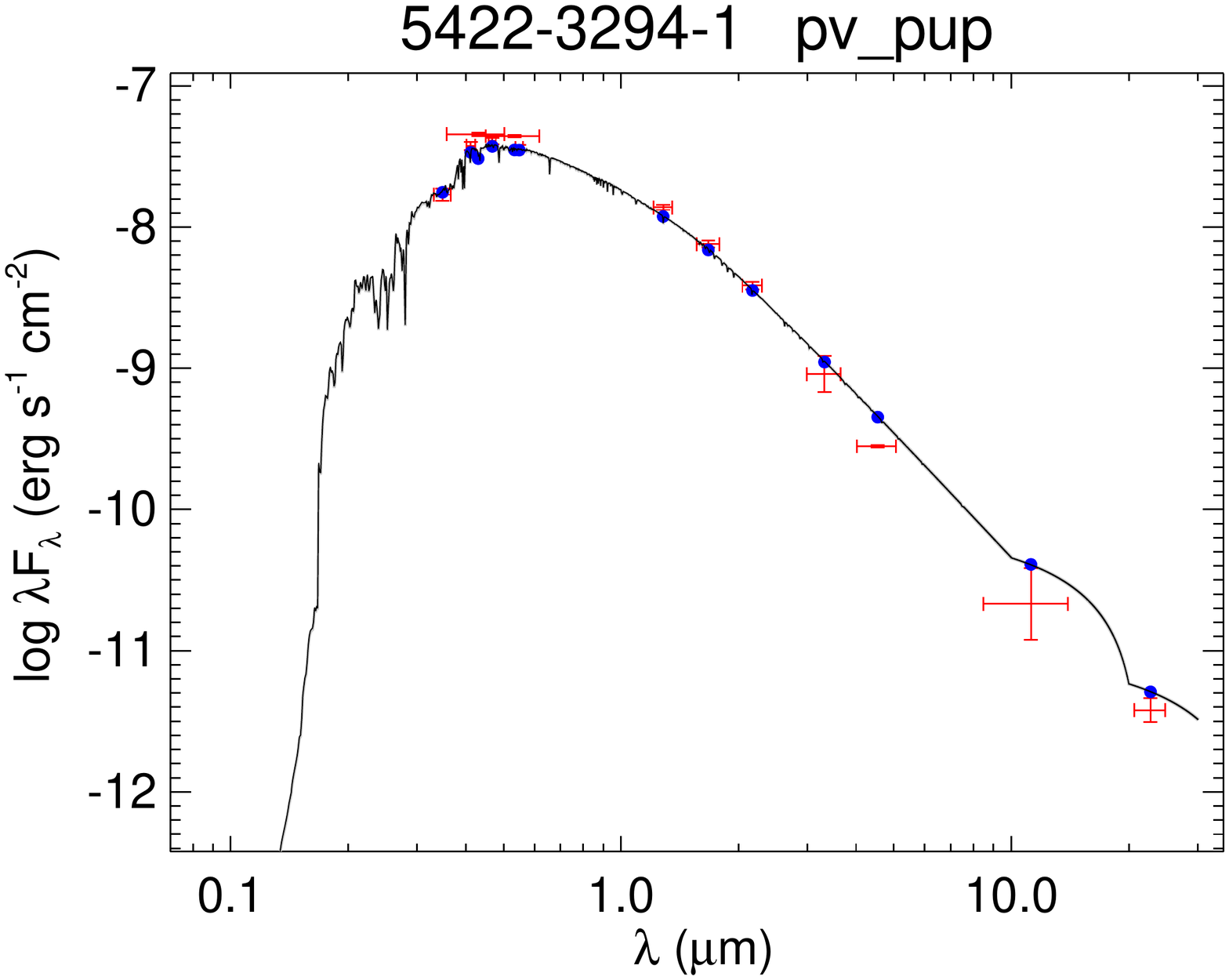}
    \caption{As in Figure~\ref{fig:rep_seds}, but for the \nbadseds\ EBs whose SED fits have unacceptably large $\chi^2$: V342~And (\teff\ = 6200 K, $R = 1.25$ \rsun, $\chi_\nu^2 = 24.2$) and PV~Pup (\teff\ = 6920 K, $R = 1.54$ \rsun, $\chi_\nu^2 = 85.1$).
    We do not consider these EBs in the rest of our analyses.}
    \label{fig:badseds}
\end{figure}

For comparison, the \nbadseds\ cases of truly unacceptable SED fits with $\chi_\nu^2 > 20$ are shown in Figure~\ref{fig:badseds}. Both of these are cases for which we adjusted the 2MASS measurements for having been observed in eclipse, but this did not improve the fits sufficiently. It is possible that alternative photometric measurements from among the many catalogs in which these EBs appear could salvage these cases. However, for the sake of consistency in methodology and in the data sources used, we have opted in this work to simply discard these \nbadseds\ cases. 

Eight of the EBs with otherwise satisfactory SED fits were flagged by {\it Hipparcos\/} as possessing sub-arcsecond companions. Such close companions could be contributing flux to the catalog photometric measurements, which typically have spatial resolutions on the order of 1 arcsec. This additional flux in the data would lead to an incorrectly high \fbol\ and thus an erroneously short inferred distance (i.e., erroneously large predicted parallax). In fact, the quality of the SED fits in these cases is in general quite good. For example, CW~Cep in Figure~\ref{fig:rep_seds} is one such EB; its $\chi_\nu^2 = 1.22$ gives no indication of problems. Nonetheless, to be conservative we have opted in the analysis that follows to disregard these eight cases. 

\begin{figure}[!ht]
    \centering
    \includegraphics[trim=0 0 0 75,clip,width=0.75\linewidth]{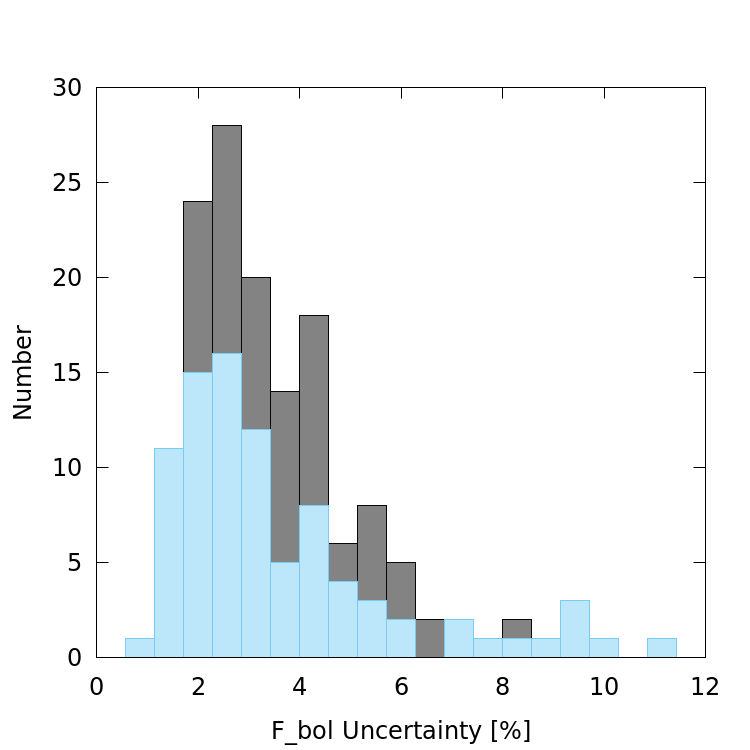}
    \caption{Distribution of \fbol\ uncertainties for the full EB sample (grey) and for the subset with {\it Hipparcos\/} parallaxes (blue). The typical uncertainty is \fbolprecper.}
    \label{fig:fbol_errors}
\end{figure}

A key empirical product of this work is the \fbol\ for each EB that results from direct summation of the SED. These \fbol\ values are tabulated in Table~\ref{tab:results}, and the distribution of their uncertainties presented in Figure~\ref{fig:fbol_errors}. The best cases have uncertainties of $\lesssim$1.5\%. The median uncertainty for the full EB sample is 3.0\%, and is better than 5\% for 90\% of the sample. This means that the uncertainty in the predicted parallaxes for the EBs will in almost all cases be dominated by the uncertainty in the EB \lbol, which is typically $\lesssim$\lbolprecper.

\subsection{Predicted Parallaxes\label{sec:eb_parallax}}
With \fbol\ and \lbol\ and in hand for each EB, we can calculate the predicted distance to each EB according to Equation \ref{eq:dist}. The uncertainty in \fbol\ comes directly from our SED fitting procedure (Section \ref{sec:fitting}). The uncertainty in \lbol\ is less straightforward, as it requires propagation of uncertainties in the stellar radii and \teff, which themselves require a proper handling of correlated uncertainties in the measured quantities from the original EB analyses. 
Our procedure for determining the uncertainties in \lbol\ is explained in Appendix \ref{sec:appendix}.

\begin{figure}[!ht]
    \centering
    \includegraphics[trim=0 0 0 75,clip,width=0.75\linewidth]{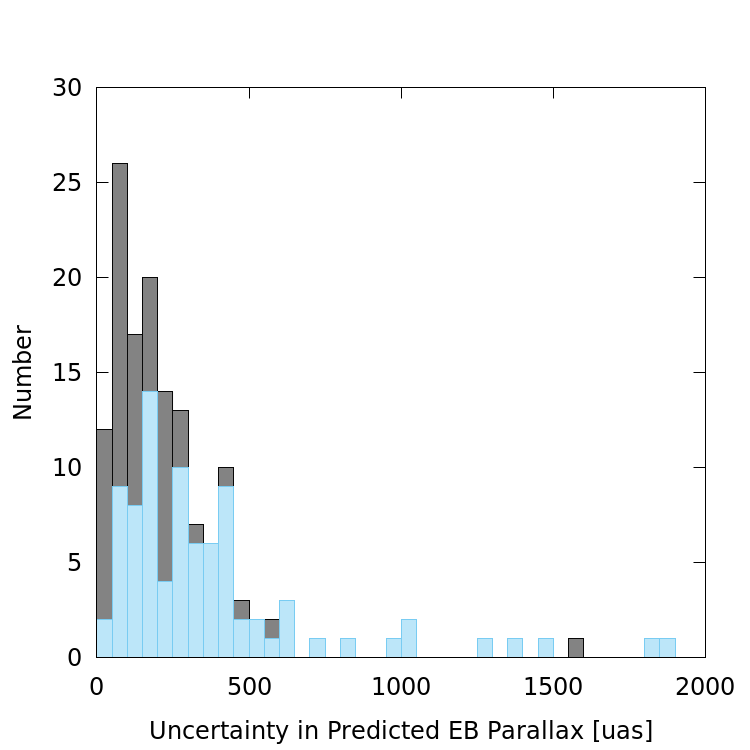}
    \caption{Distribution of uncertainties in the predicted parallaxes (in units of micro-arcseconds) for the full EB sample (grey) and for the subset with {\it Hipparcos\/} parallaxes (blue). Two EBs with parallax uncertainties larger than 2000 $\mu$as are off the plot. The typical uncertainty is \parprecwunit.}
    \label{fig:eb_parallax}
\end{figure}

The predicted distances so computed, and the parallaxes derived from them, are tabulated in Table~\ref{tab:results}. The distribution of their uncertainties is presented in Figure~\ref{fig:eb_parallax}. We note that, as explained by \cite{Bailer-Jones:2015} and others, estimating a distance from a parallax, or in our case a parallax from a distance, is not trivial when the relative errors are larger than about 20\%, and becomes sensitive to prior assumptions. The EBs in our sample all have relative errors well below 20\%, so a straightforward conversion to parallaxes is sufficient for our purposes.

The implied precision of the predicted EB parallaxes is remarkably good: The uncertainty is $\sim$30~$\mu$as in the best cases, and the median for the entire sample is 190~$\mu$as. The precision is better than $\sim$500~$\mu$as for 90\% of the sample. 

\subsection{Comparison to {\it Hipparcos}}
As just mentioned the precision of the predicted EB parallaxes is typically 4 to 5 times better than the formal uncertainties in the {\it Hipparcos\/} trigonometric parallaxes \citep{vanLeeuwen:2007}. Although the latter are therefore poorer than the EB parallaxes on an individual basis, collectively they do allow for a check on the accuracy of our results, by comparing values for the \ngoodhip\ EBs that were observed by the satellite and that have acceptable SED fits.

\begin{figure}[!ht]
    \centering
    \includegraphics[trim=0 0 0 10,clip,width=0.49\linewidth]{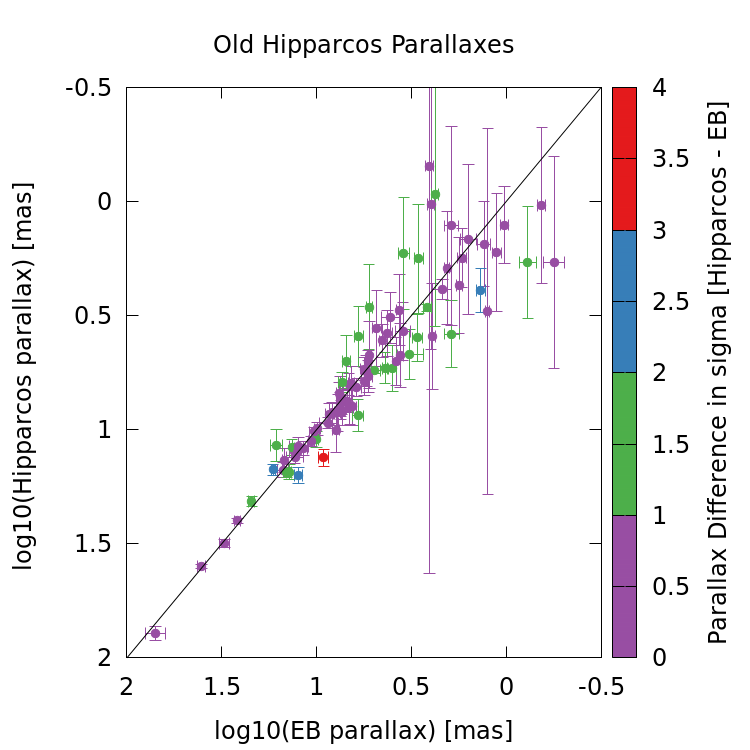}
    \includegraphics[trim=0 0 0 10,clip,width=0.49\linewidth]{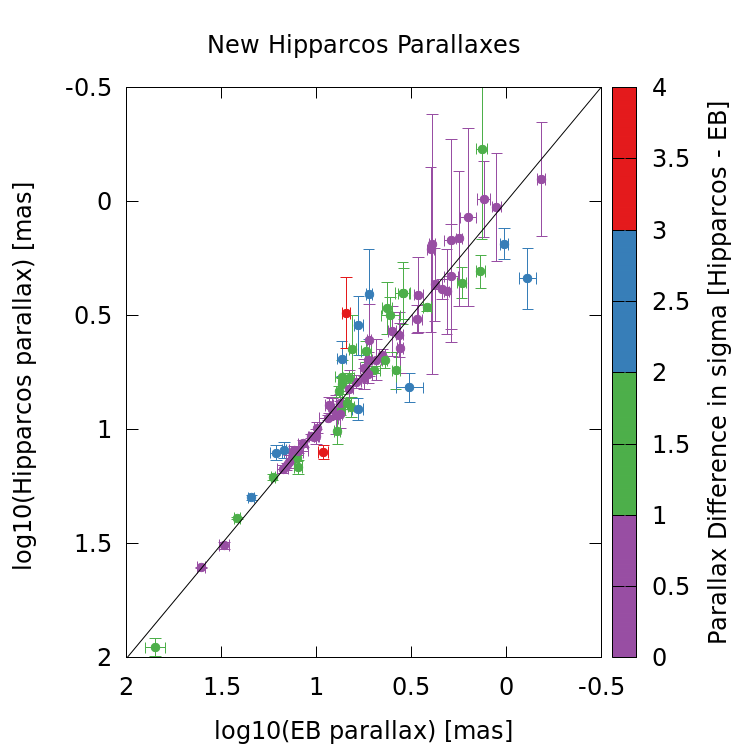}
    \caption{Comparison of predicted EB parallaxes versus their trigonometric parallaxes measured by {\it Hipparcos\/} according the old (left) and new (right) {\it Hipparcos\/} reductions. The color bar represents the difference of the {\it Hipparcos\/} parallax relative to our predicted EB parallax in units of the parallax uncertainties, $\sigma_\pi$.}
    \label{fig:hip_comp_log}
\end{figure}

Figure~\ref{fig:hip_comp_log} presents the direct comparison of our predicted EB parallaxes against the trigonometric parallaxes reported by \citet{Perryman:1997} in the original reduction (hereafter referred to as ``old Hipparcos") and as reported in the ``new Hipparcos" reduction of \citet{vanLeeuwen:2007}. In both cases the overall agreement is excellent. The EB parallaxes appear to slightly better follow the new {\it Hipparcos\/} parallaxes for the most distant EBs (i.e., the smallest parallaxes) for which the old {\it Hipparcos\/} parallaxes appear systematically smaller than the EB parallaxes. However, this small trend is well within the errors. Indeed, the old {\it Hipparcos\/} parallaxes on the whole exhibit fewer large residuals relative to the EB parallaxes, and much of this difference occurs among the smallest parallaxes for which the new {\it Hipparcos\/} reported uncertainties are much smaller than in the old {\it Hipparcos\/} reduction. 

\begin{figure}[H]
    \centering
    \includegraphics[trim=0 0 0 10,clip,width=0.6\linewidth]{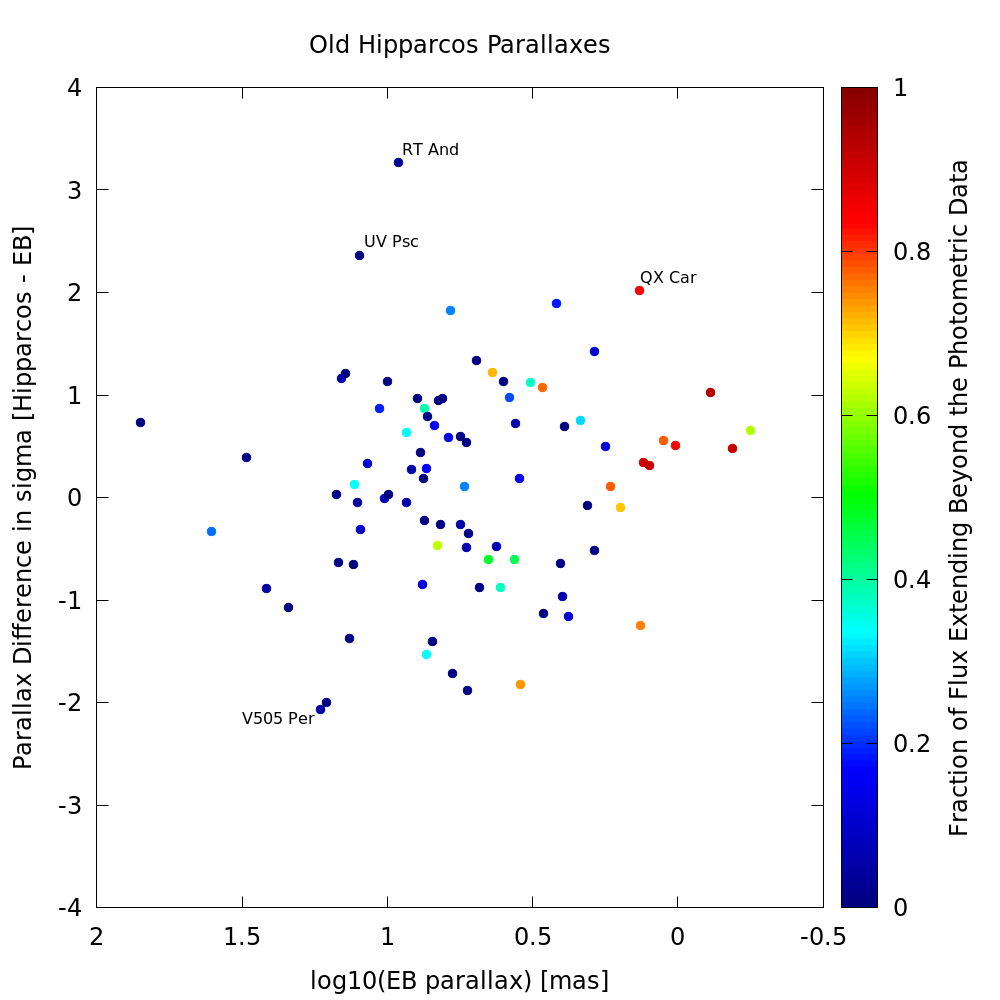}
    \includegraphics[trim=0 0 0 10,clip,width=0.6\linewidth]{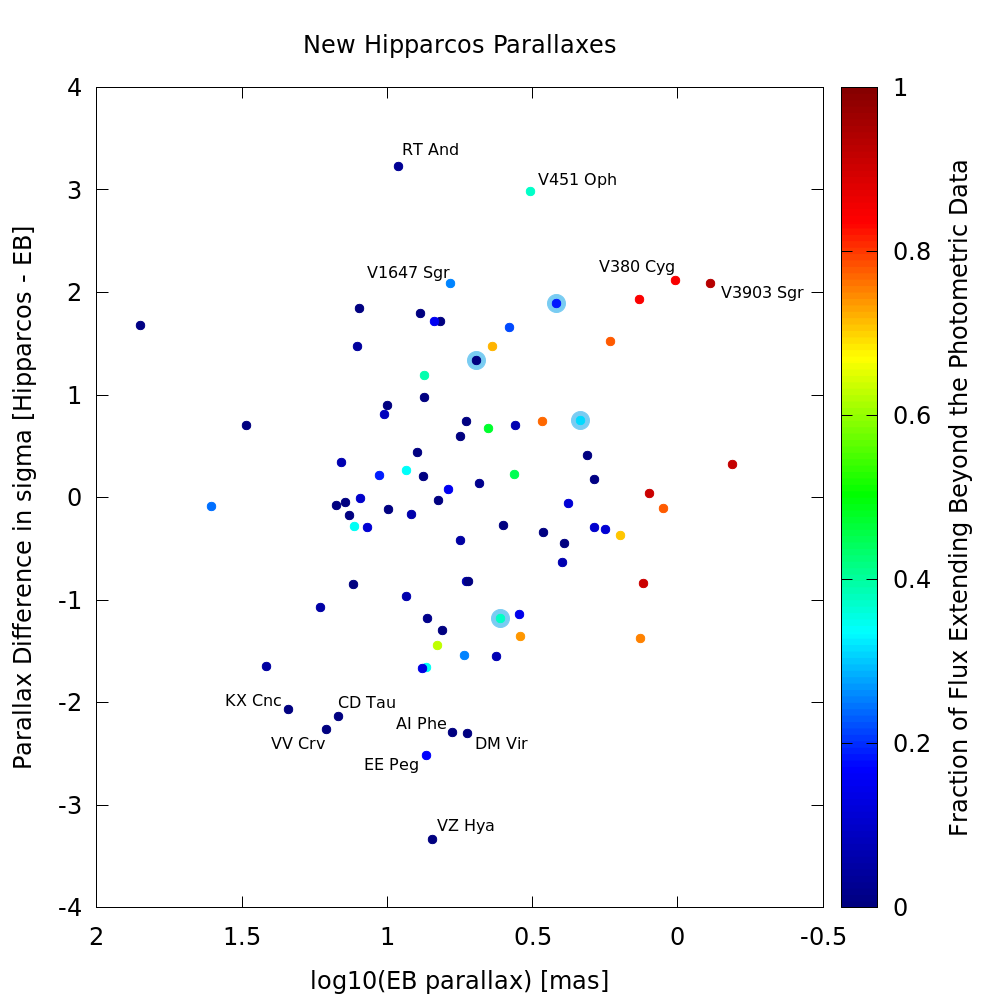}
    \caption{\small{Differences between predicted EB parallaxes and their trigonometric parallaxes measured by {\it Hipparcos}, in units of the parallax uncertainties, according the old (top) and new (bottom) {\it Hipparcos\/} reductions. The color bar represents the fraction of each EB's \fbol\ that is from the portion of the SED fit beyond the span of the photometric measurements (see Section~\ref{sec:model_depend}). Points highlighted with blue halos in the bottom panel are EBs that are members of clusters with average parallaxes reported in the new {\it Hipparcos\/} reduction (see Section~\ref{sec:data}). Specific EBs that disagree by more than 2$\sigma$ are identified; their SED fits may be inspected in Figure~\ref{fig:seds}.}}
    \label{fig:hip_comp_scatter}
\end{figure}

We show this directly in Figure~\ref{fig:hip_comp_scatter}, where we observe that while the overall distribution of parallax residuals is similar for both the old and new {\it Hipparcos\/} parallaxes, there are more outliers larger than 2$\sigma$ with the new {\it Hipparcos\/} parallaxes (12 in the new, 4 in the old). In addition, for the nearest EB in our sample (CU~Cnc), the old {\it Hipparcos\/} parallax agrees with the EB parallax within 1$\sigma$, whereas the difference is nearly 2$\sigma$ for the new {\it Hipparcos\/} parallax. 
These outliers are identified by name in Figure~\ref{fig:hip_comp_scatter} so that the SED fits may be readily compared (Fig.~\ref{fig:seds}).

We have checked for any indications of potential problems that might be common to the outliers. We checked the EBs that were reported in the original EB publications to possess tertiary companions (although any ``third light" in these cases should already be accounted for in the EB solutions from which the radii and \teff\ were derived); we checked the EBs containing metallic A (Am) stars whose metallicities are less well determined and/or anomalous (although metallicity in general has a negligible effect on the derived \fbol, see Section~\ref{sec:model_depend}); we checked the EBs flagged by {\it Hipparcos\/} as ``Variability-Induced Movers" (although this should have been accounted for in the {\it Hipparcos\/} reduction); we checked the EBs with the largest $A_{\rm V}$ values (although our SED-derived $A_{\rm V}$ values agree very well with the published values, see Fig.~\ref{fig:av_comp}); finally, we checked the EBs with large amounts of flux in the SED fits that are beyond the span of the photometric measurements (this is represented in Fig.~\ref{fig:hip_comp_scatter}). The outliers have none of these factors in common.  

It is notable that the predicted EB parallaxes compare so favorably even when the \fbol\ determination involves a large contribution from beyond the span of the photometric measurements, considering that this contribution to \fbol\ is as large as $\sim$90\% in the most extreme cases. Though this may seem surprising, it is simply a consequence of the fact that the SED fit is very stringently constrained by the stellar properties---which are in turn very accurately determined from the published EB solutions---and of the fact that the available photometry spans a sufficiently large range of wavelengths to stringently constrain $A_{\rm V}$, the only remaining free parameter. It is an extrapolation only in the sense that the model extends beyond the data, not in the sense that there is no knowledge of the nature of the SED beyond the data. 

It is also notable that the predicted EB parallaxes retain their high precision regardless of distance. This is a consequence of the fact that the accuracy of the stellar properties arising from the EB solutions does not depend on the EB distance, as long as the light curves and radial velocities used in the EB analysis are of sufficient quality. Indeed, even the EB-based distance to the Large Magellanic Cloud at $\sim$50 kpc has a demonstrated precision of $\sim$2\% \citep{Pietrzynski:2013}.

\subsection{Reliability of the EB Parallax Uncertainties\label{sec:reliability}}
The comparison of the predicted EB parallaxes to the available {\it Hipparcos\/} parallaxes suggests that our estimated EB parallax precisions are reliable. With only a few exceptions, the residuals relative to the {\it Hipparcos\/} parallaxes are distributed as expected, especially when compared to the old {\it Hipparcos\/} parallaxes. The two large outliers seen in Figure~\ref{fig:hip_comp_scatter} (top) may represent nothing more than the few $>2\sigma$ deviations expected from a normal sample of $\sim$100. 

The larger number of $>2\sigma$ outliers relative to the new {\it Hipparcos\/} parallaxes (Figure~\ref{fig:hip_comp_scatter}, bottom), however, suggests that our EB parallax uncertainties may be underestimated in some cases. 
Another way of checking our EB parallaxes and uncertainties is to use the distances to those EBs in our sample that reside in star clusters for which accurate distances have been determined. There are four such EBs in our sample, from the new {\it Hipparcos\/} parallaxes (we exclude the Pleiades for reasons discussed in Section~\ref{sec:intro}), and these are highlighted in Figure~\ref{fig:hip_comp_scatter} (bottom). 
In all four cases, the agreement between our predicted EB parallax and the {\it Hipparcos\/} cluster parallax is within 2$\sigma$. 
At the same time, the distribution of the four residuals is strictly speaking slightly broader than for a normal distribution: Two of the four EBs agree to within 1$\sigma$, whereas a normal distribution would expect three of the four to agree with 1$\sigma$. 

\begin{figure}[!ht]
    \centering
    \includegraphics[trim=0 0 0 60,clip,width=0.75\linewidth]{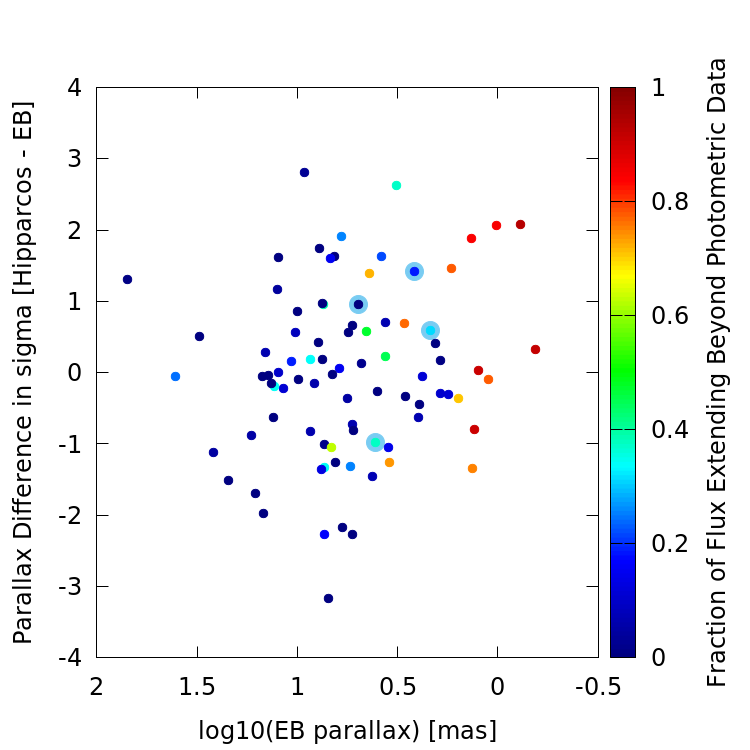}
    \caption{Same as Figure~\ref{fig:hip_comp_scatter} (bottom), except with the EB parallax uncertainties increased by 50\%.}
    \label{fig:newhip_comp_2x}
\end{figure}

It is difficult to say more than this on the basis of only four measurements. Certainly, there is not compelling evidence that the uncertainties in our predicted EB parallaxes are not reliable. However, more conservatively, these four measurements could also be interpreted as suggesting that our EB parallax uncertainties are underestimated by $\sim$50\%. This is depicted in Figure~\ref{fig:newhip_comp_2x}, where now the four EBs with cluster parallaxes are more normally distributed, although the large number of systems that deviate by more than 2$\sigma$ remains large, suggesting that perhaps it is some of the new {\it Hipparcos\/} parallax uncertainties that are underestimated. 
In any event, these comparisons allow us to conclude that the EB parallaxes are very precise, with typical errors in the range of 200--300~$\mu$as.

We can also assess the {\it accuracy} of the EB parallaxes. We have measured this by computing the mean and median parallax difference compared to {\it Hipparcos}. For the old {\it Hipparcos\/} reduction, we obtain mean and median differences of $-60$~$\mu$as and $+60$~$\mu$as (in the sense of {\it Hipparcos\/}$-$EB), respectively. For the new {\it Hipparcos\/} reduction, we obtain mean and median differences of $-40$~$\mu$as and $-55$~$\mu$as, respectively. For both the old and new {\it Hipparcos\/} reductions, these differences are $\lesssim$1/3 of our typical random errors. This is important, as it not only confirms that our EB parallaxes are highly accurate as well as very precise, but also that our estimated uncertainties are realistic and reflect any systematics inherent to the method.

\section{Discussion: Applicability to {\it Gaia} \label{sec:disc}}

The {\it Gaia\/} Mission is poised to revolutionize many areas of
astrophysics by mapping the entire sky and delivering high-precision
astrometry, photometry, and spectroscopy for up to a billion stars.
Most notably, it will provide trigonometric parallaxes that in some
cases will be as precise as a few $\mu$as, yielding useful distances
to individual objects halfway across the Galaxy. While the final
results from the nominal 5-year mission are not expected until
after 2020, interim data releases will already provide extremely
valuable information beginning with the first (DR1) slated for late
2016.

In addition to accurate positions and mean magnitudes for stars over
at least 90\% of the sky, and additional information for sources at
the ecliptic poles, DR1 will contain parallaxes for up to 2.5 million
stars based on the Tycho-Gaia Astrometric Solution
\citep[TGAS;][]{Michalik:2015}. This full-sky solution takes advantage
of the {\it Tycho-2\/} astrometry \citep{Hog:2000} gathered 20 years ago
during the {\it Hipparcos\/} Mission, and combines it with the first six
months of {\it Gaia\/} data to solve for the positions, proper motions,
and parallaxes of the majority of the {\it Tycho-2\/} stars. The
precision expected for these parallaxes is comparable to or better
than that of {\it Hipparcos\/} for most stars, with typical nominal
errors below a milli-arc second, and somewhat poorer precision up to
about 3~mas along the ecliptic. The data set should be almost complete
down to $V \approx 11.5$, or about 3 to 4 magnitudes deeper than {\it                    
Hipparcos}.

Even if it is preliminary, this large collection of parallaxes
featuring 20 times more stars than {\it Hipparcos\/} will have numerous
scientific applications. In particular, it will significantly improve
the stellar characterization of targets of exoplanet searches,
including transit surveys, many of which observe relatively bright
stars that are contained in the {\it Tycho-2\/} catalog. This, in turn,
should have an immediate impact on the precision and accuracy of the
derived planetary properties. NASA's Transiting Exoplanet Survey
Satellite Mission \citep[TESS;][]{Ricker:2015} will benefit enormously
as well, as {\it Gaia\/} DR1 will supply parallaxes for a large fraction
of the bright nearby stars being considered for observation, enabling
a better selection of the optimal targets in preparation for launch at
the end of 2017.

\begin{figure}[!ht]
    \centering
    \includegraphics[trim=0 0 0 75,clip,width=0.75\linewidth]{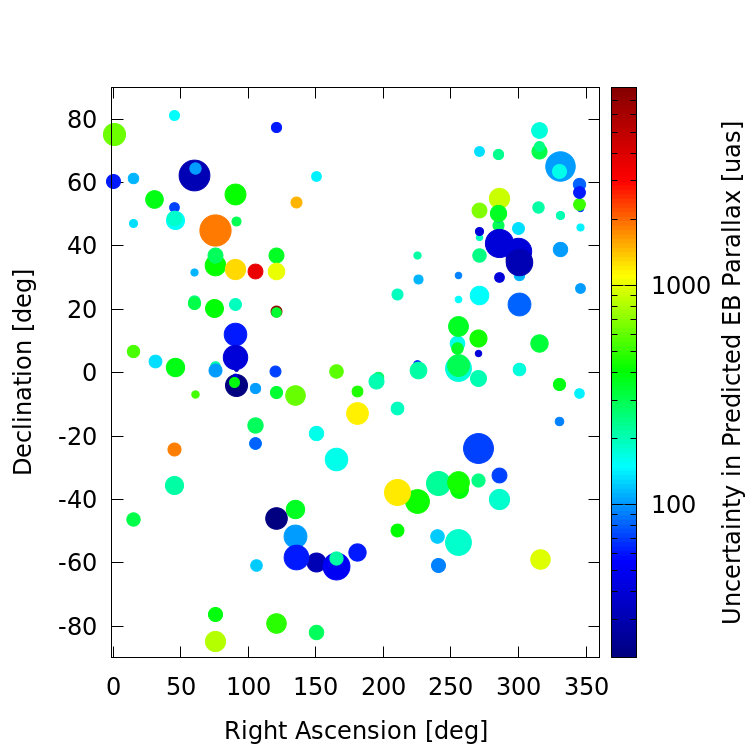}
    \caption{Map in equatorial coordinates of the EBs in our sample. Color represents the uncertainty in our predicted parallaxes. Symbol size is proportional to EB brightness, with the largest symbols corresponding to $V\approx2$ and the smallest corresponding to $V\approx12$.}
    \label{fig:eb_map}
\end{figure}

While the external \emph{accuracy} of the {\it Gaia\/} DR1 parallaxes is
expected to be very good, and perhaps similar to that of {\it                            
Hipparcos\/} according to simulations by \citep{Michalik:2015},
independent checks are highly desirable as a means to validate the new
astrometric results. The set of predicted EB parallaxes derived in this work,
obtained in a completely different way, offers an excellent opportunity
for this test given that our typical \parprecwunit\ precision is at
least as good as, and often better than, that expected of {\it Gaia\/} DR1. We
note also that our parallax errors do not tend to increase as stars
get fainter or more distant. Our faintest system, CoRoT~102918586 ($V = 12.4$), has a
predicted parallax with an uncertainty of 30~$\mu$as; our most distant system,
V467~Vel, is more than 5~kpc away and has a correspondingly well-determined
parallax under 0.2~mas also with an uncertainty of only 20--30~$\mu$as.
Furthermore, as shown in Figure~\ref{fig:eb_map},
our stars are distributed over the entire sky, and span
more than a 10 magnitude range in brightness ($V = 1.9$--12.4)
potentially allowing the discovery of magnitude-dependent parallax
discrepancies, should they be present.

\section{Summary and Conclusions\label{sec:summary}}

Eclipsing binaries with well-measured physical properties allow the determination of distances that are essentially model-independent, and can be both highly accurate and highly precise. They are therefore ideal as benchmarks for validating other methods of establishing distances, and can often be used out to many kiloparsecs without loss of precision. They have been employed to great advantage even in external galaxies such as the LMC, the SMC, and others.

In this paper we have assembled an all-sky list of \nebs\ EBs contained in the {\it Tycho-2\/} catalog with high-quality determinations of the component radii and effective temperatures from the literature, and combined this information with constrained SED fits using existing photometric measurements over a wide range of wavelengths. The distances calculated from the accurate absolute stellar luminosities and bolometric fluxes lead to predicted parallaxes having typical precisions of \parprecwunit\ (\parprecper\ relative errors), which are 4--5 times better than the trigonometric parallaxes from {\it Hipparcos}. We find excellent overall agreement between our results and those from {\it Hipparcos\/}. To the extent that our EB parallaxes represent a test of the {\it Hipparcos\/} parallaxes, we find no obvious systematic deviations of the sort that appear to have affected the Pleiades, at least for this particular sample. In any case, the good agreement between our EB parallaxes and the {\it Hipparcos\/} parallaxes supports the accuracy of our measurements, and other tests suggest that our precision estimates are also realistic.

The quality of our predicted parallaxes also compares very favorably with that expected for the {\it Gaia\/} parallaxes from the Tycho-Gaia Astrometric Solution \citep{Michalik:2015}, soon to be delivered as part of the mission's first Data Release. Our results will therefore serve as an important external check on the spacecraft's astrometric performance early on in the mission. While subsequent {\it Gaia\/} releases will feature steadily increasing parallax precision for larger numbers of stars as the time base lengthens, we anticipate that predicted parallaxes from EBs will continue to provide a valuable reference that is completely independent of astrometry and whose precision does not degrade with increasing distance or diminishing brightness.

\acknowledgments
K.G.S.\ acknowledges partial support from NSF PAARE grant AST-1358862. G.T.\ acknowledges partial support from NSF grant AST-1509375. This work made extensive use of the Filtergraph data visualization system \citep{Burger:2013} at {\tt \url{filtergraph.vanderbilt.edu}}.

\clearpage
\begin{deluxetable}{l c c c c c c c c c}
\rotate
\tabletypesize{\footnotesize}
\tablewidth{0pt}
\tablecolumns{10}
\tablecaption{Stellar properties of our sample of EBs\label{tab:sample}}
\tablehead{
\colhead{} & \colhead{{\it Tycho}} & \colhead{{\it Hipparcos}} & \colhead{$R_1$} & \colhead{$R_2$} & \colhead{Primary} & \colhead{Secondary} & \colhead{$E(B-V)$} & \colhead{} & \colhead{}  \\
\colhead{System} & \colhead{ID} & \colhead{ID} & \colhead{($R_{\odot}$)} & \colhead{($R_{\odot}$)} & \colhead{$T_{\rm eff}$ (K)} & \colhead{$T_{\rm eff}$ (K)} & \colhead{(mag)} & \colhead{[Fe/H]} & \colhead{Source}
}
\startdata
  UV Psc                      &  0026-0577-1  &    5980   &   1.110~$\pm$~0.023   &   0.835~$\pm$~0.018   &   5780~$\pm$~100   &   4750~$\pm$~80     &  \nodata        &   \nodata      &  1 \\
  XY Cet                      &  0051-0832-1  &   13937   &   1.873~$\pm$~0.035   &   1.773~$\pm$~0.029   &   7870~$\pm$~115   &   7620~$\pm$~125    &   0.04~$\pm$~0.02    &   Am / Am      &  2 \\
  V1130 Tau                   &  0066-1108-1  &   17988   &   1.782~$\pm$~0.011   &   1.489~$\pm$~0.010   &   6625~$\pm$~70    &   6650~$\pm$~70     &  0.000~$\pm$~0.011   & $-$0.25~$\pm$~0.10  &  3 \\
  EW Ori                      &  0104-1206-1  &  \nodata  &   1.133~$\pm$~0.011   &   1.083~$\pm$~0.011   &   5970~$\pm$~100   &   5780~$\pm$~100    &  0.014~$\pm$~0.012   &   \nodata      &  1 \\
  V578 Mon                    &  0154-2528-1  &  \nodata  &   5.149~$\pm$~0.091   &    4.21~$\pm$~0.10    &  30000~$\pm$~740   &  26400~$\pm$~600    &  0.455~$\pm$~0.029   & $-$0.30~$\pm$~0.13  &  1 \\
  AI Hya                      &  0196-0626-1  &  \nodata  &   3.916~$\pm$~0.031   &   2.767~$\pm$~0.019   &   6700~$\pm$~60    &   7100~$\pm$~65     &  0              &   Am / --      &  1 \\
  FM Leo                      &  0263-0727-1  &   54766   &   1.648~$\pm$~0.043   &   1.511~$\pm$~0.049   &   6316~$\pm$~240   &   6190~$\pm$~211    &  0.043          &   \nodata      &  4 \\
  AQ Ser                      &  0340-0588-1  &  \nodata  &   2.281~$\pm$~0.014   &   2.451~$\pm$~0.027   &   6430~$\pm$~100   &   6340~$\pm$~100    &  0.029~$\pm$~0.010   & $-$0.19~$\pm$~0.15  &  5 \\
\enddata
\end{deluxetable}

\clearpage
\begin{deluxetable}{l c c c c c c c}
\rotate
\tabletypesize{\footnotesize}
\tablewidth{0pt}
\tablecolumns{8}
\tablecaption{Results\label{tab:results}}
\tablehead{
\colhead{} & 
\colhead{{\it Tycho}} & 
\colhead{$V$} & 
\colhead{} &
\colhead{$F_{\rm bol}$} & 
\colhead{$A_V$} & 
\colhead{Distance} & 
\colhead{Parallax}
\\
\colhead{System} & 
\colhead{ID} & 
\colhead{(mag)} & 
\colhead{$\chi^2_{\nu}$} & 
\colhead{(erg s$^{-1}$ cm$^{-2}$)} & 
\colhead{(mag)} & 
\colhead{(pc)} & 
\colhead{(mas)}
}
\startdata
 UV Psc                     &  0026-0577-1 &   9.01   &    2.72      &   7.748e-09~$\pm$~(3.4e-10/3.2e-10)  &   0.01~$\pm$~(0.00/0.01)  &    80.2~$\pm$~(  3.6/   3.5)  &   12.47~$\pm$~(0.57/0.53)    \\  
 XY Cet                     &  0051-0832-1 &   8.75   &    1.60      &   9.072e-09~$\pm$~(3.9e-10/3.2e-10)  &   0.11~$\pm$~(0.07/0.05)  &   276.5~$\pm$~( 10.2/  10.7)  &    3.62~$\pm$~(0.14/0.13)    \\  
 V1130 Tau                  &  0066-1108-1 &   6.66   &    1.34      &   6.202e-08~$\pm$~(1.4e-09/1.7e-09)  &   0.03~$\pm$~(0.00/0.02)  &    69.7~$\pm$~(  1.9/   1.7)  &   14.35~$\pm$~(0.36/0.37)    \\  
 EW Ori                     &  0104-1206-1 &   9.78   &    0.71      &   2.945e-09~$\pm$~(6.3e-11/6.1e-11)  &   0.02~$\pm$~(0.02/0.00)  &   169.6~$\pm$~(  6.3/   6.1)  &    5.90~$\pm$~(0.22/0.21)    \\  
 V578 Mon                   &  0154-2528-1 &   8.55   &    0.65 *    &   4.922e-07~$\pm$~(1.3e-08/1.1e-08)  &   1.37~$\pm$~(0.02/0.02)  &  1327.2~$\pm$~( 71.9/  70.9)  &    0.75~$\pm$~(0.04/0.04)    \\  
 AI Hya                     &  0196-0626-1 &   9.35   &    7.02      &   4.826e-09~$\pm$~(3.1e-10/2.9e-10)  &   0.21~$\pm$~(0.03/0.00)  &   548.4~$\pm$~( 20.5/  19.6)  &    1.82~$\pm$~(0.07/0.07)    \\  
 FM Leo                     &  0263-0727-1 &   8.45   &    3.49      &   1.175e-08~$\pm$~(5.3e-10/5.0e-10)  &   0.11~$\pm$~(0.02/0.00)  &   137.2~$\pm$~( 11.7/  11.1)  &    7.29~$\pm$~(0.64/0.57)    \\  
 AQ Ser                     &  0340-0588-1 &  10.65   &    1.89      &   1.576e-09~$\pm$~(3.6e-11/3.2e-11)  &   0.12~$\pm$~(0.00/0.01)  &   583.2~$\pm$~( 20.0/  20.0)  &    1.71~$\pm$~(0.06/0.06)    \\
\enddata
\tablenotetext{a}{Systems flagged with an asterisk have $T_{\rm eff} > 15,000$~K and a large fraction of their flux extrapolated from a blackbody distribution on the blue side.
Those marked with an ``X'' are considered to have poor SED fits.}

\end{deluxetable}

\clearpage

\appendix

\section{Spectral Energy Distribution Measurements and Fits for EB Study Sample\label{sec:sed_appendix}}
\begin{figure}[!ht]
\centering
\includegraphics[trim=70 70 70 50,clip,width=\linewidth]{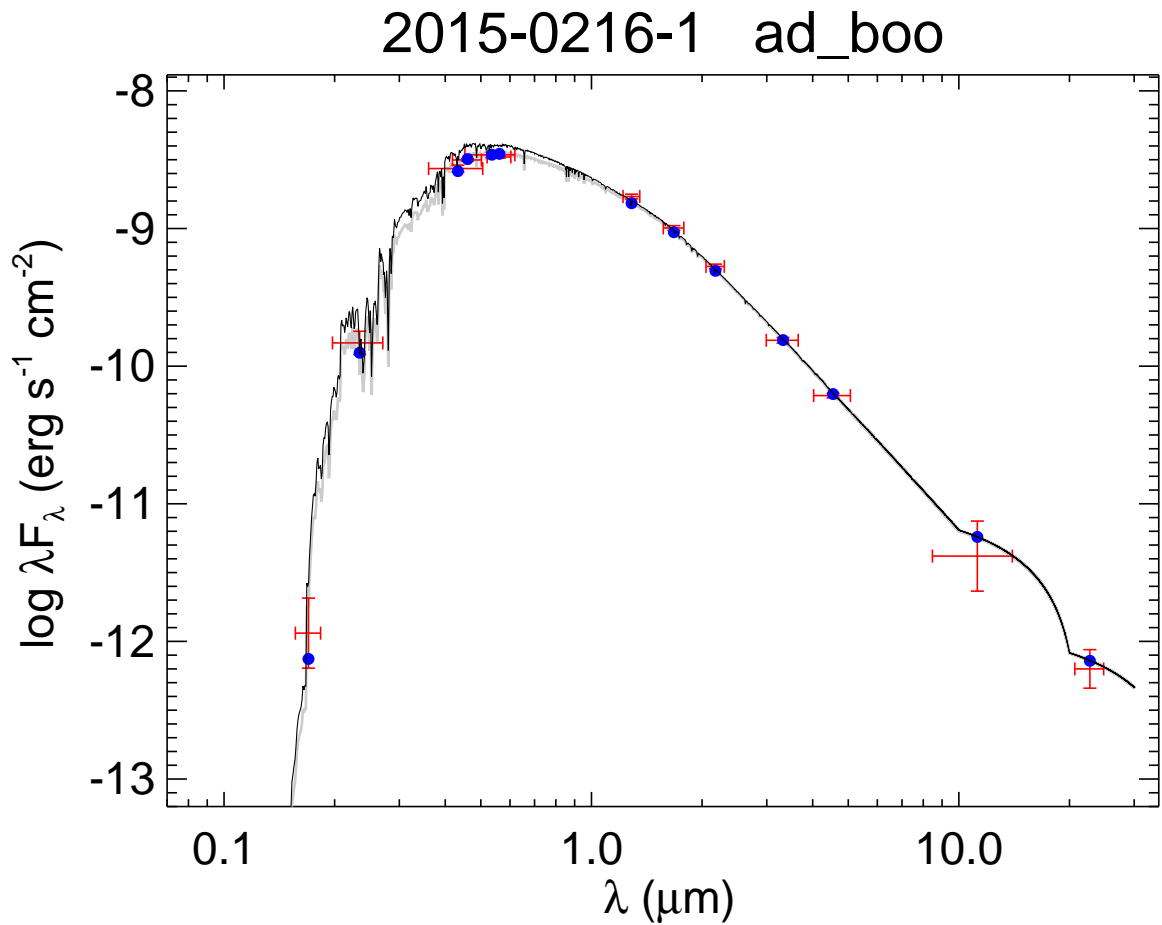}
\caption{AD Boo is shown as example of the figure set. Each panel in the Figure Set is labeled at top by the {\it Tycho-2} ID and name of the EB, and shows the observed fluxes (in units of erg cm$^{-2}$ s$^{-1}$) versus wavelength (in \micron) as red error bars, where the vertical error bar represents the uncertainty in the measurement and the horizontal ``error" bar represents the effective width of the passband. Also in each figure is the fitted SED model including extinction (light gray curve), on which is shown the model passband fluxes as blue dots. The discontinuity in the model SED at 0.1 \micron\ among the hot stars is due to the blackbody extrapolation for $\lambda < 0.1$ \micron\ for stars with \teff\ $> 15,000$~K (see the text). The corresponding un-extincted SED model is also shown (dark black curve); the reported \fbol\ is the sum over all wavelengths of this un-extincted model (see the text). The full figure set is displayed in Figures \ref{fig:seds_1}--\ref{fig:seds_27}.}
\label{fig:seds}
\end{figure}

In Figure Set \ref{fig:seds} we present the observed and fitted spectral energy distributions of the \nebs\ EBs in our study sample (Table~\ref{tab:sample}). 

\section{The uncertainties in the total bolometric luminosity\label{sec:appendix}}

The uncertainty in the total bolometric luminosity $L_{\rm bol}$ of
each our EBs, which factors into the final error in the distances and
parallaxes, depends on the radius and temperature uncertainties of
both components.  A naive approach to error propagation would consider
the published primary and secondary temperature errors to be
independent, and the radius errors to also be independent, whereas in
reality they are not. For example, light curve analyses typically
yield highly precise temperature \emph{ratios} from the relative depth
of the eclipses such that the secondary temperature essentially scales
with the primary temperature (a positive correlation), and most of the
uncertainty in the secondary value is inherited from the primary. The
primary temperature and its error are usually determined from external
information (e.g., spectroscopically) and held fixed in the
photometric analysis.  Similarly, light curves can often determine the
radius \emph{sum} very precisely from the total duration of the
eclipses, whereas the radius \emph{ratio} is less well constrained, as
are the individual radii if solved for separately. This results in the
primary and secondary radius errors being highly anti-correlated.

Given these correlations, propagating the temperature errors for the
primary and secondary as if they were independent will cause the
uncertainty in $L_{\rm bol}$ to be underestimated, and doing the same
with the radius errors will tend to give an $L_{\rm bol}$ error that
is too high, partially offsetting the temperature bias. However, as
temperature errors dominate due to the fourth power dependence in the
Stefan-Boltzmann law, and because they are typically larger
(fractionally) compared to the radius errors, the net result will be
an underestimate of the total luminosity error.

A more sensible approach would be to recast the published
uncertainties in the secondary temperature $T_{\rm eff,2}$ in terms of
the errors in the primary value and in the temperature ratio $t \equiv
T_{\rm eff,2}/T_{\rm eff,1}$, which are usually uncorrelated, and the
radius uncertainties in terms of those of their sum $r \equiv R_1+R_2$
and their ratio $k \equiv R_2/R_1$, which are also uncorrelated, and
to then propagate these new uncertainties independently to infer the
error in $L_{\rm bol}$. This should lead to more realistic errors for
$L_{\rm bol}$.  However, in practice it is not possible to recover the
uncertainties in $t$, $r$, and $k$ accurately without access to the
details of each light curve analysis, which are not always published,
and the issue is further complicated by the sometimes subjective
assignment of radius and/or temperature uncertainties in some of the
original publications. For this work we have nevertheless attempted to
estimate $\sigma_t$, $\sigma_r$, and $\sigma_k$ in a statistical sense
to match the ensemble of reported errors for the individual
temperatures and radii as closely as possible. We proceed as follows.

Beginning with the radii, the individual component values may be
expressed in terms of their sum and ratio as
\begin{equation}
R_1 = \left(\frac{1}{1+k}\right) r~~~,~~~R_2 = \left(\frac{k}{1+k}\right) r
\end{equation}
\citep[see][]{Torres:2000} in which the correlation between $r$ and $k$ is
generally weak. Standard error propagation then gives
\begin{equation}
\label{eq:sigmaR}
\sigma_{R_1} = \frac{1}{1+k}\left[\left(\frac{r}{1+k}\right)^2\sigma_k^2 + \sigma_r^2\right]^{1/2}~~,~~
\sigma_{R_2} = \frac{1}{1+k}\left[\left(\frac{r}{1+k}\right)^2\sigma_k^2 + k^2\sigma_r^2\right]^{1/2}~,
\end{equation}
from which it can be seen that if the secondary is a smaller star,
its uncertainty should in principle also be smaller. As mentioned
above, many systems with smaller secondaries have reported
$\sigma_{R_2}$ values that are in fact larger than $\sigma_{R_1}$,
which prevents one from solving Eqs.[\ref{eq:sigmaR}] directly for
$\sigma_r$, and $\sigma_k$. Since the expectation is that the radius
sum should be better determined than the radius ratio, we introduce
this condition by defining
\begin{equation}
\label{eq:fr}
f_R \equiv \frac{\sigma_r/r}{\sigma_k/k}
\end{equation}
where $f_R$, the ratio of the fractional uncertainties in $r$ and $k$,
should be smaller than unity. Solving for $\sigma_k$ and inserting it
in Eqs.[\ref{eq:sigmaR}] leads to two expressions for $\sigma_r$ in
terms of the published errors, one based on $\sigma_{R_1}$ and the
other on $\sigma_{R_2}$:
\begin{equation}
\label{eq:sigrsum}
\sigma_{r,1} = \sigma_{R_1} (1+k) \left[\left(\frac{k}{1+k}\right)^2 f_R^{-2}+1\right]^{-1/2}~~,~~
\sigma_{r,2} = \sigma_{R_2} (1+k) \left[\left(\frac{k}{1+k}\right)^2 f_R^{-2}+k^2\right]^{-1/2}~.
\end{equation}
We adopt here a simple average of these two estimates, $\sigma_r =
(\sigma_{r,1}+\sigma_{r,2})/2$. We then derive the uncertainty in $k$
from Eq.[\ref{eq:fr}] as $\sigma_k = \sigma_r (k/r) f_R^{-1}$.

We apply a similar reasoning to the temperatures, assuming the
uncertainties in the primary temperature and in the temperature ratio
are largely independent. In that case, since $T_{\rm eff,2} = t\,T_{\rm
eff,1}$, the uncertainty $\sigma_2$ in $T_{\rm eff,2}$ is expressed as
\begin{equation}
\label{eq:secT}
\sigma_2^2 = T_{\rm eff,1}^2 \sigma_t^2 + t^2 \sigma_1^2~,
\end{equation}
and as before we may define
\begin{equation}
\label{eq:fT}
f_T \equiv \frac{\sigma_t/t}{\sigma_1/T_{\rm eff,1}}
\end{equation}
with the expectation that the fractional error in the temperature
ratio will typically be smaller than that of the primary
temperature. Solving for $\sigma_1$ and inserting it in
Eq.[\ref{eq:secT}] leads to the required estimate of the uncertainty
in the temperature ratio,
\begin{equation}
\label{eq:sigtrat}
\sigma_t = \sigma_2 T_{\rm eff,1}^{-1} (1+f_T^{-2})^{-1/2}~.
\end{equation}

\begin{figure}[!ht]
%
\centering
\includegraphics[width=0.65\linewidth]{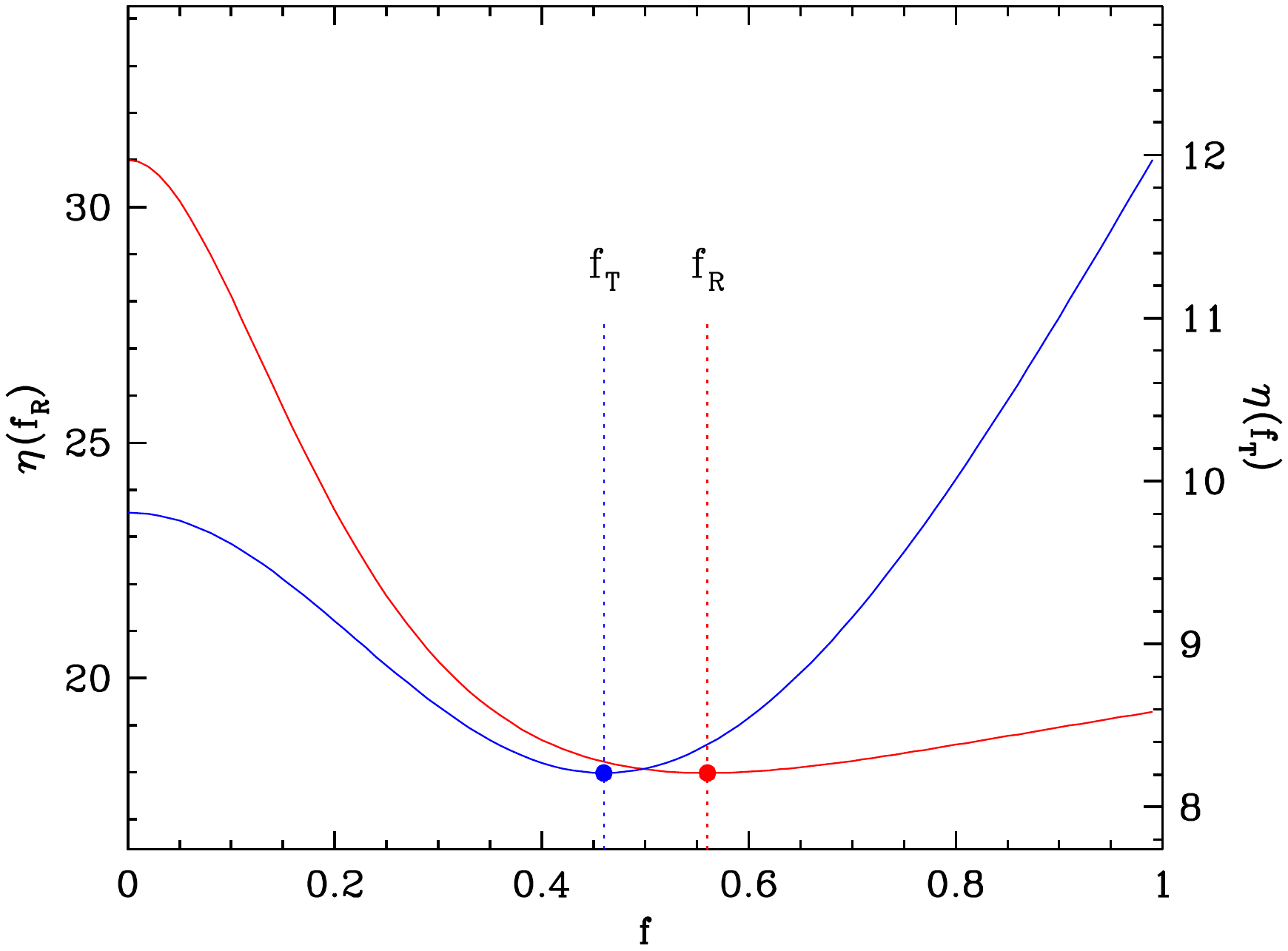}
\caption{Illustration of how we estimated the scale factors
  $f_R$ and $f_T$ that provide the best overall fit to the ensemble of
  published errors in the radii and temperatures for the \ngoodseds\ EBs in
  our sample. The best-fit values that minimize the respective figures
  of merit $\eta$ are indicated.\label{fig:f}}
\end{figure}

Because of the ways in which individual temperature errors have
sometimes been assigned to our eclipsing binaries, it is not always
possible to use the above formalism to infer a value of $\sigma_t$ or
$f_T$ for each system directly from the published uncertainties
$\sigma_1$ and $\sigma_2$. This is similar to the difficulty mentioned
earlier for the radii.  We have therefore chosen to adopt a single
value of $f_R$ and $f_T$ for the entire sample, and to use the
ensemble of published measurement errors for all \ngoodseds\ systems to tell
us what the optimal values should be, such that when inserted into the
expressions for $\sigma_r$, $\sigma_k$, and $\sigma_t$ for each EB we
obtain the closest match to the published individual radius and
temperature errors using Eqs.[\ref{eq:sigmaR}] and
Eq.[\ref{eq:secT}]. Denoting the individual radius and temperature
errors predicted by our prescription for given values of $f_R$ and
$f_T$ as $\sigma_{R_1}^{\prime}$, $\sigma_{R_2}^{\prime}$,
$\sigma_1^{\prime}$, and $\sigma_2^{\prime}$, we construct a figure of
merit $\eta$ for $R$ and $T_{\rm eff}$ as follows:
\begin{equation}
\label{eq:chi2}
\eta(f_R) = \sum \left(1-\frac{\sigma_{R_1}^{\prime}}{\sigma_{R_1}}\right)^2 +
\sum \left(1-\frac{\sigma_{R_2}^{\prime}}{\sigma_{R_2}}\right)^2~~,~~
\eta(f_T) = \sum \left(1-\frac{\sigma_2^{\prime}}{\sigma_2}\right)^2~.
\end{equation}
Both statistics have a single absolute minimum yielding best-fit
values $f_R = 0.56$ and $f_T = 0.46$ (see Figure~\ref{fig:f}). 
These factors are both smaller than unity,
supporting our assumption that the radius sum is usually better
determined than the radius ratio, and that the temperature ratio is
typically more precise than the fractional error in the primary
temperature. The fact that the $f$ factors are not much smaller, as
one might expect for canonical light curve solutions, is a reflection
of the inhomogeneous nature of the published errors.

We used the above prescription to estimate the uncertainty in $L_{\rm
  bol}$ for each system by propagating the errors in $r$ and $k$
through a Monte Carlo procedure as if they were independent, and
similarly with $T_{\rm eff,1}$ and $t$.  Compared to the more
simplistic approach in which the individual temperature errors and
individual radius errors are considered to be uncorrelated, our
approach results in typical luminosity uncertainties up to 50\%
larger, which we expect to be more realistic.


\clearpage

\begin{figure}[H]
  \centering
  \includegraphics[trim=60 60 60 60,clip,width=0.49\linewidth]{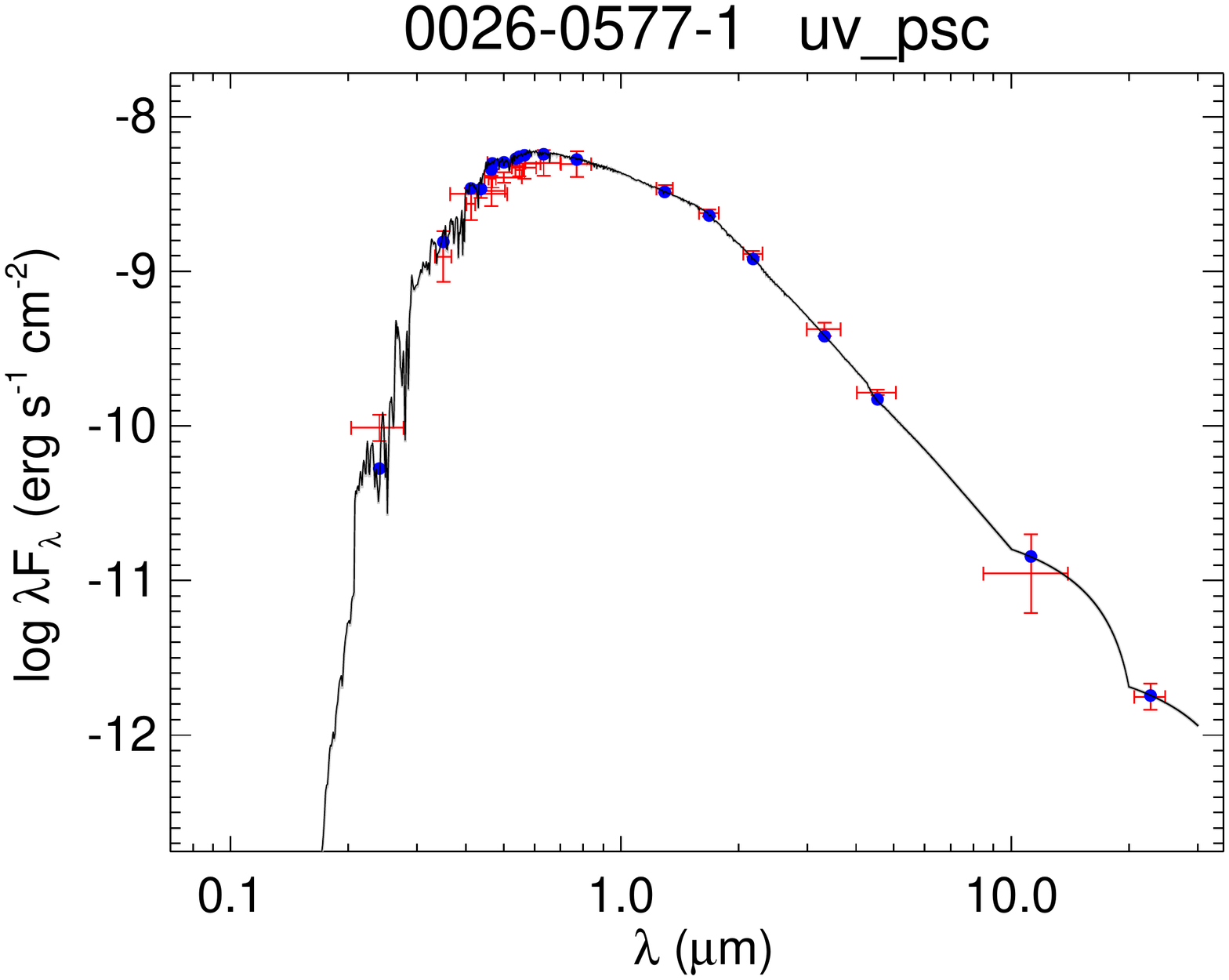}
  \includegraphics[trim=60 60 60 60,clip,width=0.49\linewidth]{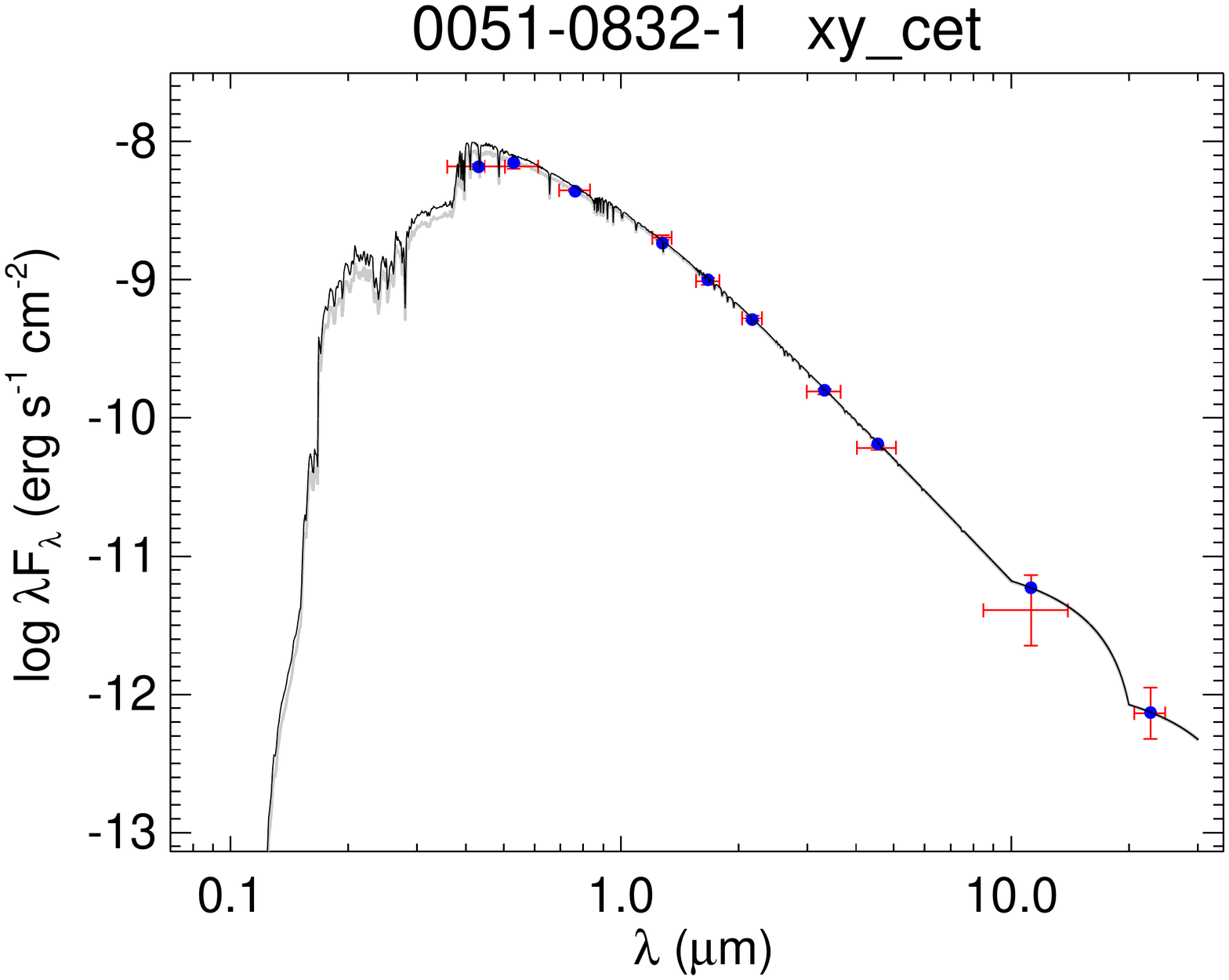}
  \includegraphics[trim=60 60 60 60,clip,width=0.49\linewidth]{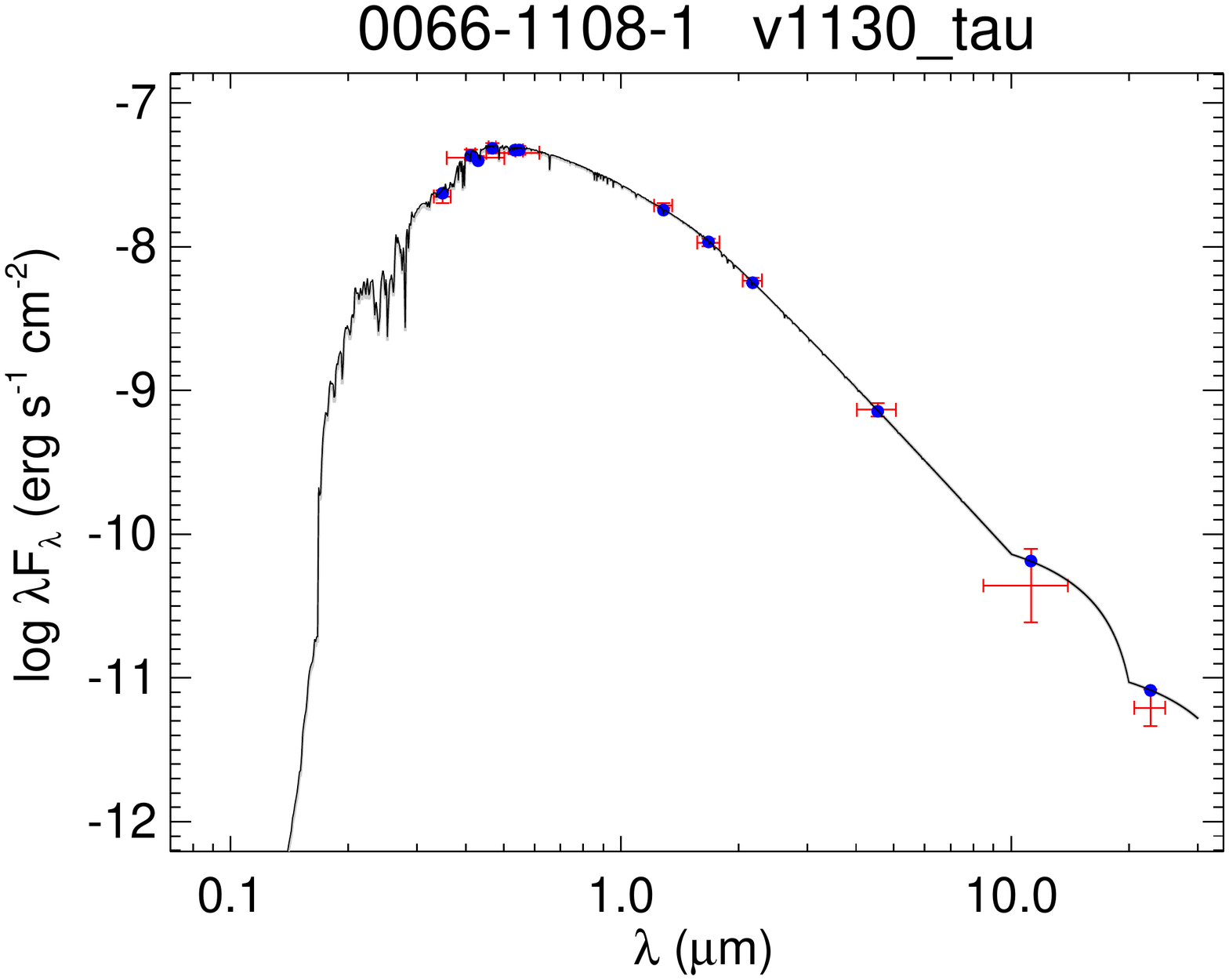}
  \includegraphics[trim=60 60 60 60,clip,width=0.49\linewidth]{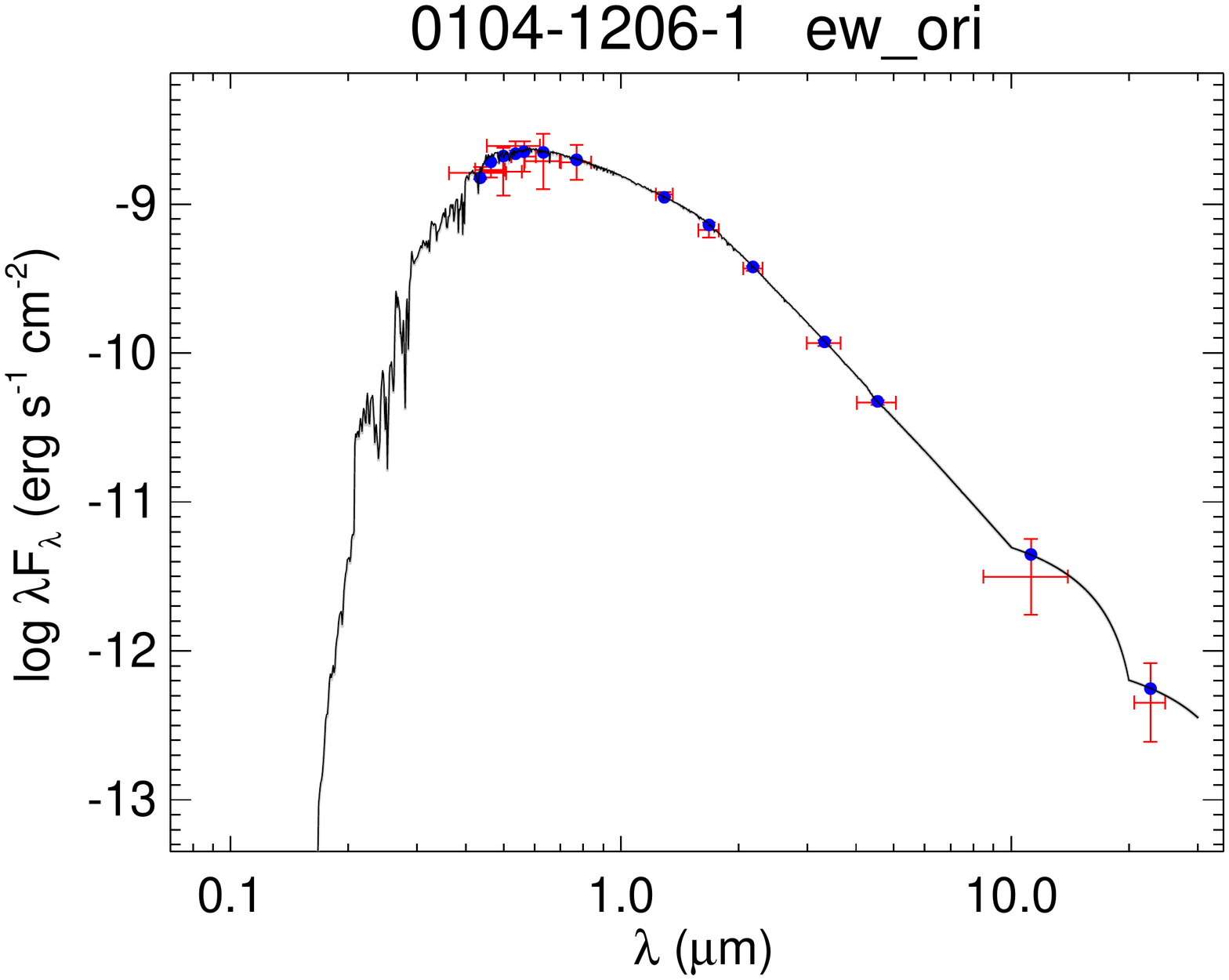}
  \includegraphics[trim=60 60 60 60,clip,width=0.49\linewidth]{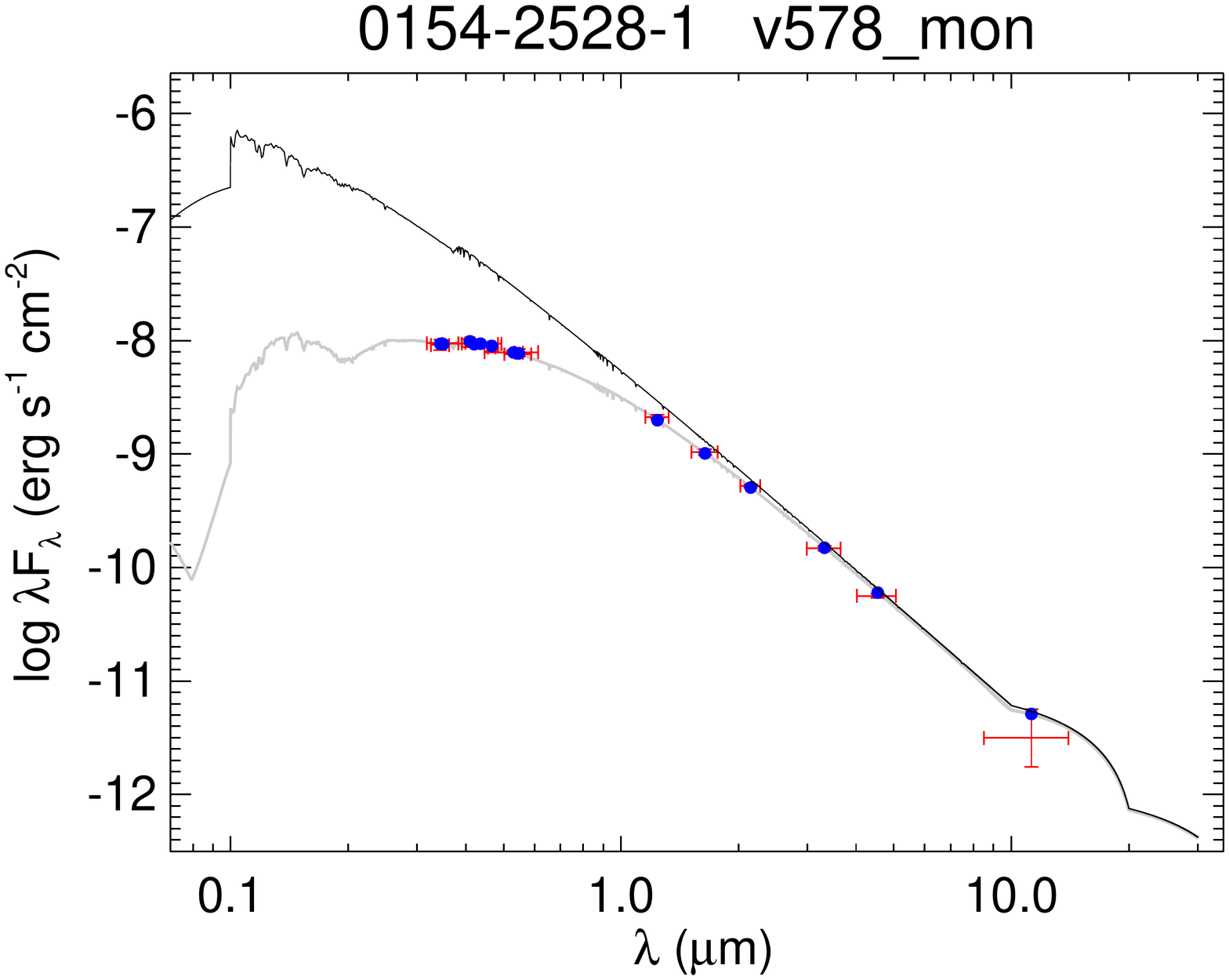}
  \includegraphics[trim=60 60 60 60,clip,width=0.49\linewidth]{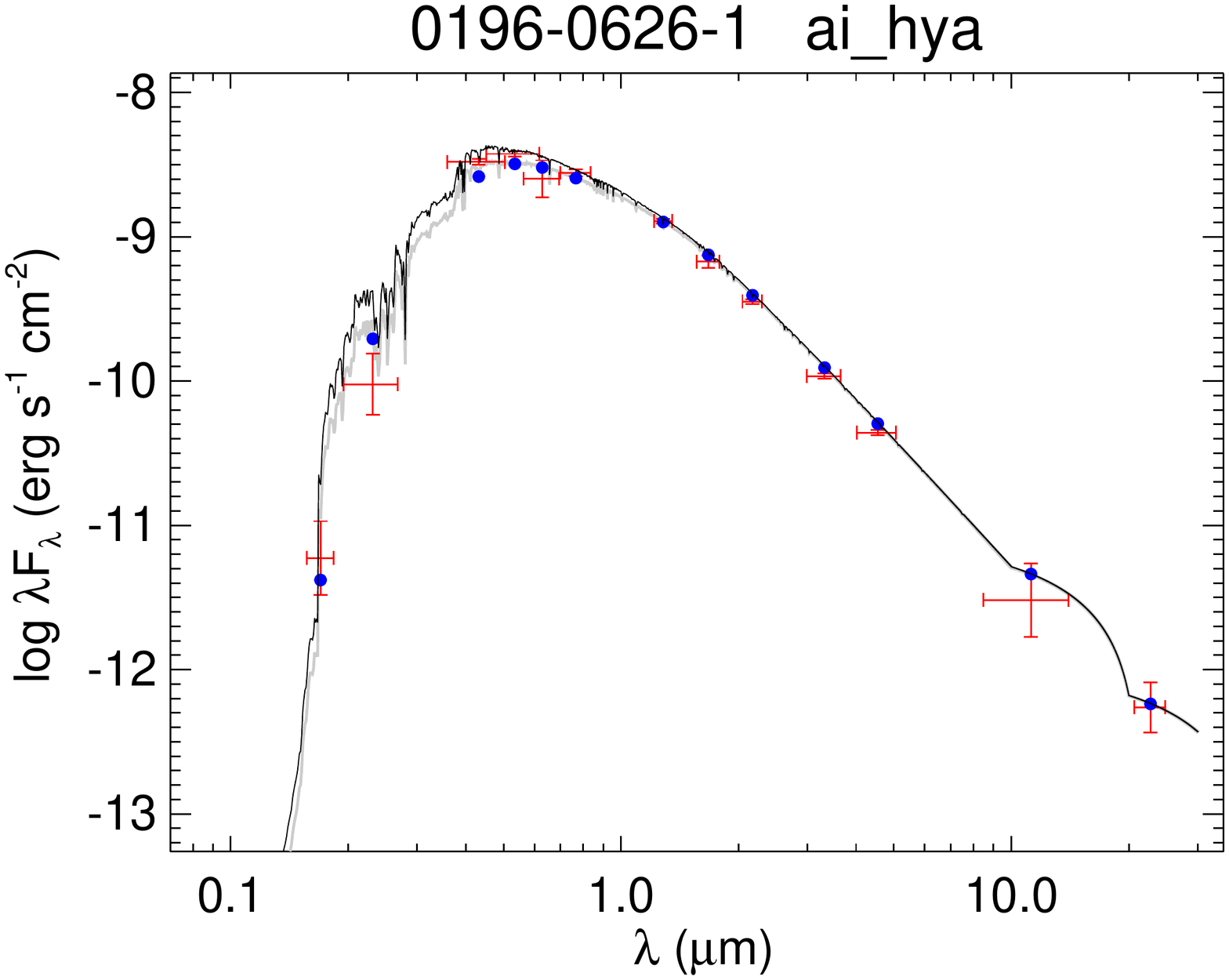}
  \caption{All labels, lines, symbols, and colors as in Figure \ref{fig:seds}.}
  \label{fig:seds_1}
\end{figure}

\begin{figure}[H]
  \centering
  \includegraphics[trim=60 60 60 60,clip,width=0.49\linewidth]{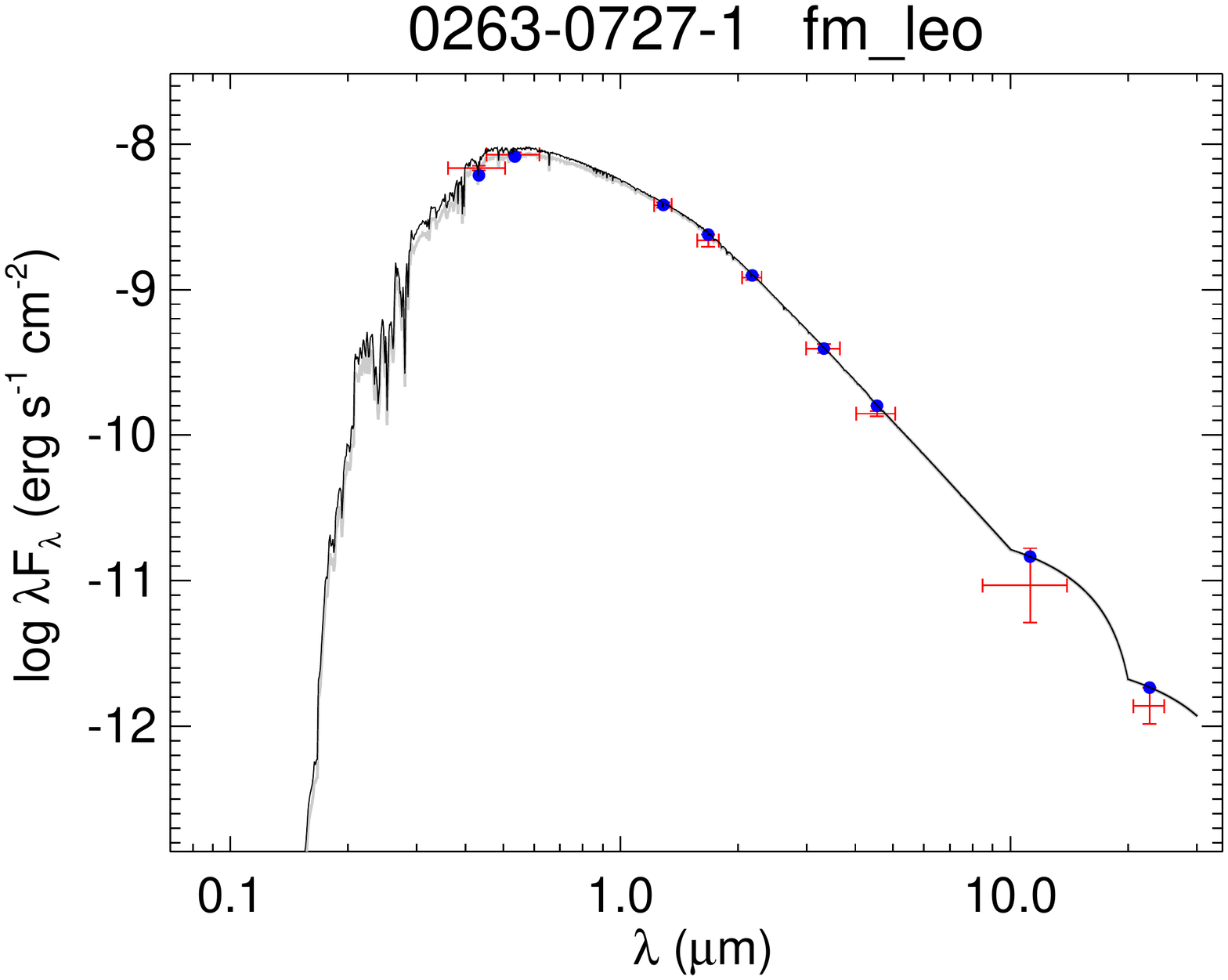}
  \includegraphics[trim=60 60 60 60,clip,width=0.49\linewidth]{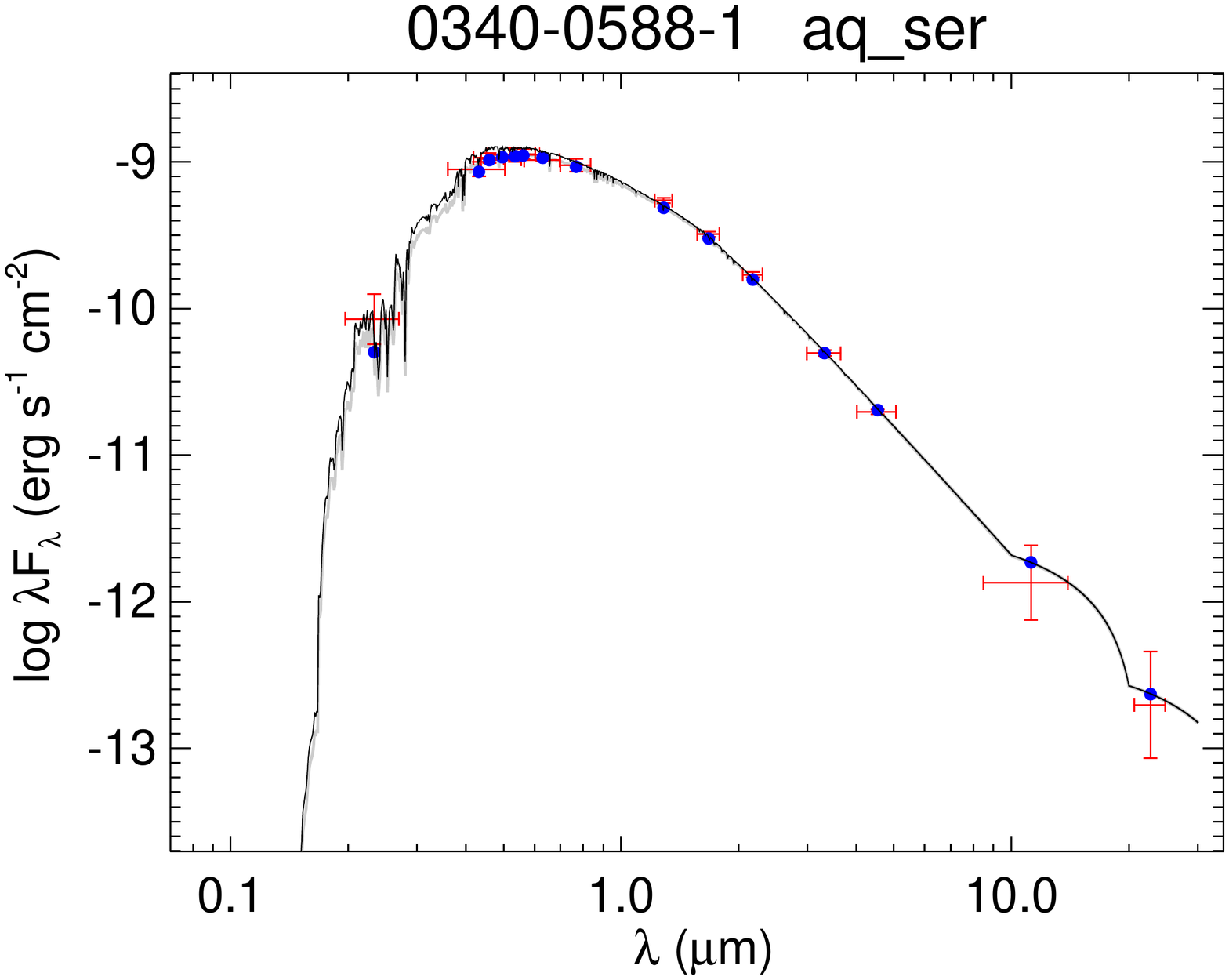}
  \includegraphics[trim=60 60 60 60,clip,width=0.49\linewidth]{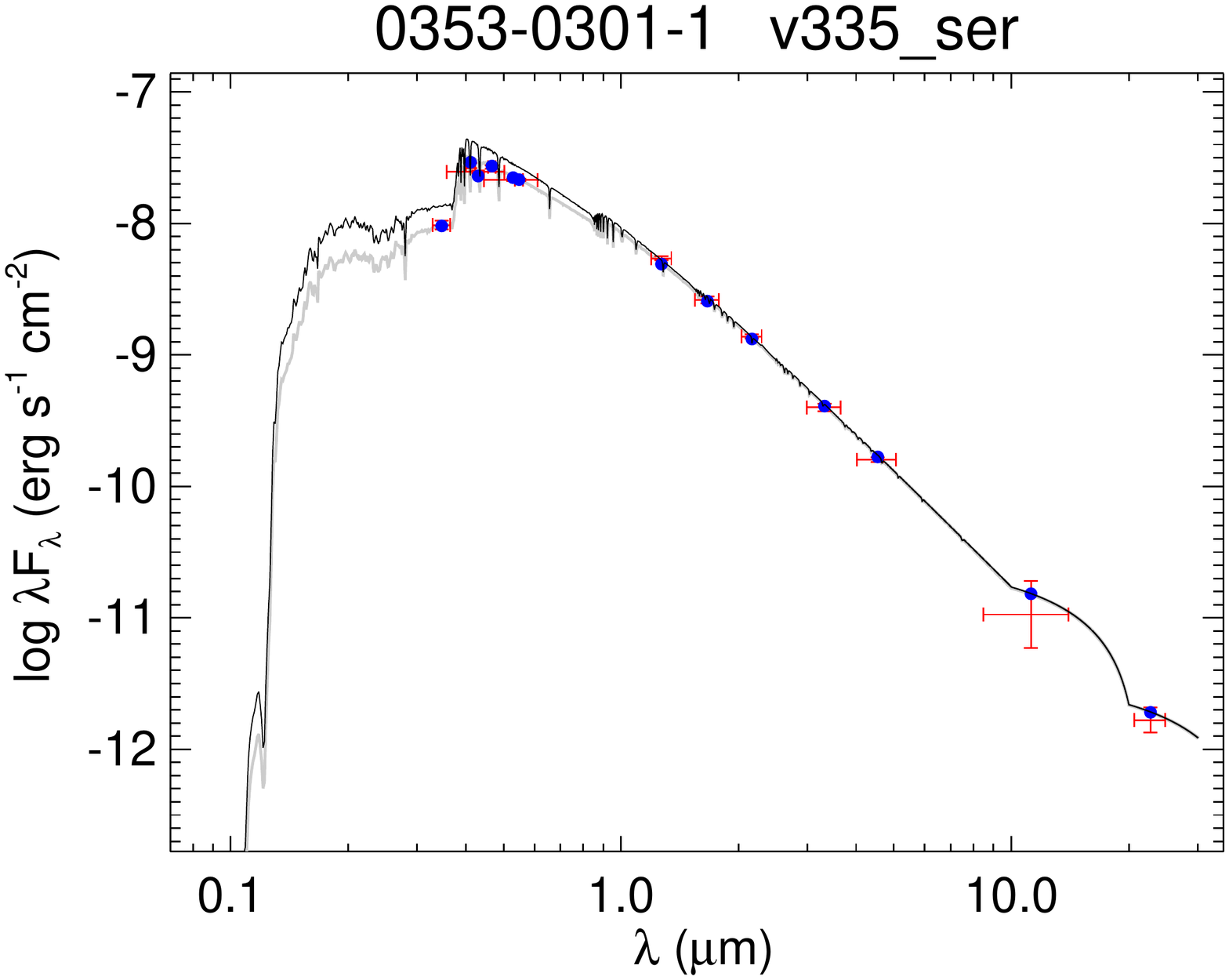}
  \includegraphics[trim=60 60 60 60,clip,width=0.49\linewidth]{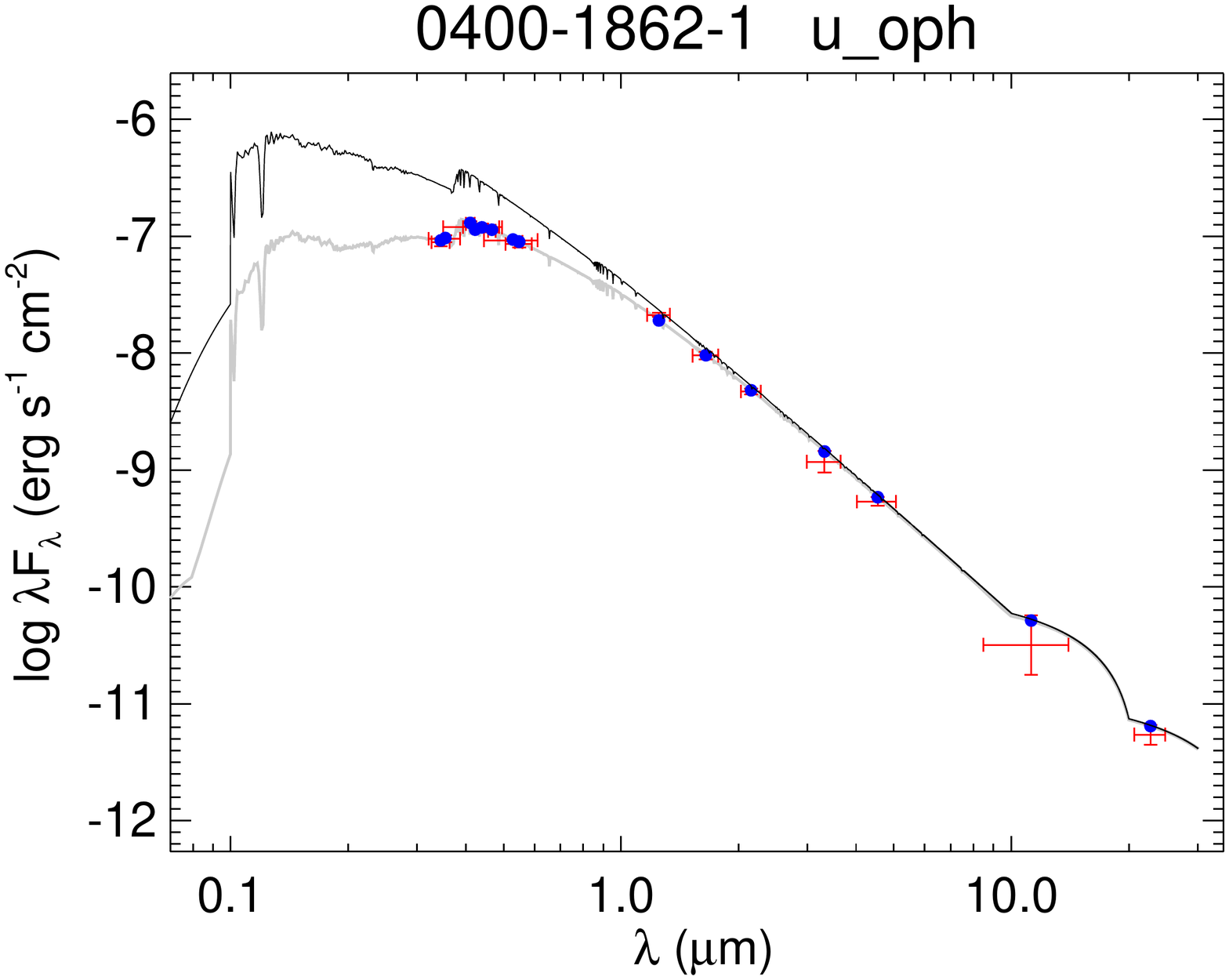}
  \includegraphics[trim=60 60 60 60,clip,width=0.49\linewidth]{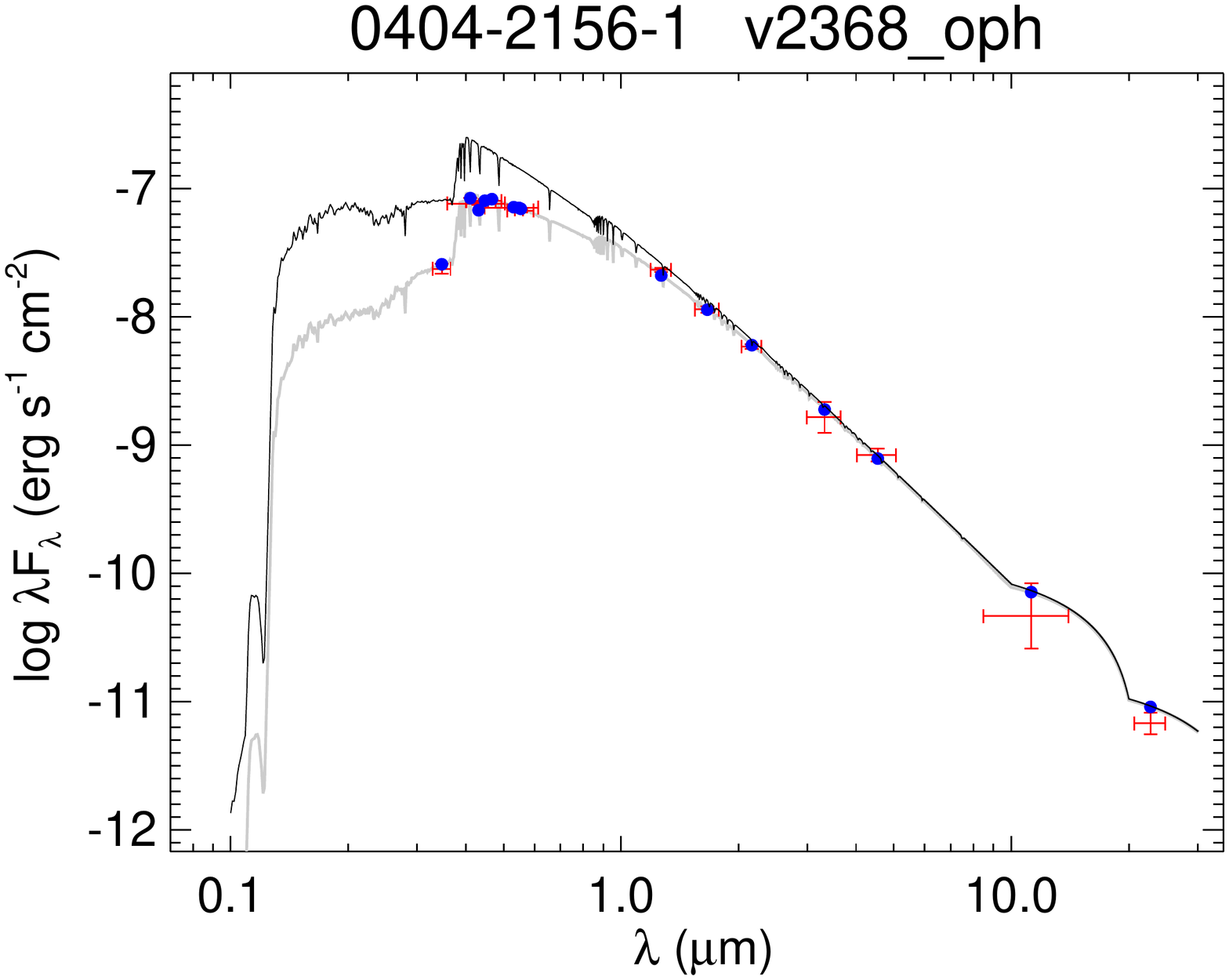}
  \includegraphics[trim=60 60 60 60,clip,width=0.49\linewidth]{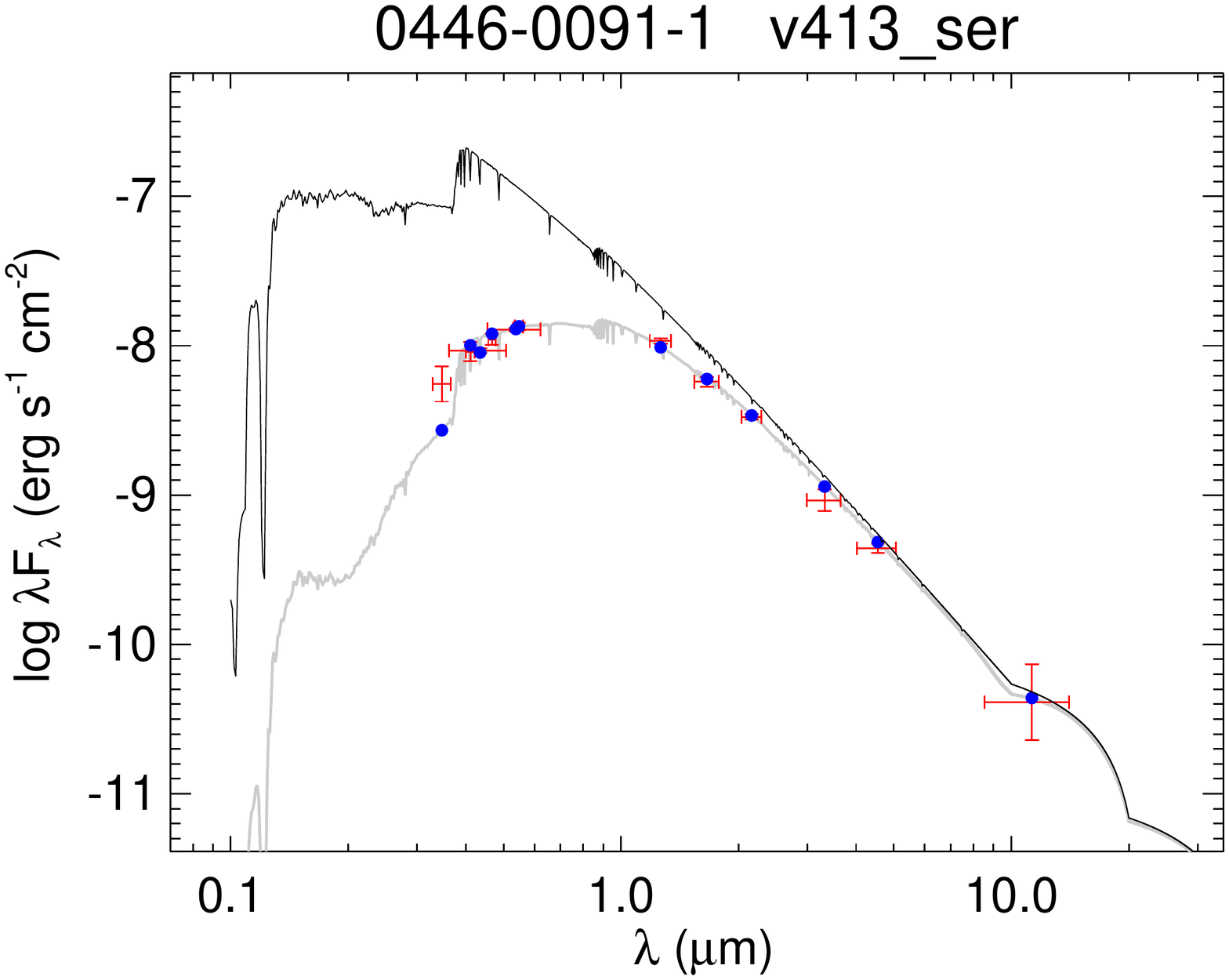}
  \caption{All labels, lines, symbols, and colors as in Figure \ref{fig:seds}.}
  \label{fig:seds_2}
\end{figure}

\begin{figure}[H]
  \centering
  \includegraphics[trim=60 60 60 60,clip,width=0.49\linewidth]{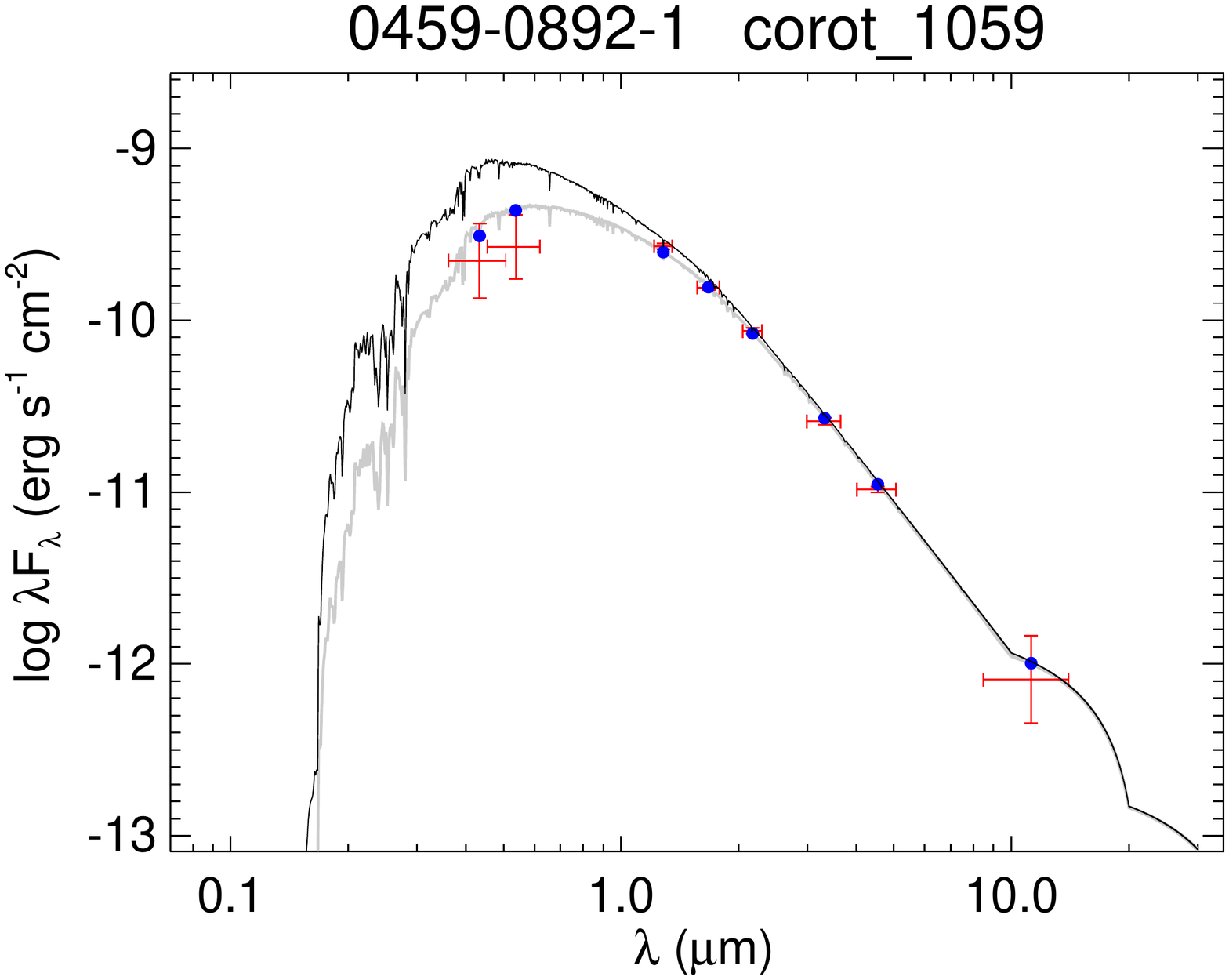}
  \includegraphics[trim=60 60 60 60,clip,width=0.49\linewidth]{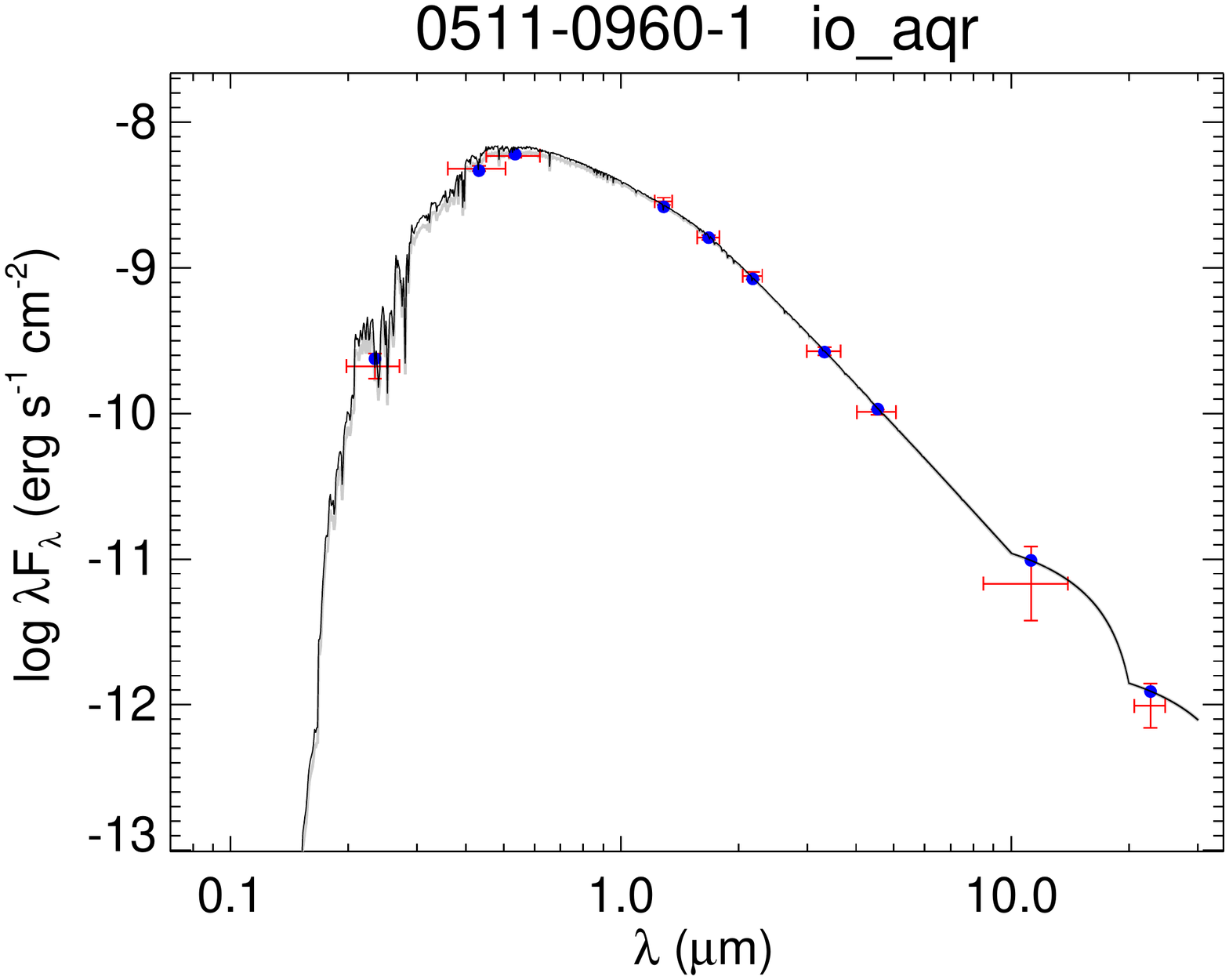}
  \includegraphics[trim=60 60 60 60,clip,width=0.49\linewidth]{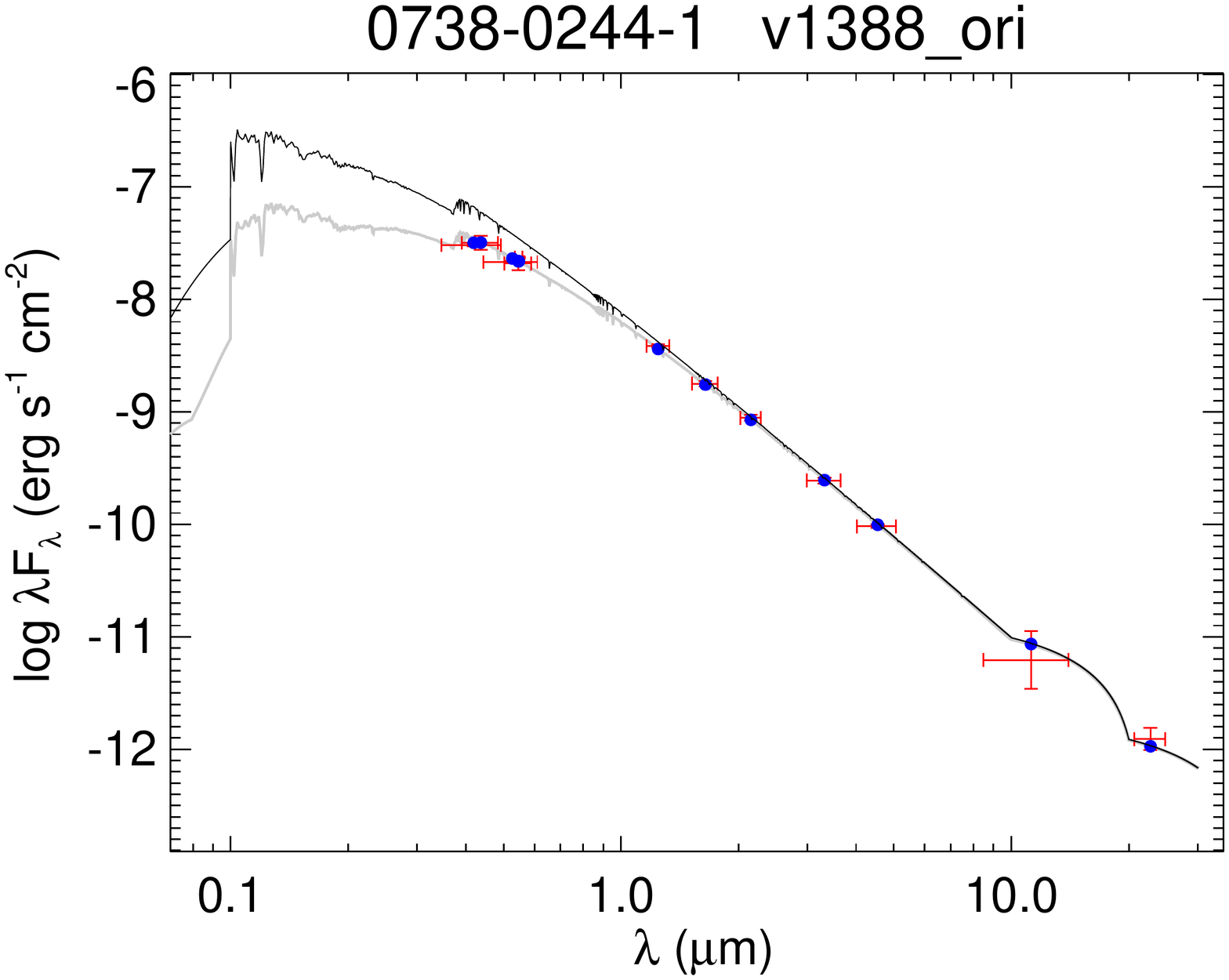}
  \includegraphics[trim=60 60 60 60,clip,width=0.49\linewidth]{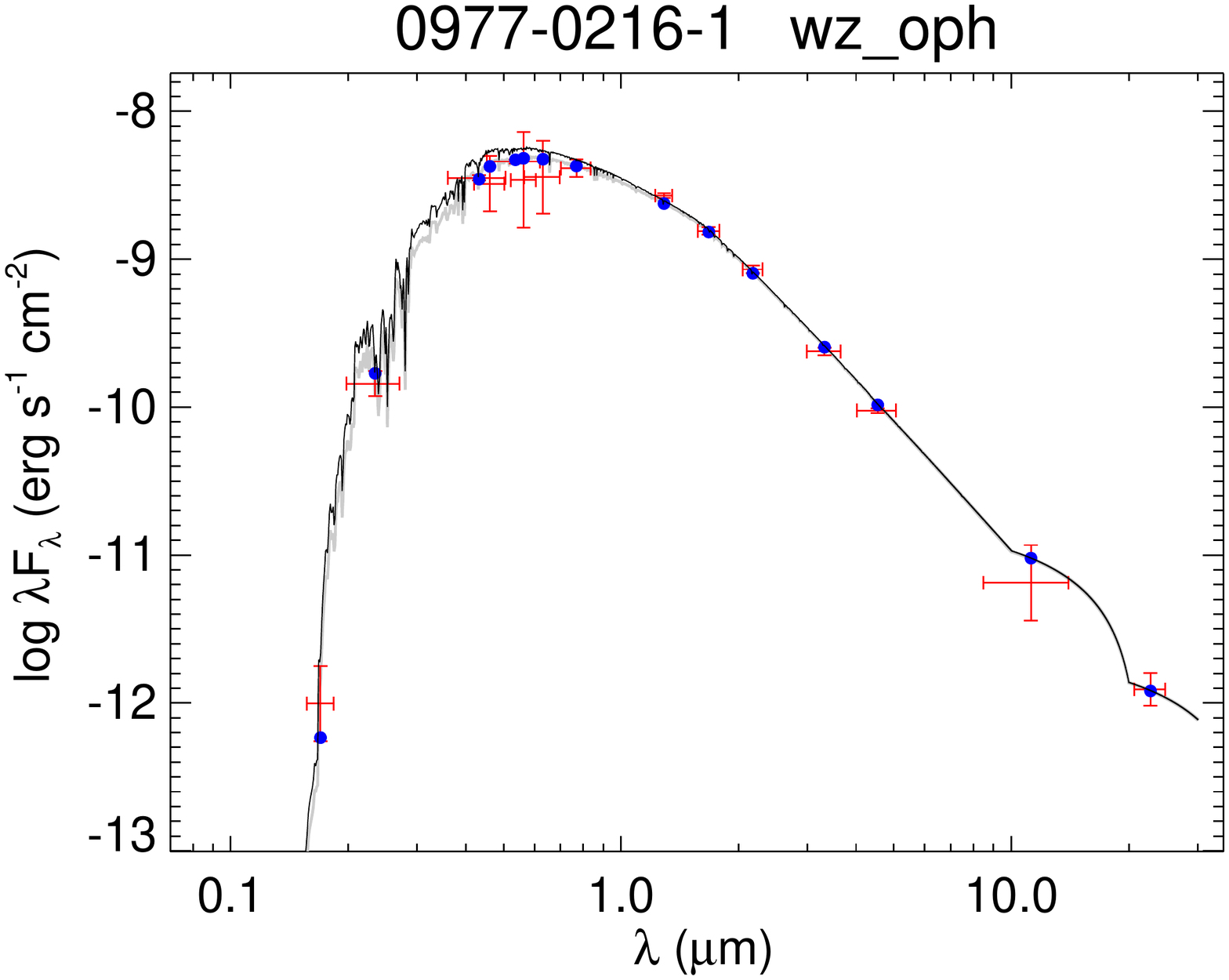}
  \includegraphics[trim=60 60 60 60,clip,width=0.49\linewidth]{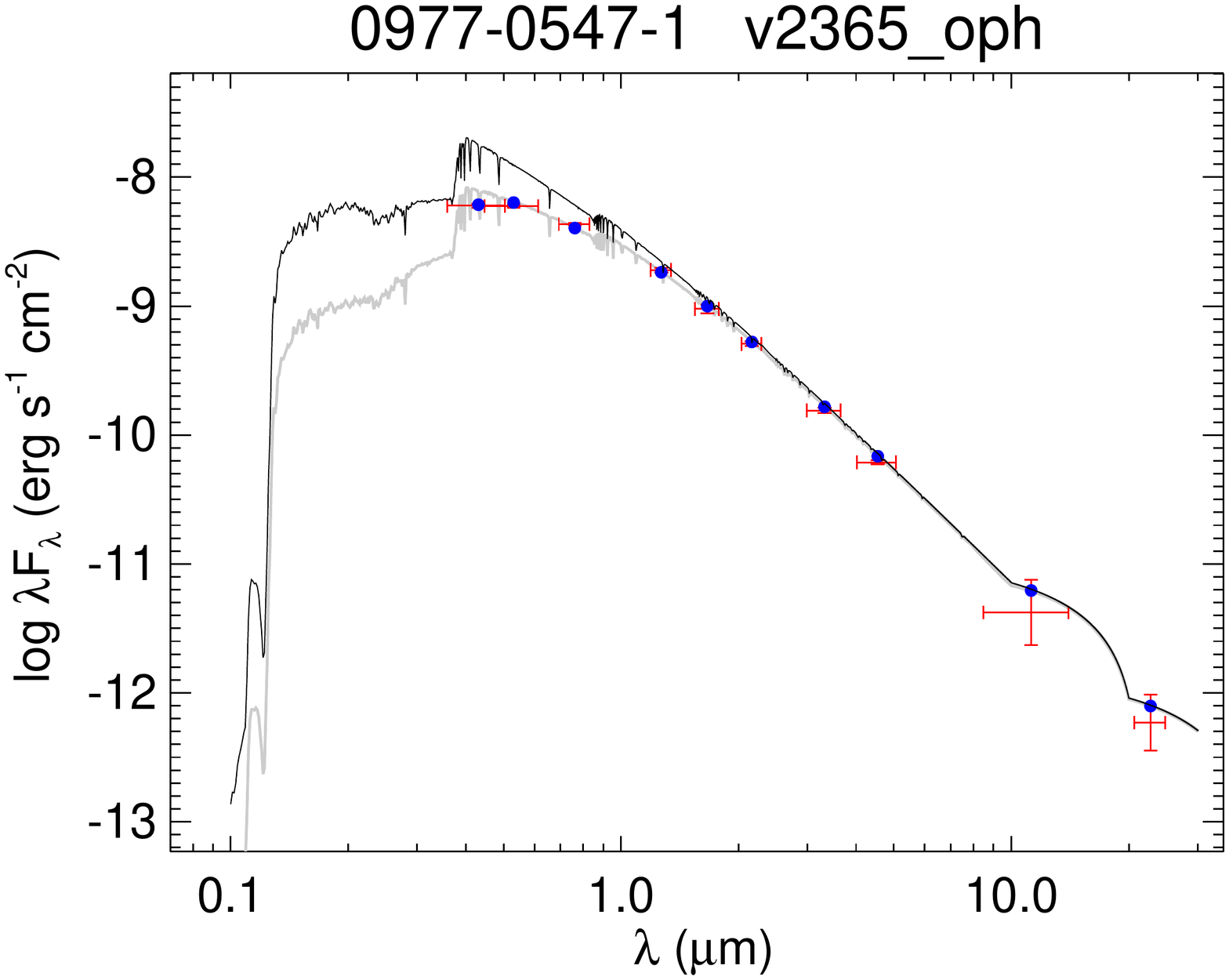}
  \includegraphics[trim=60 60 60 60,clip,width=0.49\linewidth]{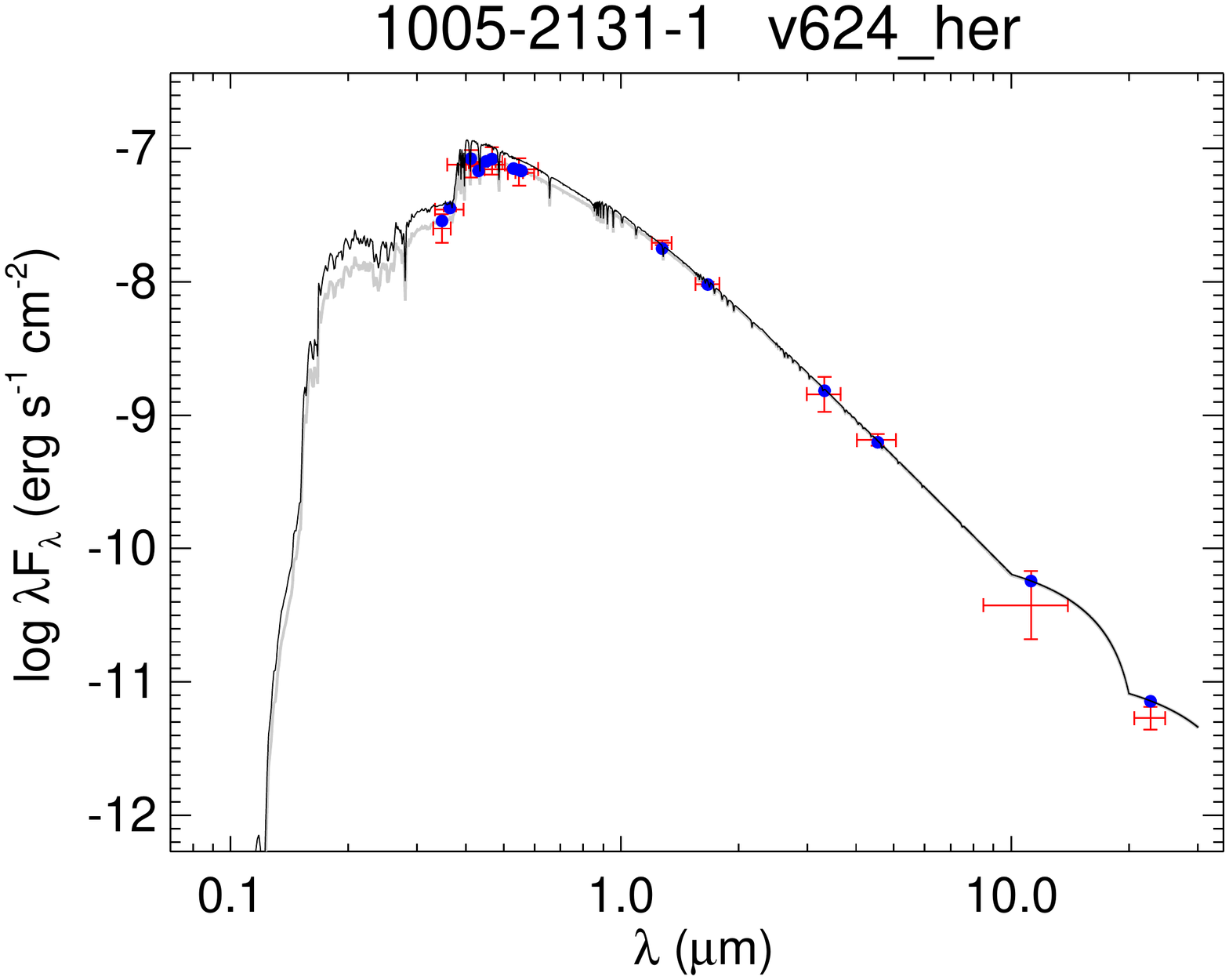}
  \caption{All labels, lines, symbols, and colors as in Figure \ref{fig:seds}.}
  \label{fig:seds_3}
\end{figure}

\begin{figure}[H]
  \centering
  \includegraphics[trim=60 60 60 60,clip,width=0.49\linewidth]{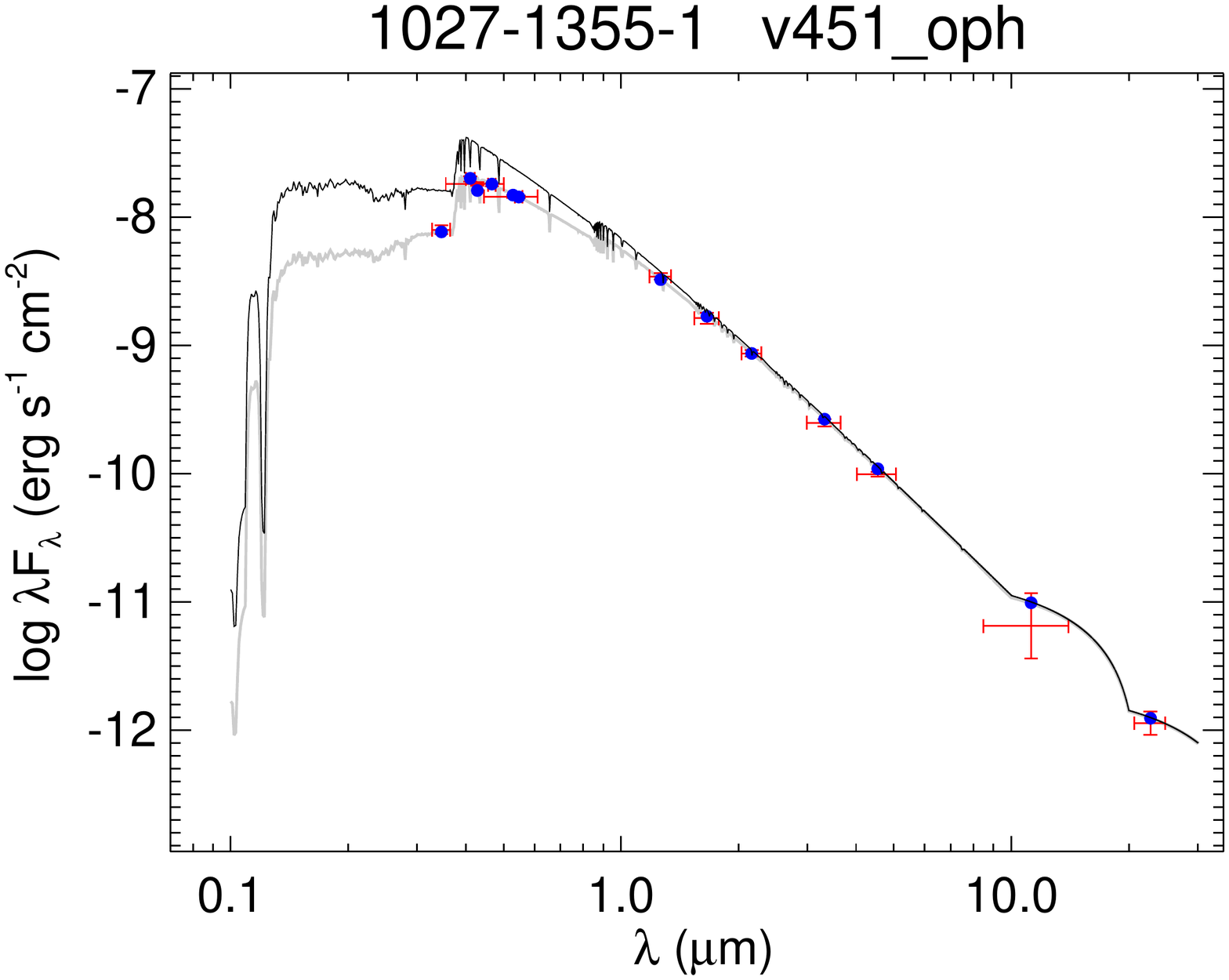}
  \includegraphics[trim=60 60 60 60,clip,width=0.49\linewidth]{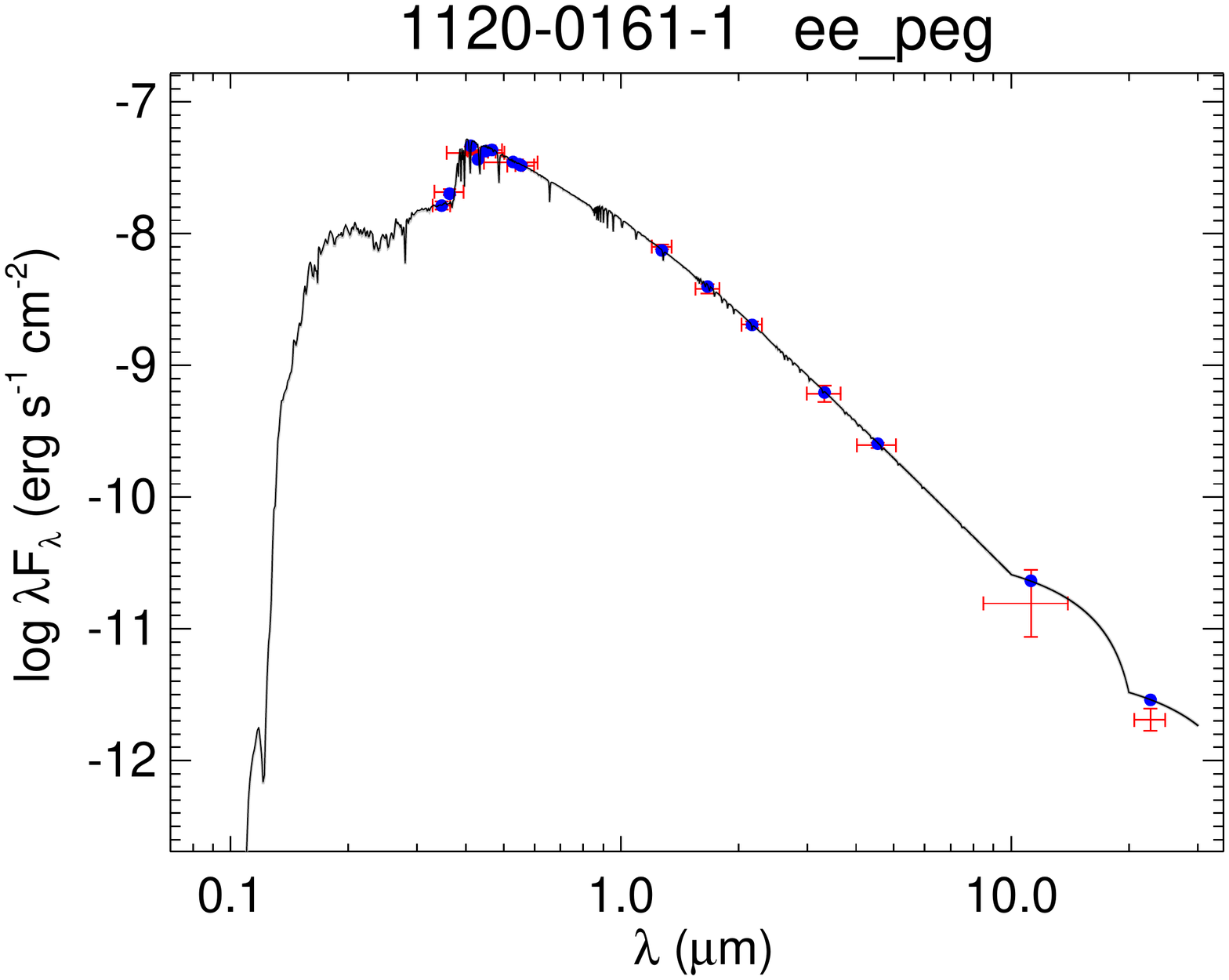}
  \includegraphics[trim=60 60 60 60,clip,width=0.49\linewidth]{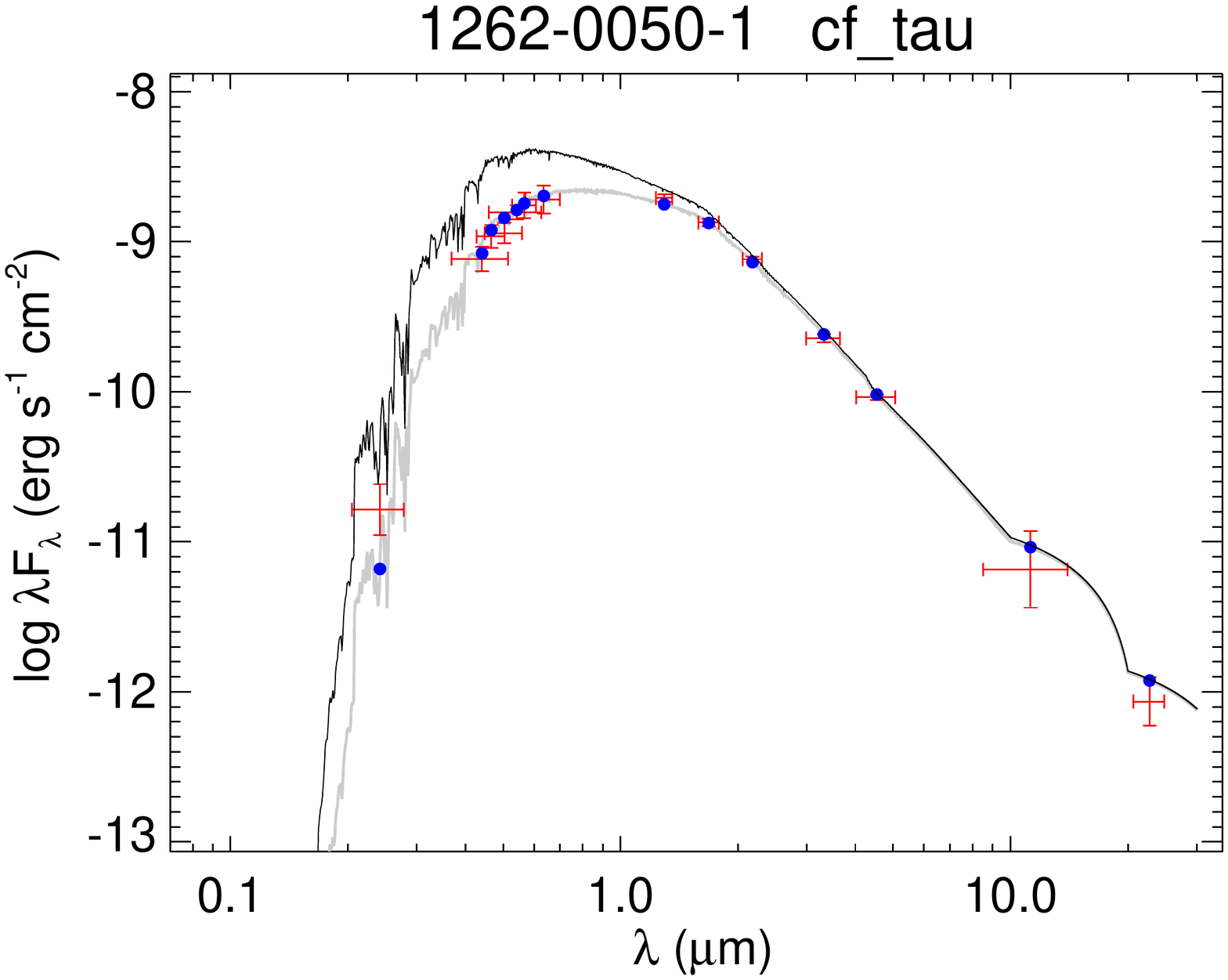}
  \includegraphics[trim=60 60 60 60,clip,width=0.49\linewidth]{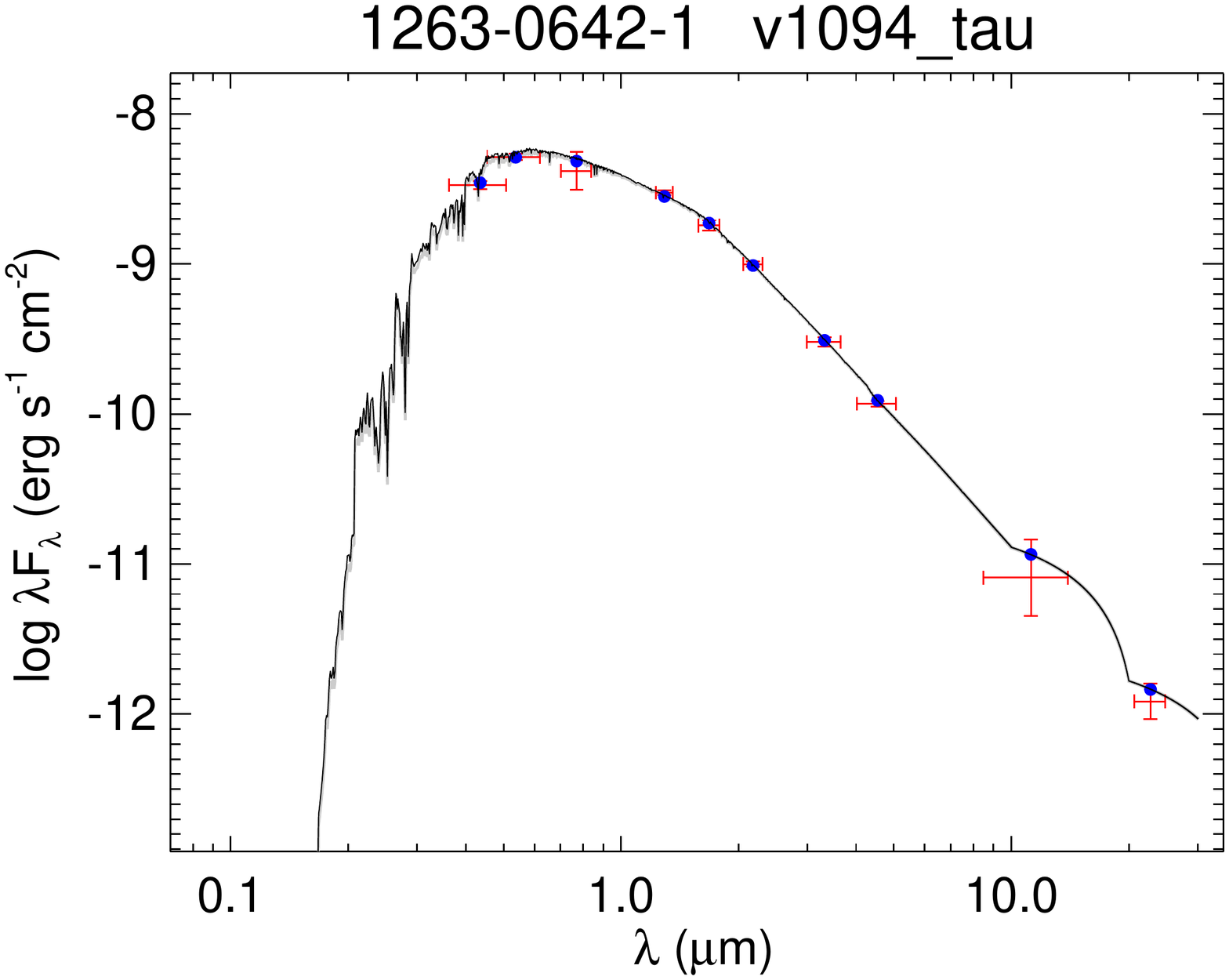}
  \includegraphics[trim=60 60 60 60,clip,width=0.49\linewidth]{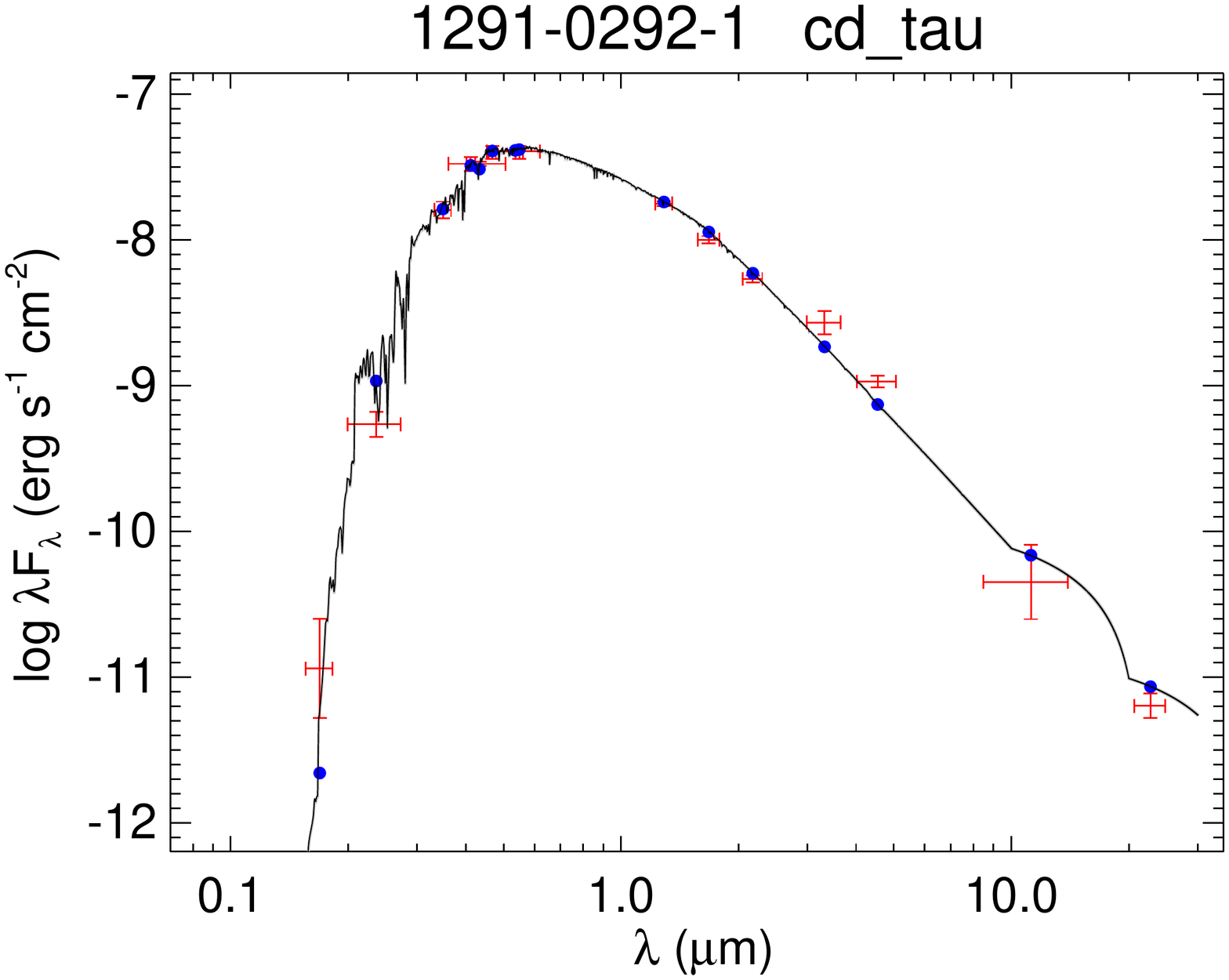}
  \includegraphics[trim=60 60 60 60,clip,width=0.49\linewidth]{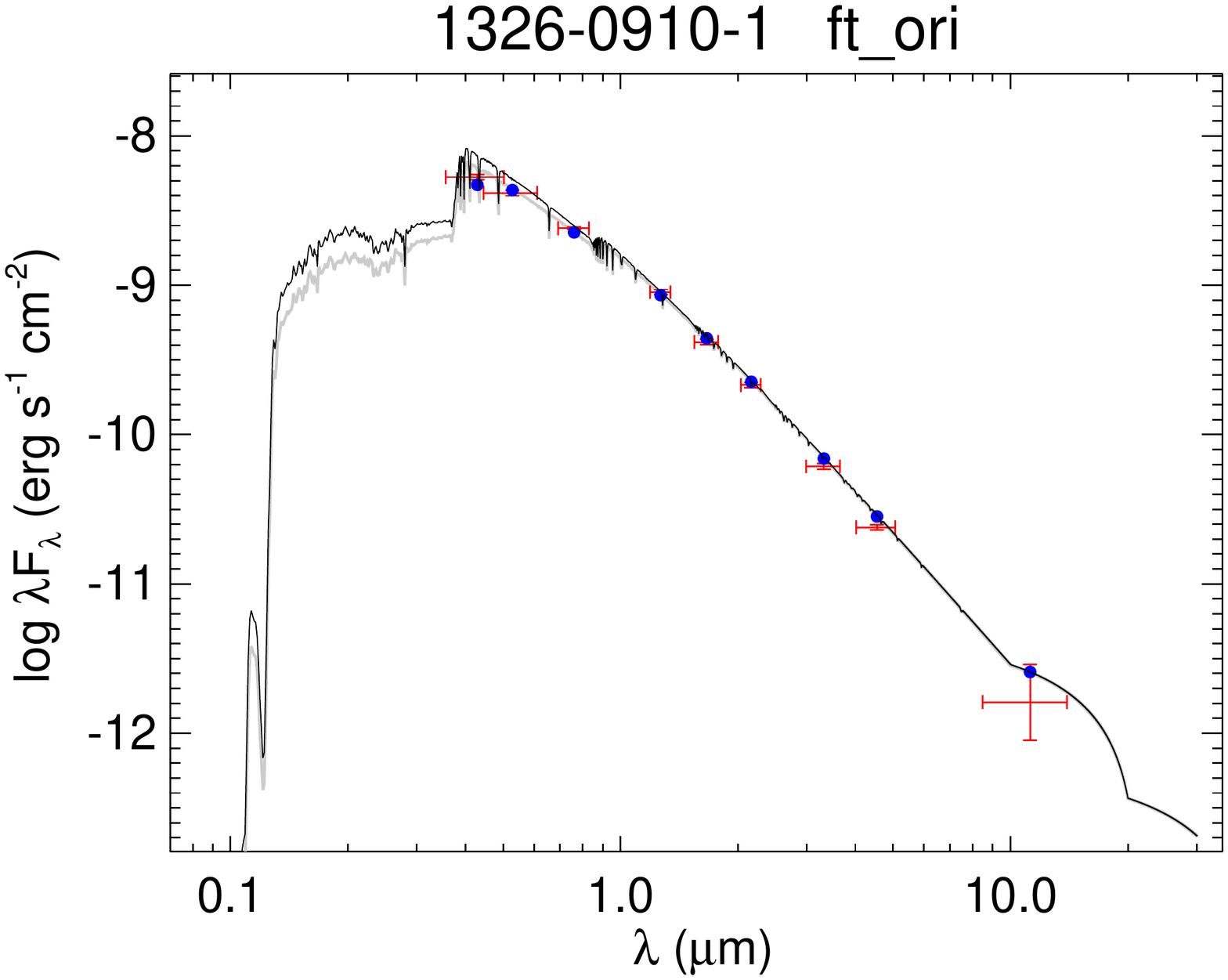}
  \caption{All labels, lines, symbols, and colors as in Figure \ref{fig:seds}.}
  \label{fig:seds_4}
\end{figure}

\begin{figure}[H]
  \centering
  \includegraphics[trim=60 60 60 60,clip,width=0.49\linewidth]{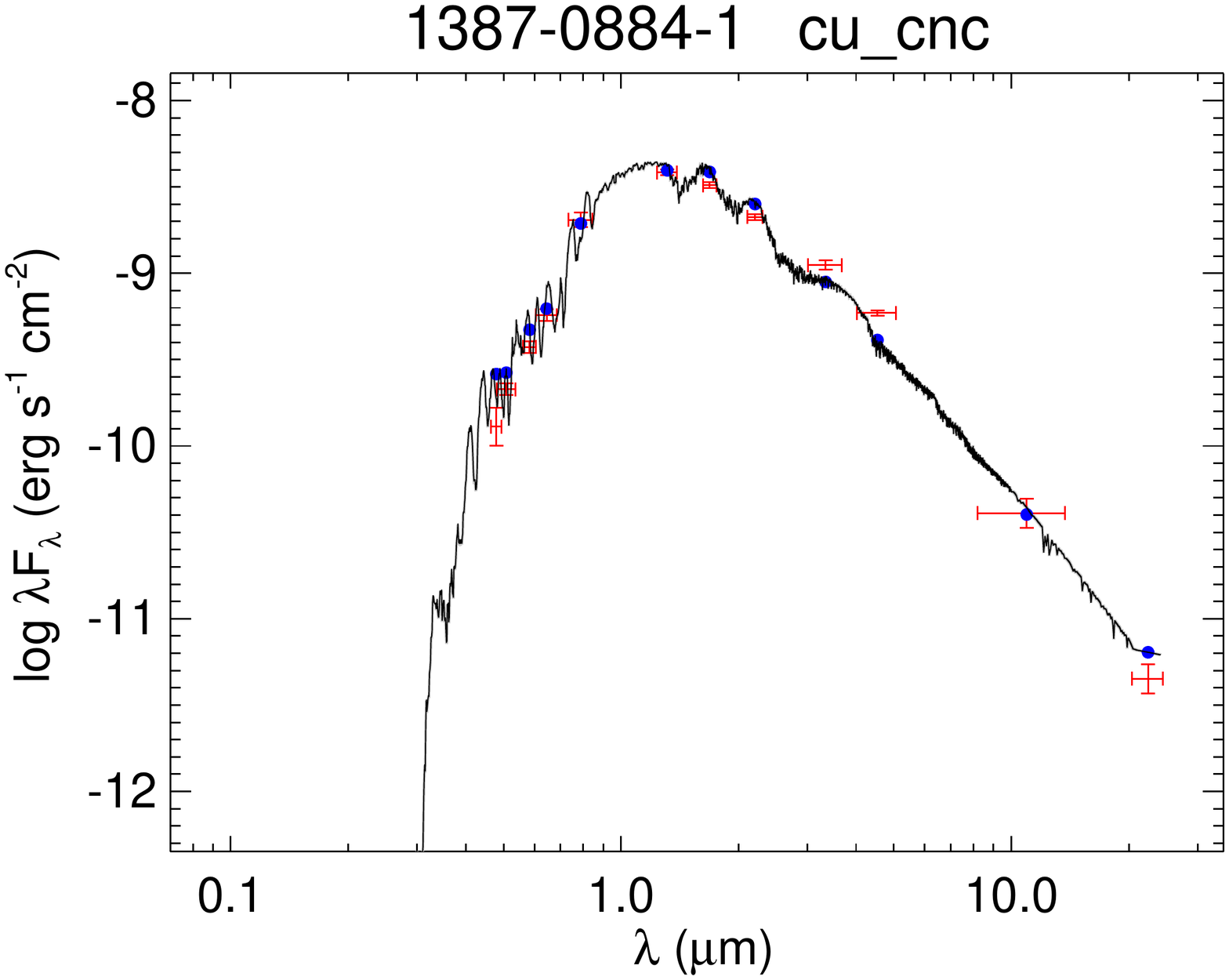}
  \includegraphics[trim=60 60 60 60,clip,width=0.49\linewidth]{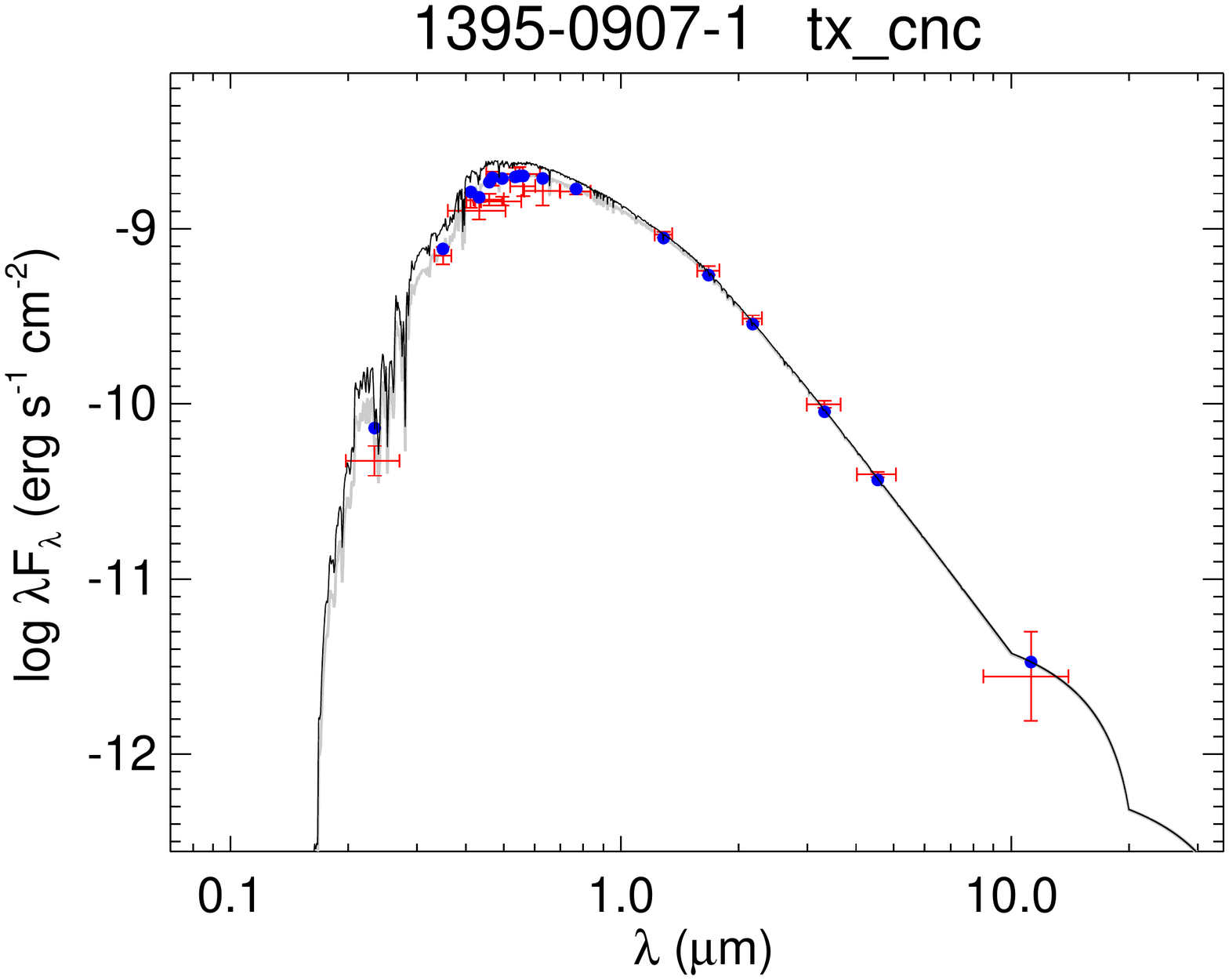}
  \includegraphics[trim=60 60 60 60,clip,width=0.49\linewidth]{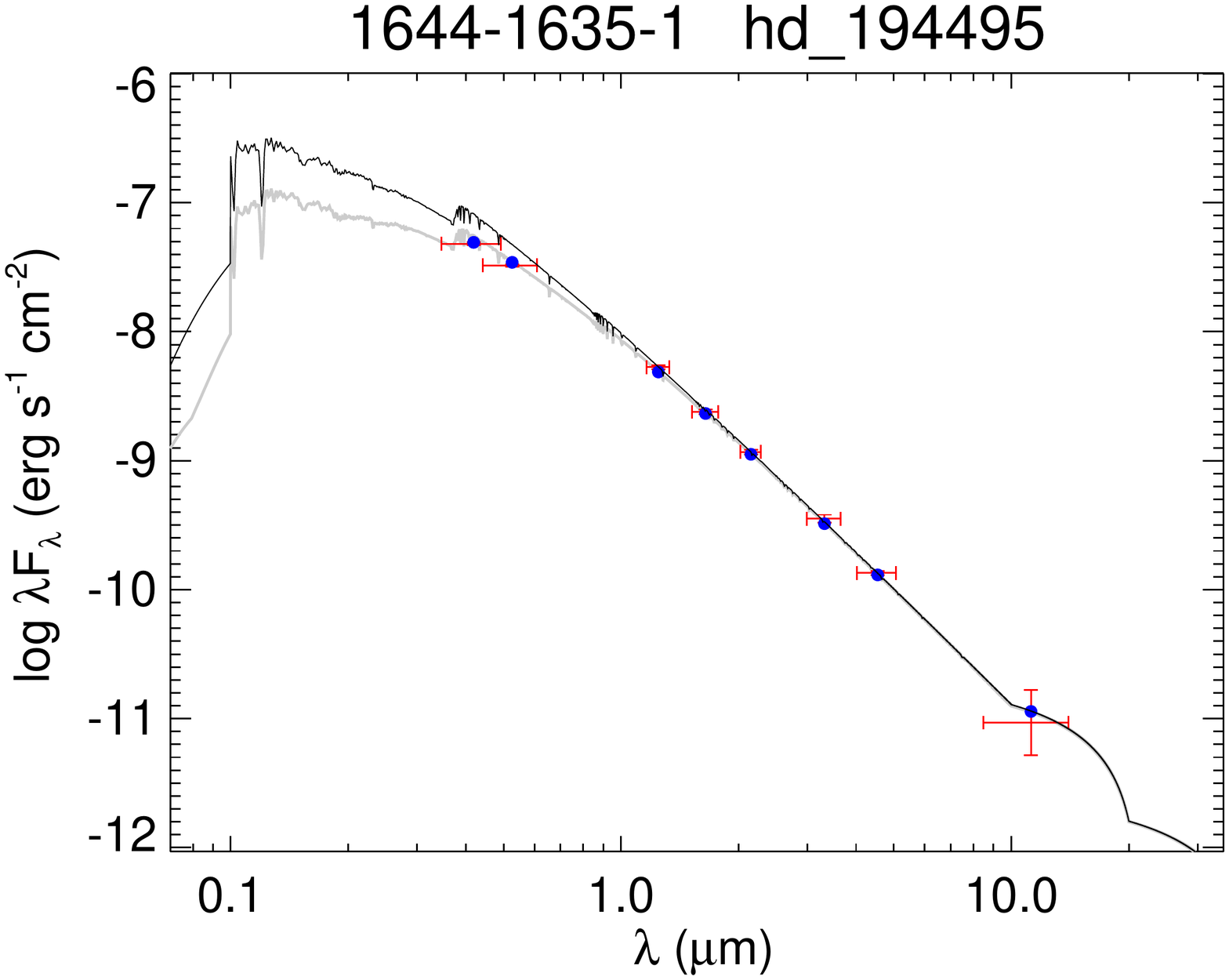}
  \includegraphics[trim=60 60 60 60,clip,width=0.49\linewidth]{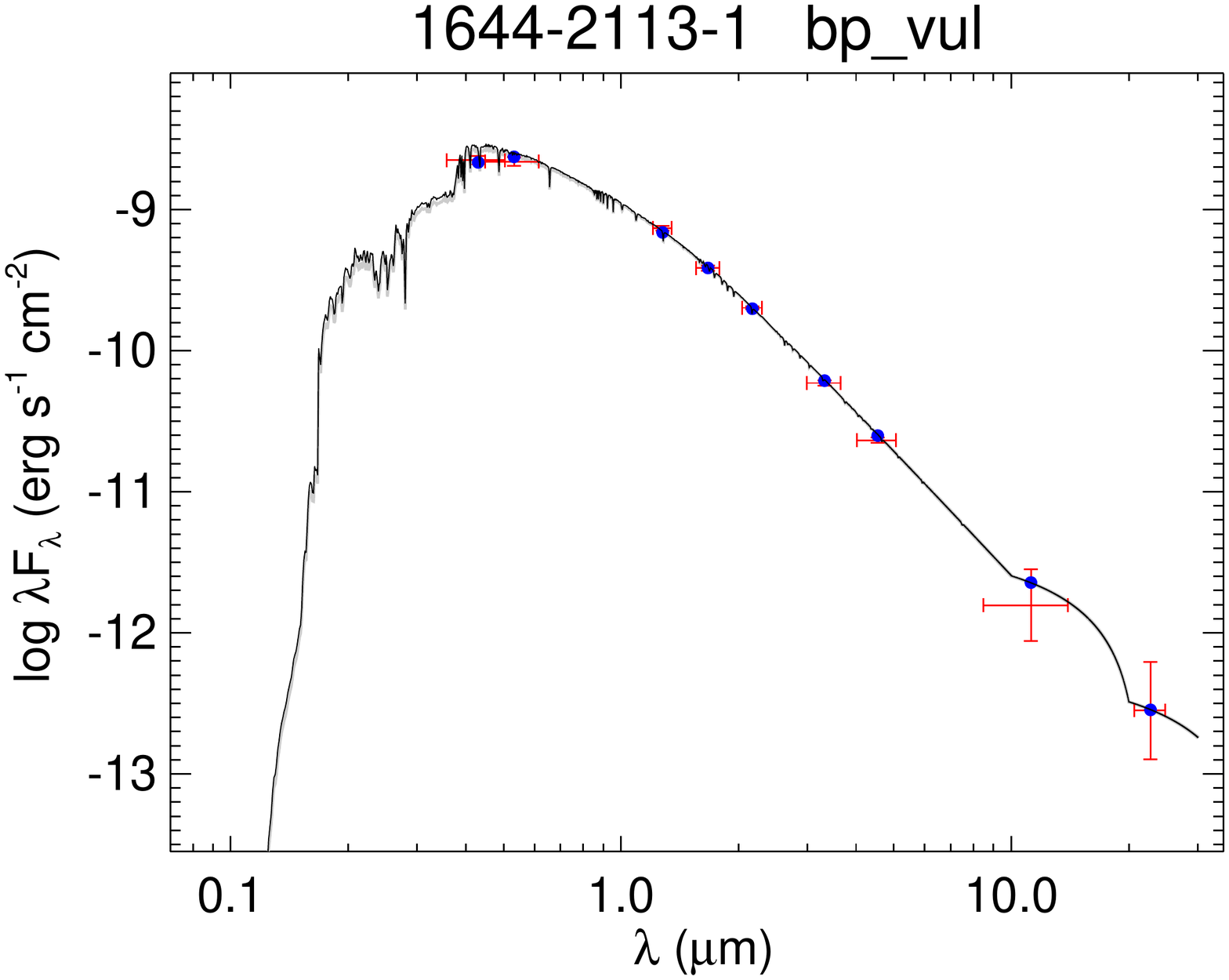}
  \includegraphics[trim=60 60 60 60,clip,width=0.49\linewidth]{sedfigs/ad_boo.pdf}
  \includegraphics[trim=60 60 60 60,clip,width=0.49\linewidth]{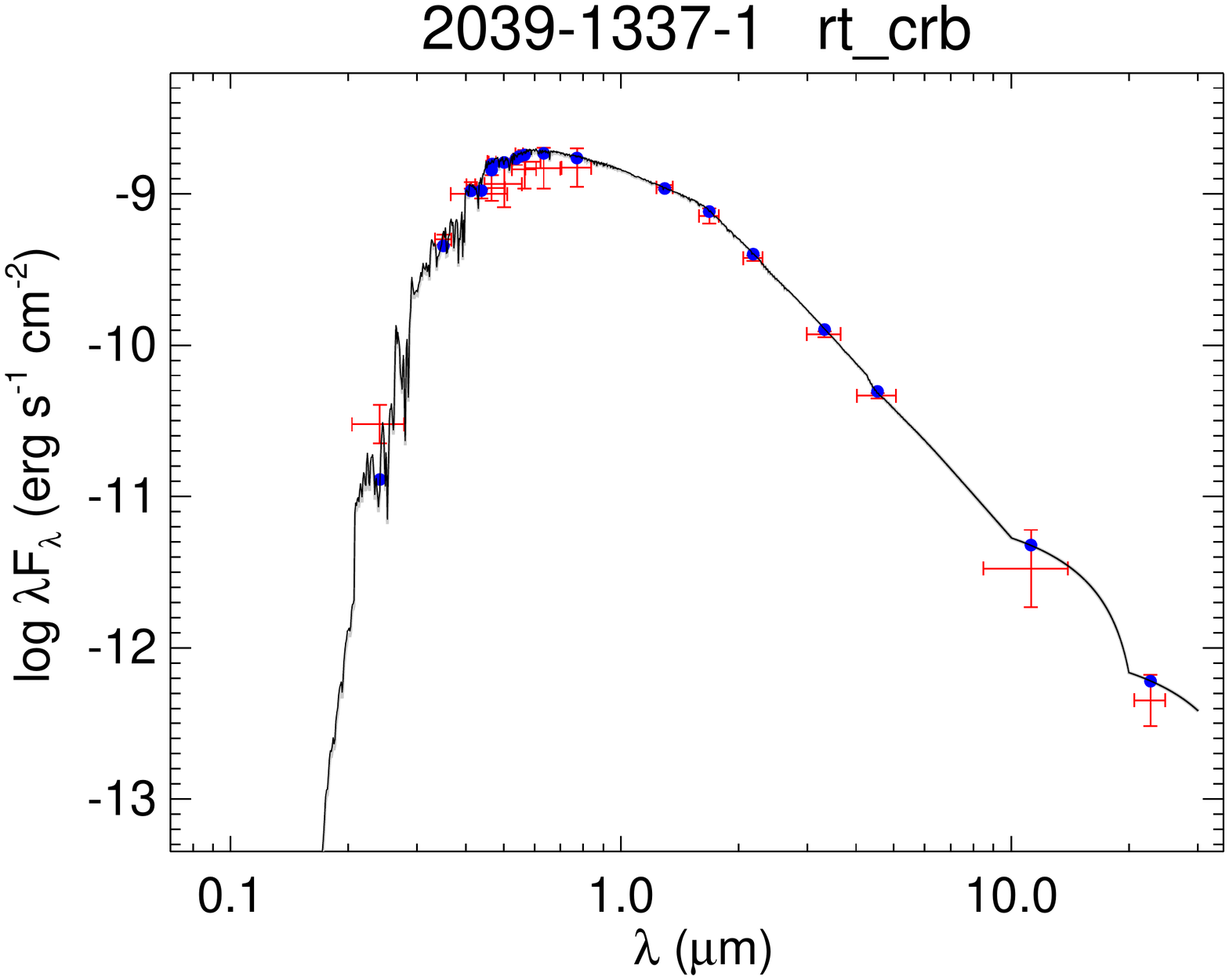}
  \caption{All labels, lines, symbols, and colors as in Figure \ref{fig:seds}.}
  \label{fig:seds_5}
\end{figure}

\begin{figure}[H]
  \centering
  \includegraphics[trim=60 60 60 60,clip,width=0.49\linewidth]{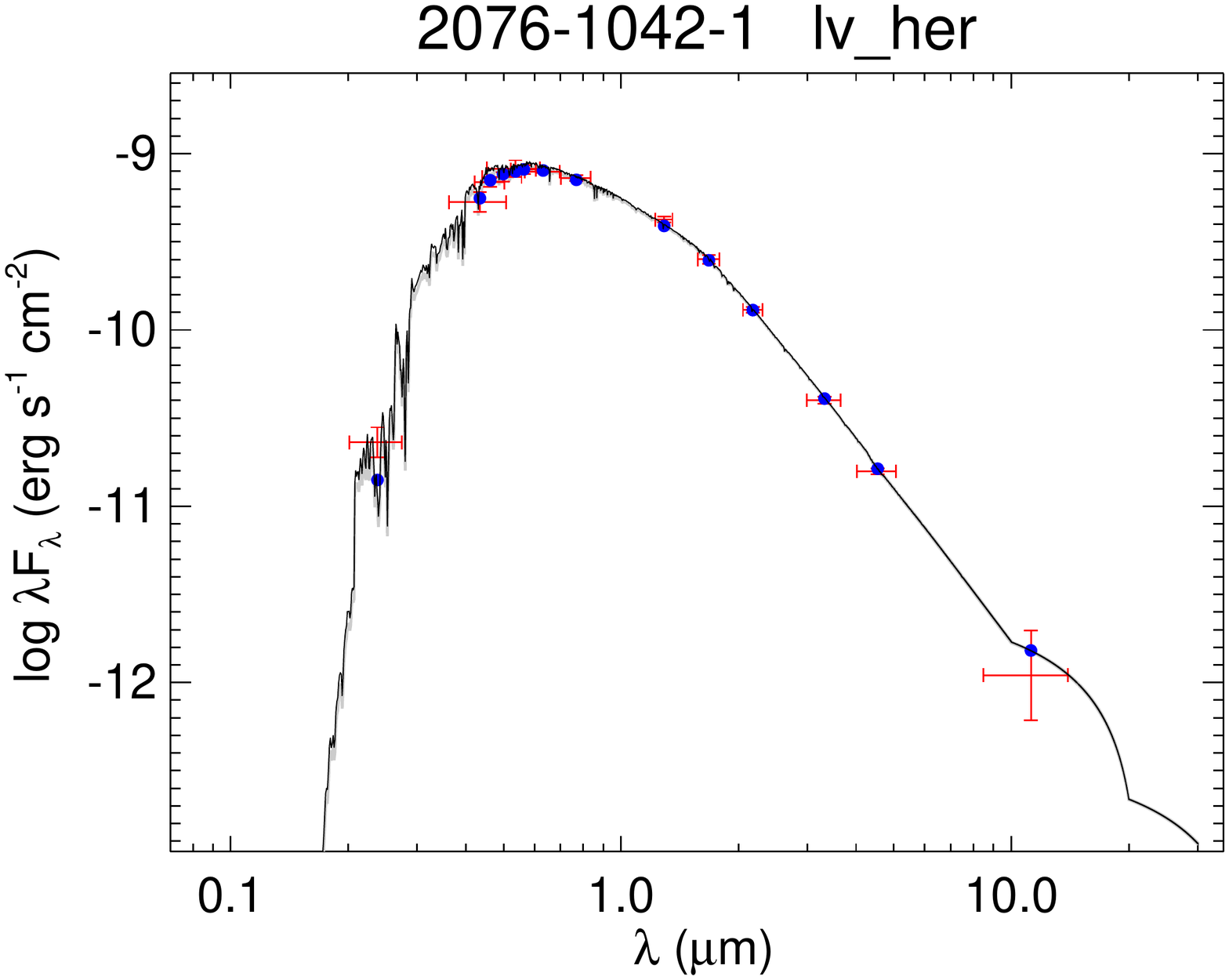}
  \includegraphics[trim=60 60 60 60,clip,width=0.49\linewidth]{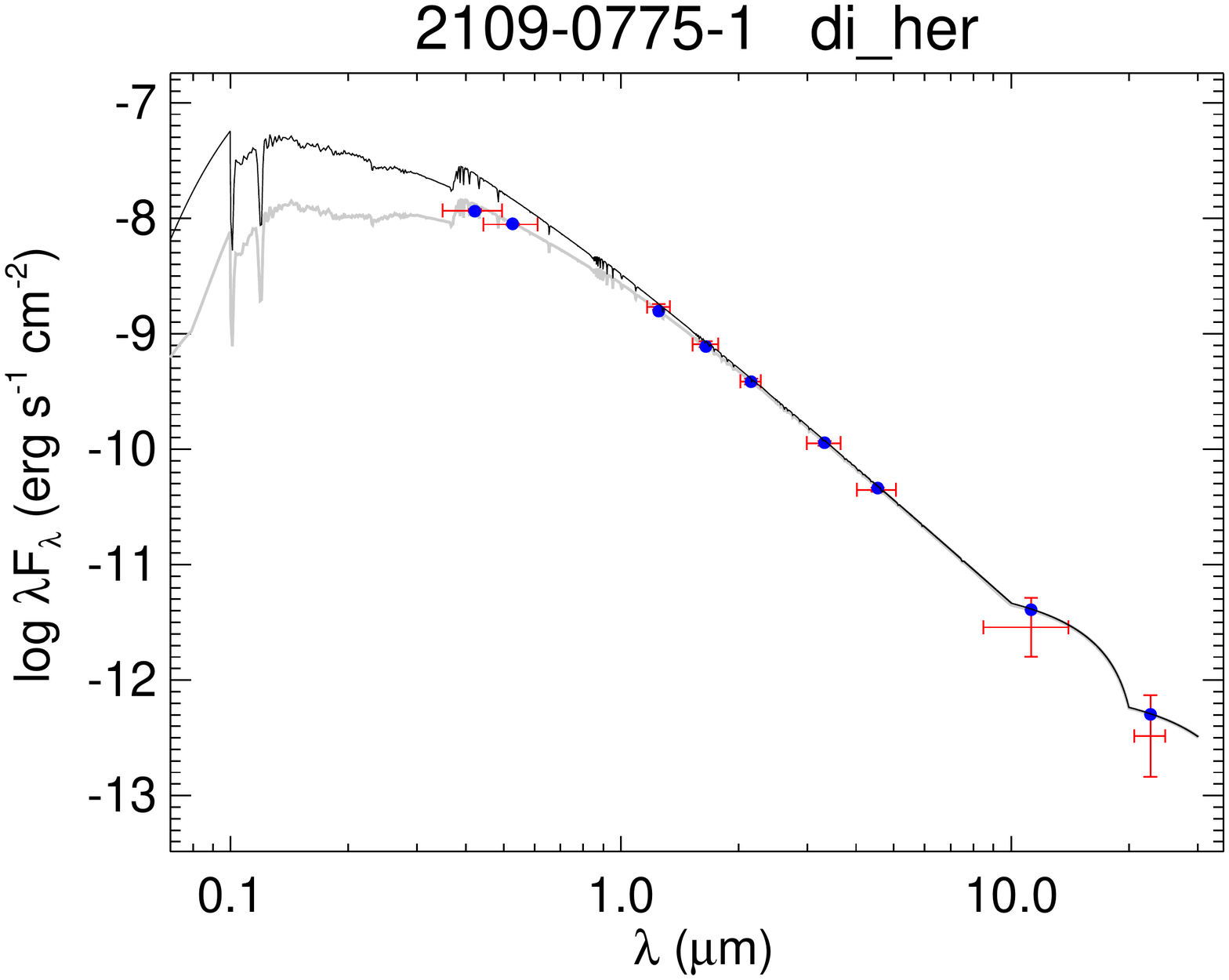}
  \includegraphics[trim=60 60 60 60,clip,width=0.49\linewidth]{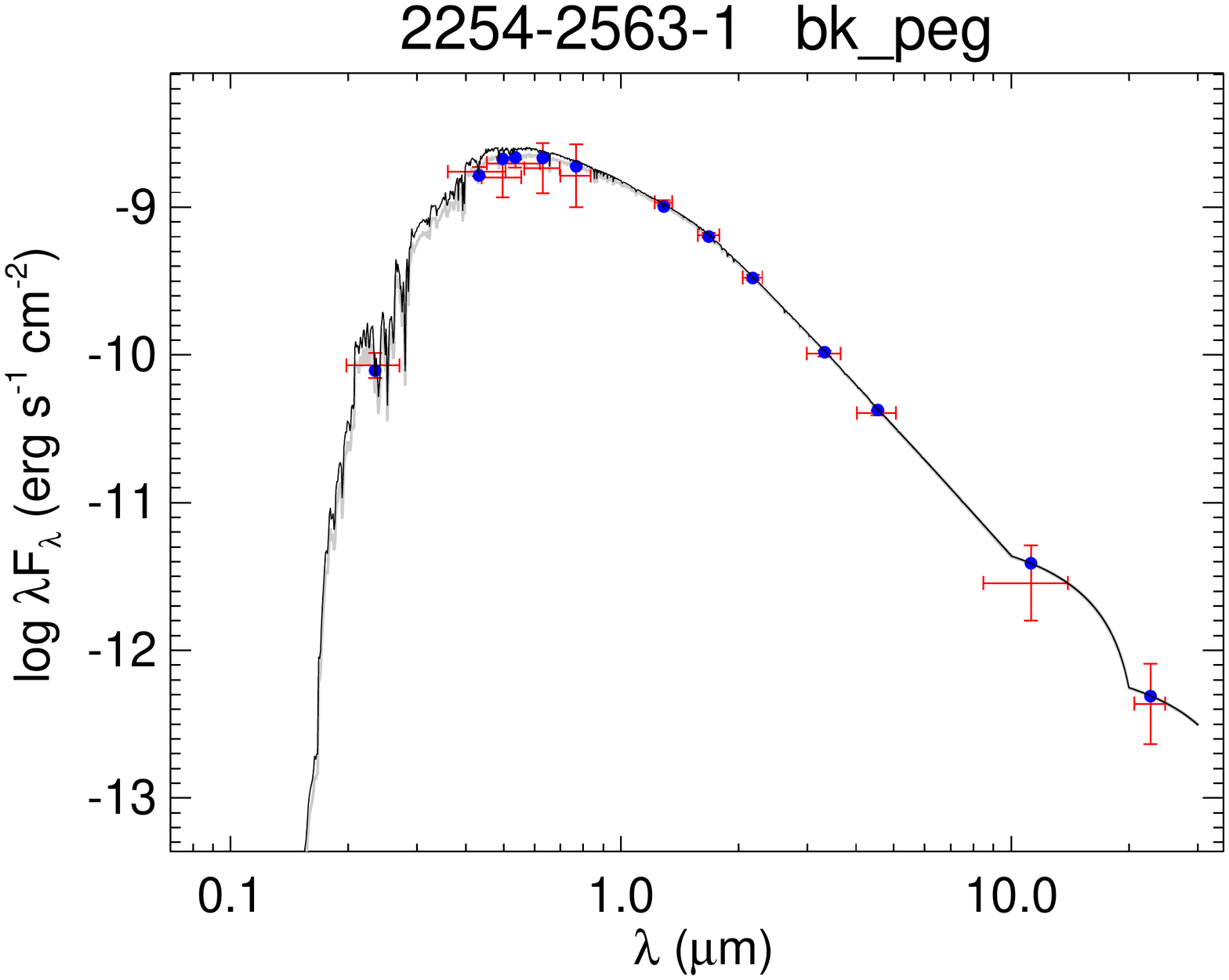}
  \includegraphics[trim=60 60 60 60,clip,width=0.49\linewidth]{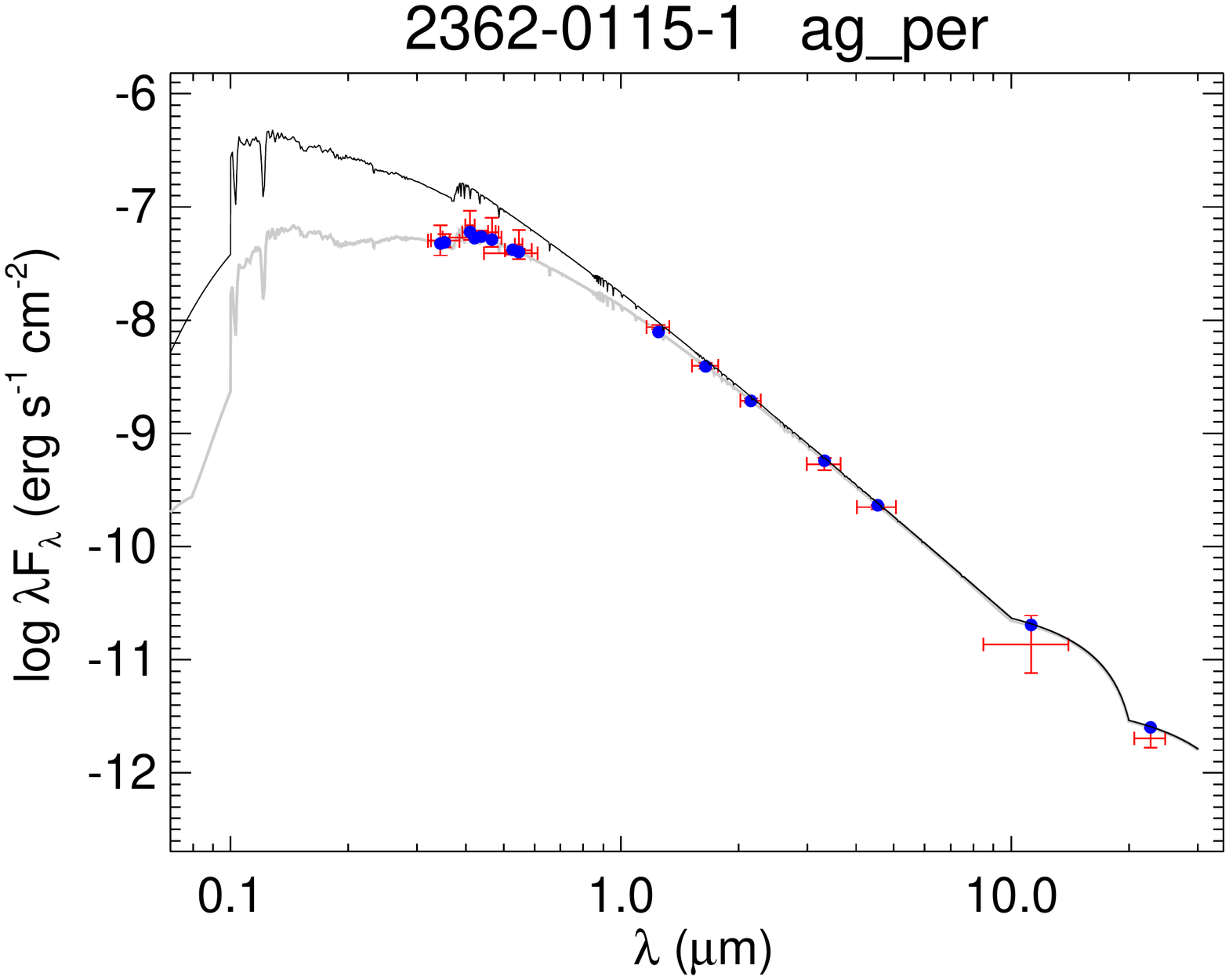}
  \includegraphics[trim=60 60 60 60,clip,width=0.49\linewidth]{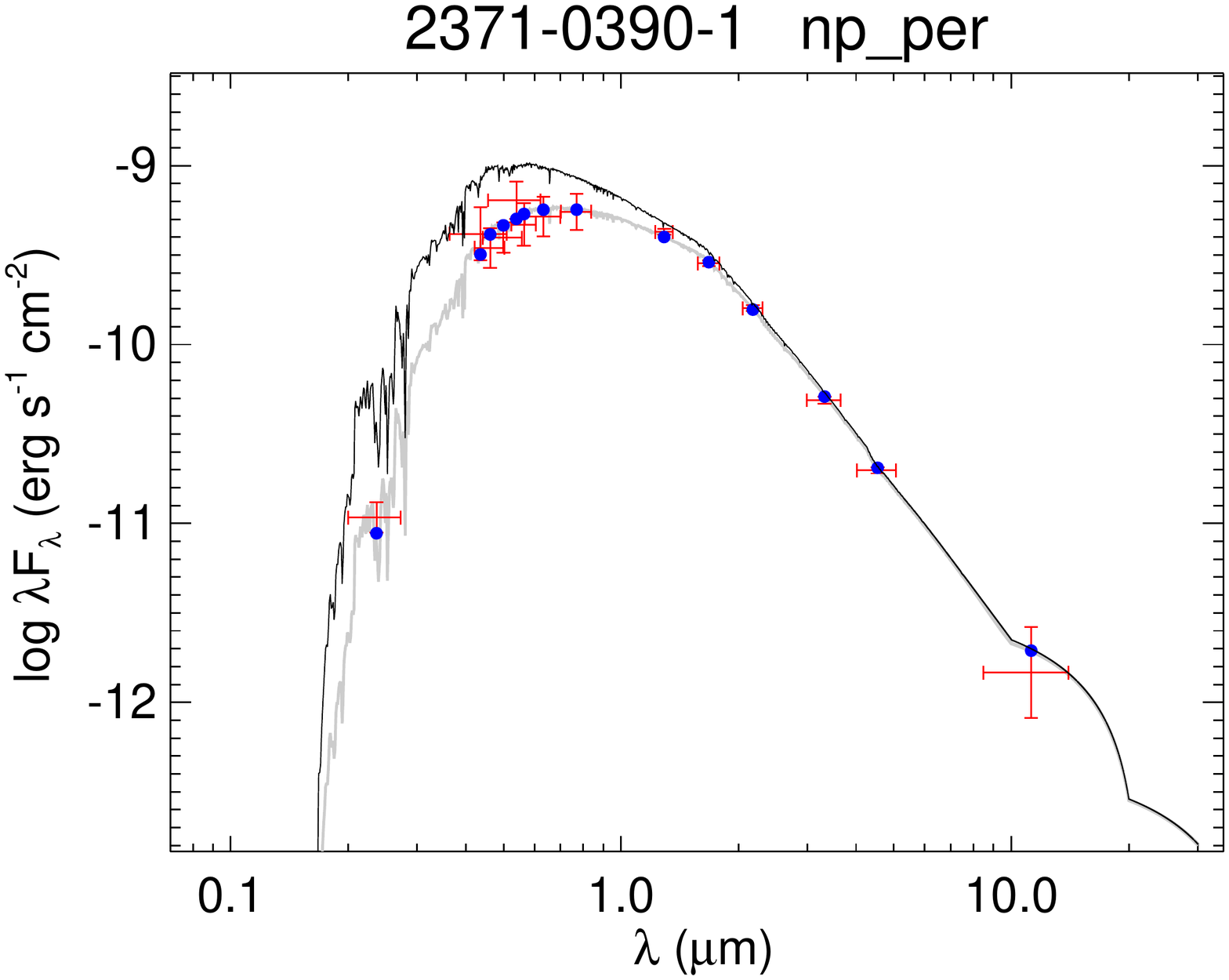}
  \includegraphics[trim=60 60 60 60,clip,width=0.49\linewidth]{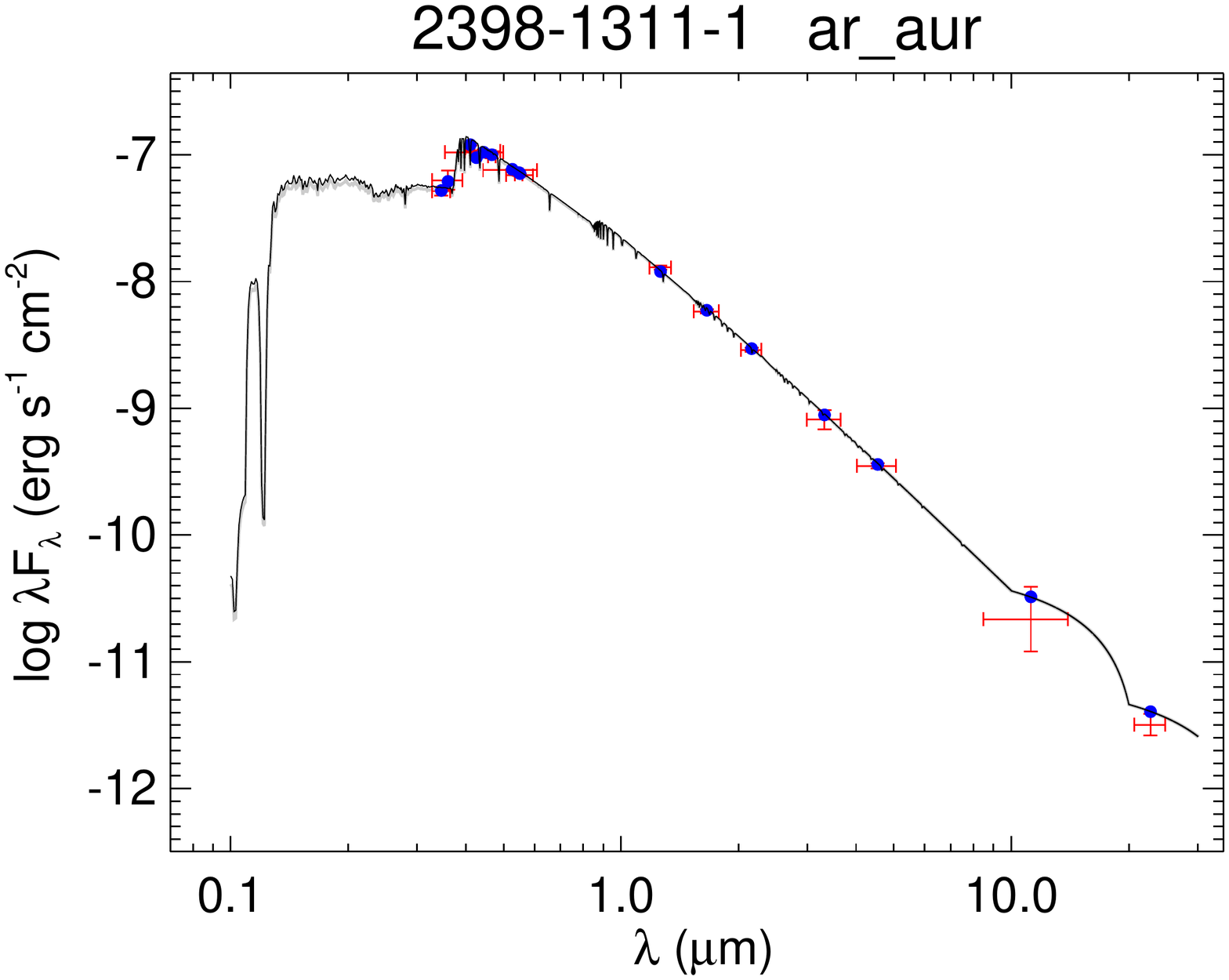}
  \caption{All labels, lines, symbols, and colors as in Figure \ref{fig:seds}.}
  \label{fig:seds_6}
\end{figure}

\begin{figure}[H]
  \centering
  \includegraphics[trim=60 60 60 60,clip,width=0.49\linewidth]{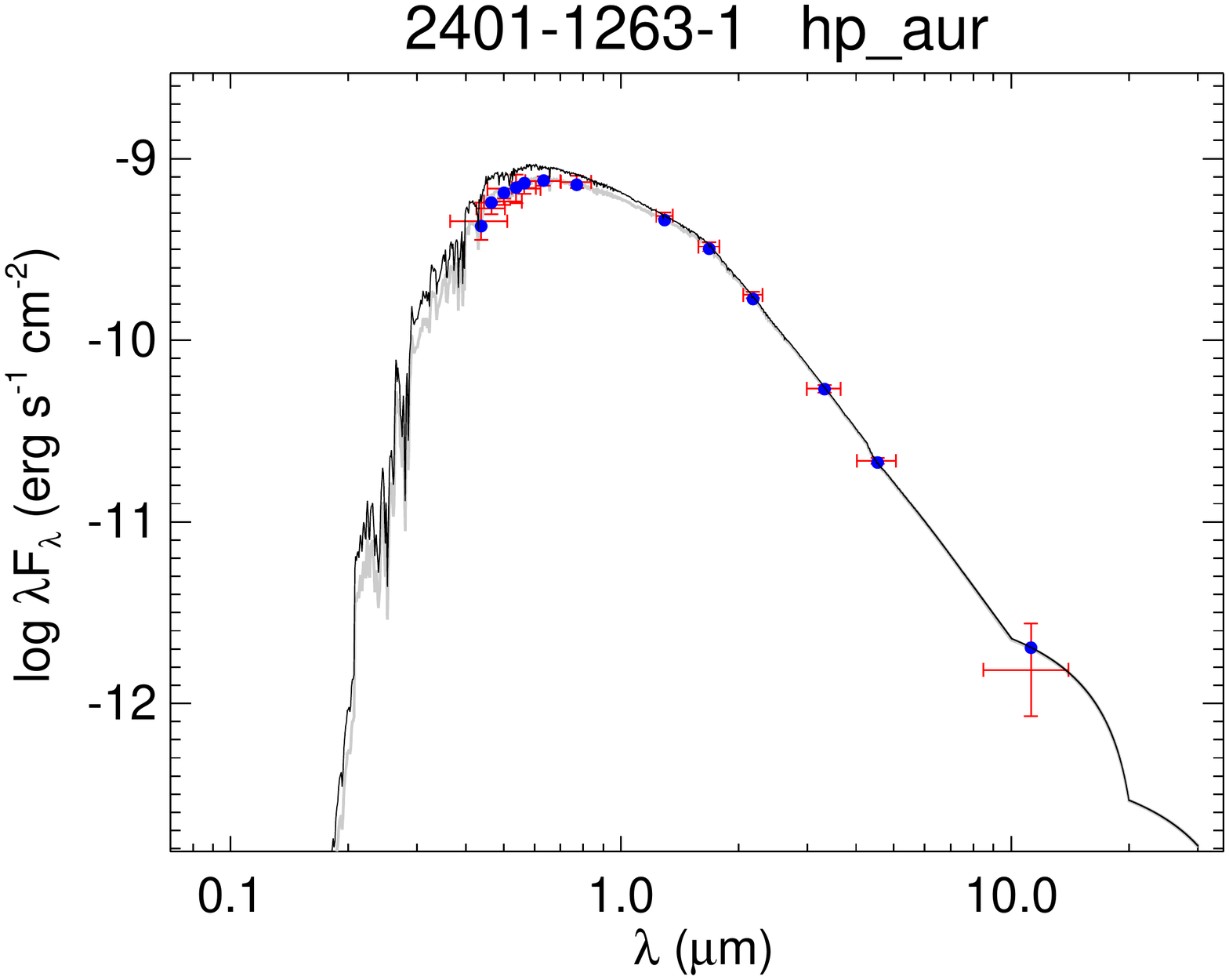}
  \includegraphics[trim=60 60 60 60,clip,width=0.49\linewidth]{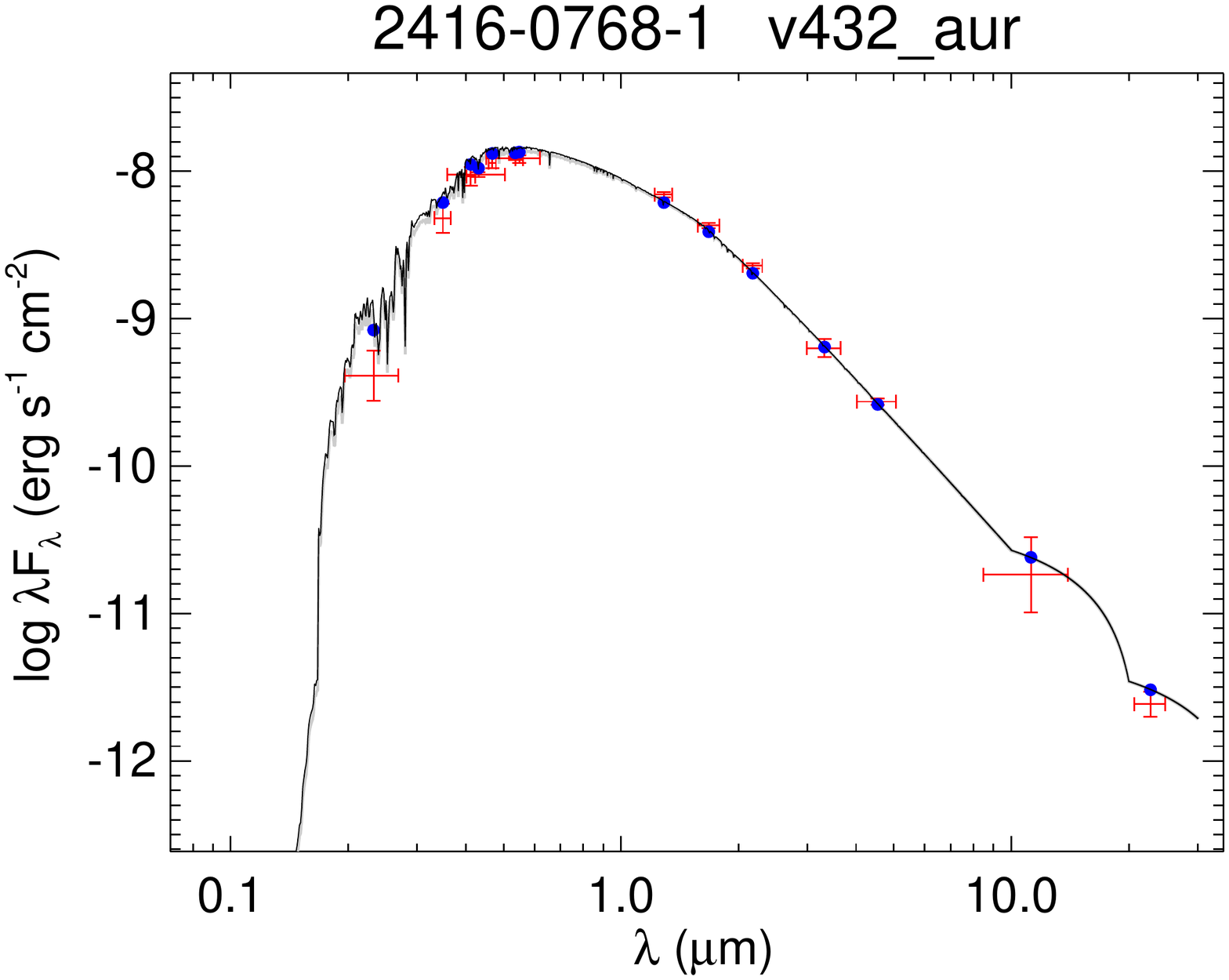}
  \includegraphics[trim=60 60 60 60,clip,width=0.49\linewidth]{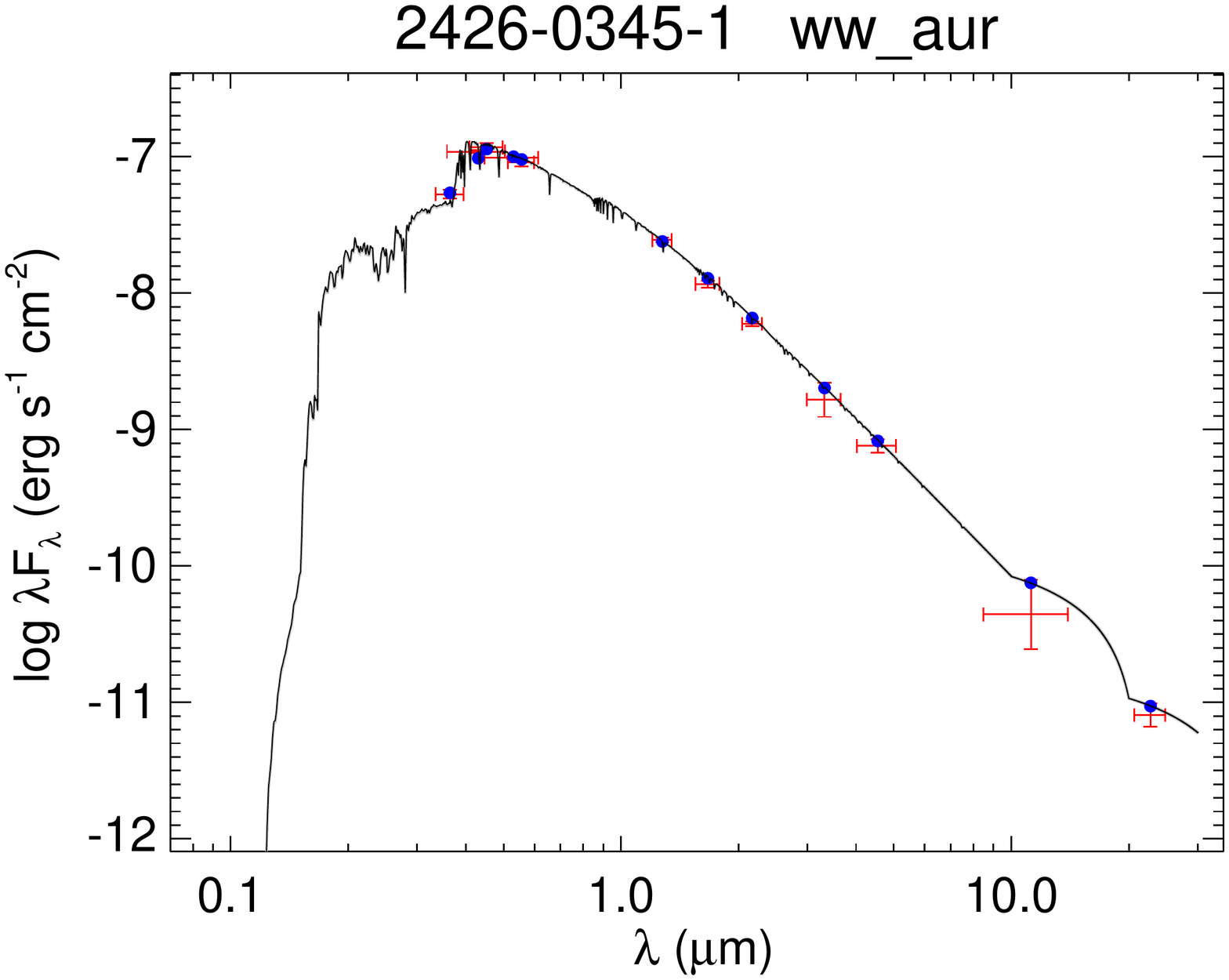}
  \includegraphics[trim=60 60 60 60,clip,width=0.49\linewidth]{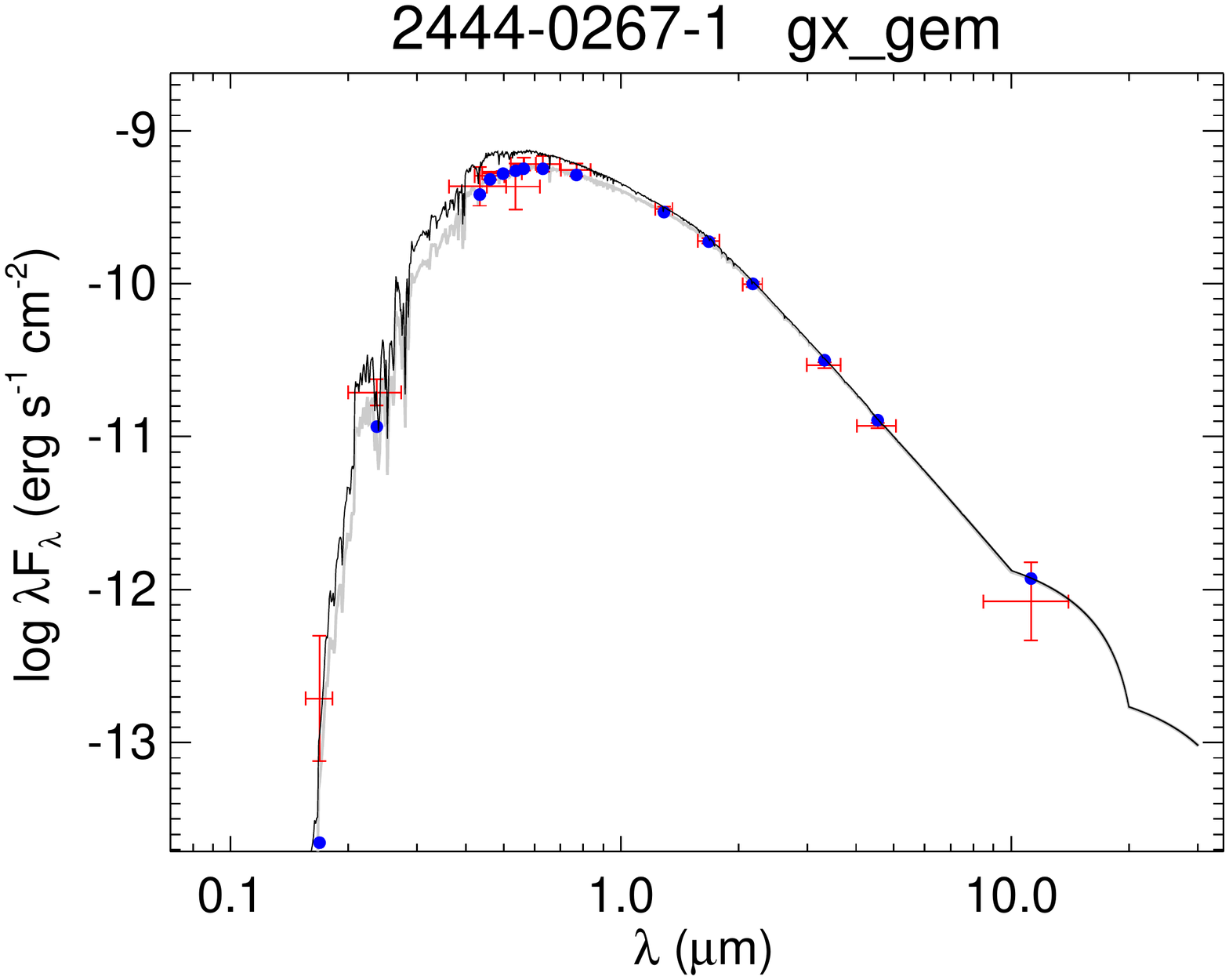}
  \includegraphics[trim=60 60 60 60,clip,width=0.49\linewidth]{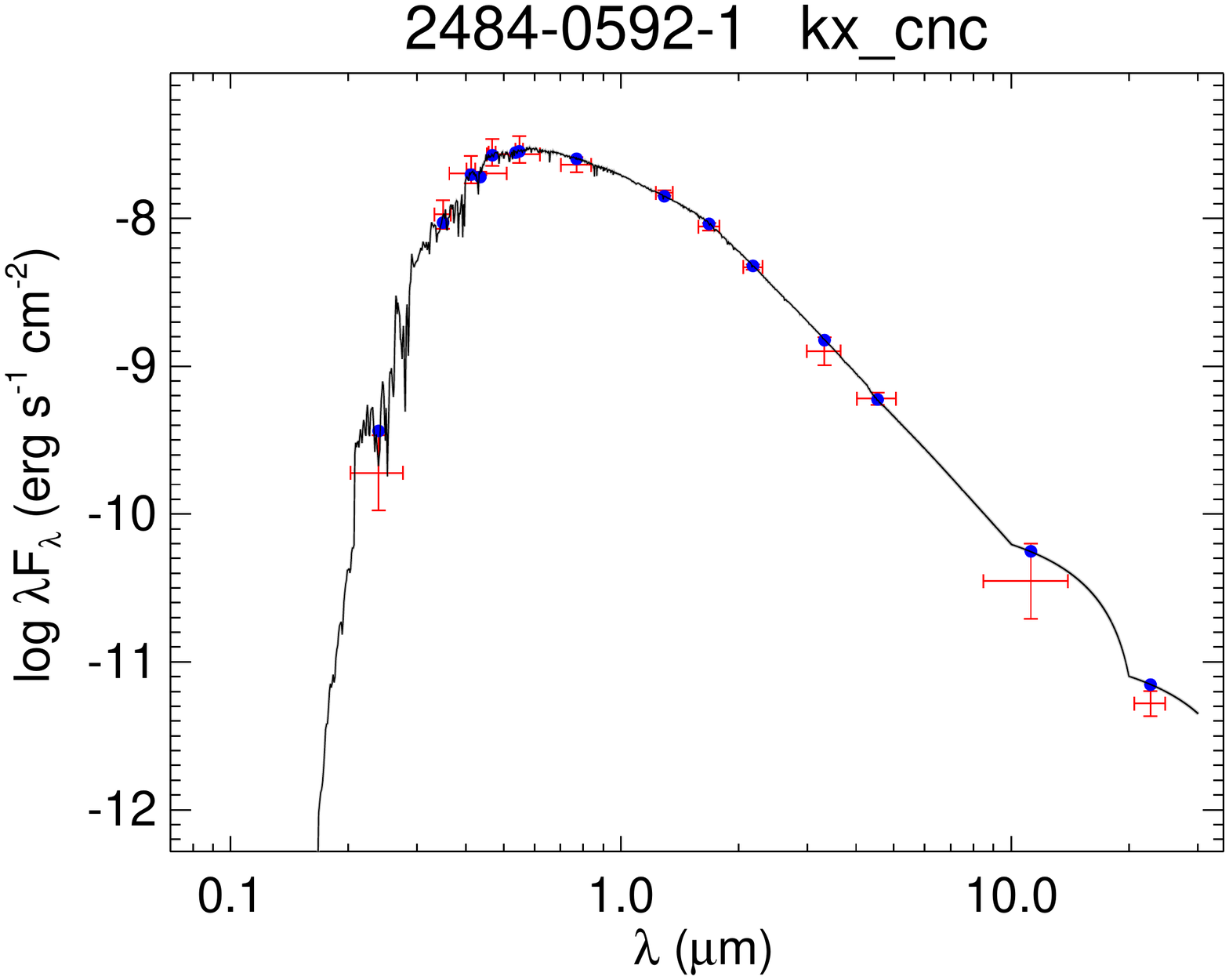}
  \includegraphics[trim=60 60 60 60,clip,width=0.49\linewidth]{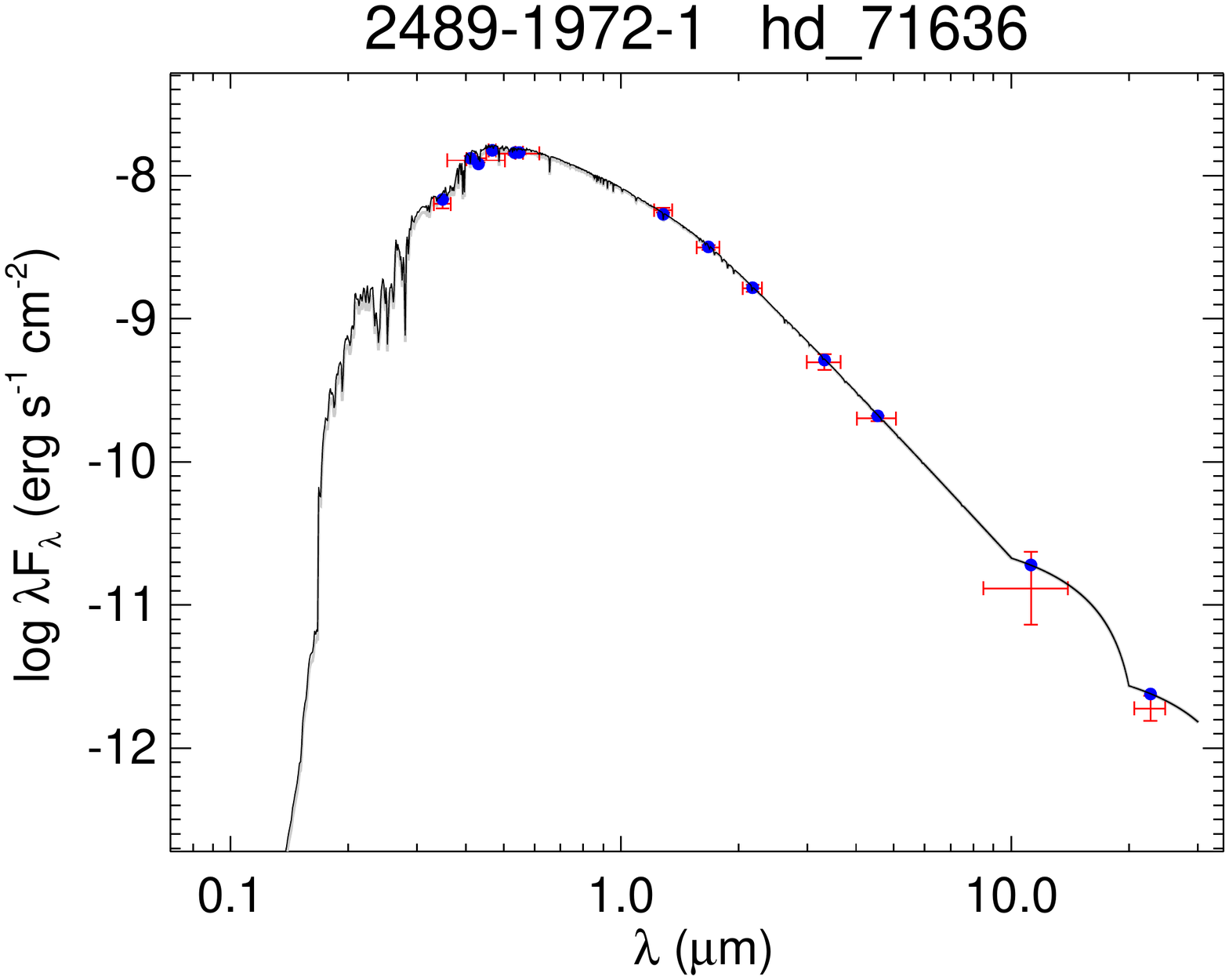}
  \caption{All labels, lines, symbols, and colors as in Figure \ref{fig:seds}.}
  \label{fig:seds_7}
\end{figure}

\begin{figure}[H]
  \centering
  \includegraphics[trim=60 60 60 60,clip,width=0.49\linewidth]{sedfigs/yy_gem.pdf}
  \includegraphics[trim=60 60 60 60,clip,width=0.49\linewidth]{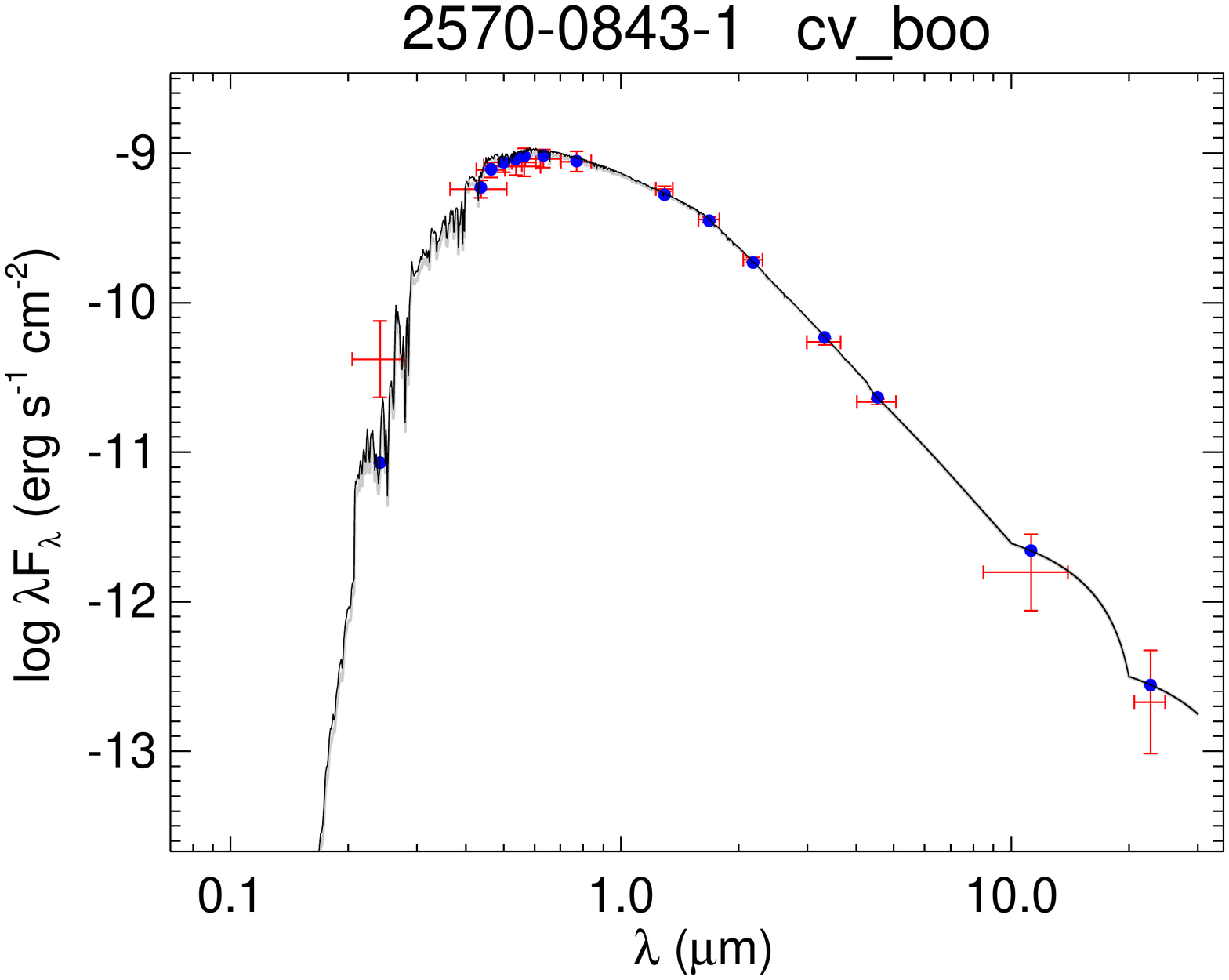}
  \includegraphics[trim=60 60 60 60,clip,width=0.49\linewidth]{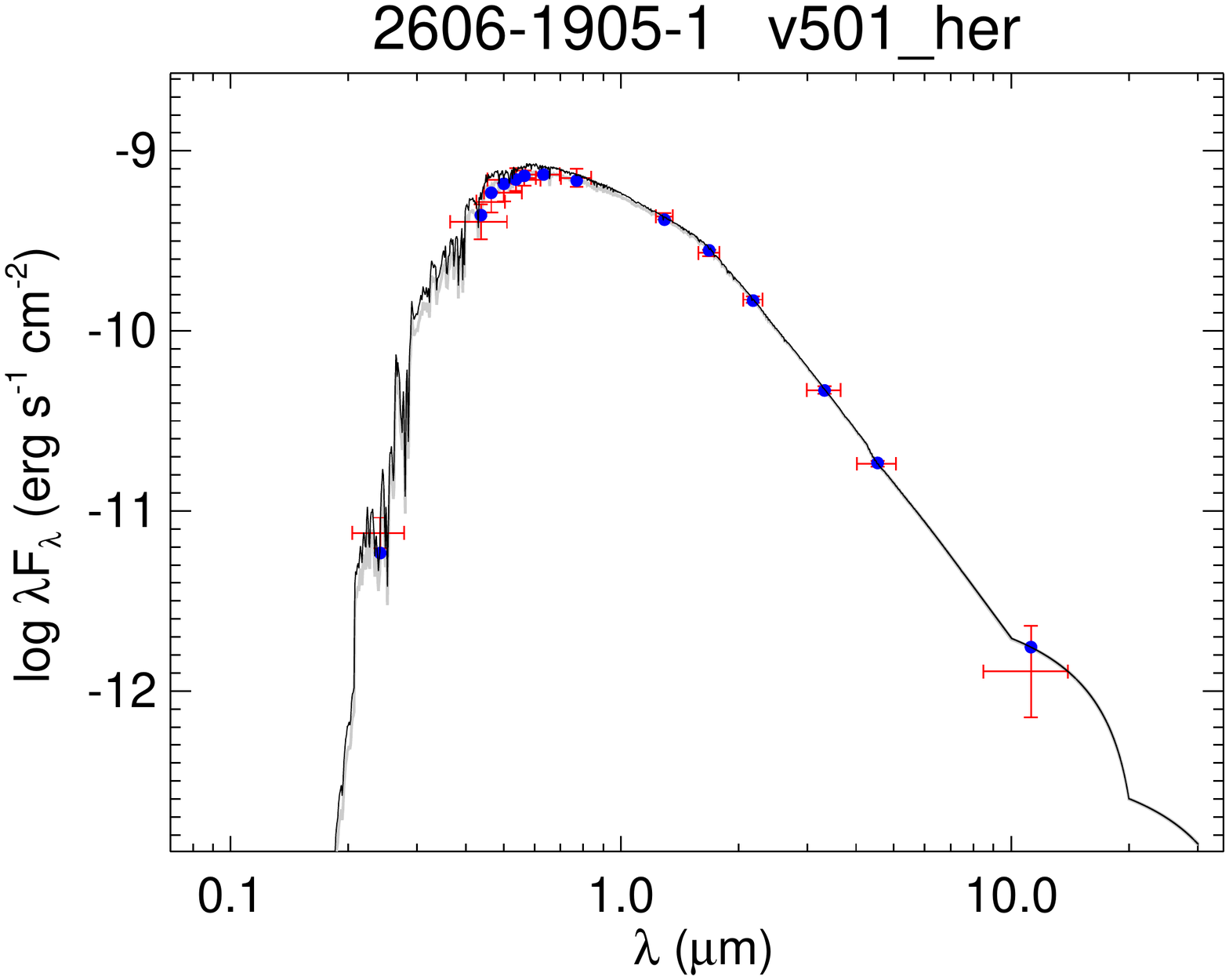}
  \includegraphics[trim=60 60 60 60,clip,width=0.49\linewidth]{sedfigs/bd+36_3317.pdf}
  \includegraphics[trim=60 60 60 60,clip,width=0.49\linewidth]{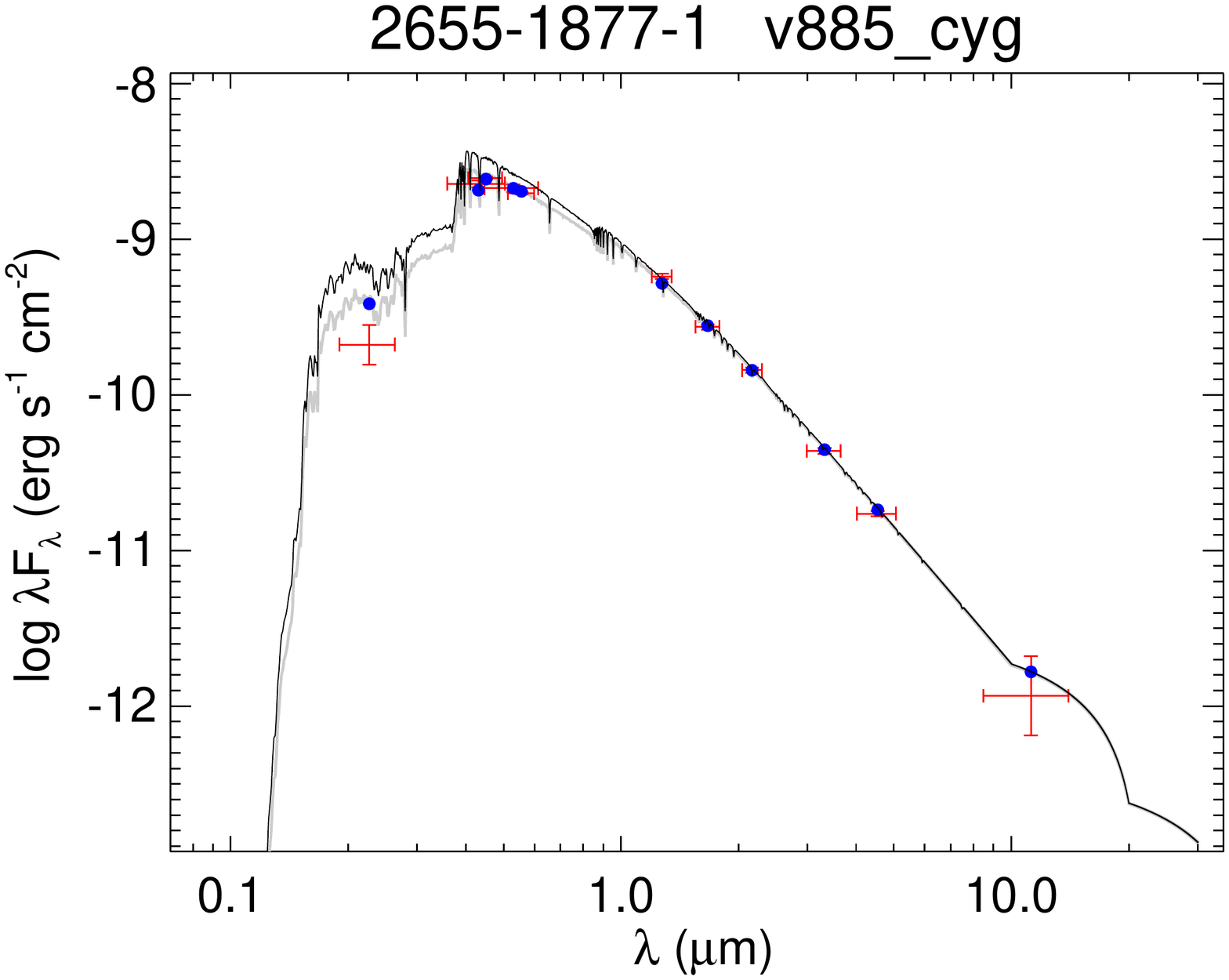}
  \includegraphics[trim=60 60 60 60,clip,width=0.49\linewidth]{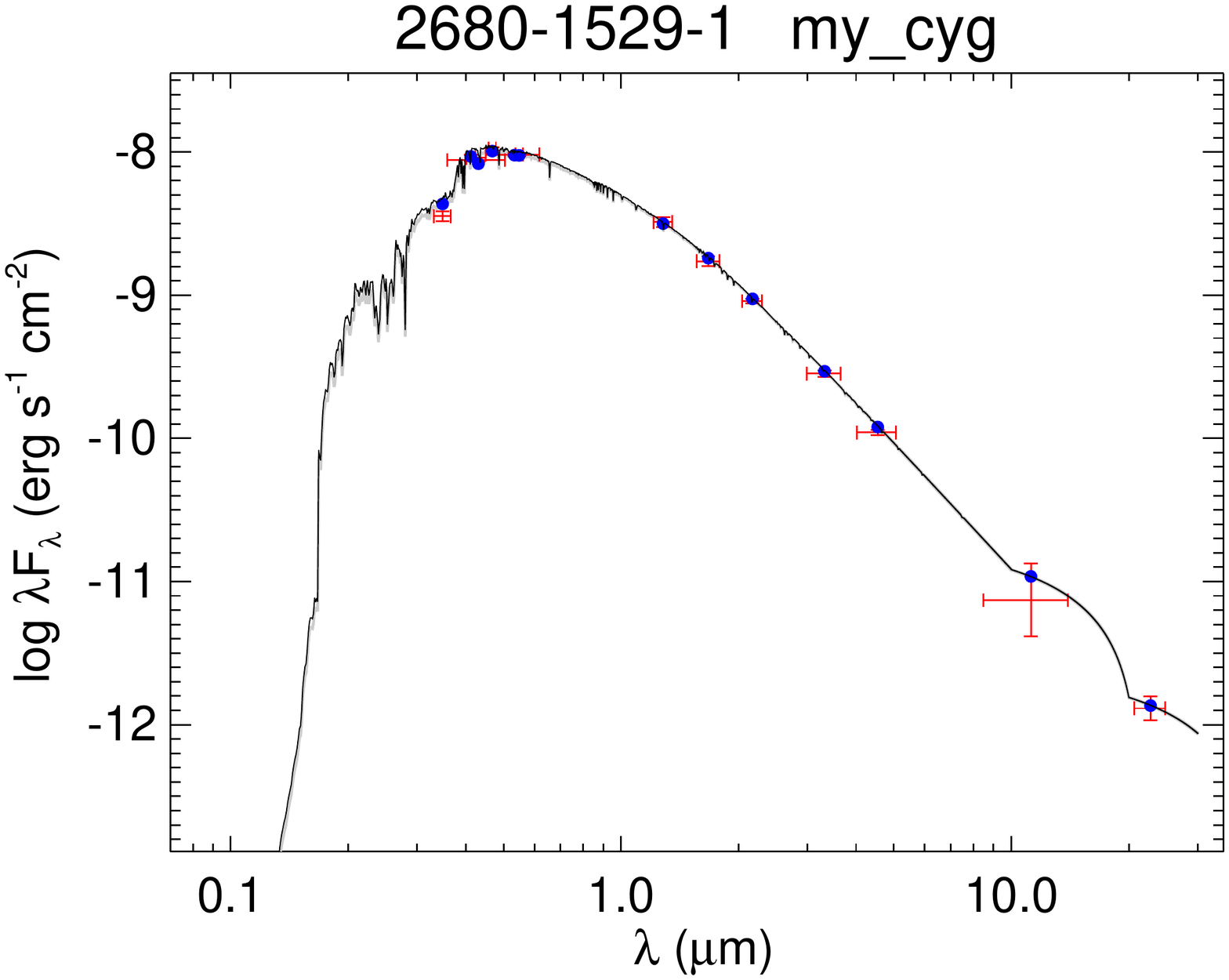}
  \caption{All labels, lines, symbols, and colors as in Figure \ref{fig:seds}.}
  \label{fig:seds_8}
\end{figure}

\begin{figure}[H]
  \centering
  \includegraphics[trim=60 60 60 60,clip,width=0.49\linewidth]{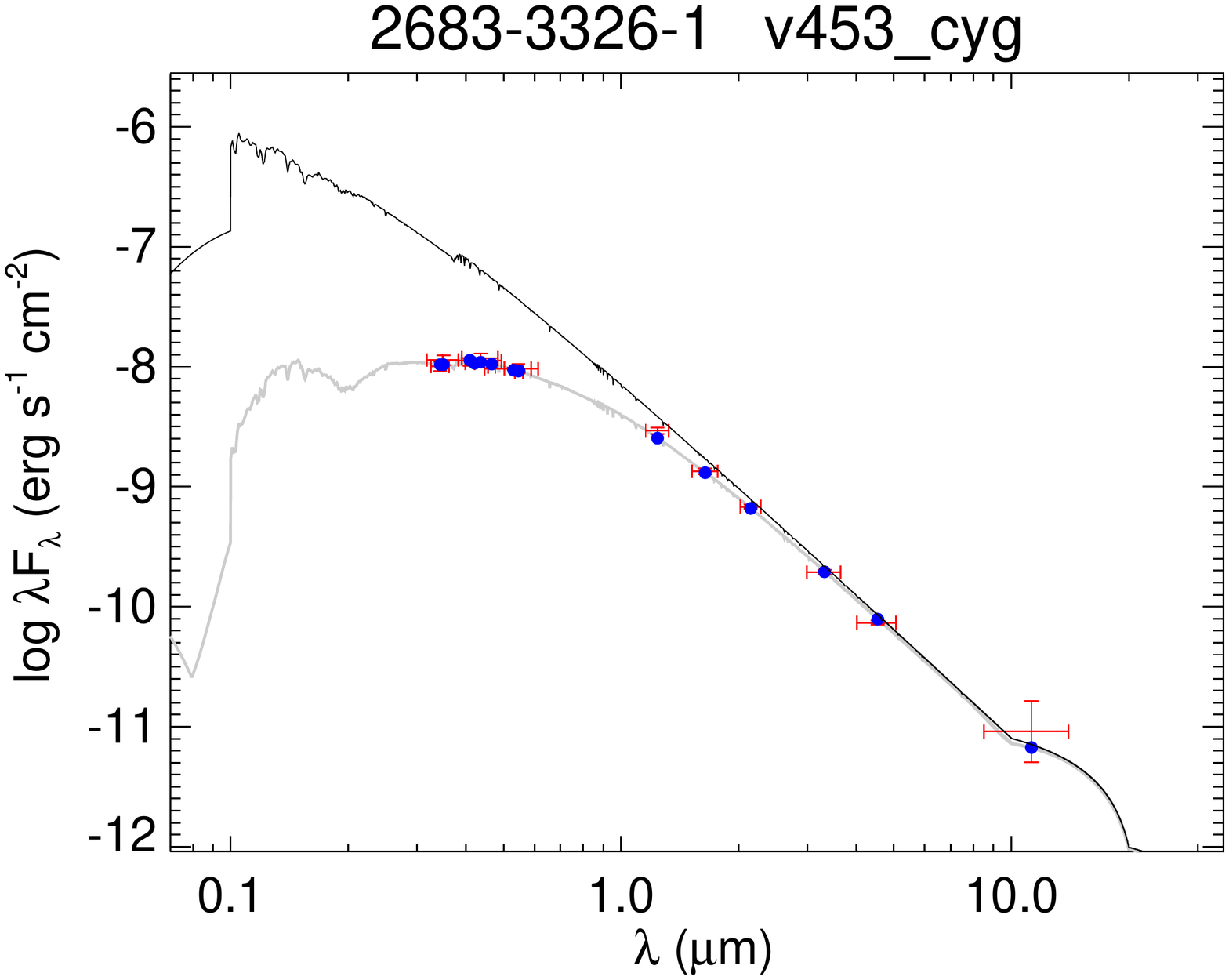}
  \includegraphics[trim=60 60 60 60,clip,width=0.49\linewidth]{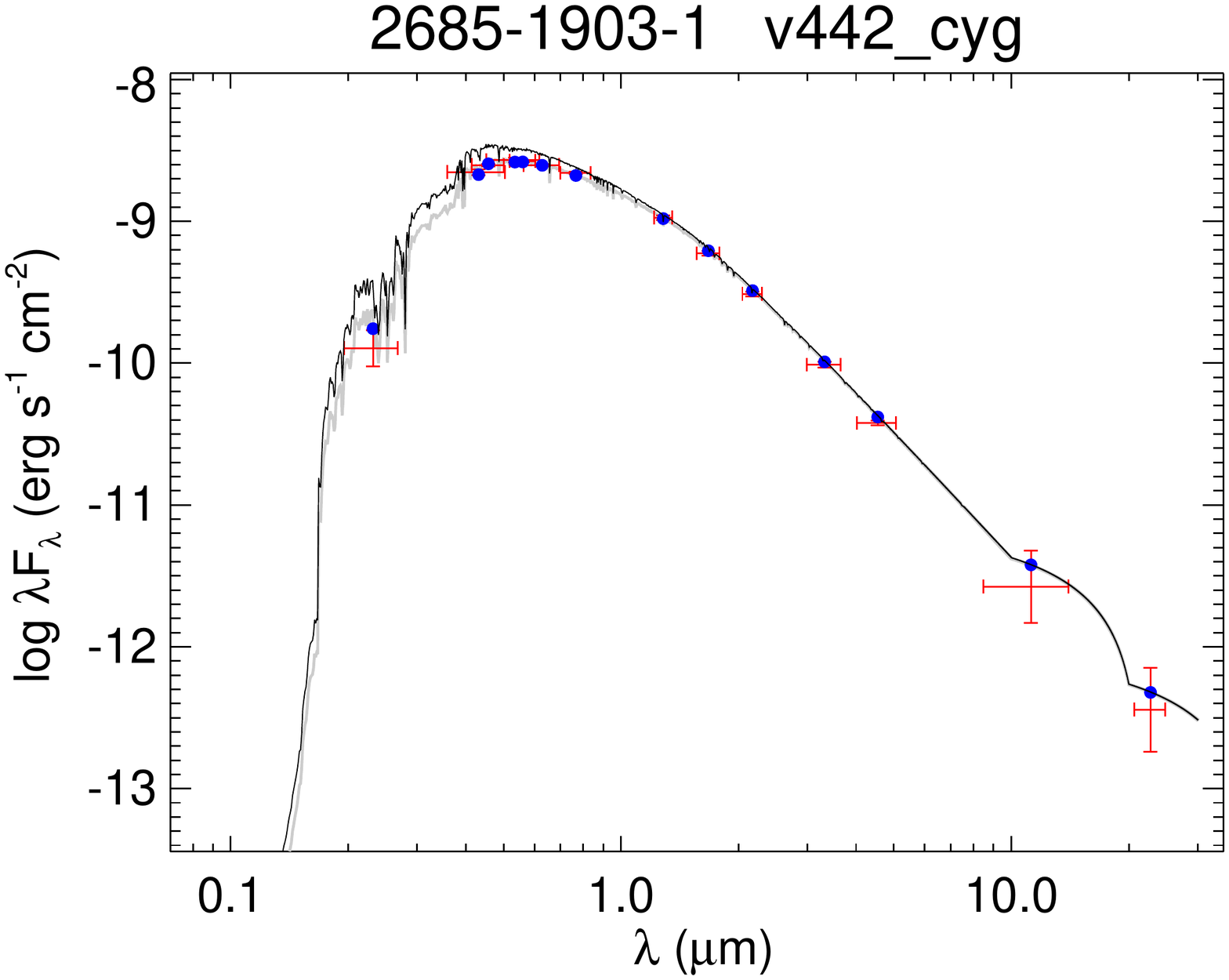}
  \includegraphics[trim=60 60 60 60,clip,width=0.49\linewidth]{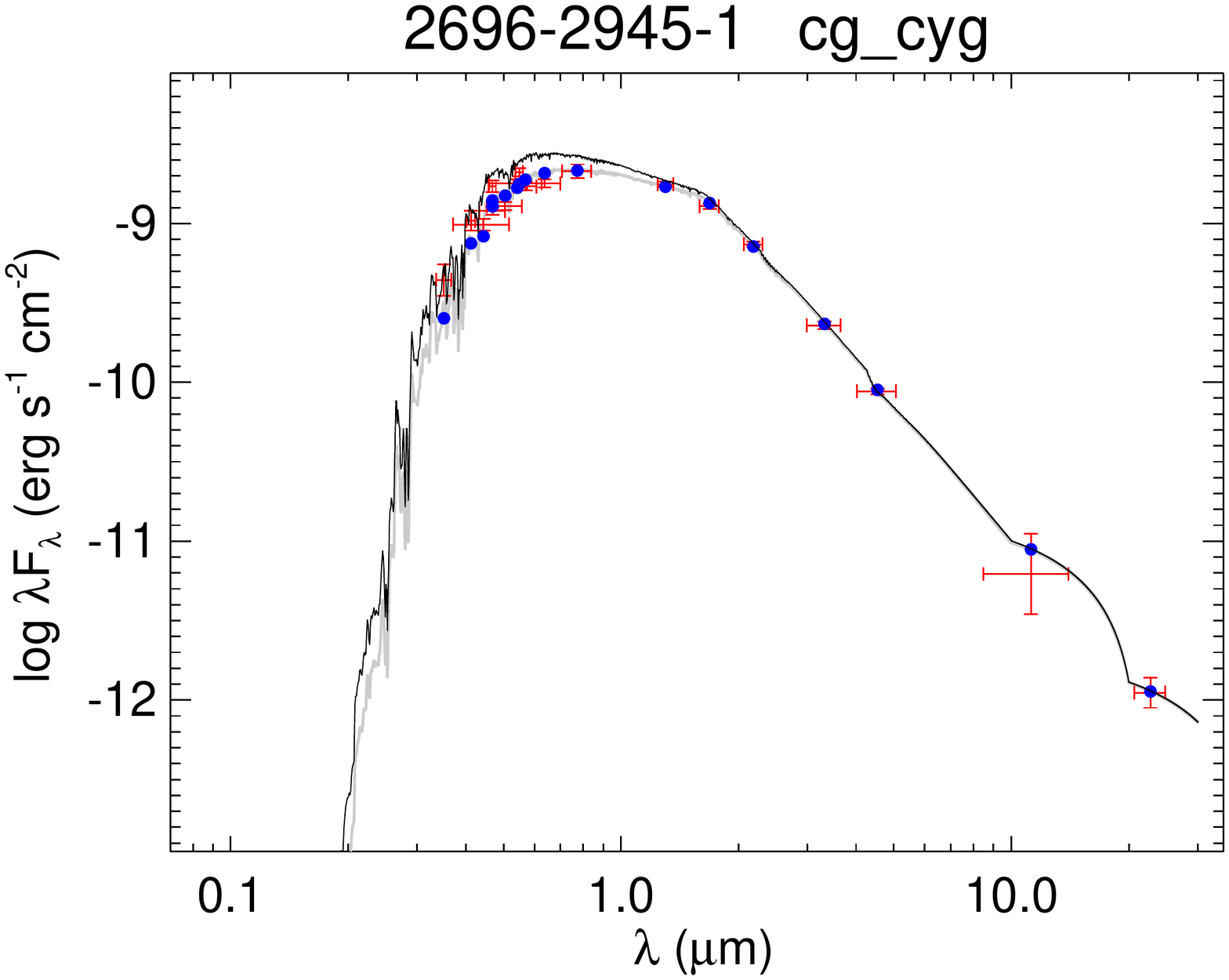}
  \includegraphics[trim=60 60 60 60,clip,width=0.49\linewidth]{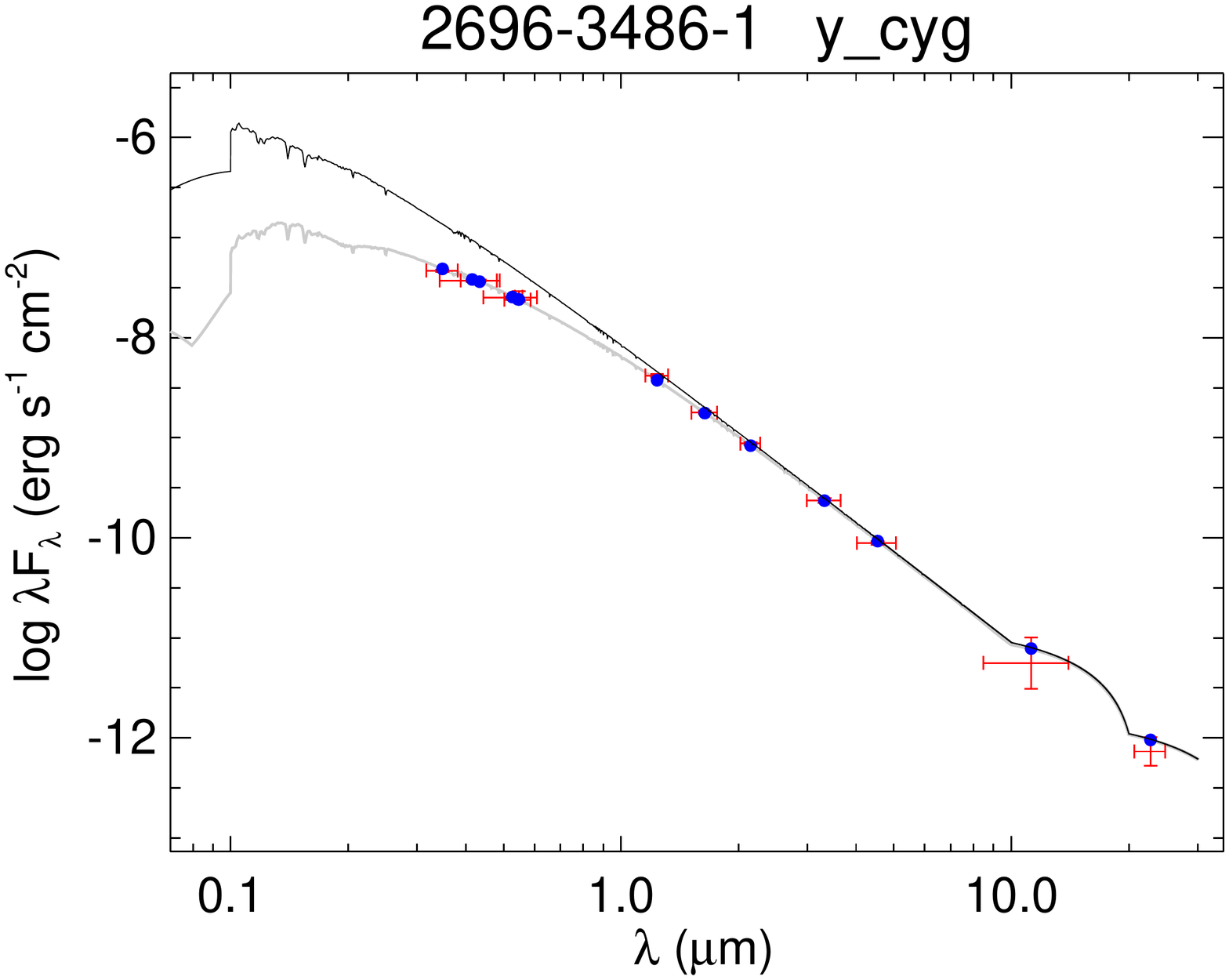}
  \includegraphics[trim=60 60 60 60,clip,width=0.49\linewidth]{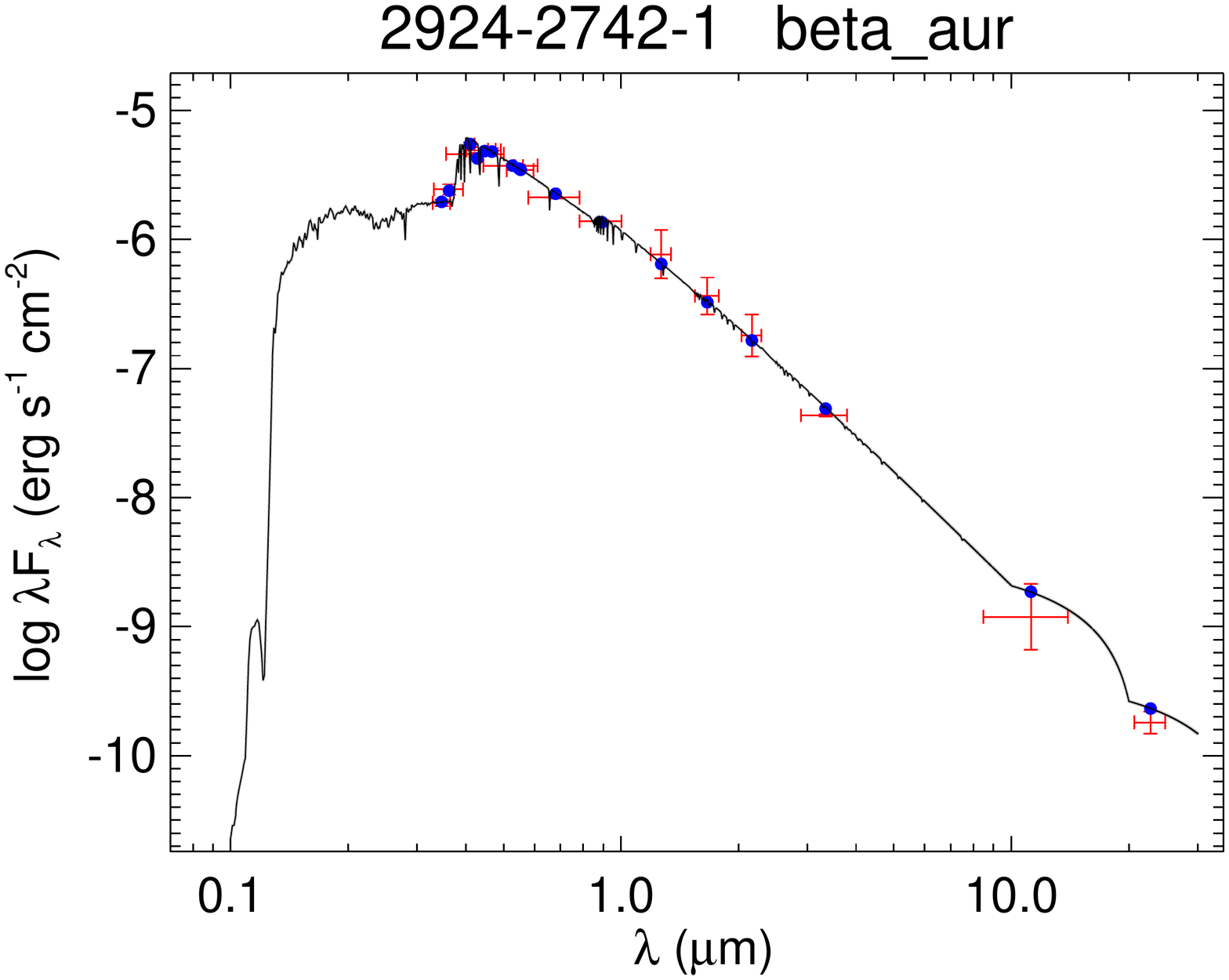}
  \includegraphics[trim=60 60 60 60,clip,width=0.49\linewidth]{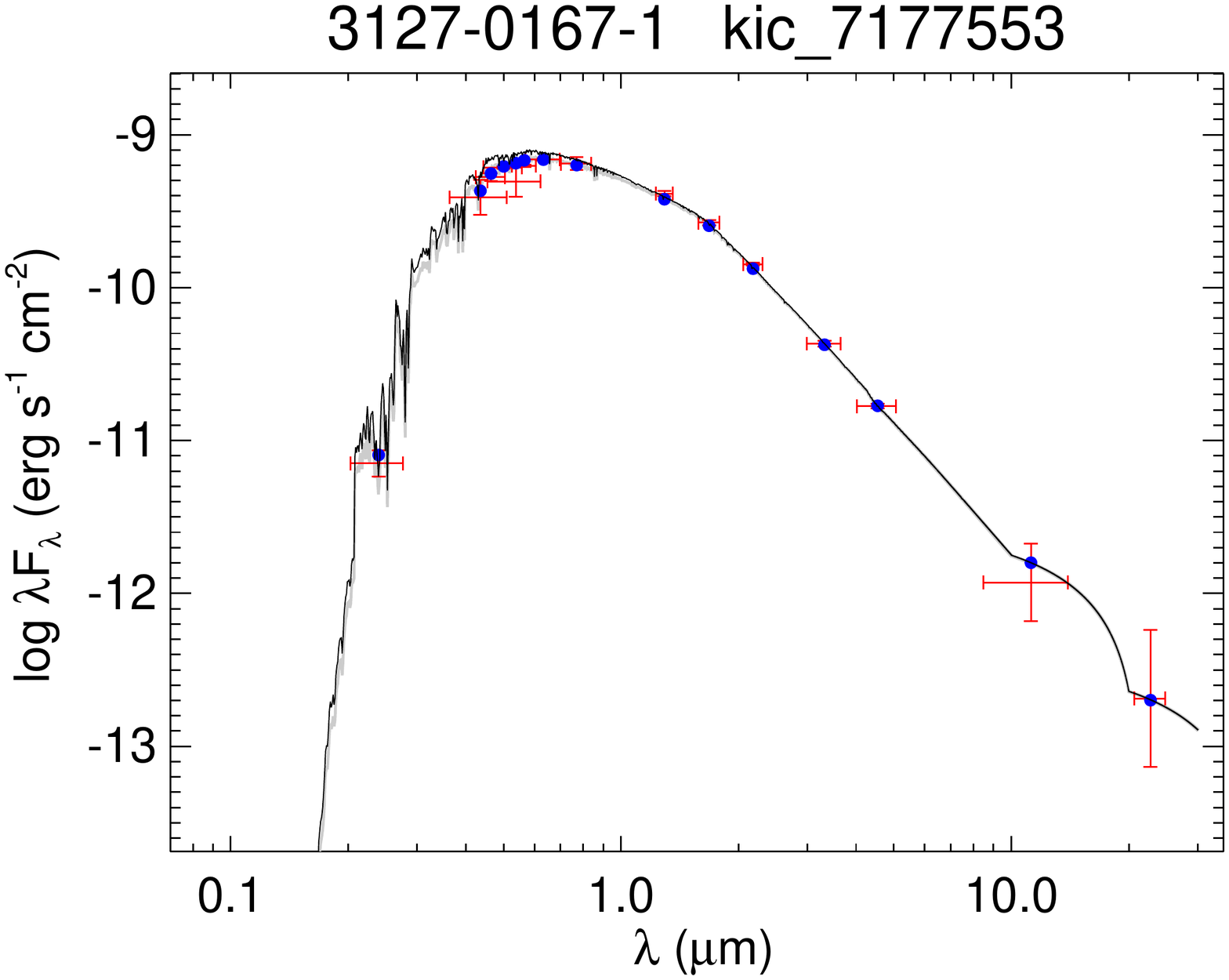}
  \caption{All labels, lines, symbols, and colors as in Figure \ref{fig:seds}.}
  \label{fig:seds_9}
\end{figure}

\begin{figure}[H]
  \centering
  \includegraphics[trim=60 60 60 60,clip,width=0.49\linewidth]{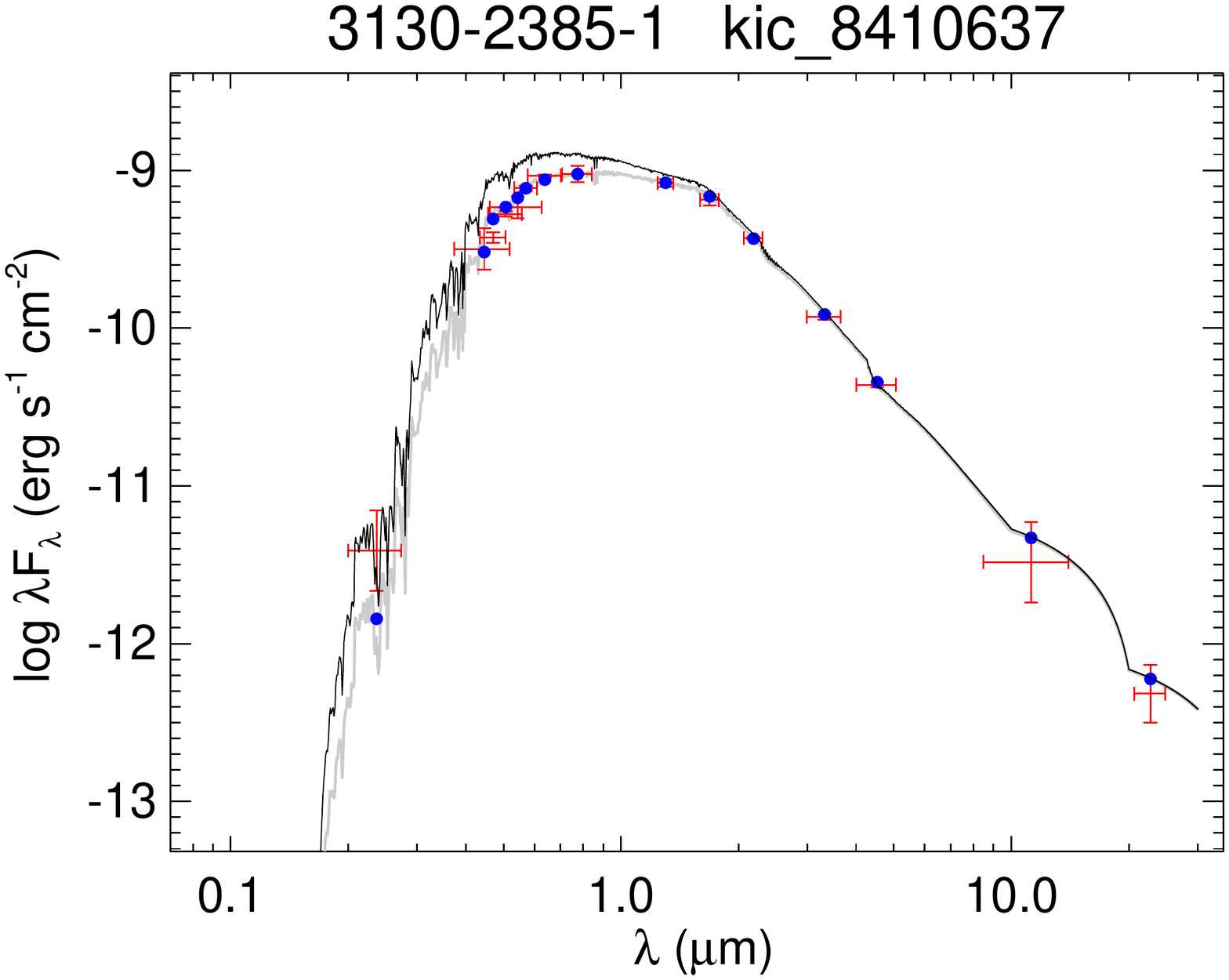}
  \includegraphics[trim=60 60 60 60,clip,width=0.49\linewidth]{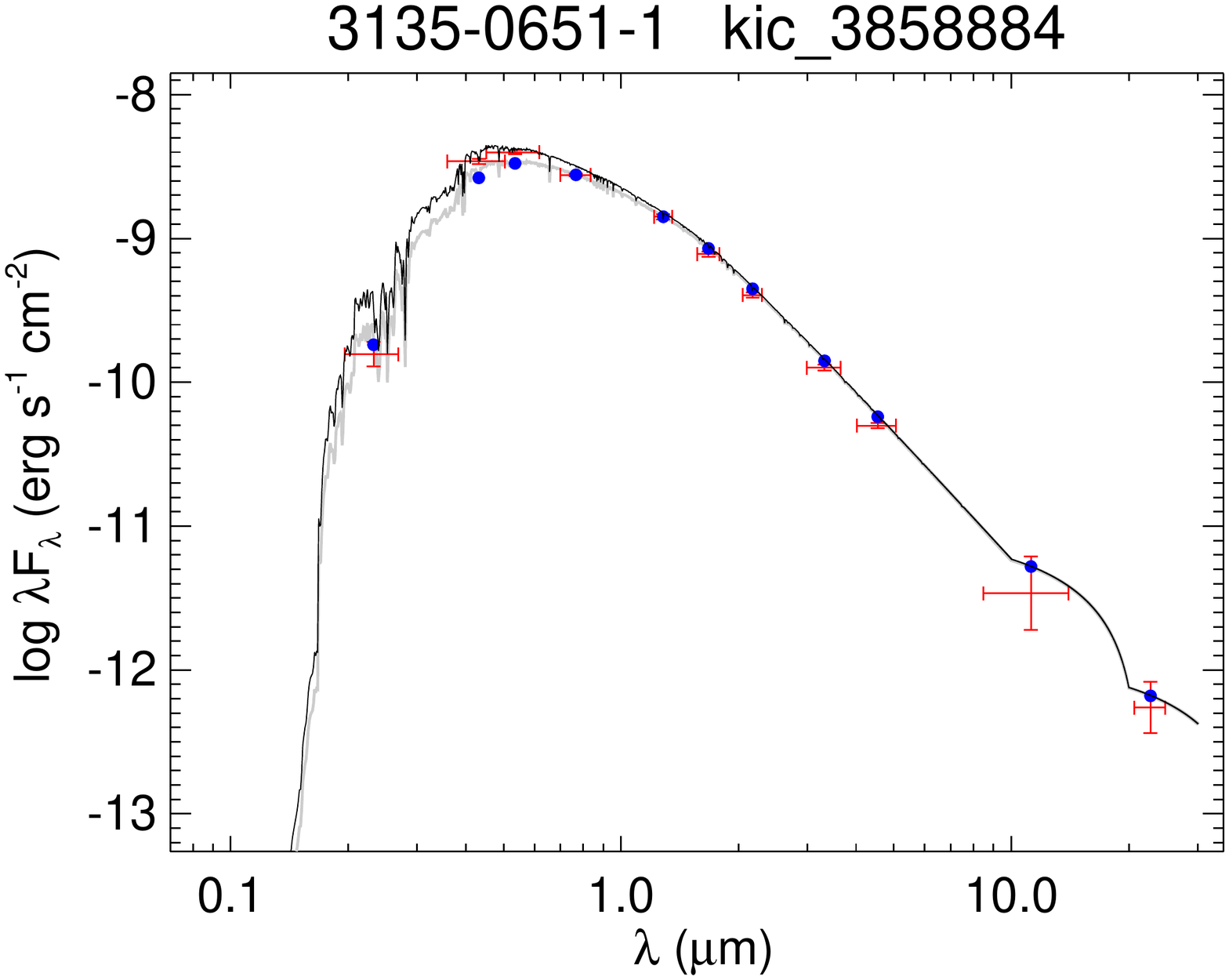}
  \includegraphics[trim=60 60 60 60,clip,width=0.49\linewidth]{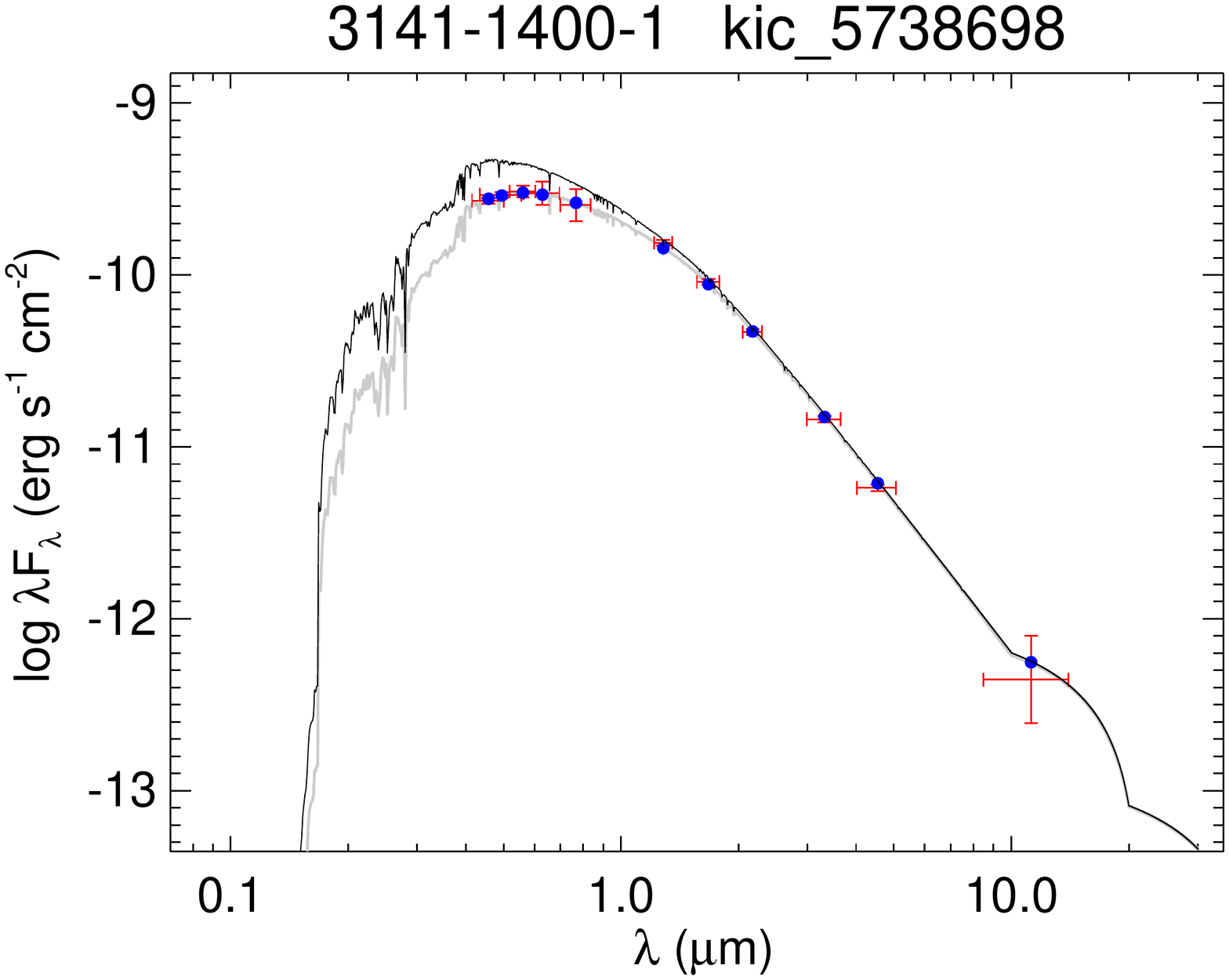}
  \includegraphics[trim=60 60 60 60,clip,width=0.49\linewidth]{sedfigs/v380_cyg.pdf}
  \includegraphics[trim=60 60 60 60,clip,width=0.49\linewidth]{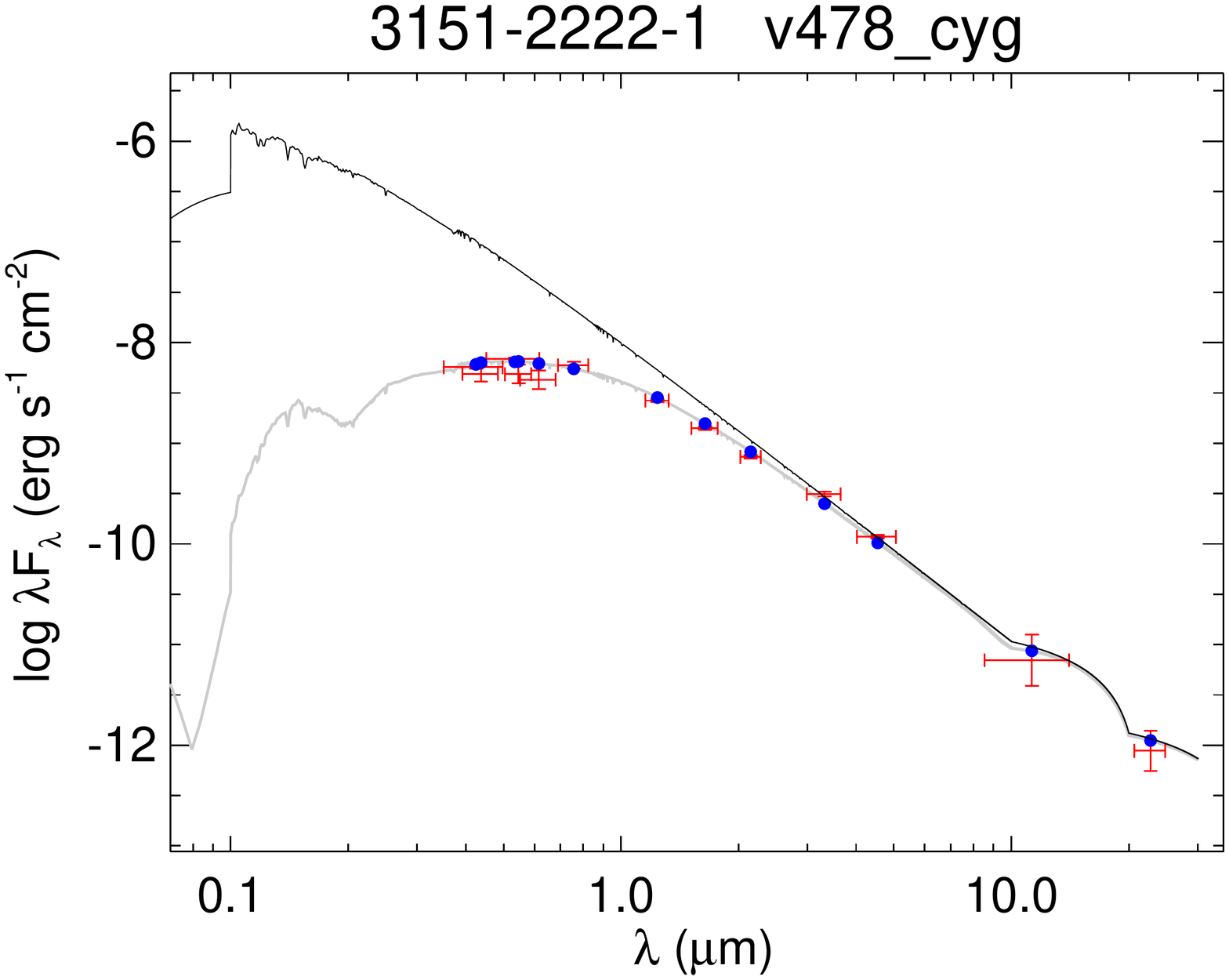}
  \includegraphics[trim=60 60 60 60,clip,width=0.49\linewidth]{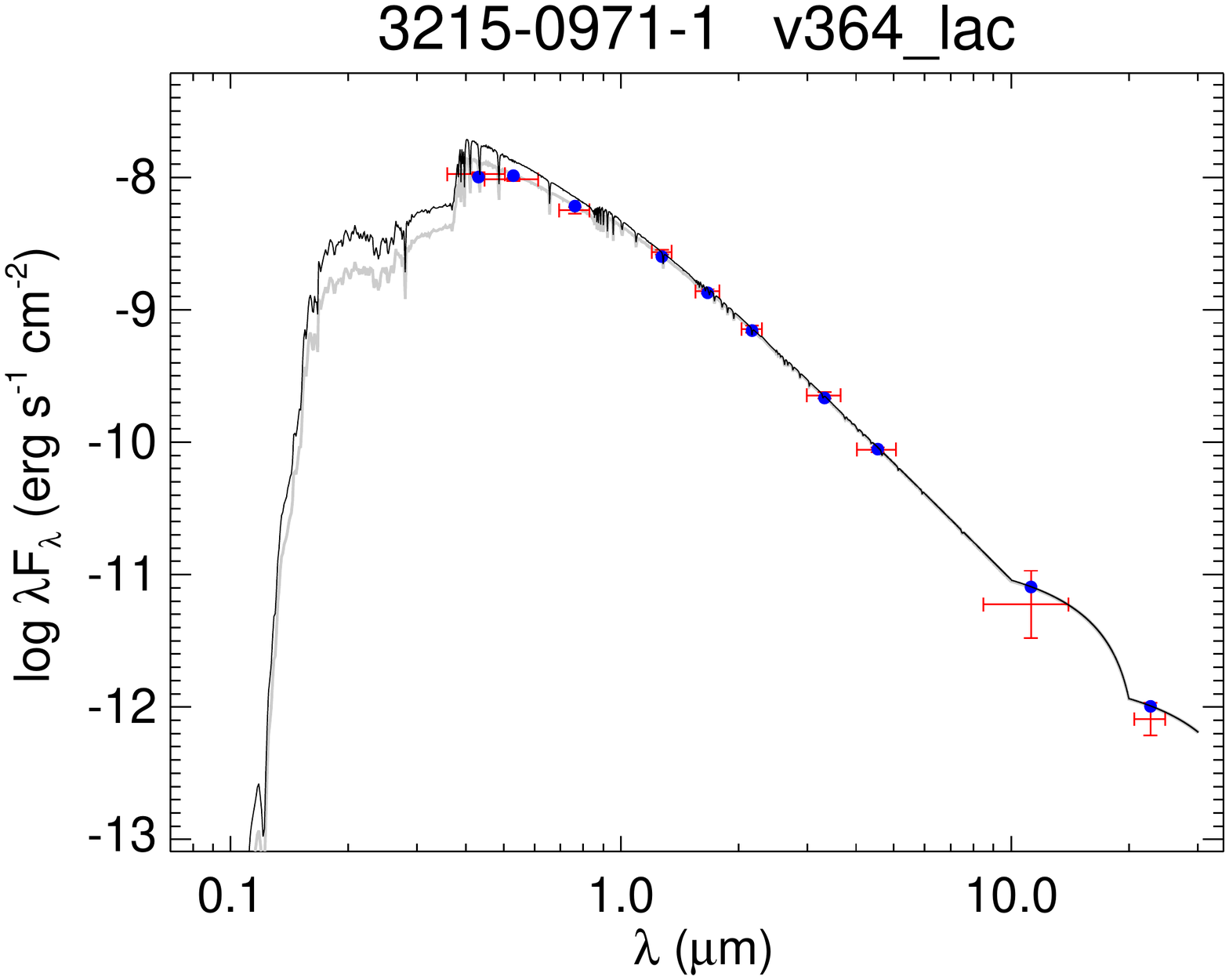}
  \caption{All labels, lines, symbols, and colors as in Figure \ref{fig:seds}.}
  \label{fig:seds_10}
\end{figure}

\begin{figure}[H]
  \centering
  \includegraphics[trim=60 60 60 60,clip,width=0.49\linewidth]{sedfigs/v342_and.pdf}
  \includegraphics[trim=60 60 60 60,clip,width=0.49\linewidth]{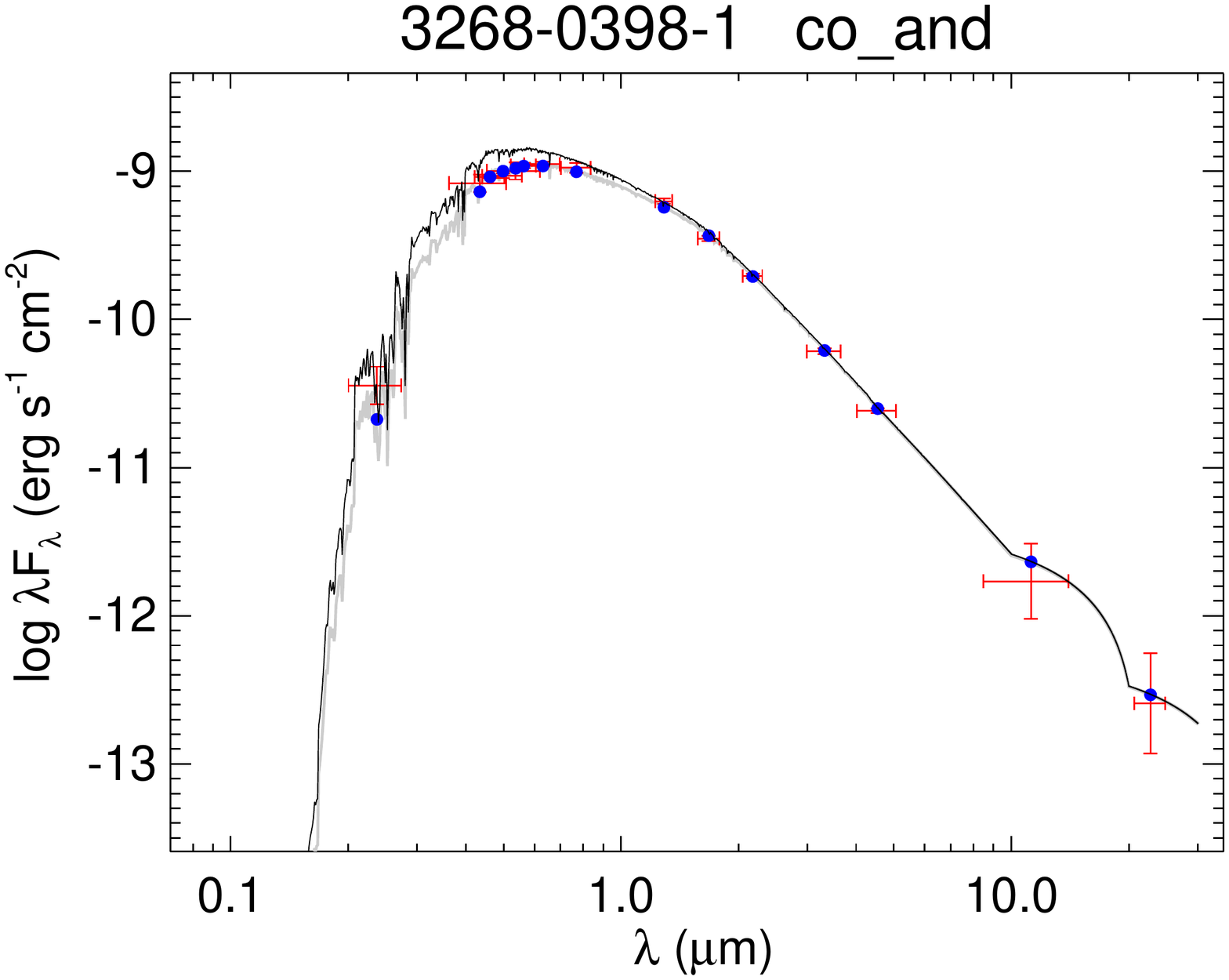}
  \includegraphics[trim=60 60 60 60,clip,width=0.49\linewidth]{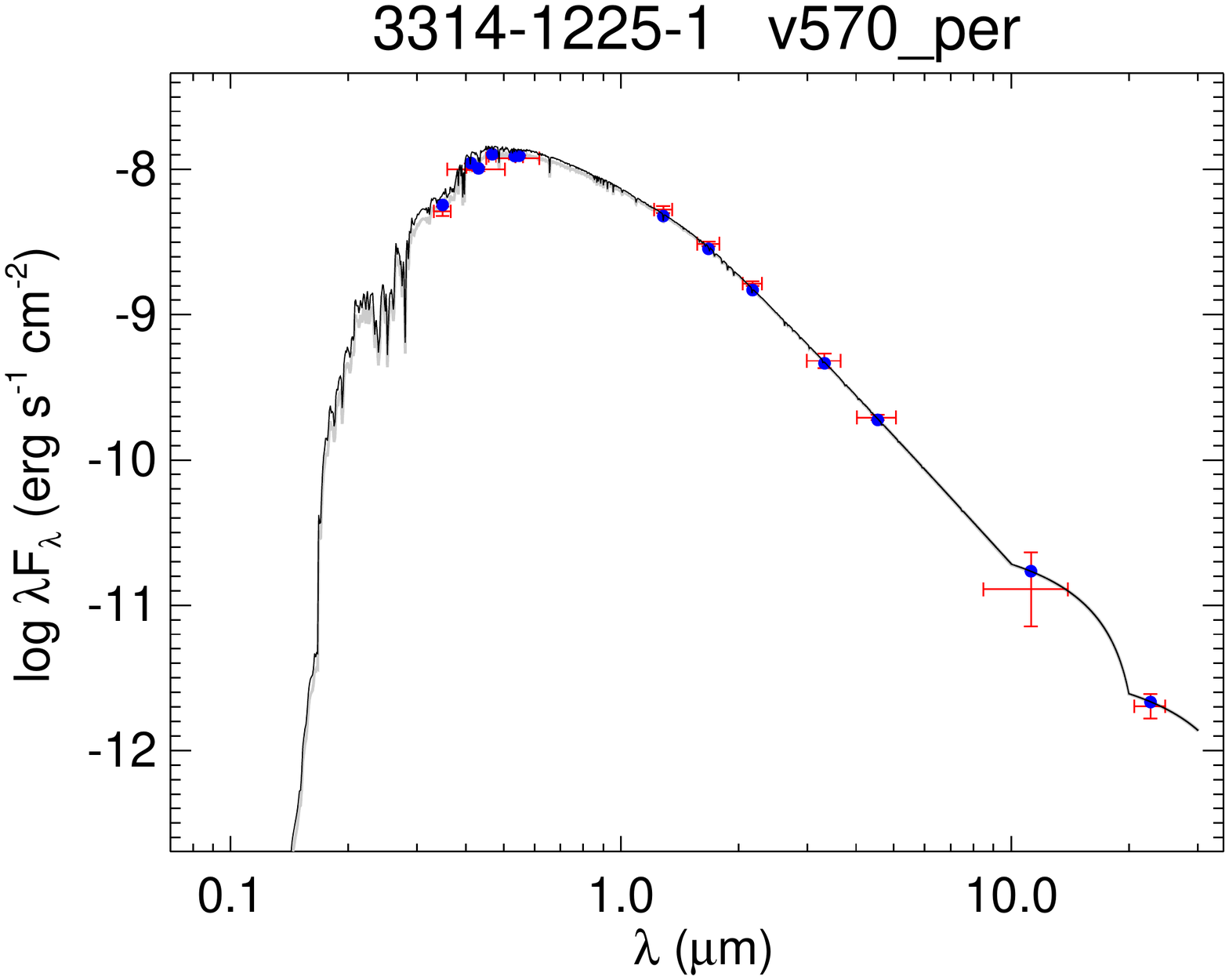}
  \includegraphics[trim=60 60 60 60,clip,width=0.49\linewidth]{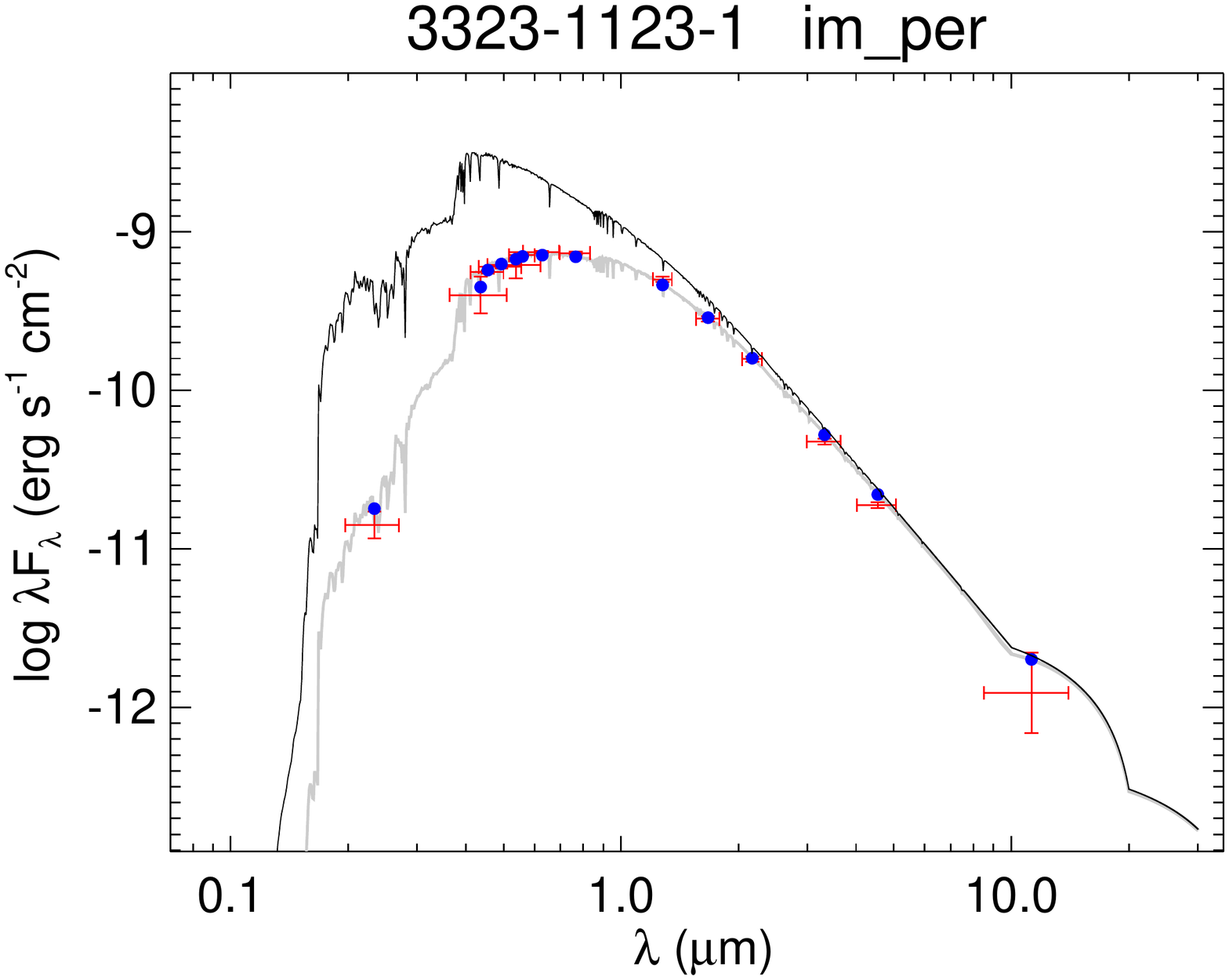}
  \includegraphics[trim=60 60 60 60,clip,width=0.49\linewidth]{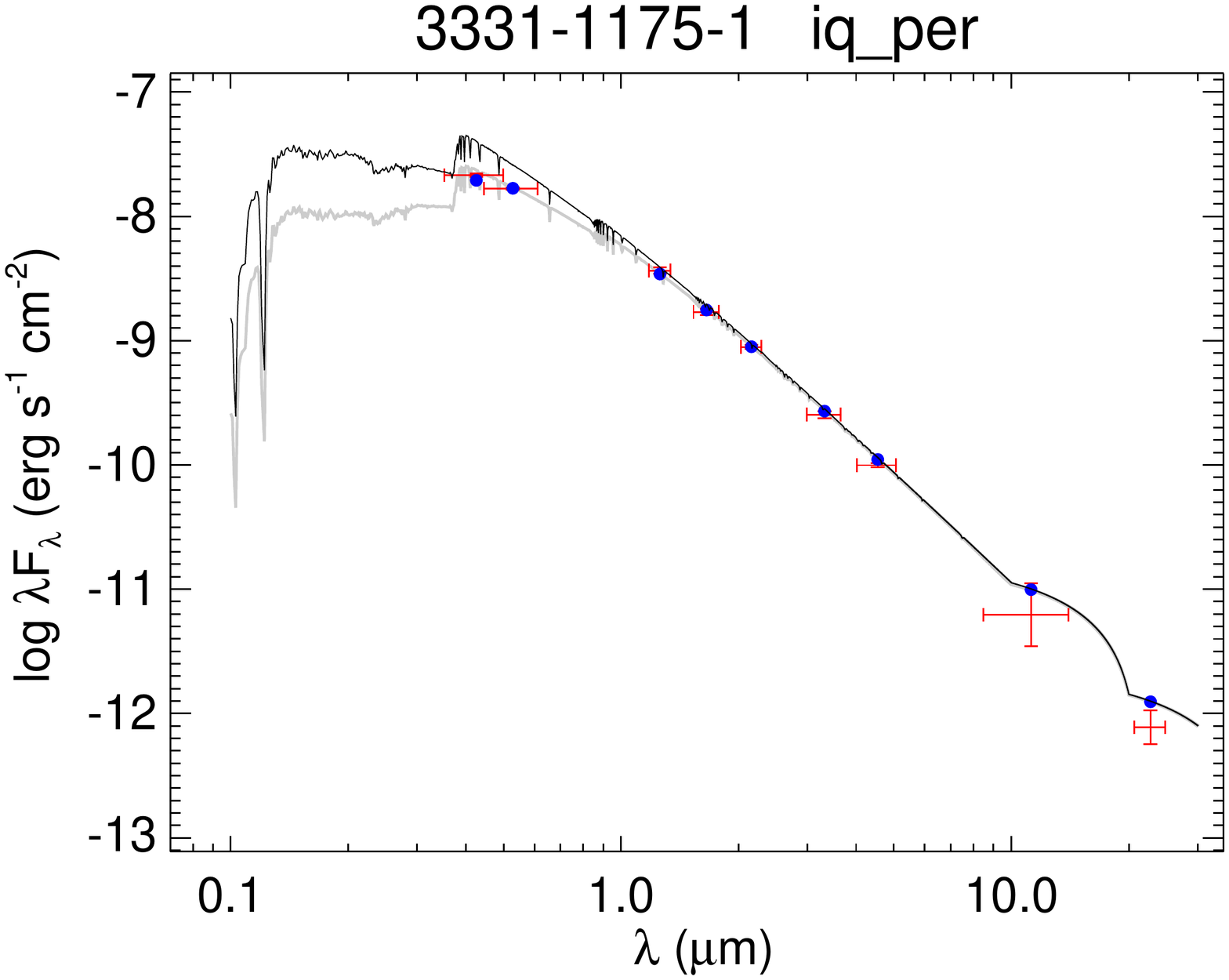}
  \includegraphics[trim=60 60 60 60,clip,width=0.49\linewidth]{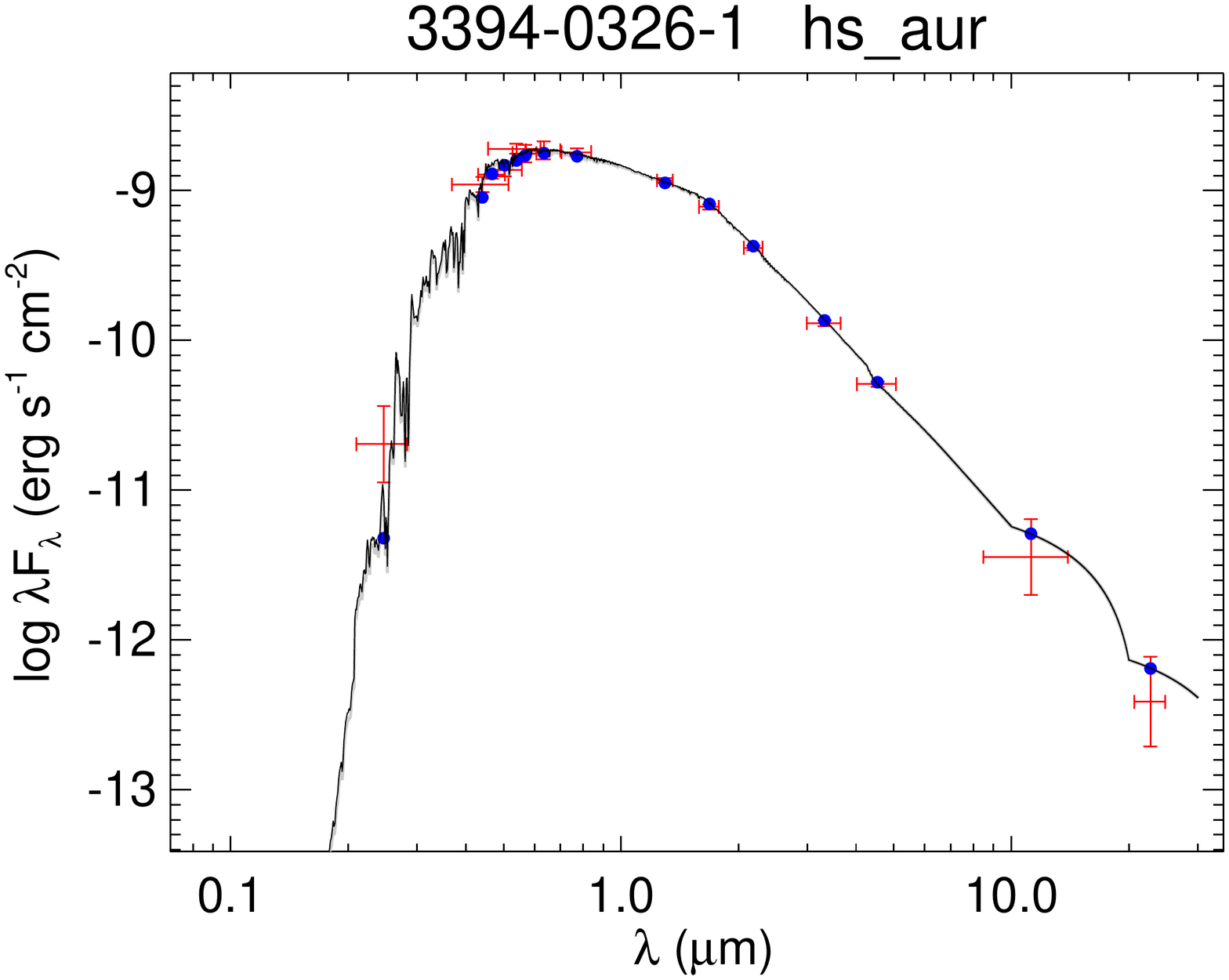}
  \caption{All labels, lines, symbols, and colors as in Figure \ref{fig:seds}.}
  \label{fig:seds_11}
\end{figure}

\begin{figure}[H]
  \centering
  \includegraphics[trim=60 60 60 60,clip,width=0.49\linewidth]{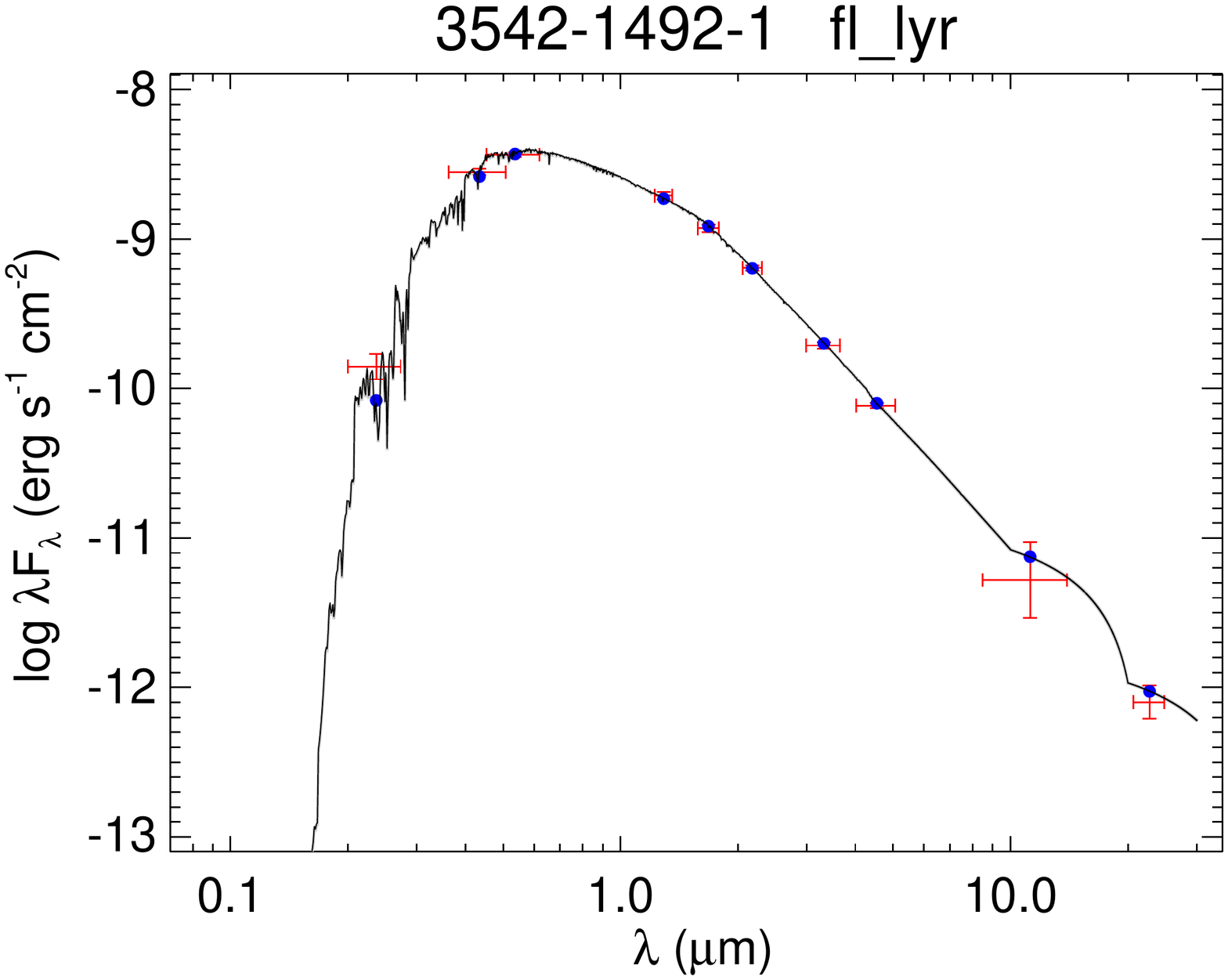}
  \includegraphics[trim=60 60 60 60,clip,width=0.49\linewidth]{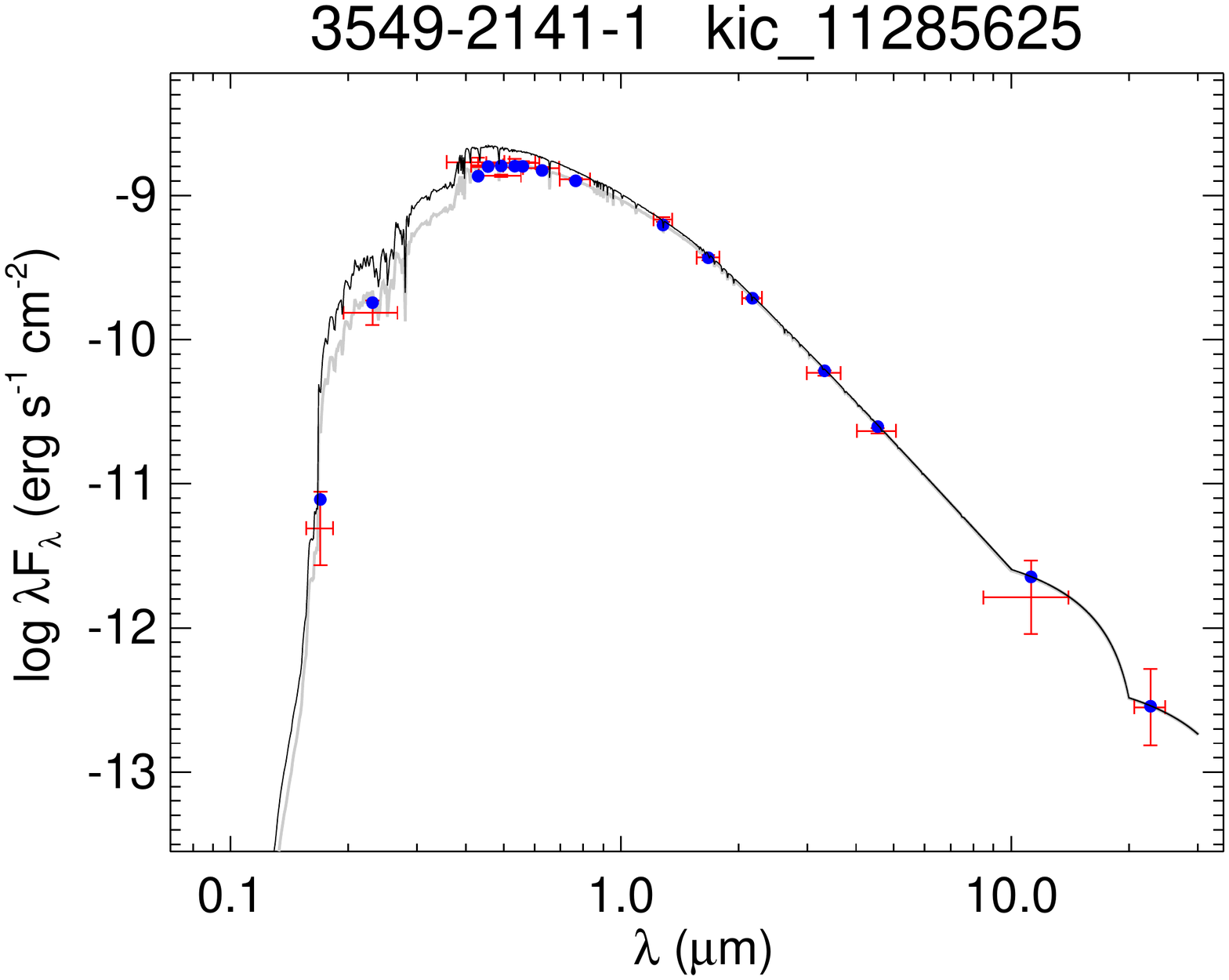}
  \includegraphics[trim=60 60 60 60,clip,width=0.49\linewidth]{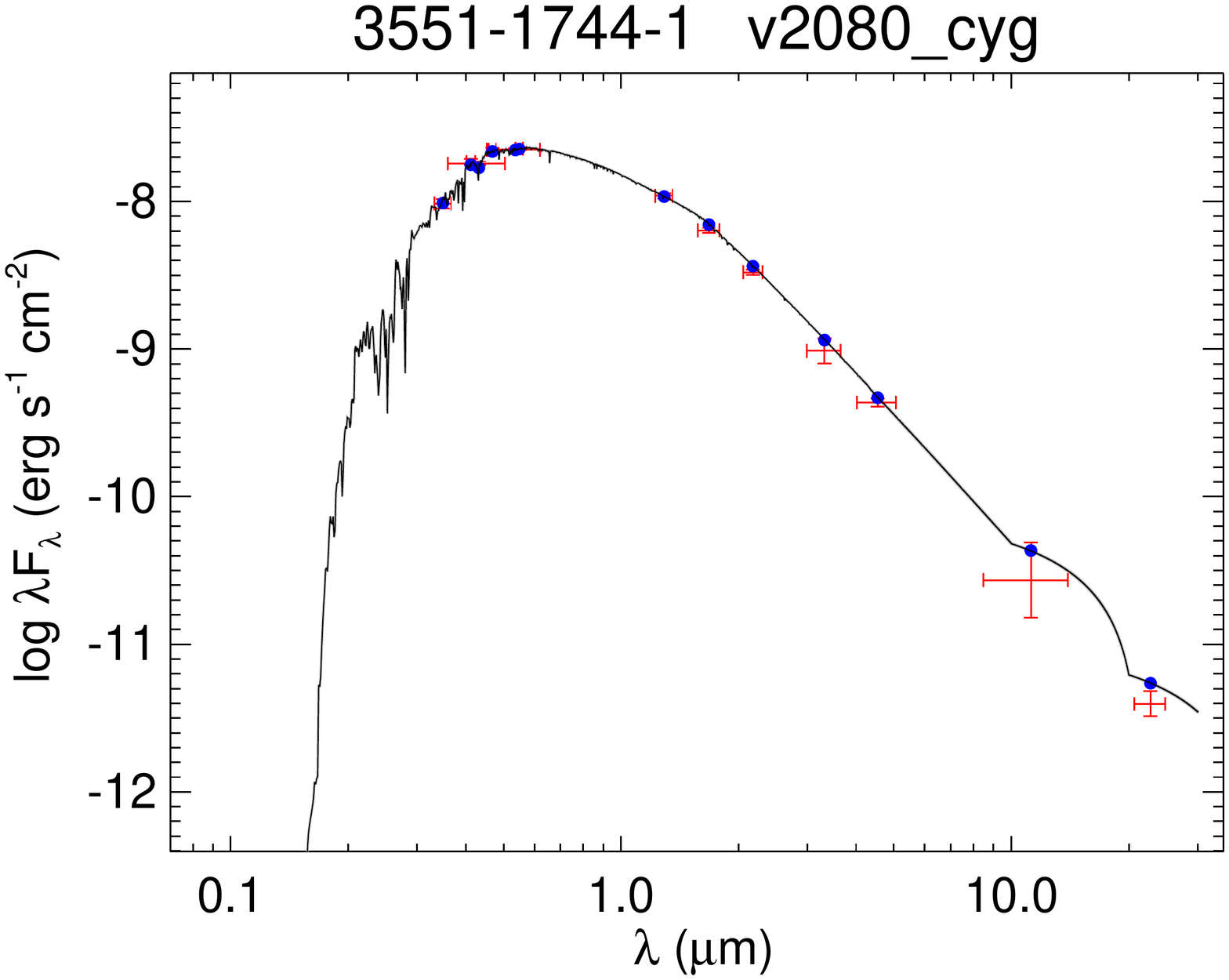}
  \includegraphics[trim=60 60 60 60,clip,width=0.49\linewidth]{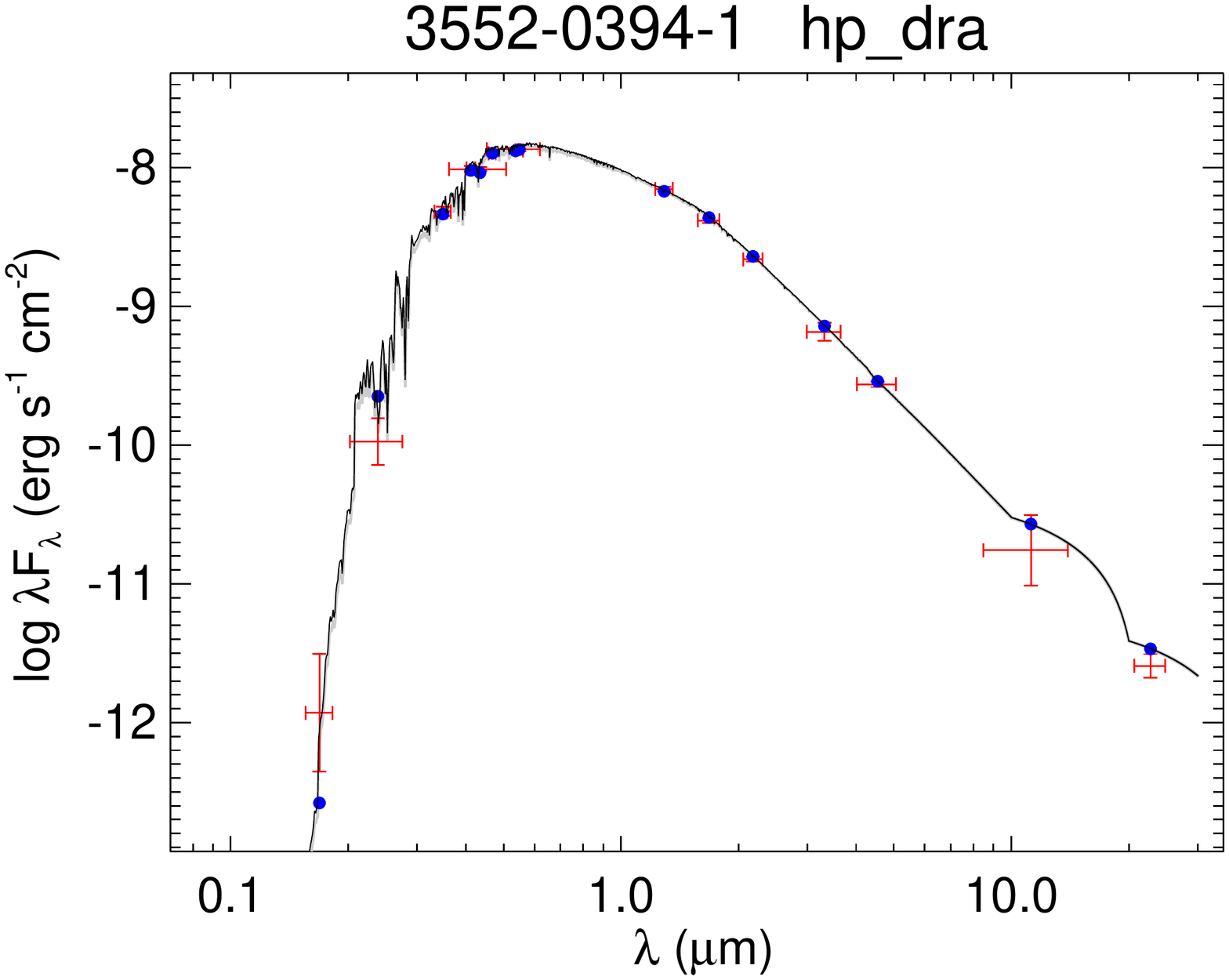}
  \includegraphics[trim=60 60 60 60,clip,width=0.49\linewidth]{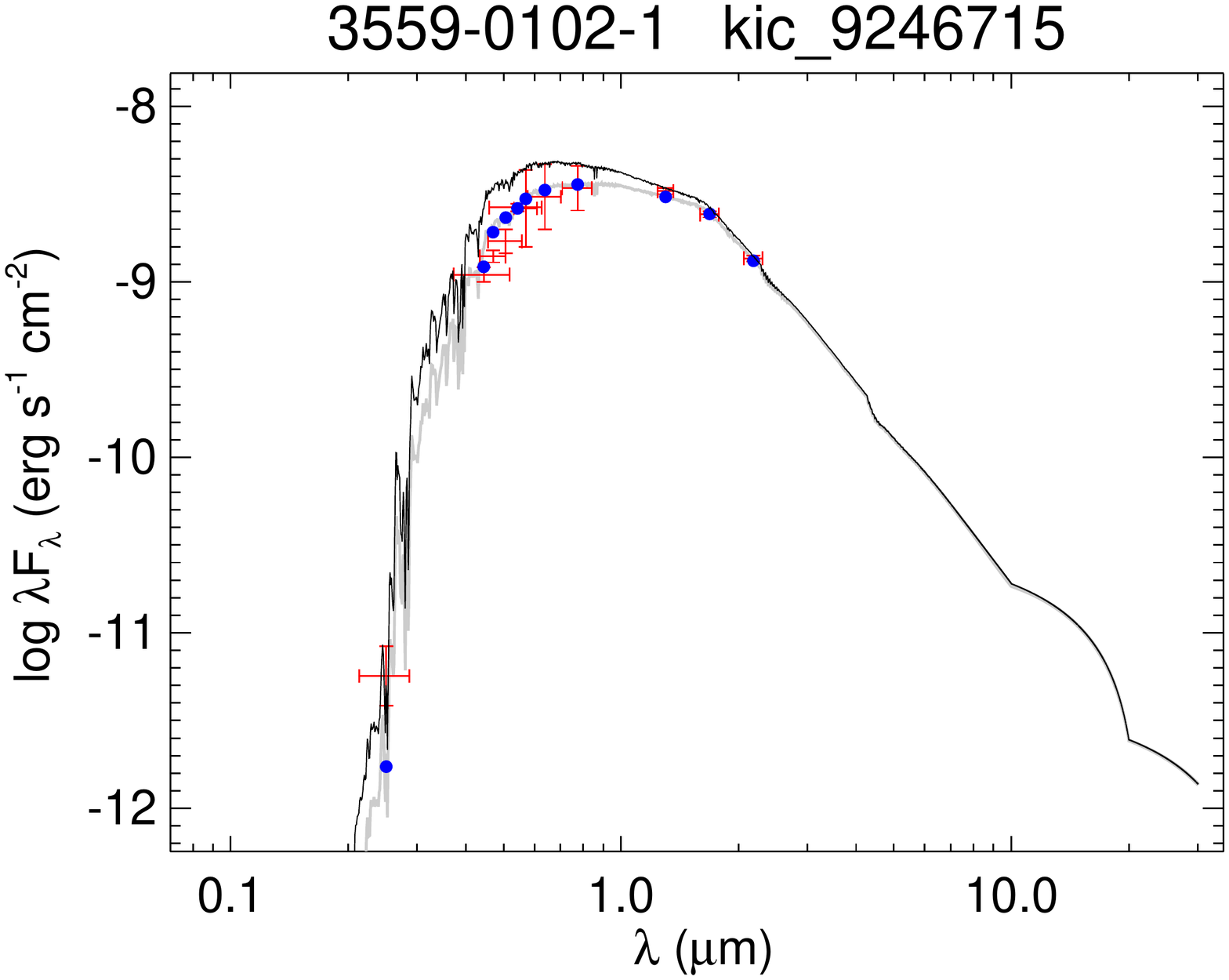}
  \includegraphics[trim=60 60 60 60,clip,width=0.49\linewidth]{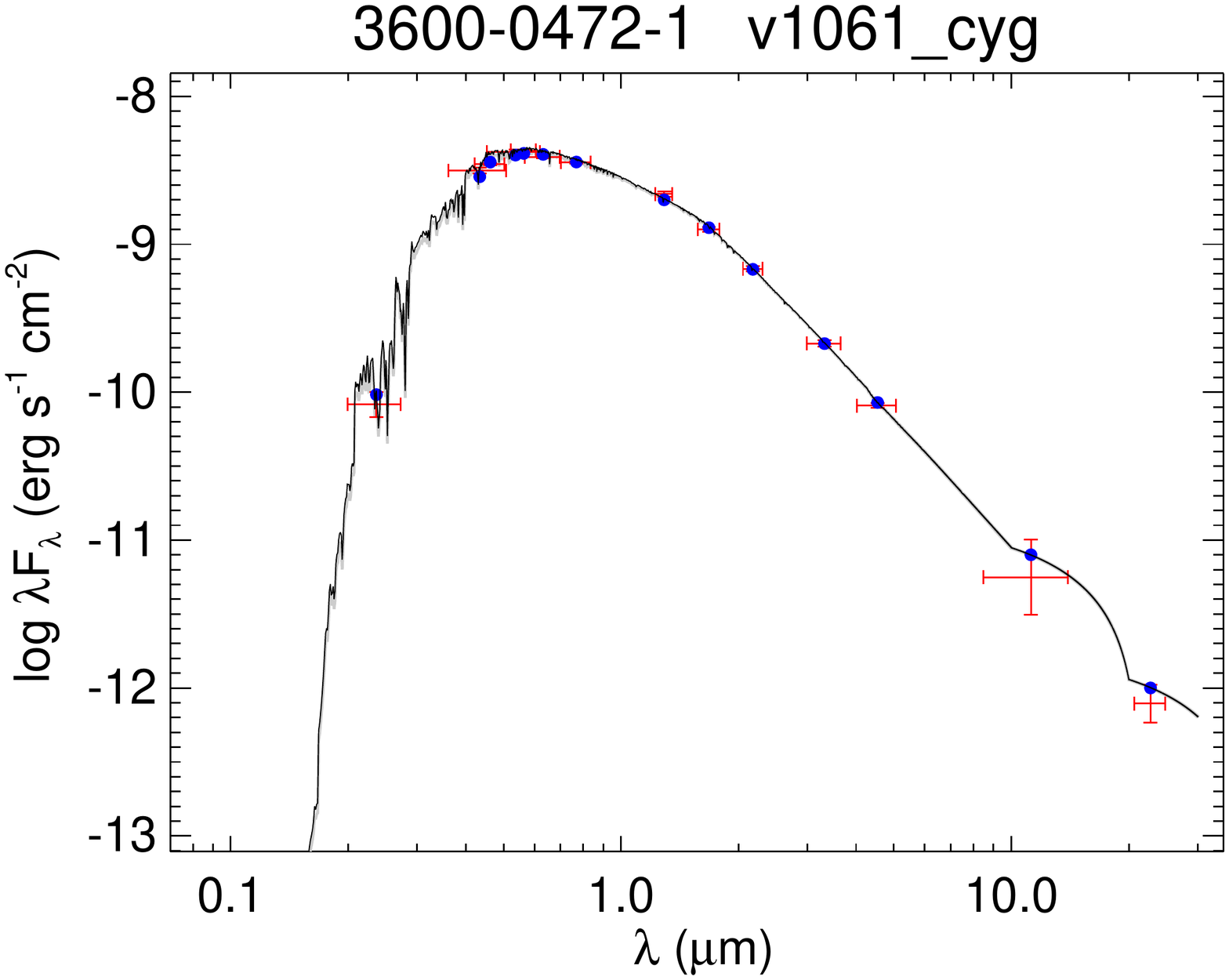}
  \caption{All labels, lines, symbols, and colors as in Figure \ref{fig:seds}.}
  \label{fig:seds_12}
\end{figure}

\begin{figure}[H]
  \centering
  \includegraphics[trim=60 60 60 60,clip,width=0.49\linewidth]{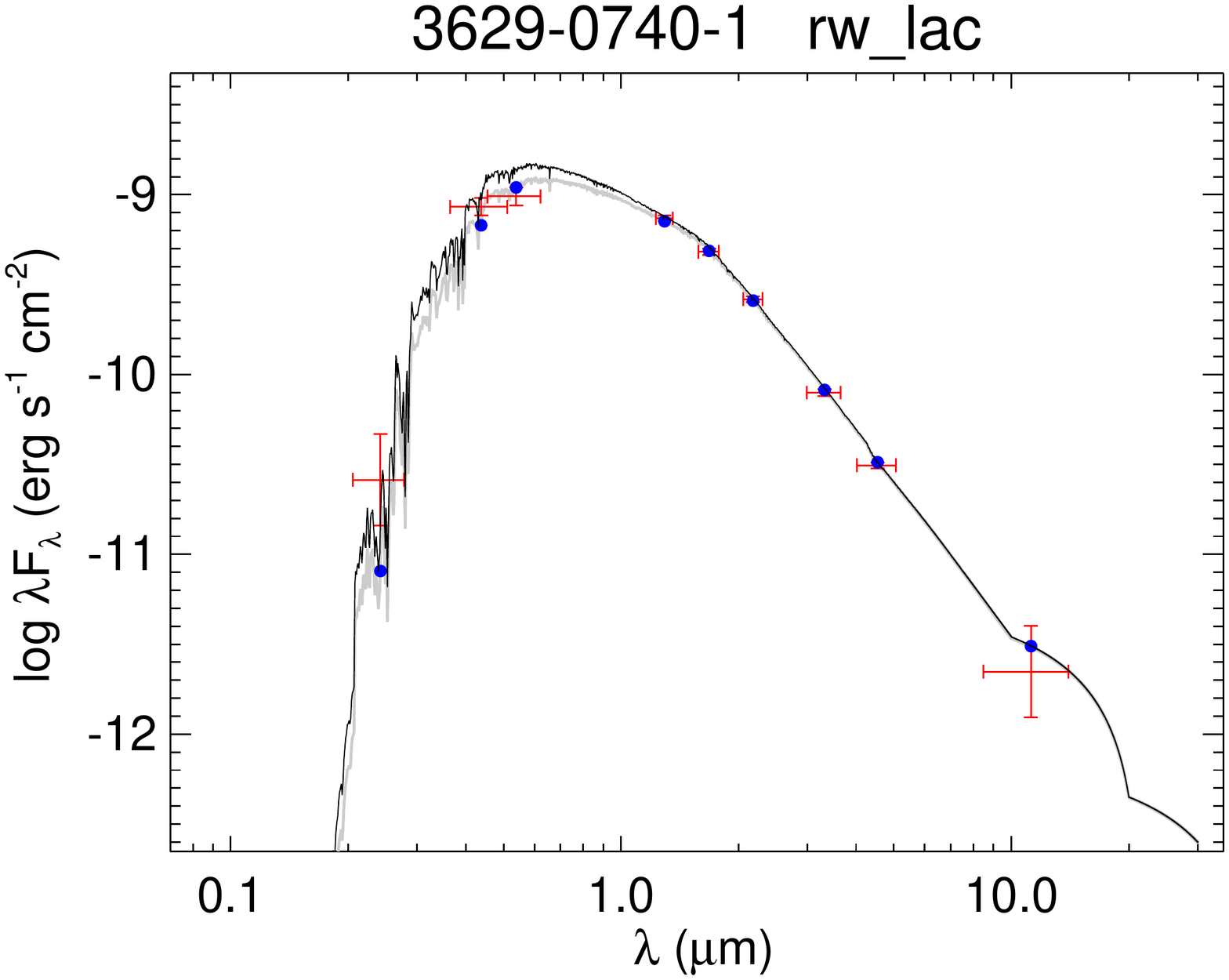}
  \includegraphics[trim=60 60 60 60,clip,width=0.49\linewidth]{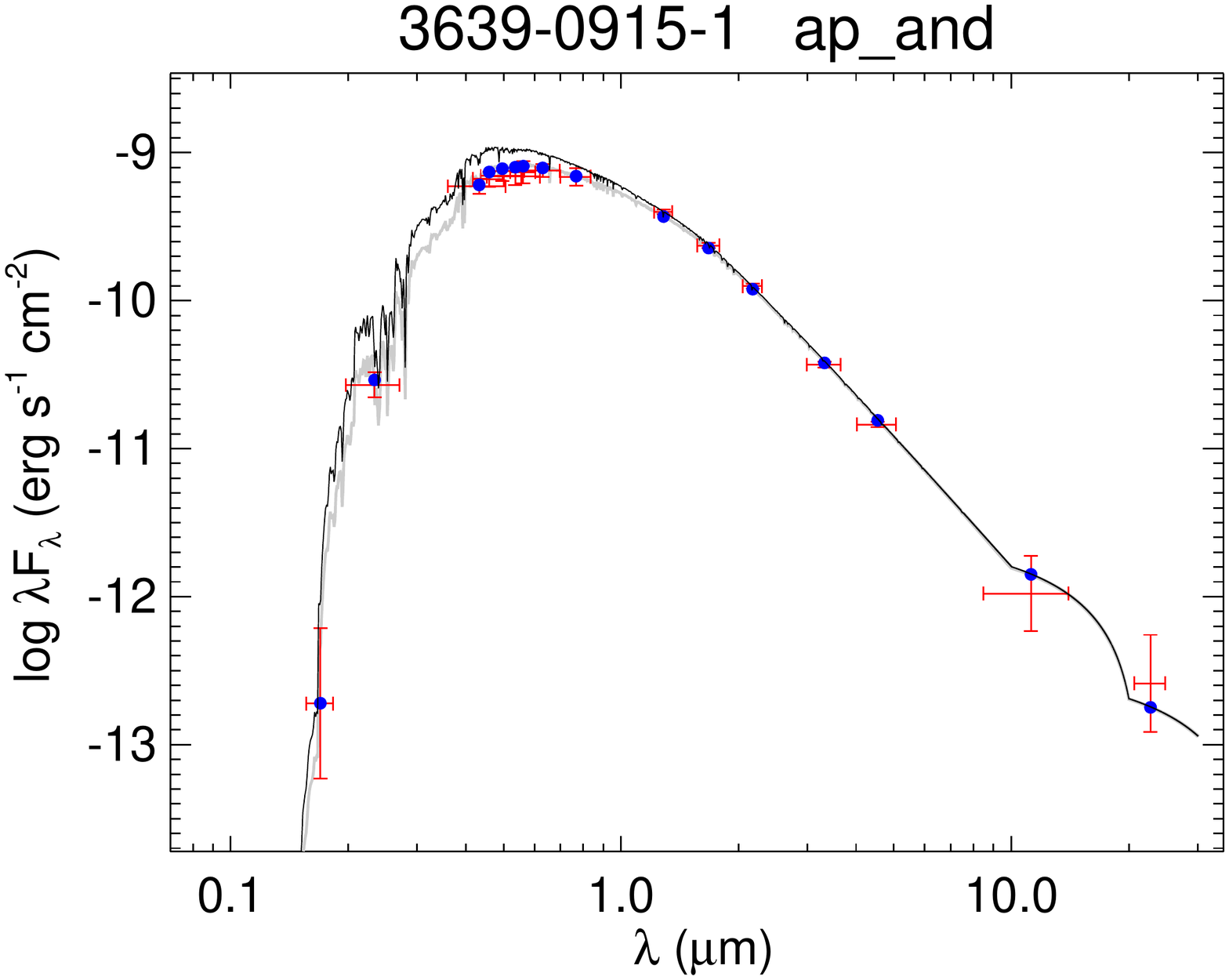}
  \includegraphics[trim=60 60 60 60,clip,width=0.49\linewidth]{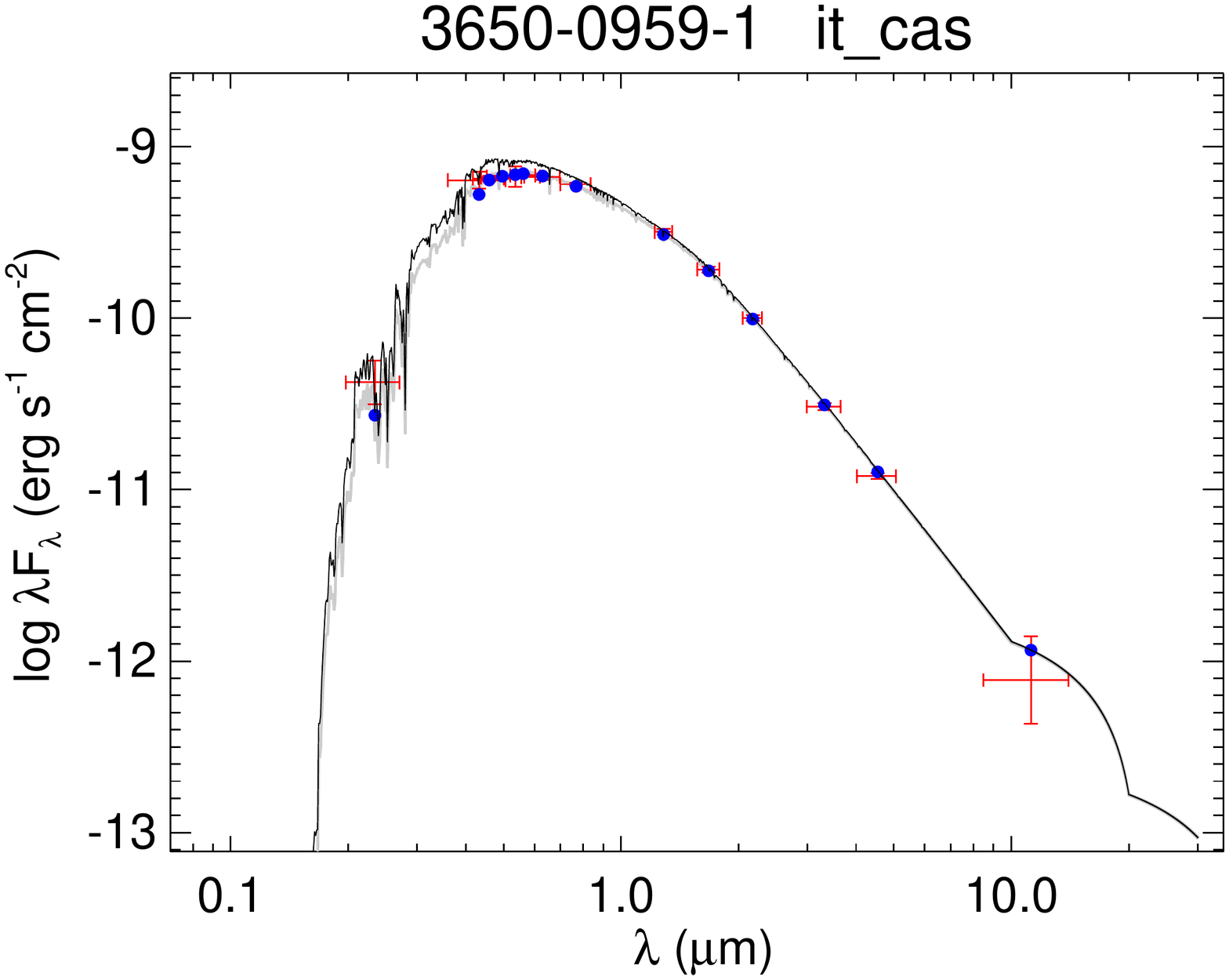}
  \includegraphics[trim=60 60 60 60,clip,width=0.49\linewidth]{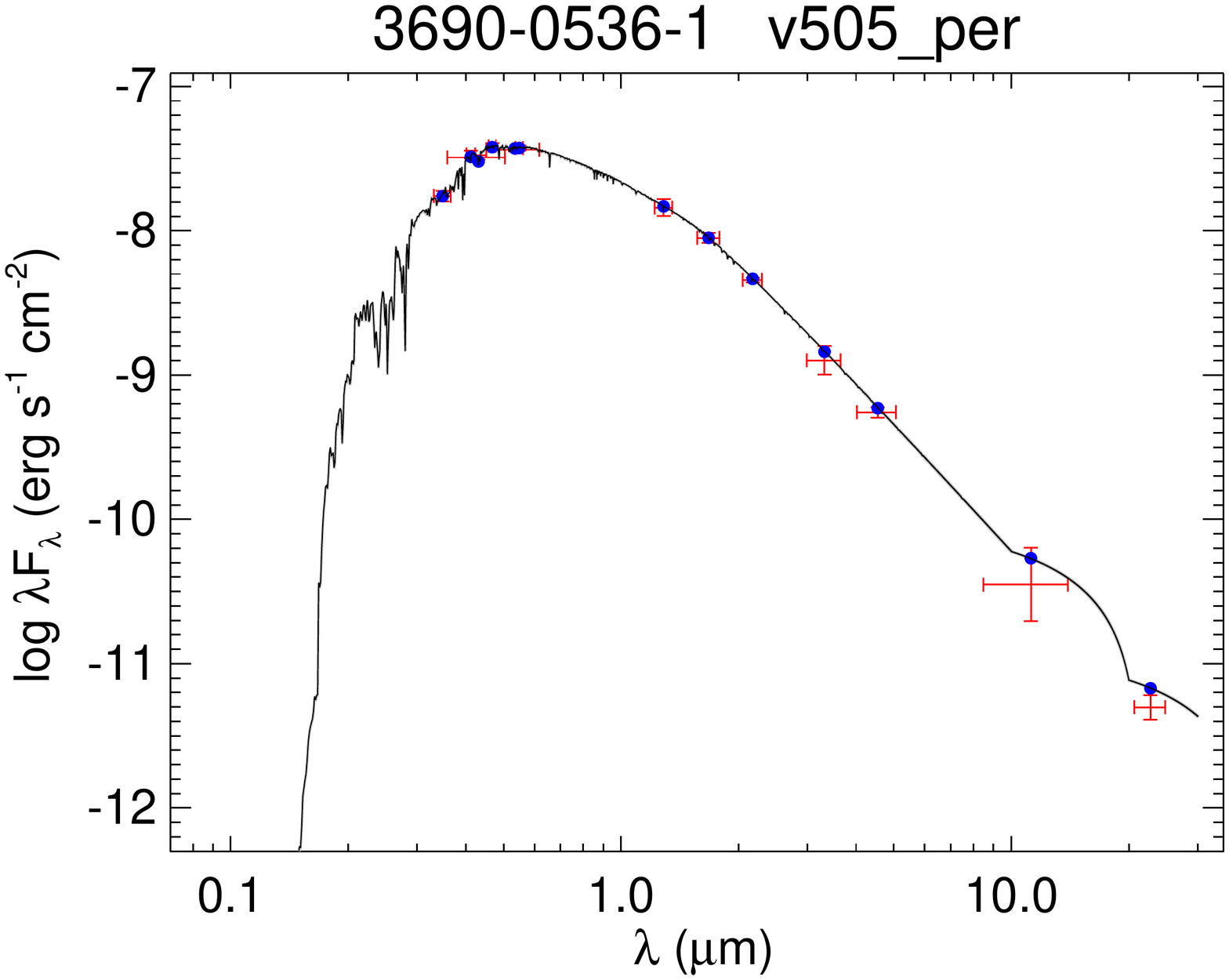}
  \includegraphics[trim=60 60 60 60,clip,width=0.49\linewidth]{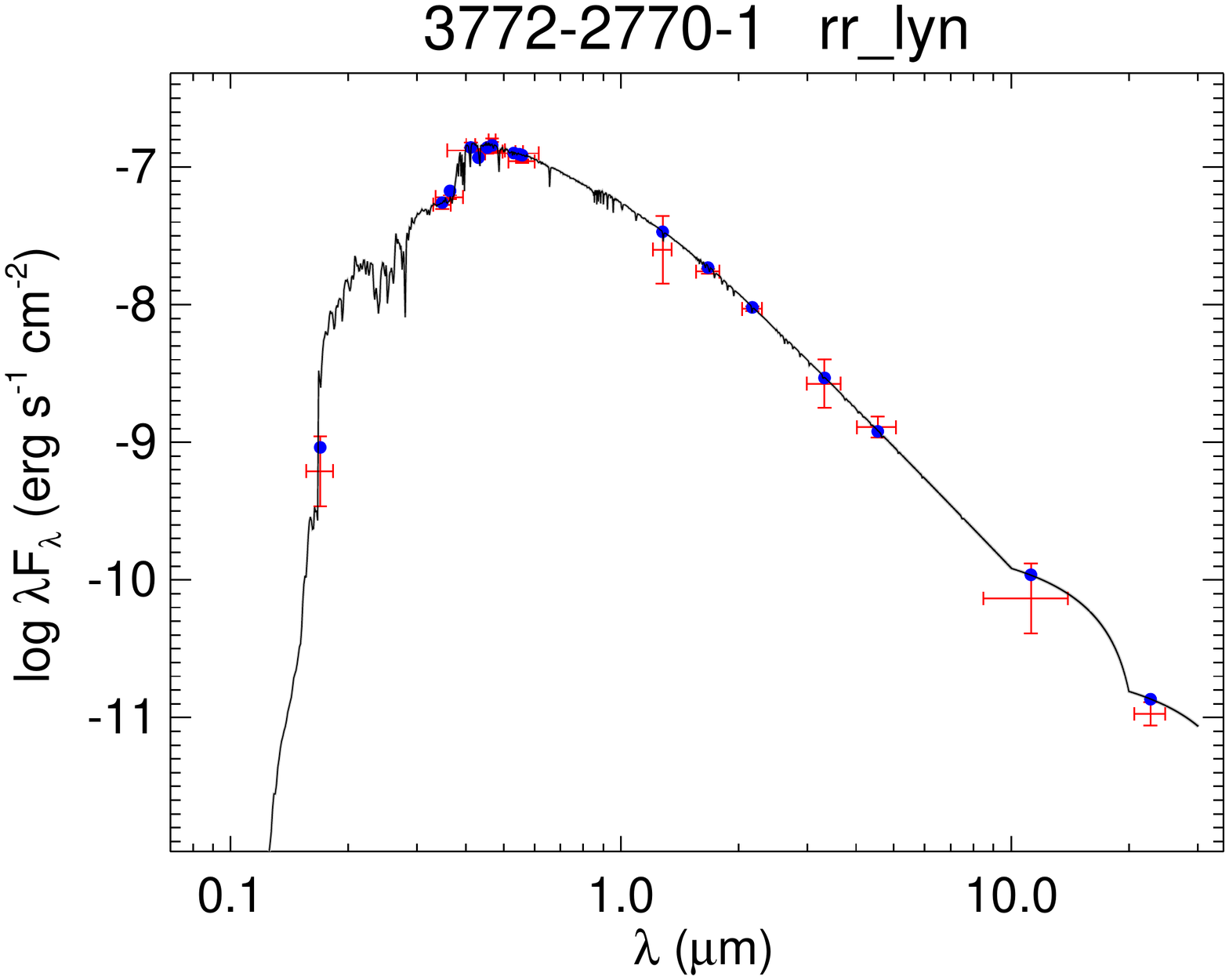}
  \includegraphics[trim=60 60 60 60,clip,width=0.49\linewidth]{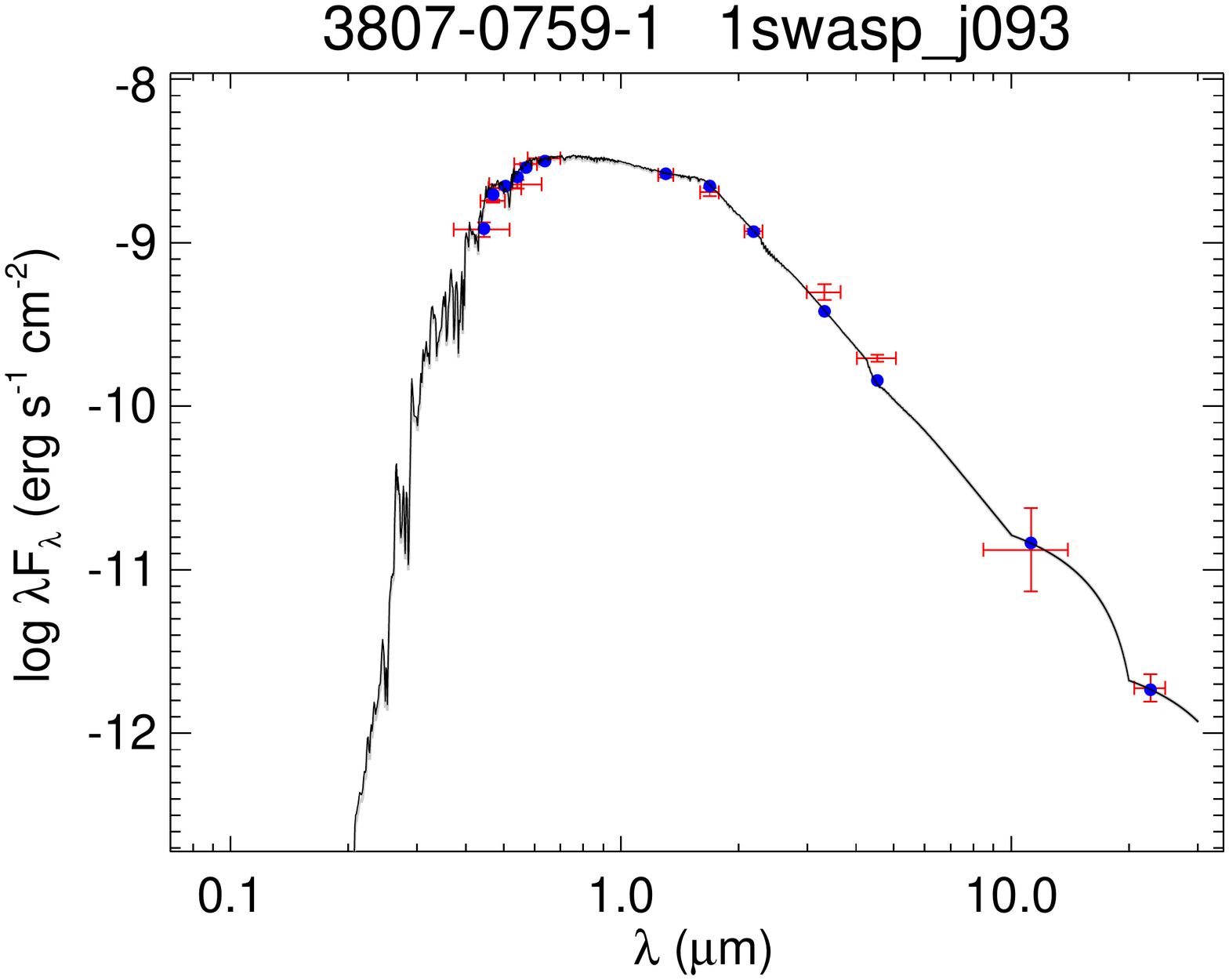}
  \caption{All labels, lines, symbols, and colors as in Figure \ref{fig:seds}.}
  \label{fig:seds_13}
\end{figure}

\begin{figure}[H]
  \centering
  \includegraphics[trim=60 60 60 60,clip,width=0.49\linewidth]{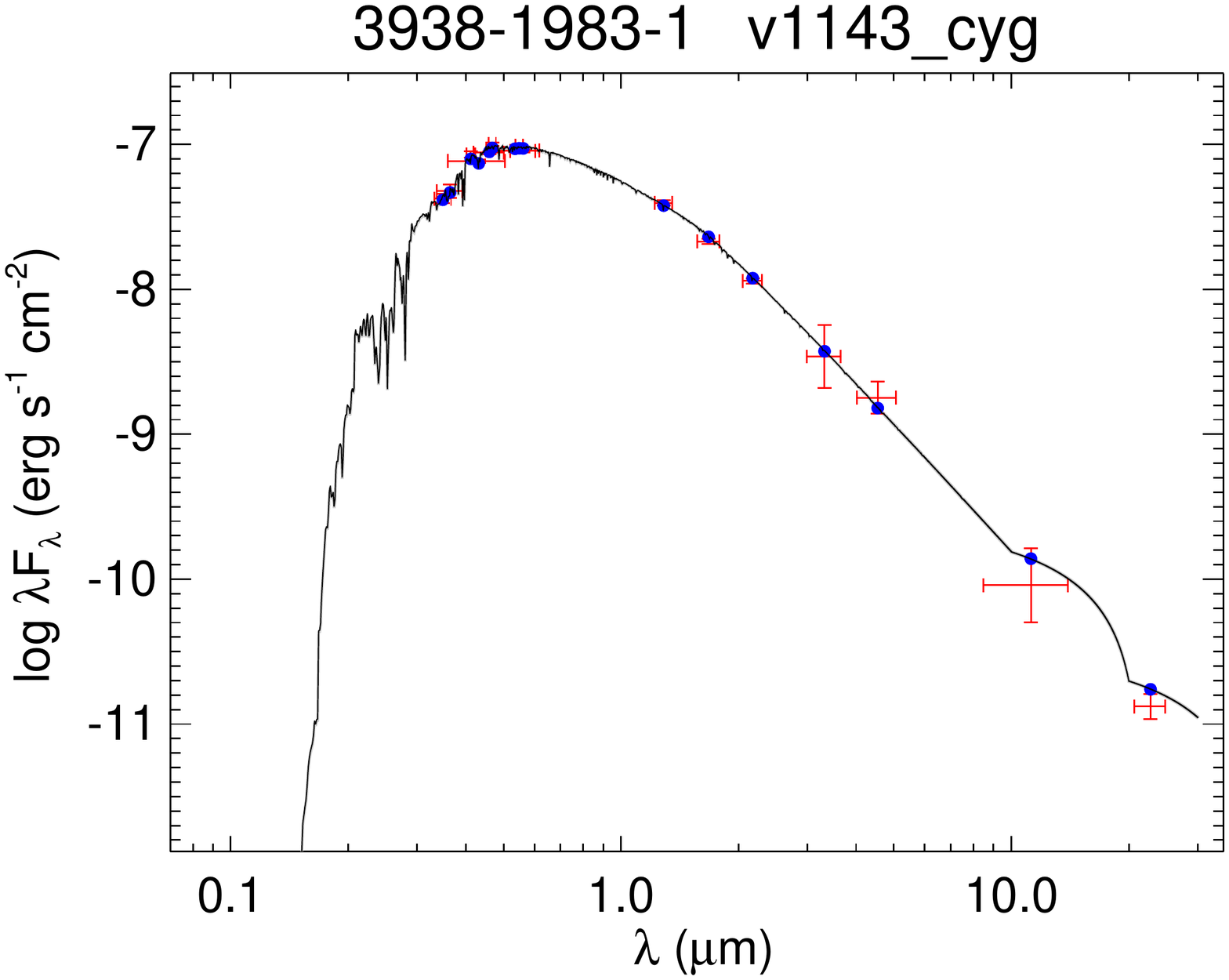}
  \includegraphics[trim=60 60 60 60,clip,width=0.49\linewidth]{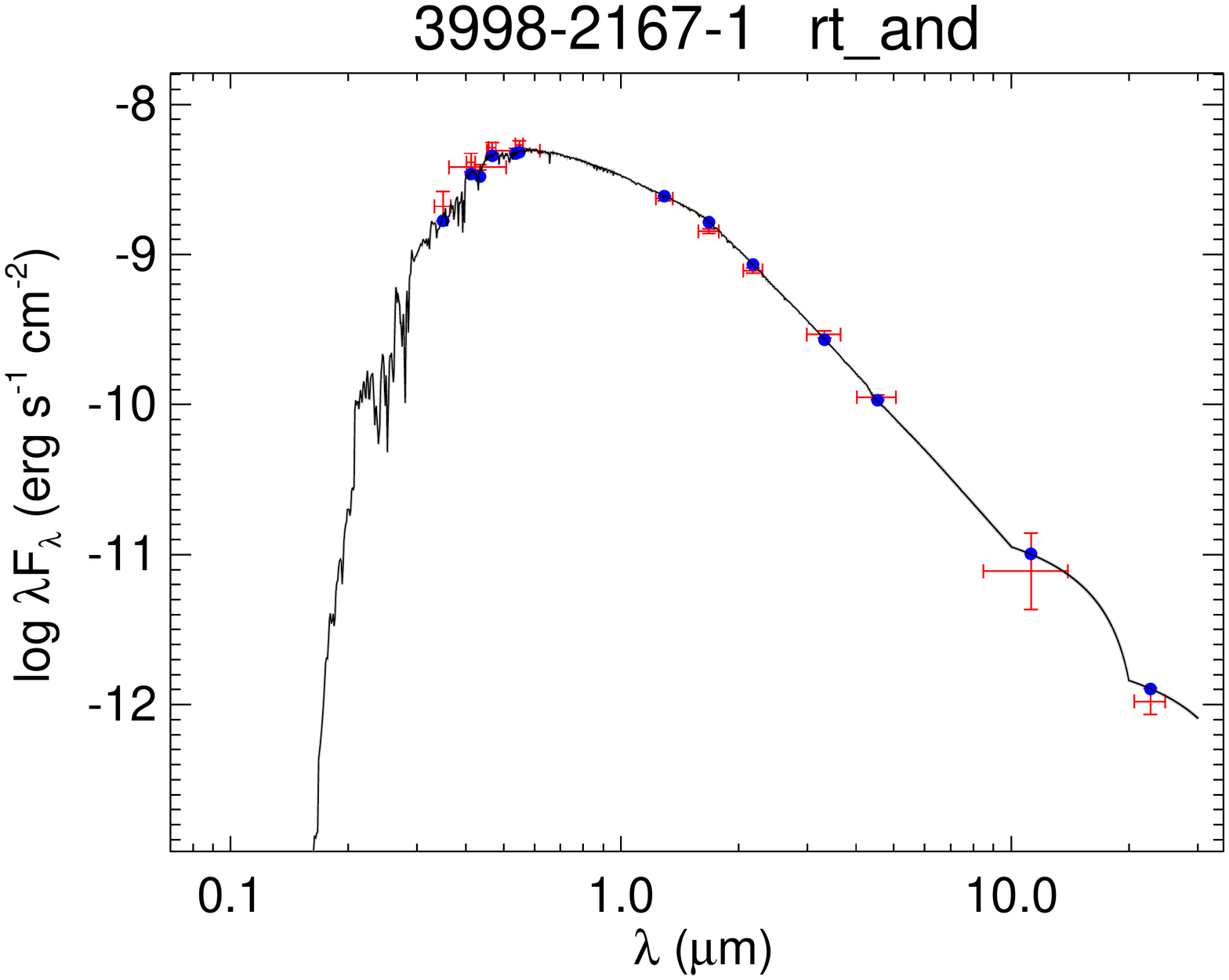}
  \includegraphics[trim=60 60 60 60,clip,width=0.49\linewidth]{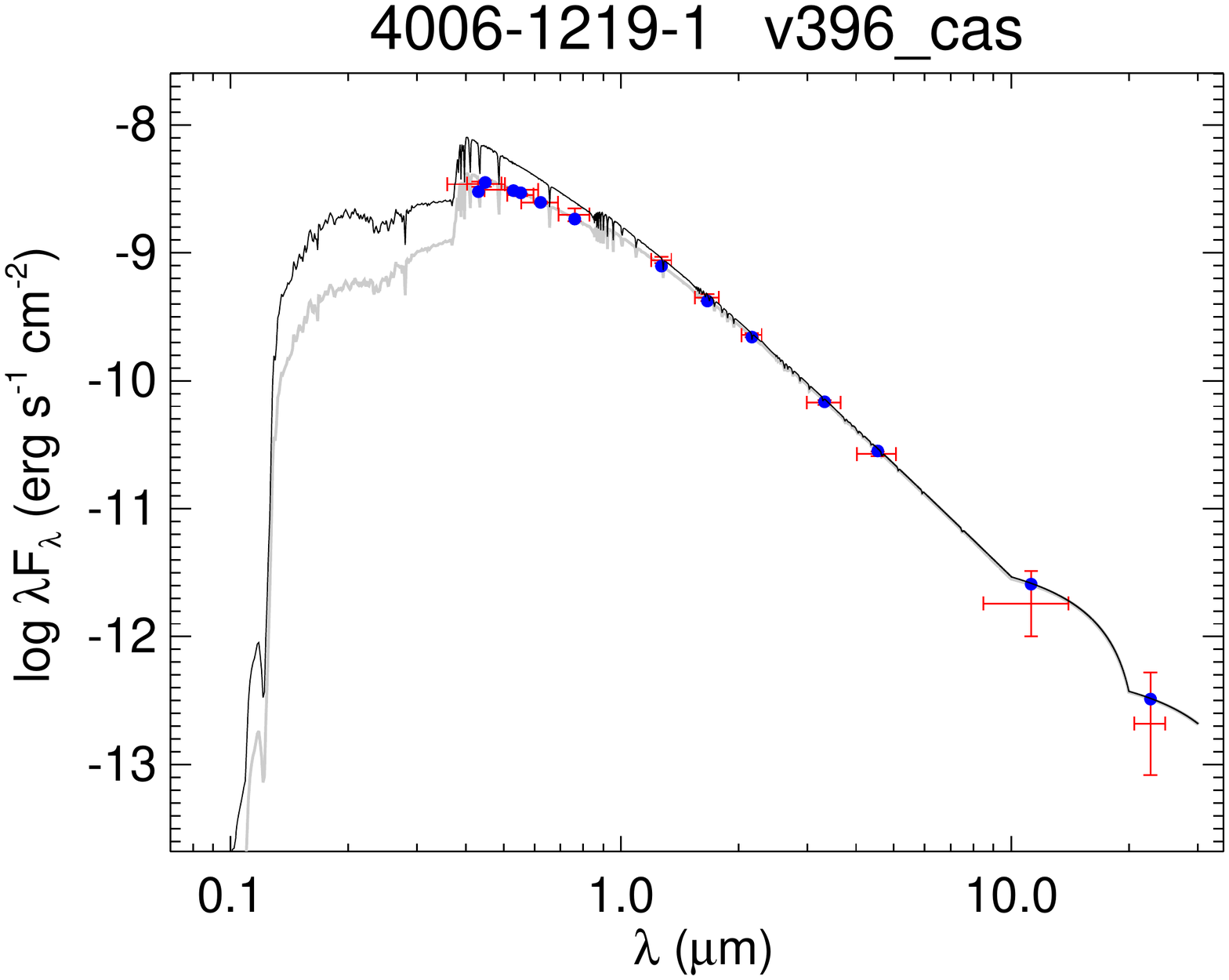}
  \includegraphics[trim=60 60 60 60,clip,width=0.49\linewidth]{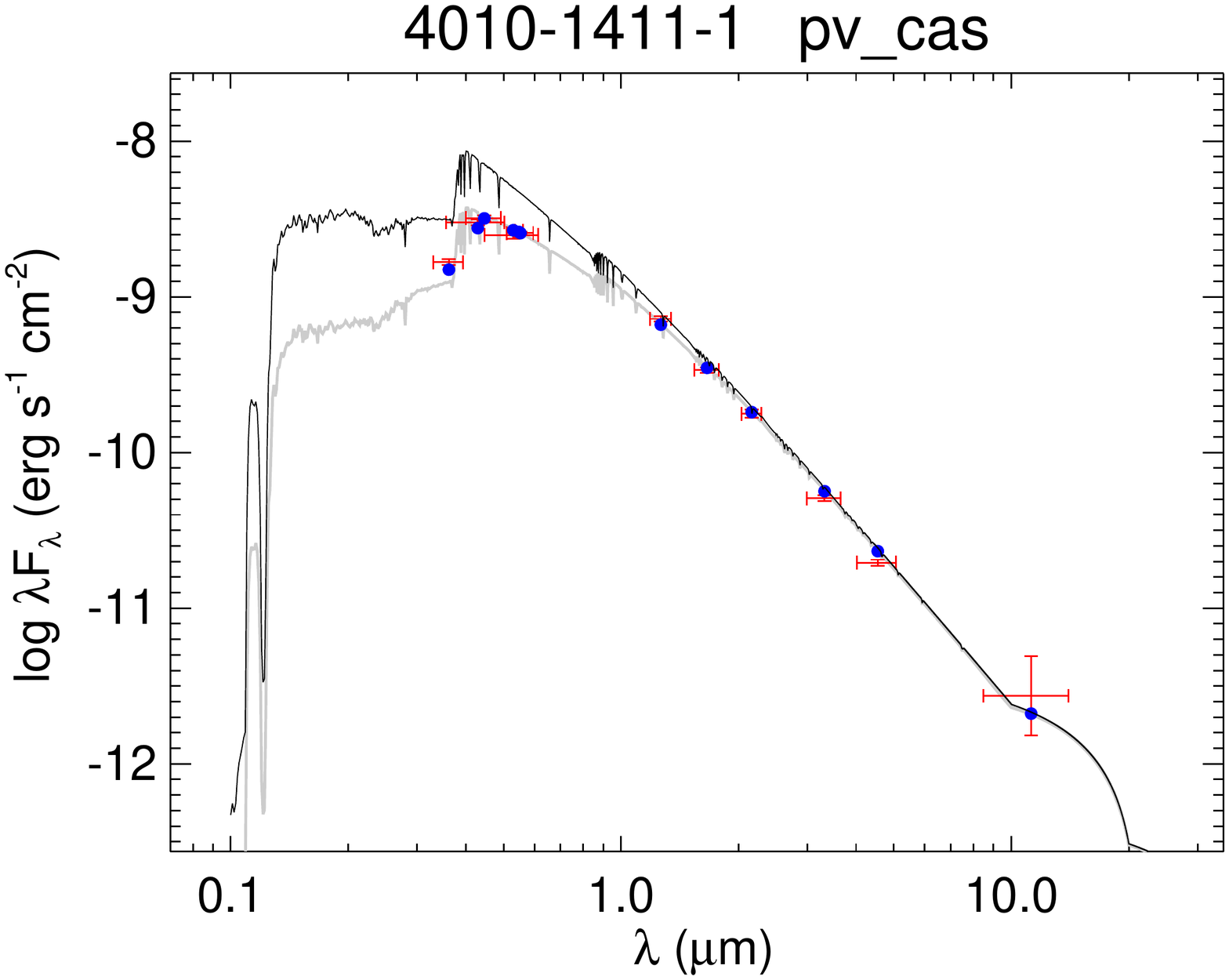}
  \includegraphics[trim=60 60 60 60,clip,width=0.49\linewidth]{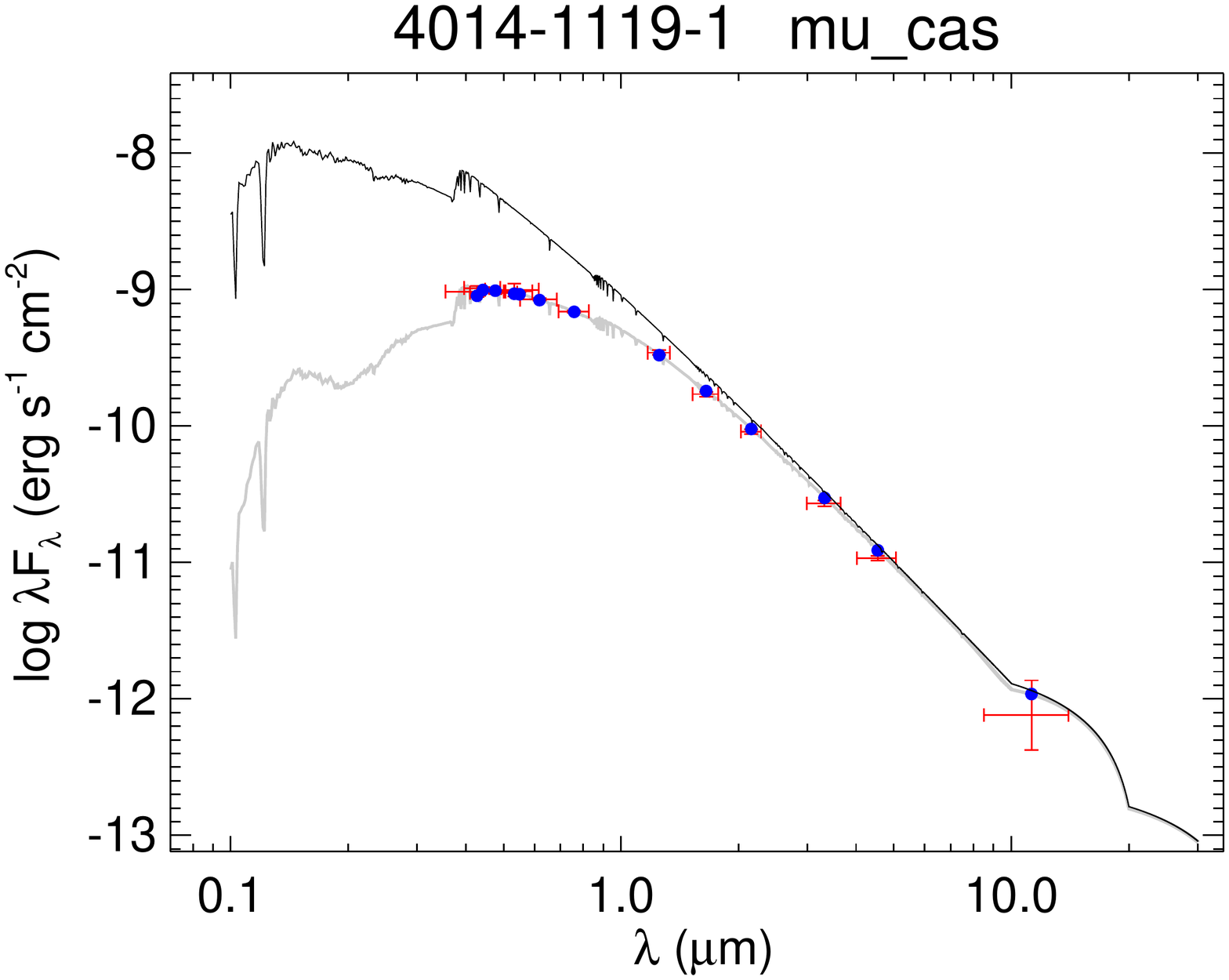}
  \includegraphics[trim=60 60 60 60,clip,width=0.49\linewidth]{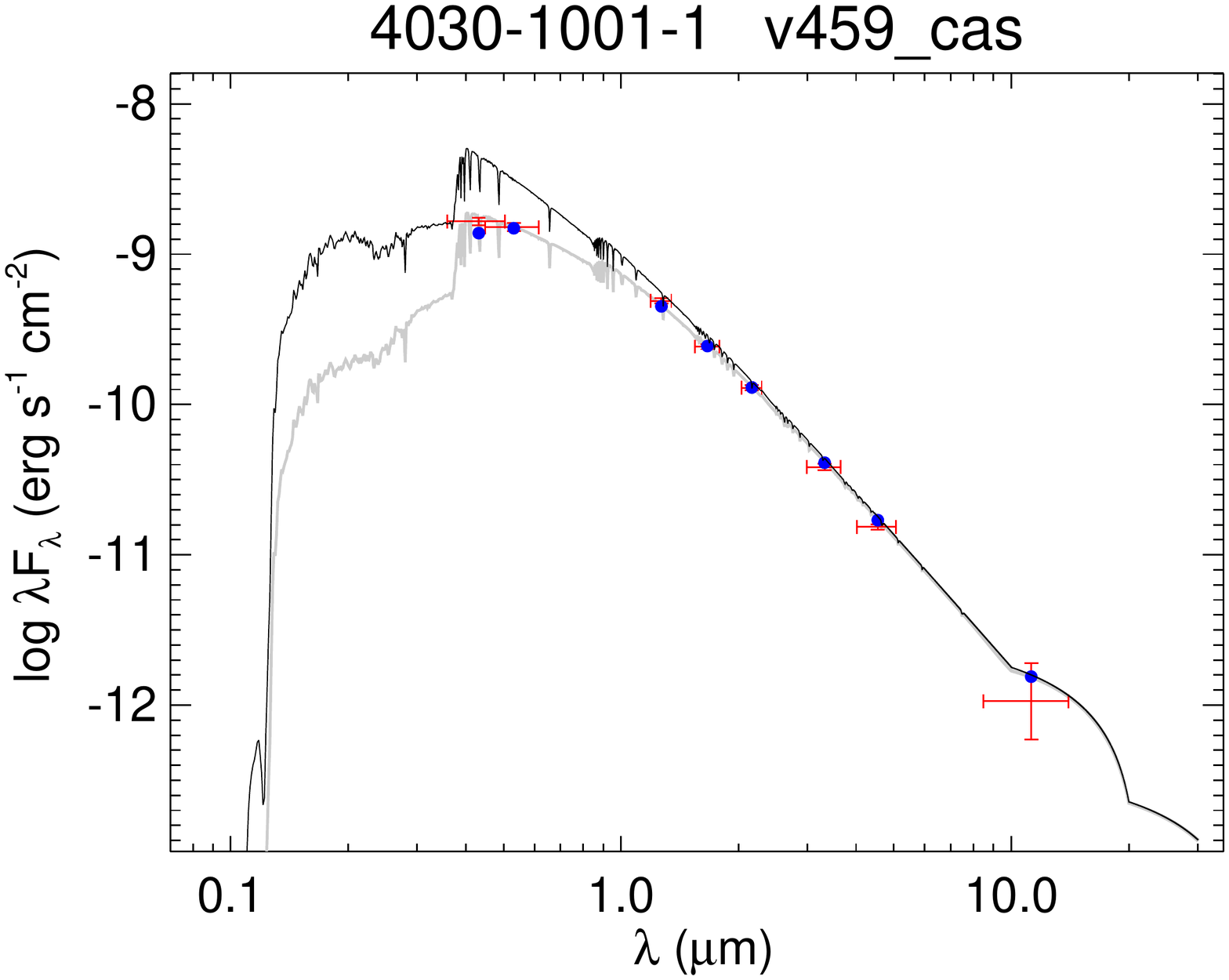}
  \caption{All labels, lines, symbols, and colors as in Figure \ref{fig:seds}.}
  \label{fig:seds_14}
\end{figure}

\begin{figure}[H]
  \centering
  \includegraphics[trim=60 60 60 60,clip,width=0.49\linewidth]{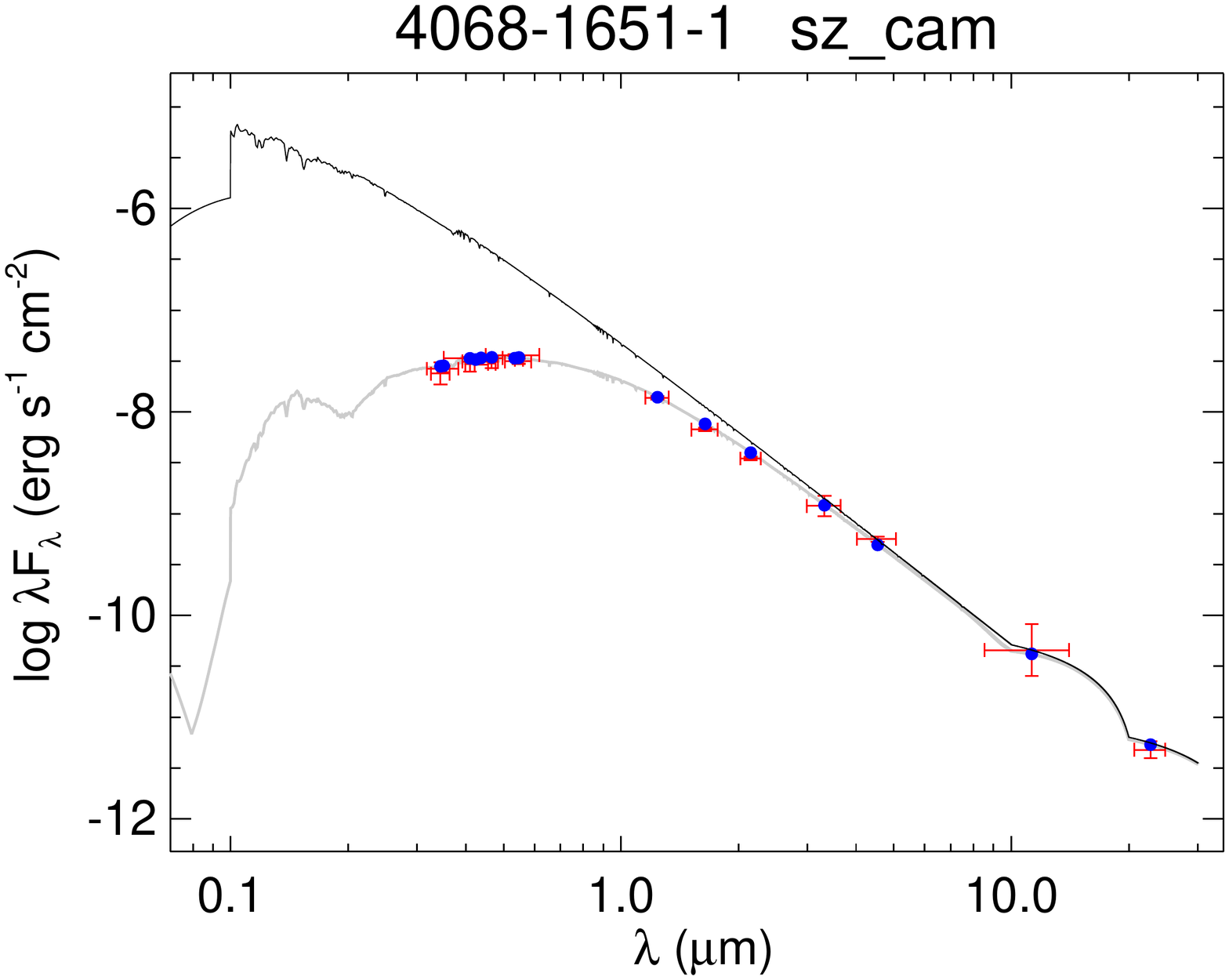}
  \includegraphics[trim=60 60 60 60,clip,width=0.49\linewidth]{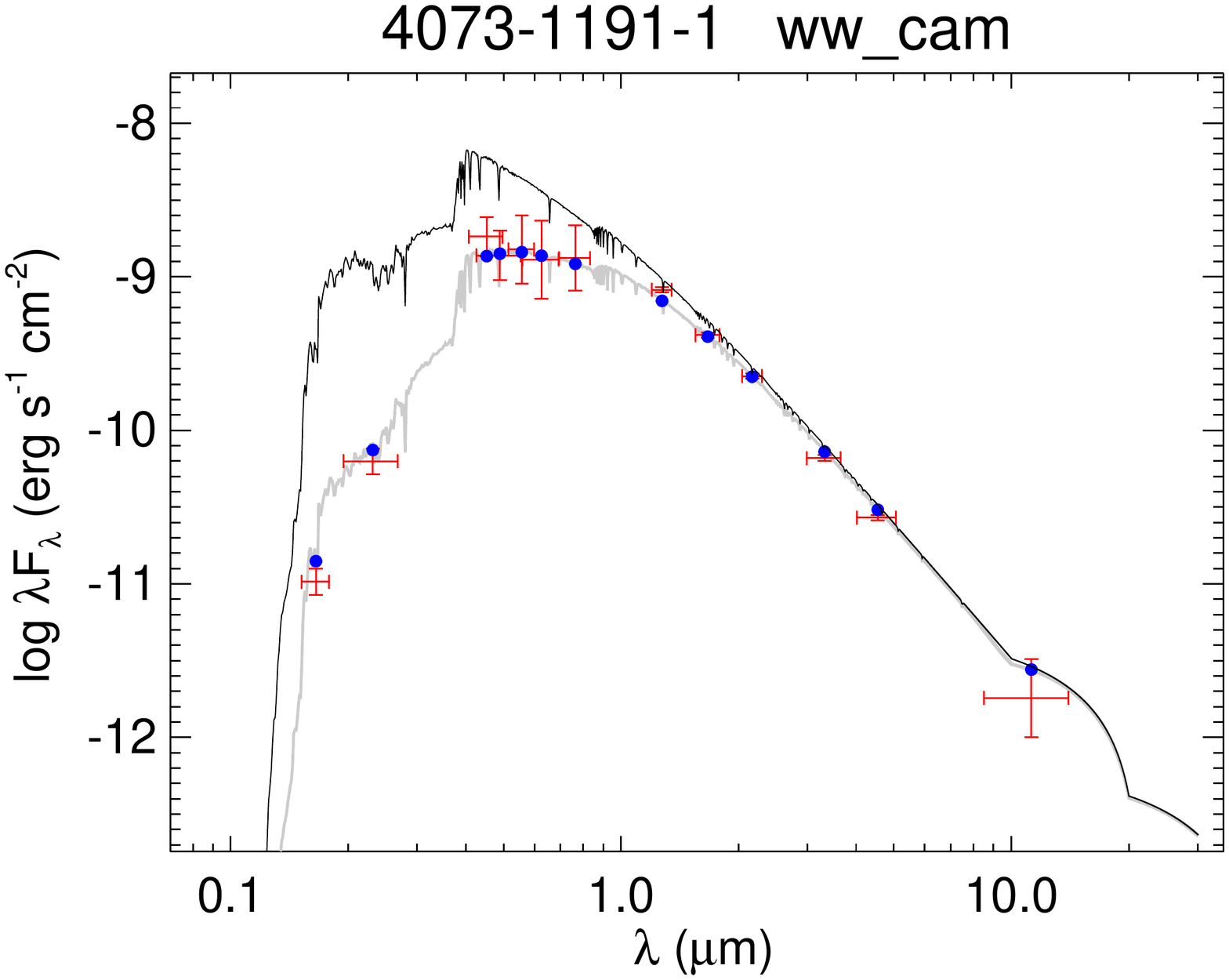}
  \includegraphics[trim=60 60 60 60,clip,width=0.49\linewidth]{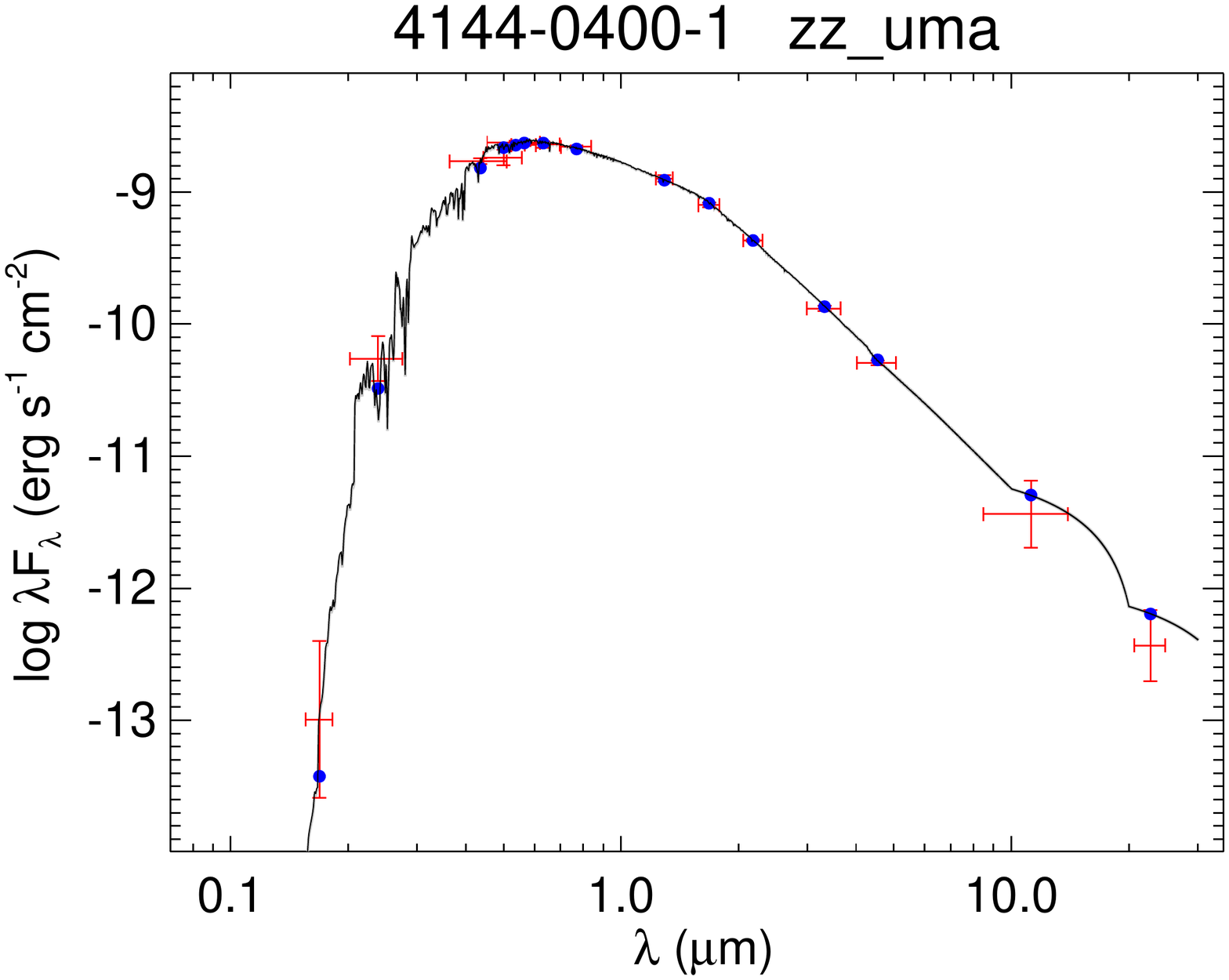}
  \includegraphics[trim=60 60 60 60,clip,width=0.49\linewidth]{sedfigs/wx_cep.pdf}
  \includegraphics[trim=60 60 60 60,clip,width=0.49\linewidth]{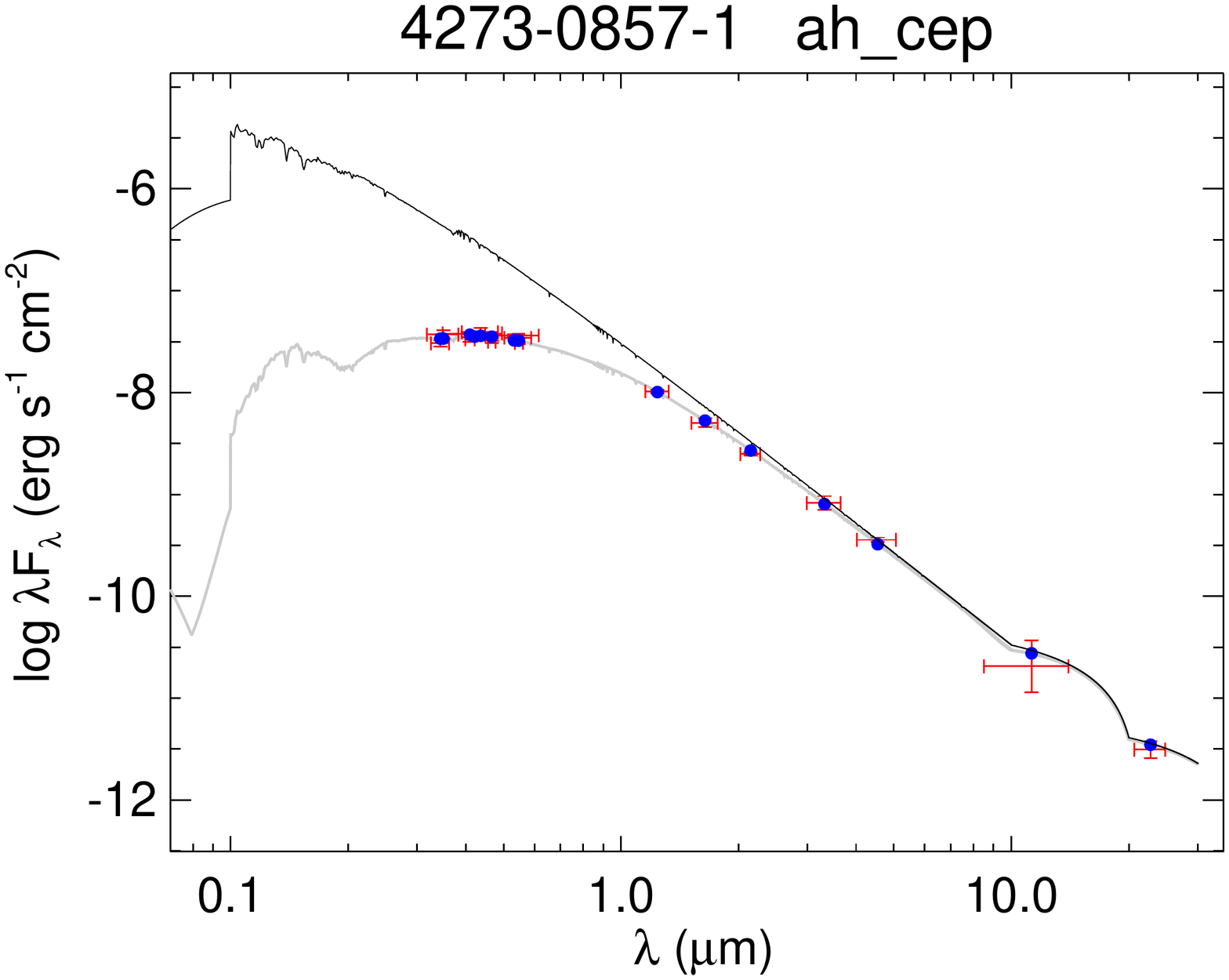}
  \includegraphics[trim=60 60 60 60,clip,width=0.49\linewidth]{sedfigs/cw_cep.pdf}
  \caption{All labels, lines, symbols, and colors as in Figure \ref{fig:seds}.}
  \label{fig:seds_15}
\end{figure}

\begin{figure}[H]
  \centering
  \includegraphics[trim=60 60 60 60,clip,width=0.49\linewidth]{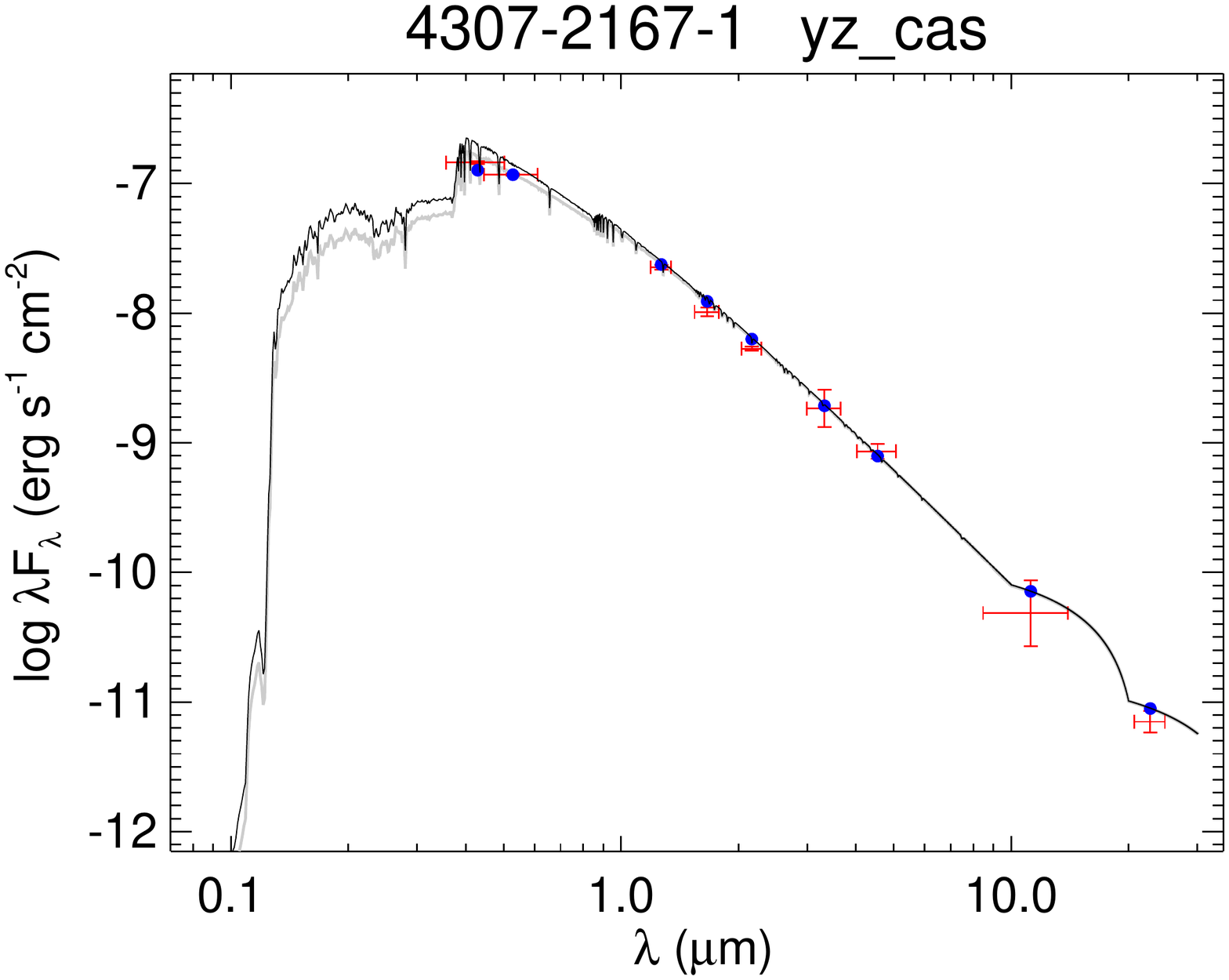}
  \includegraphics[trim=60 60 60 60,clip,width=0.49\linewidth]{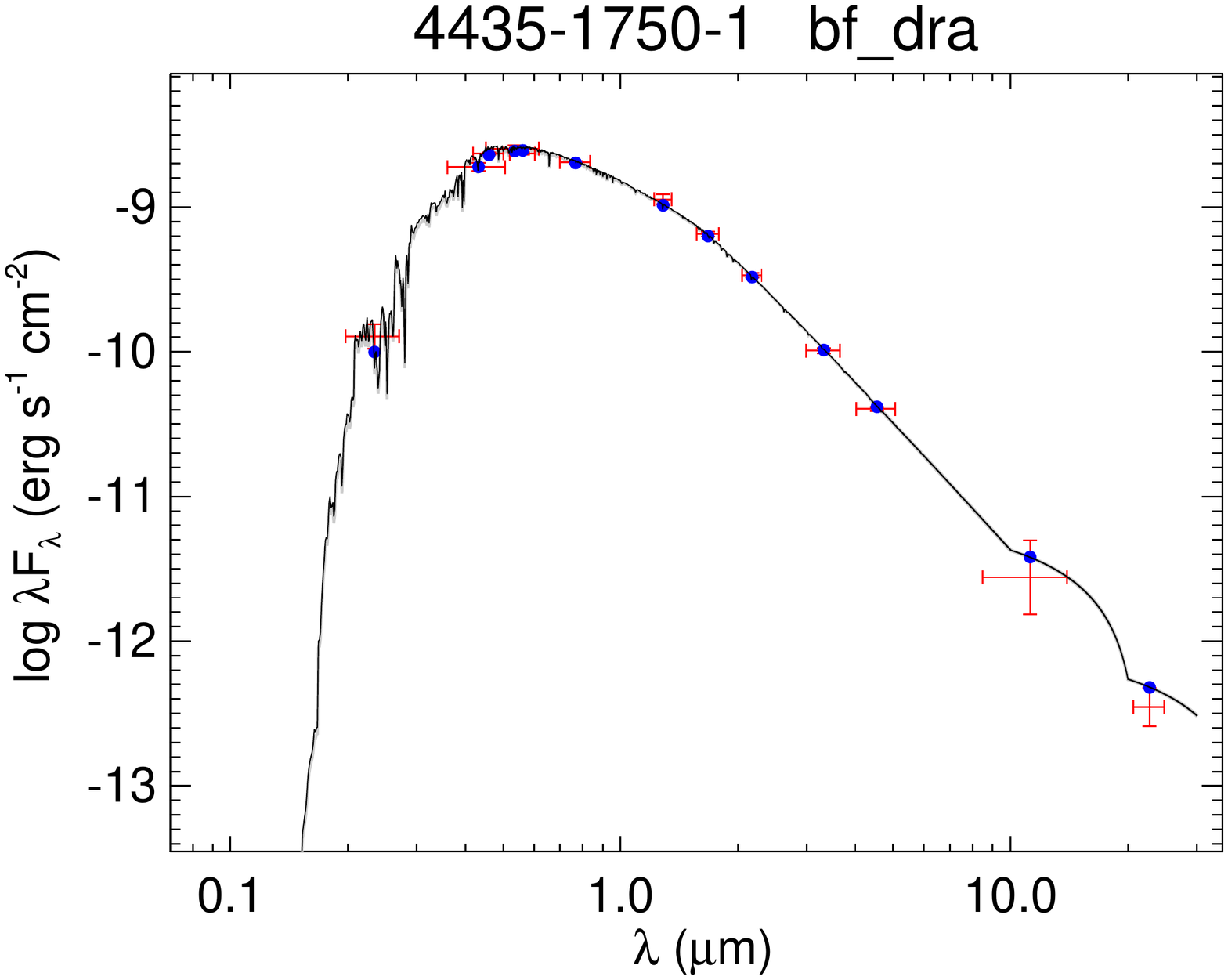}
  \includegraphics[trim=60 60 60 60,clip,width=0.49\linewidth]{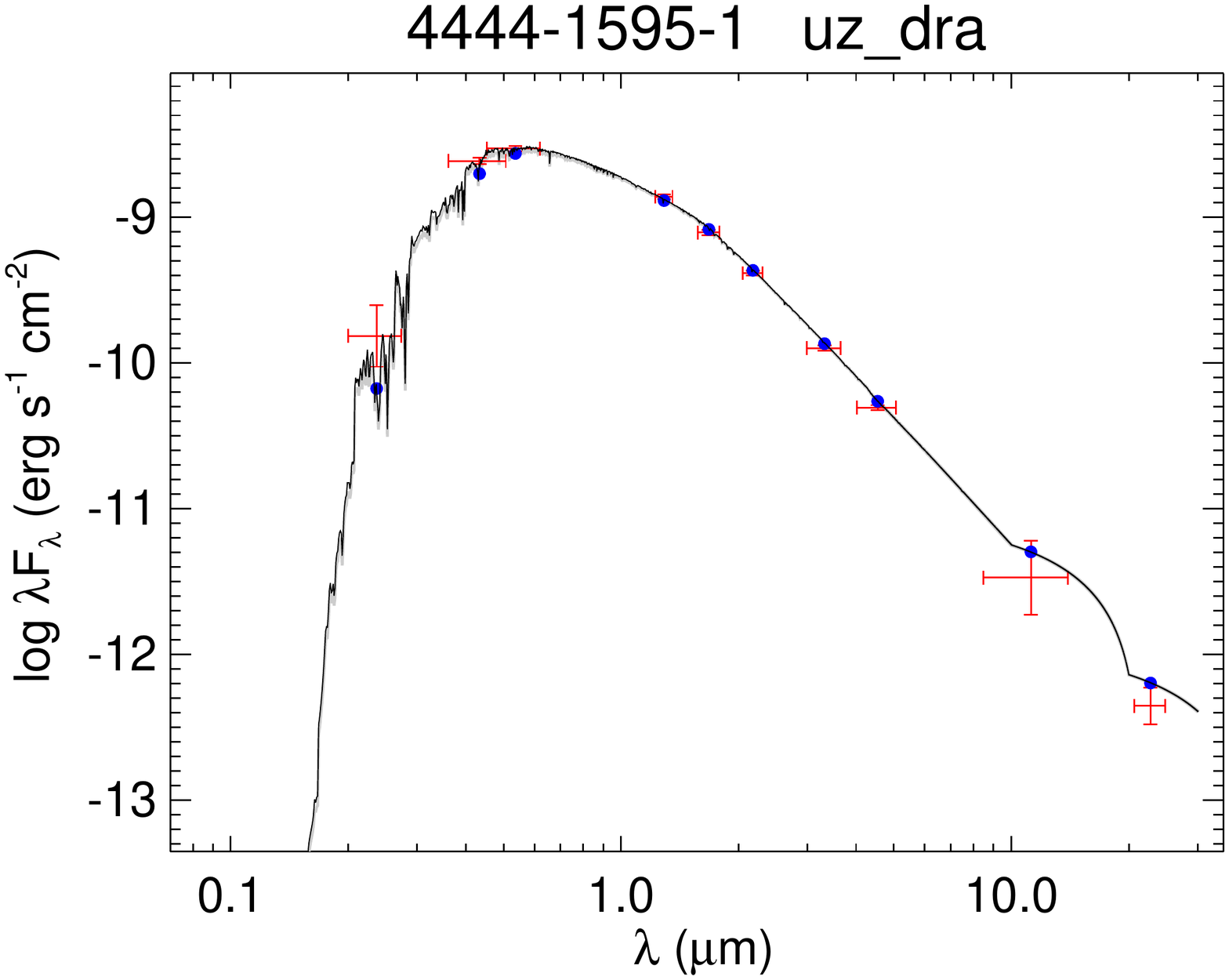}
  \includegraphics[trim=60 60 60 60,clip,width=0.49\linewidth]{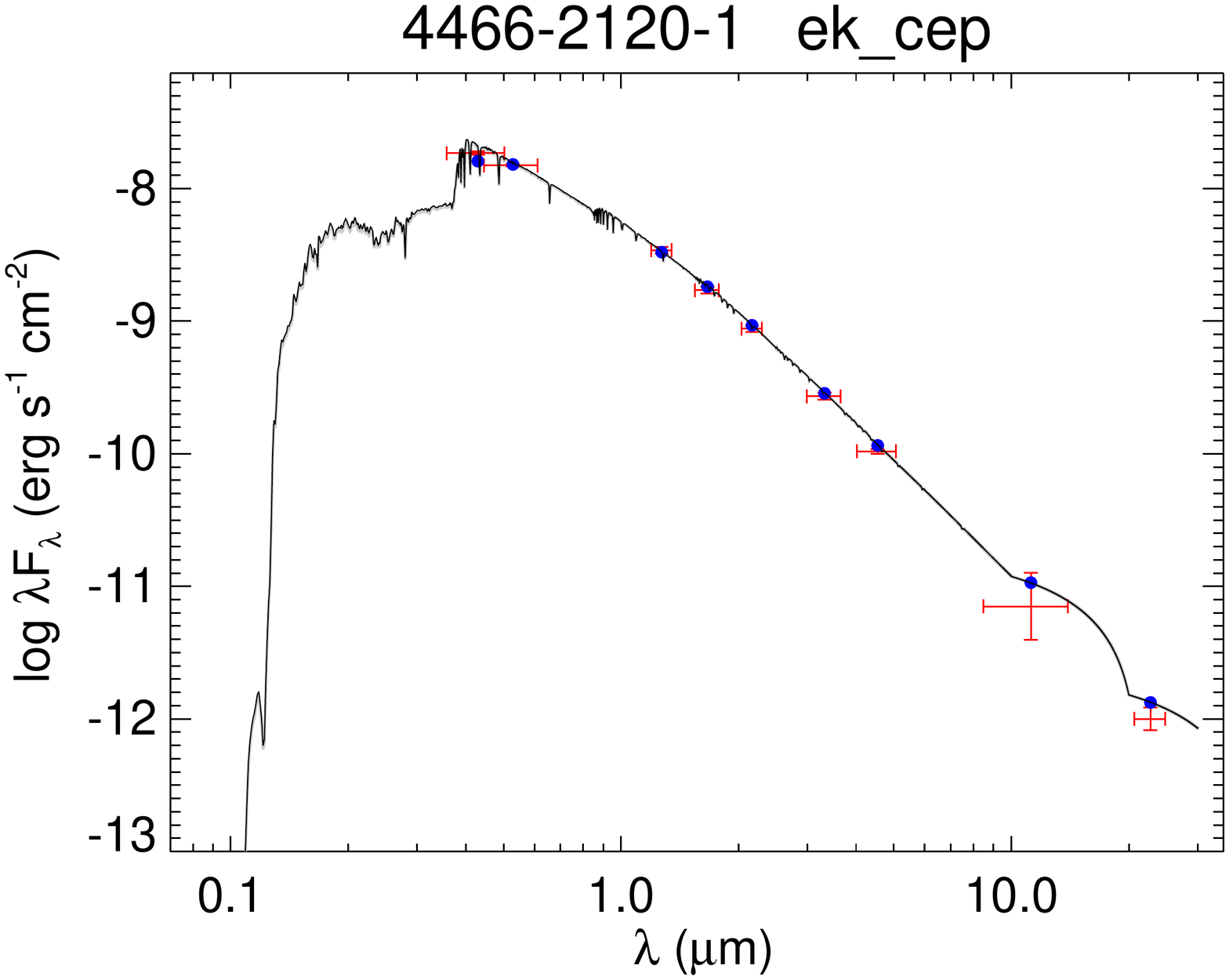}
  \includegraphics[trim=60 60 60 60,clip,width=0.49\linewidth]{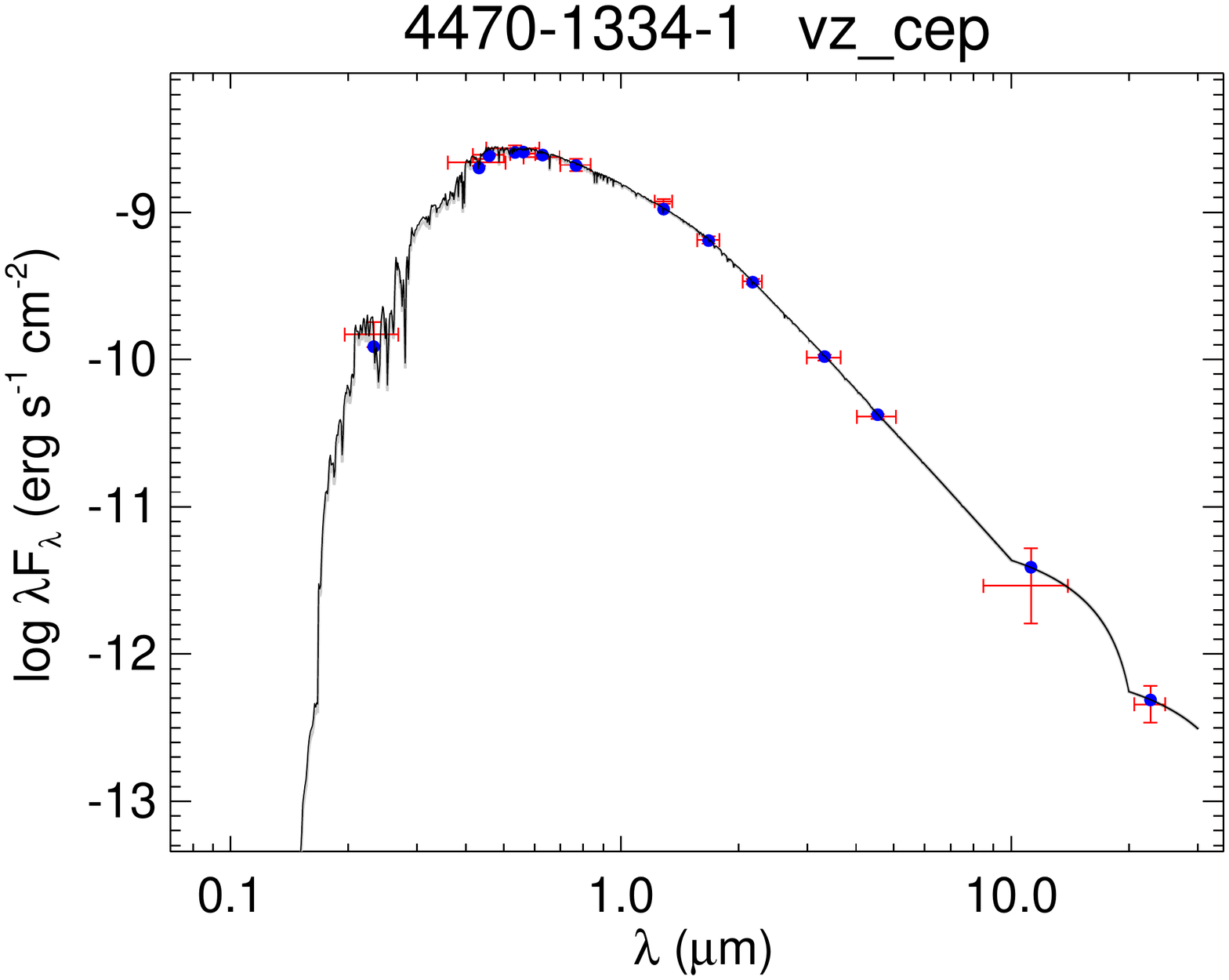}
  \includegraphics[trim=60 60 60 60,clip,width=0.49\linewidth]{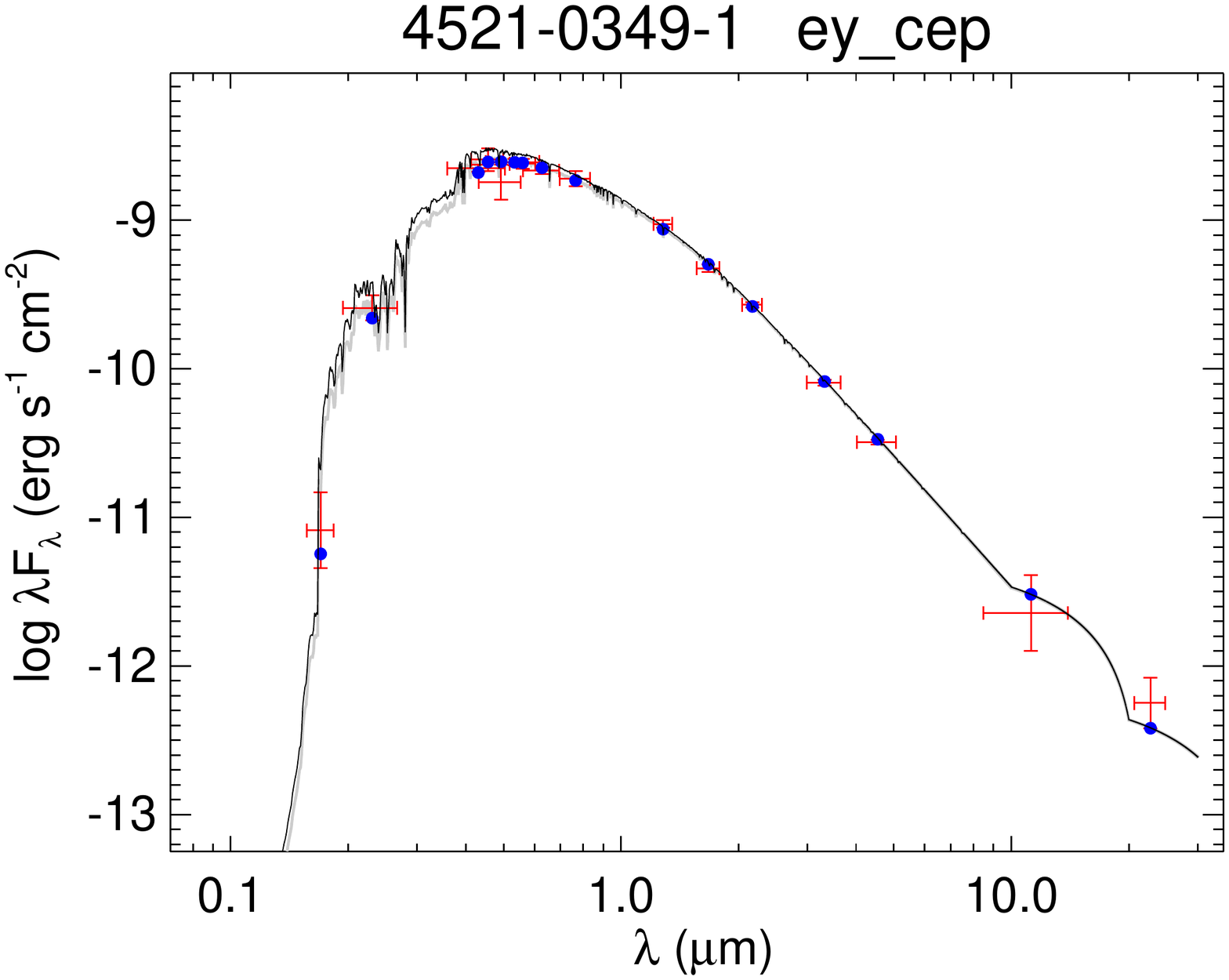}
  \caption{All labels, lines, symbols, and colors as in Figure \ref{fig:seds}.}
  \label{fig:seds_16}
\end{figure}

\begin{figure}[H]
  \centering
  \includegraphics[trim=60 60 60 60,clip,width=0.49\linewidth]{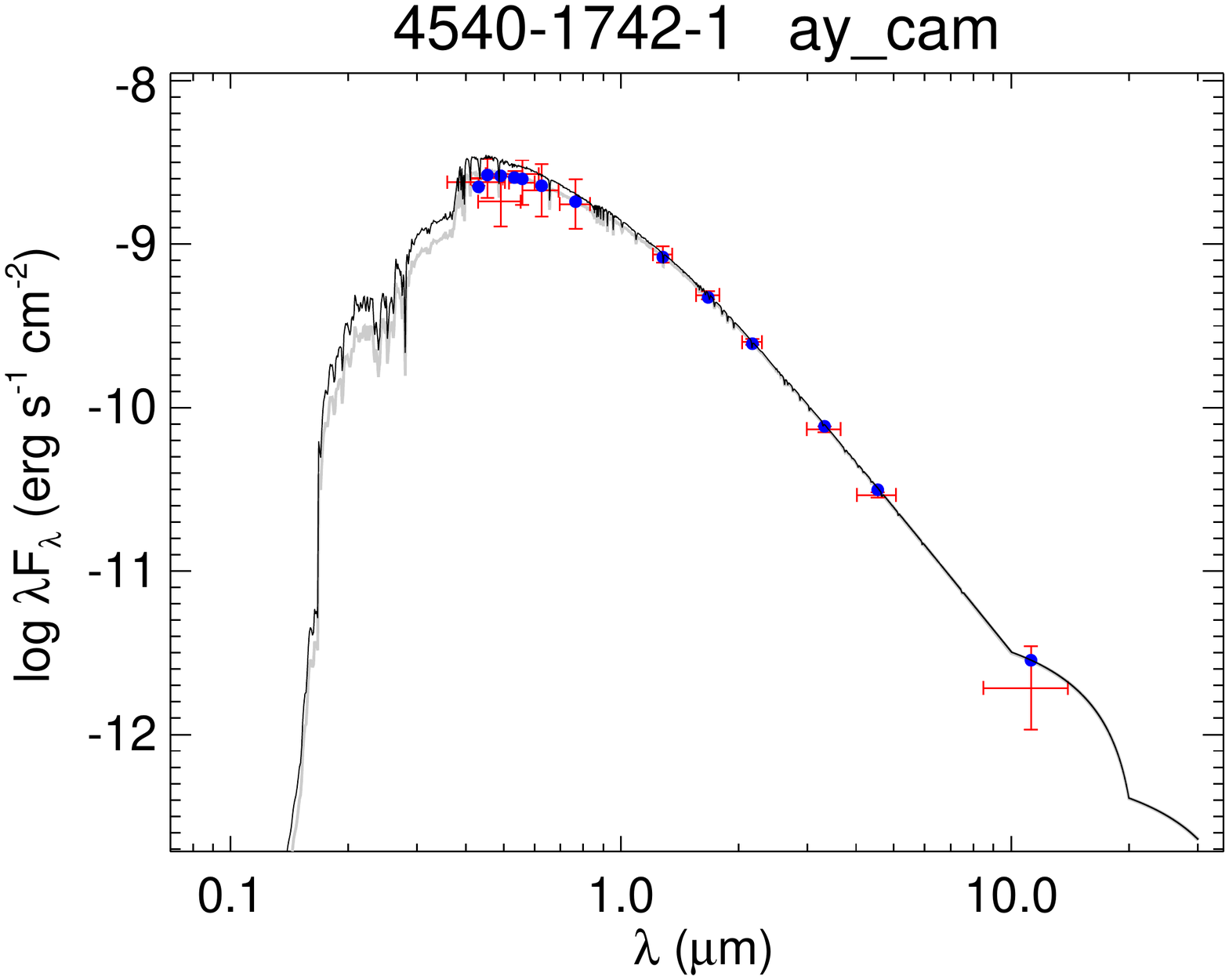}
  \includegraphics[trim=60 60 60 60,clip,width=0.49\linewidth]{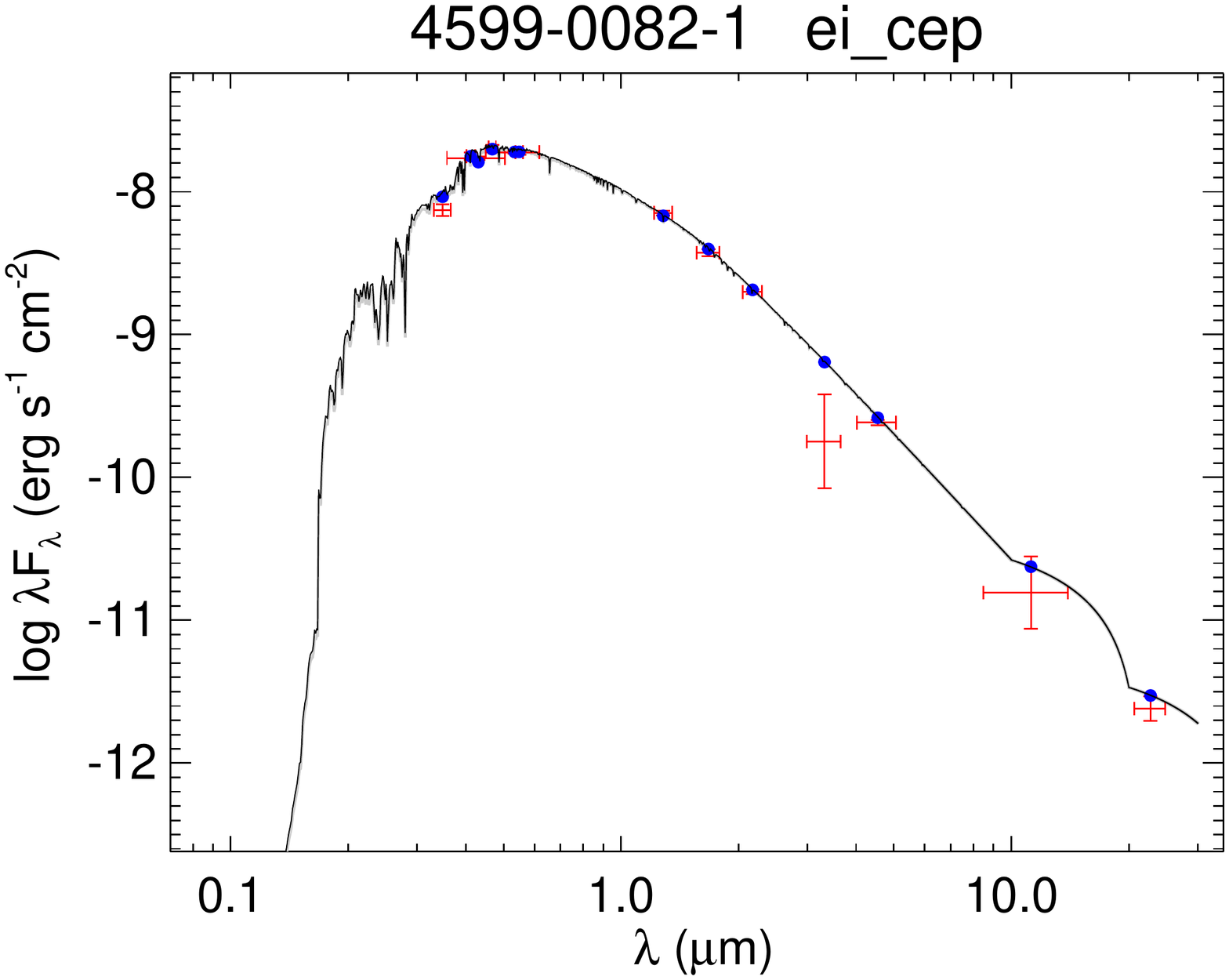}
  \includegraphics[trim=60 60 60 60,clip,width=0.49\linewidth]{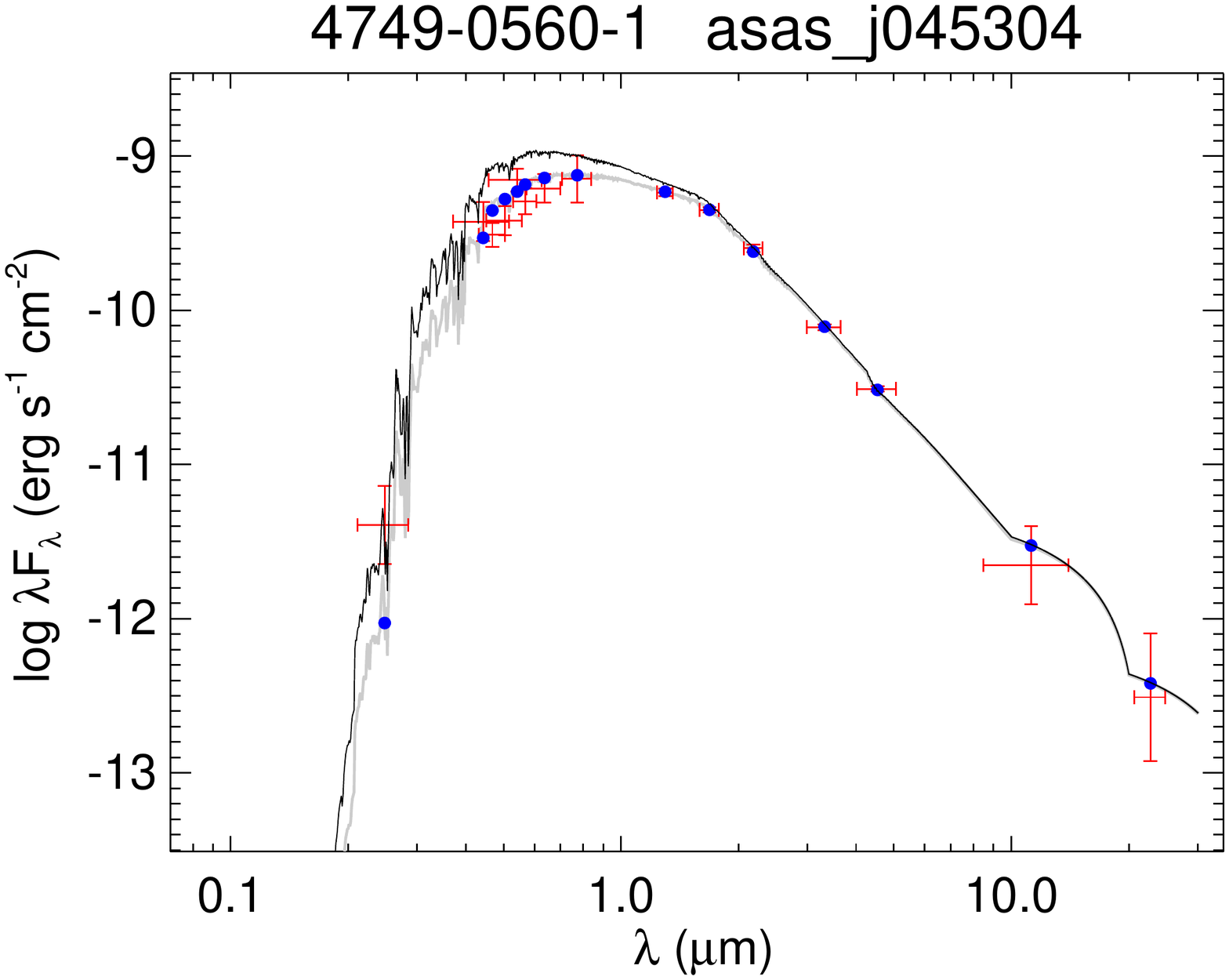}
  \includegraphics[trim=60 60 60 60,clip,width=0.49\linewidth]{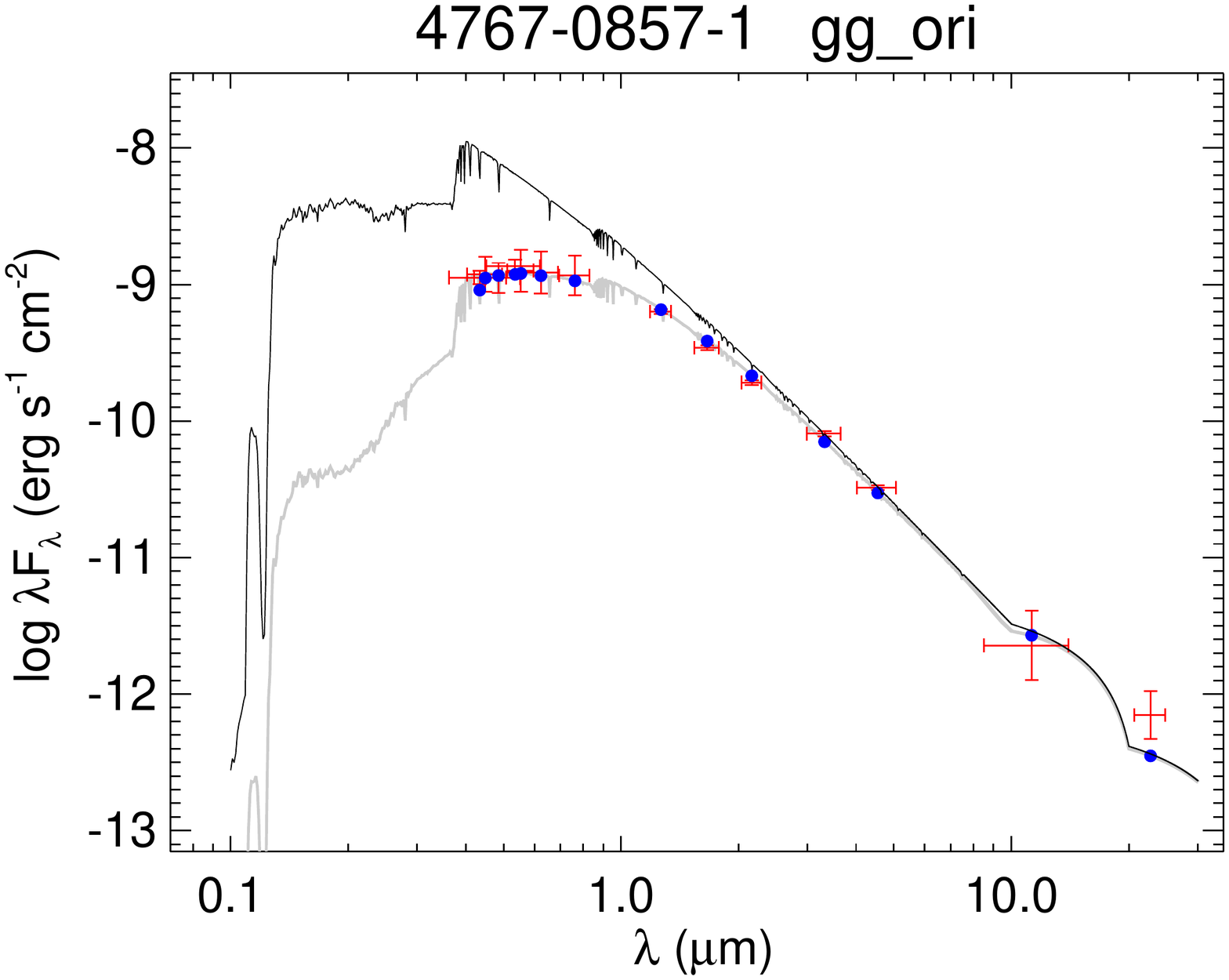}
  \includegraphics[trim=60 60 60 60,clip,width=0.49\linewidth]{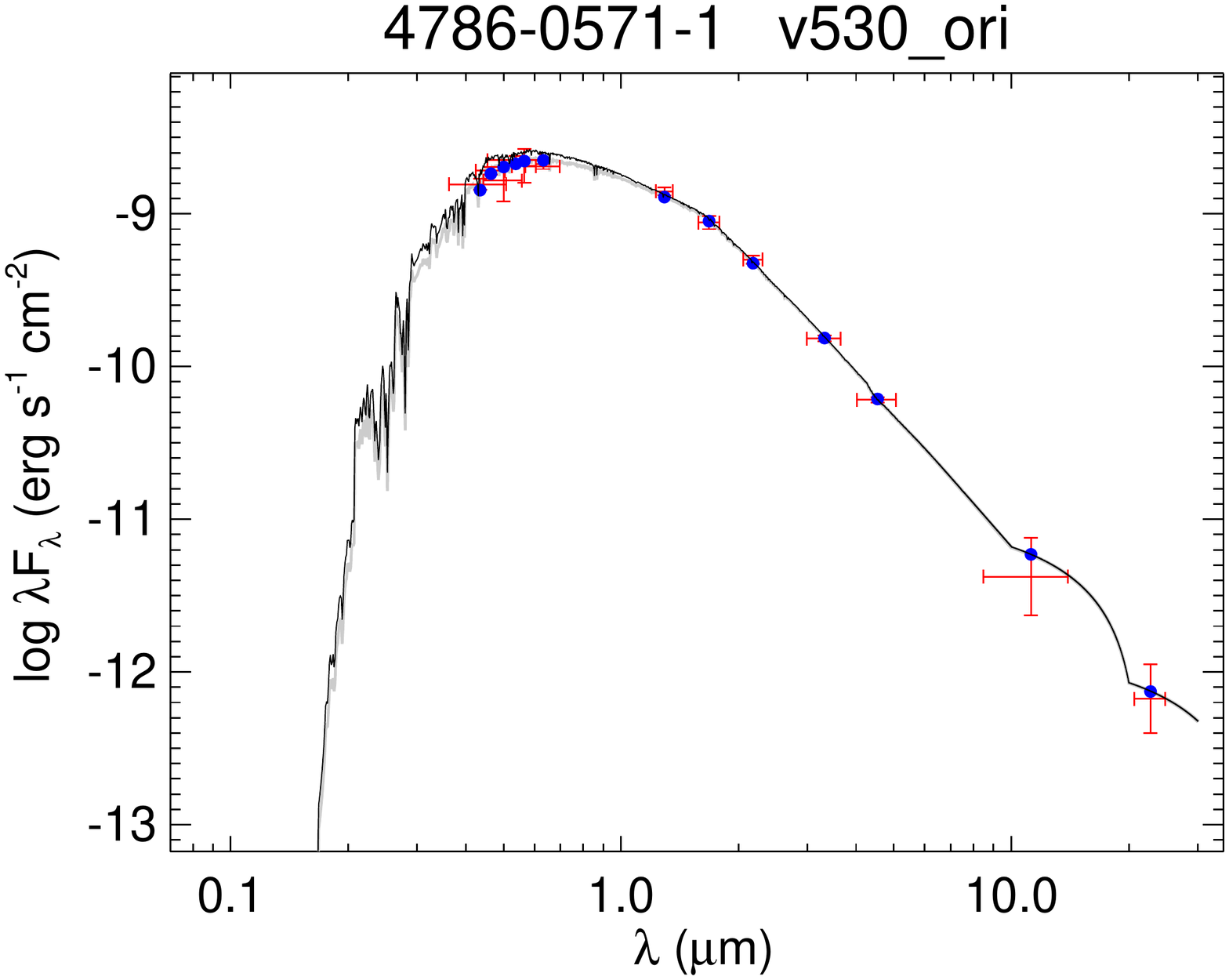}
  \includegraphics[trim=60 60 60 60,clip,width=0.49\linewidth]{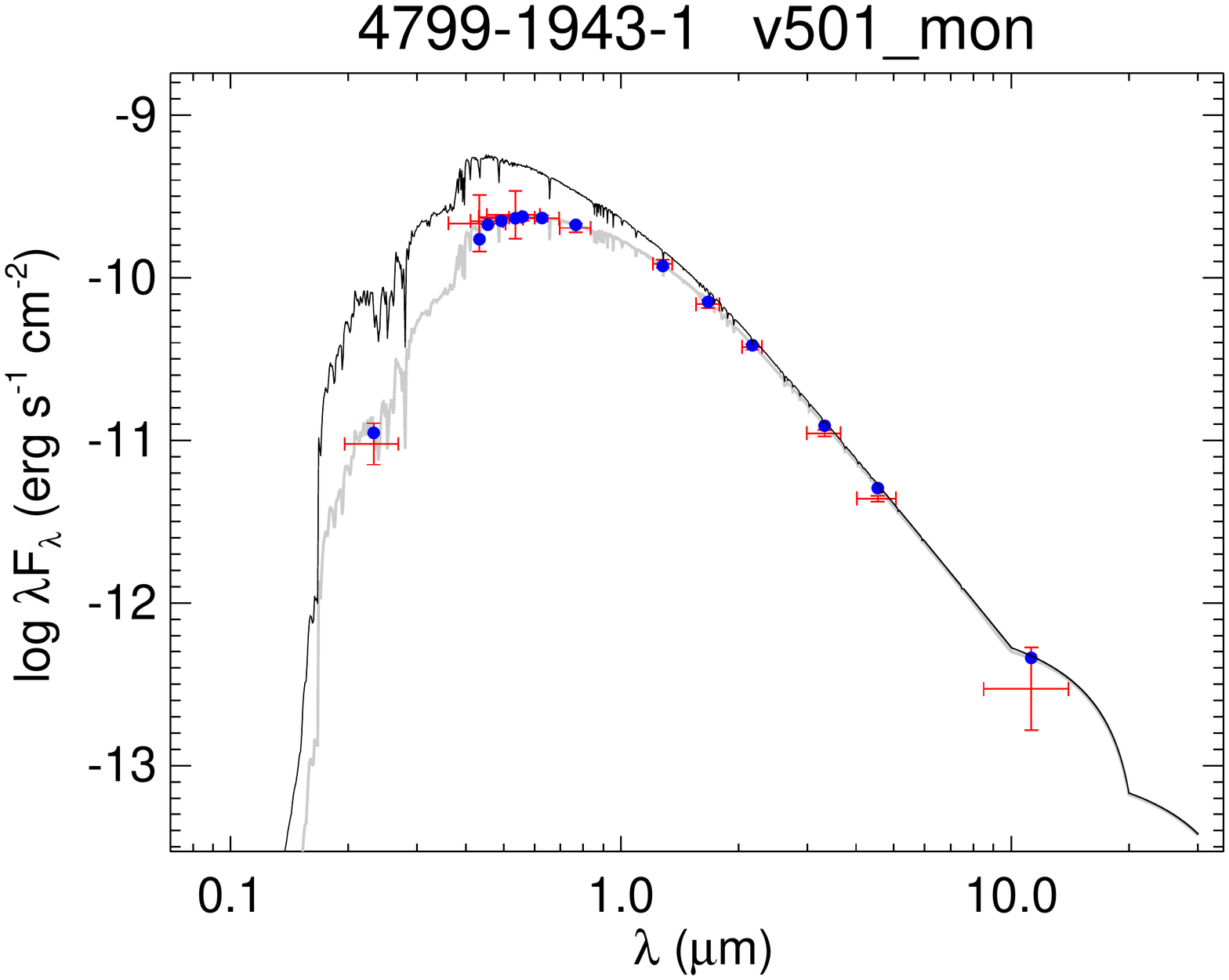}
  \caption{All labels, lines, symbols, and colors as in Figure \ref{fig:seds}.}
  \label{fig:seds_17}
\end{figure}

\begin{figure}[H]
  \centering
  \includegraphics[trim=60 60 60 60,clip,width=0.49\linewidth]{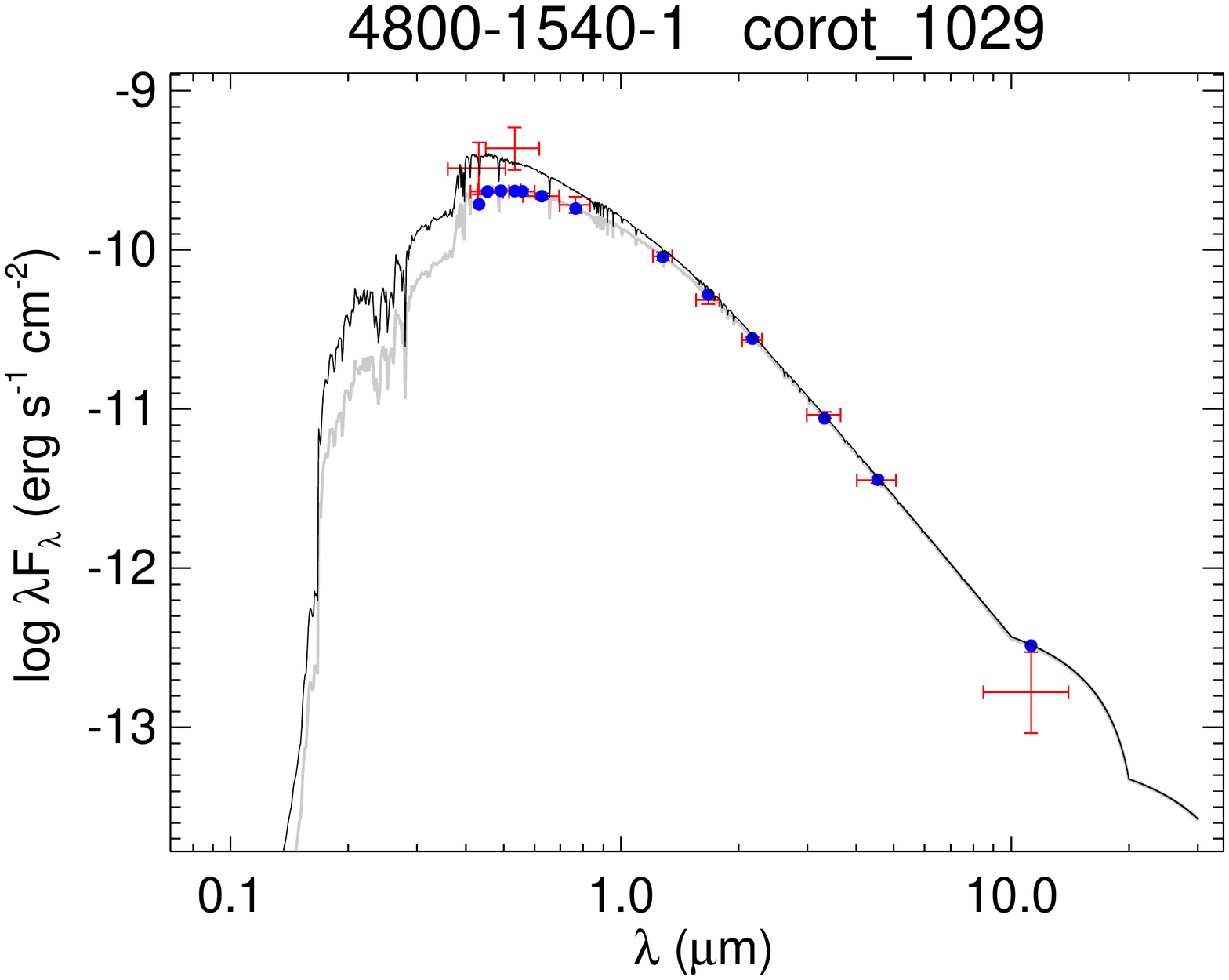}
  \includegraphics[trim=60 60 60 60,clip,width=0.49\linewidth]{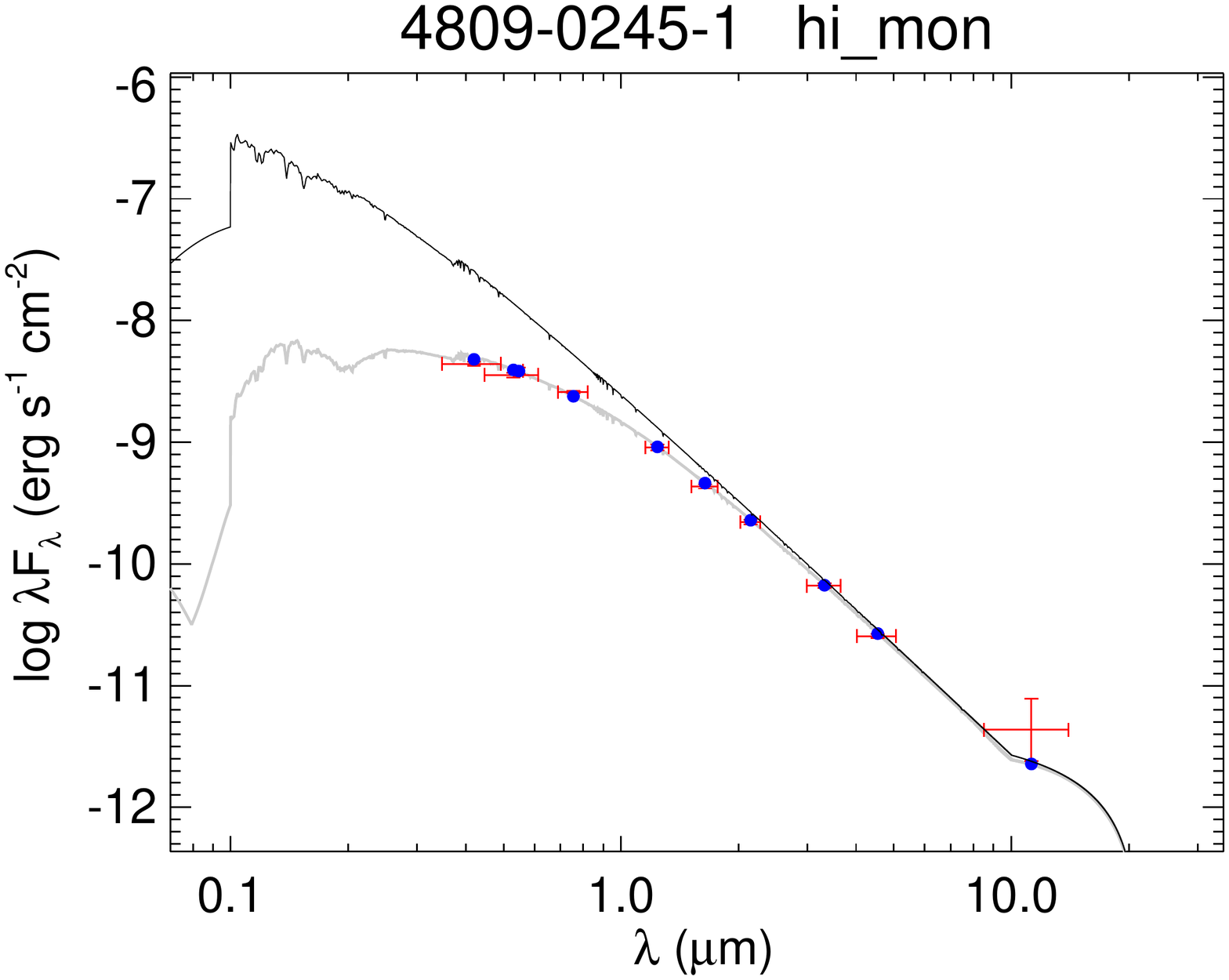}
  \includegraphics[trim=60 60 60 60,clip,width=0.49\linewidth]{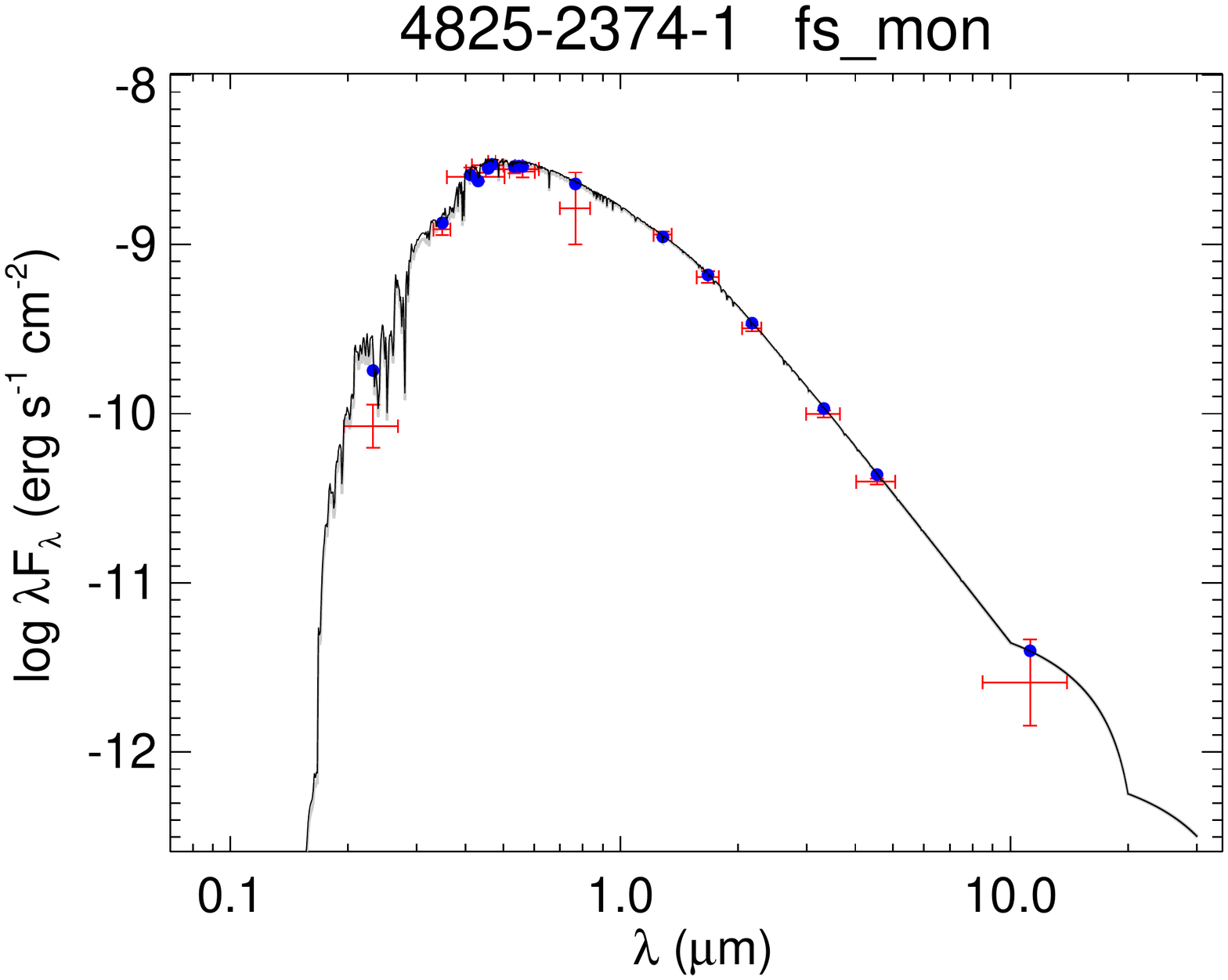}
  \includegraphics[trim=60 60 60 60,clip,width=0.49\linewidth]{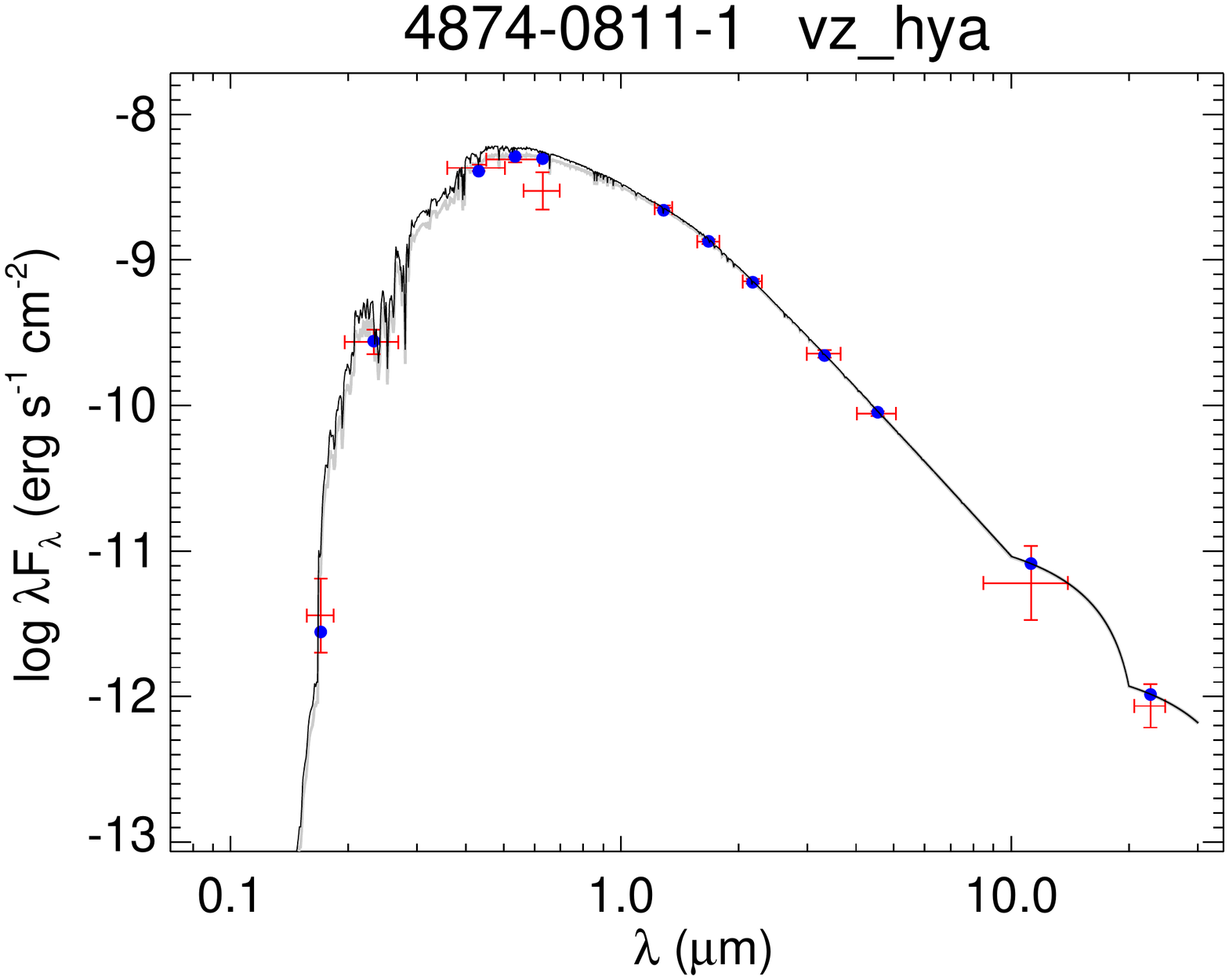}
  \includegraphics[trim=60 60 60 60,clip,width=0.49\linewidth]{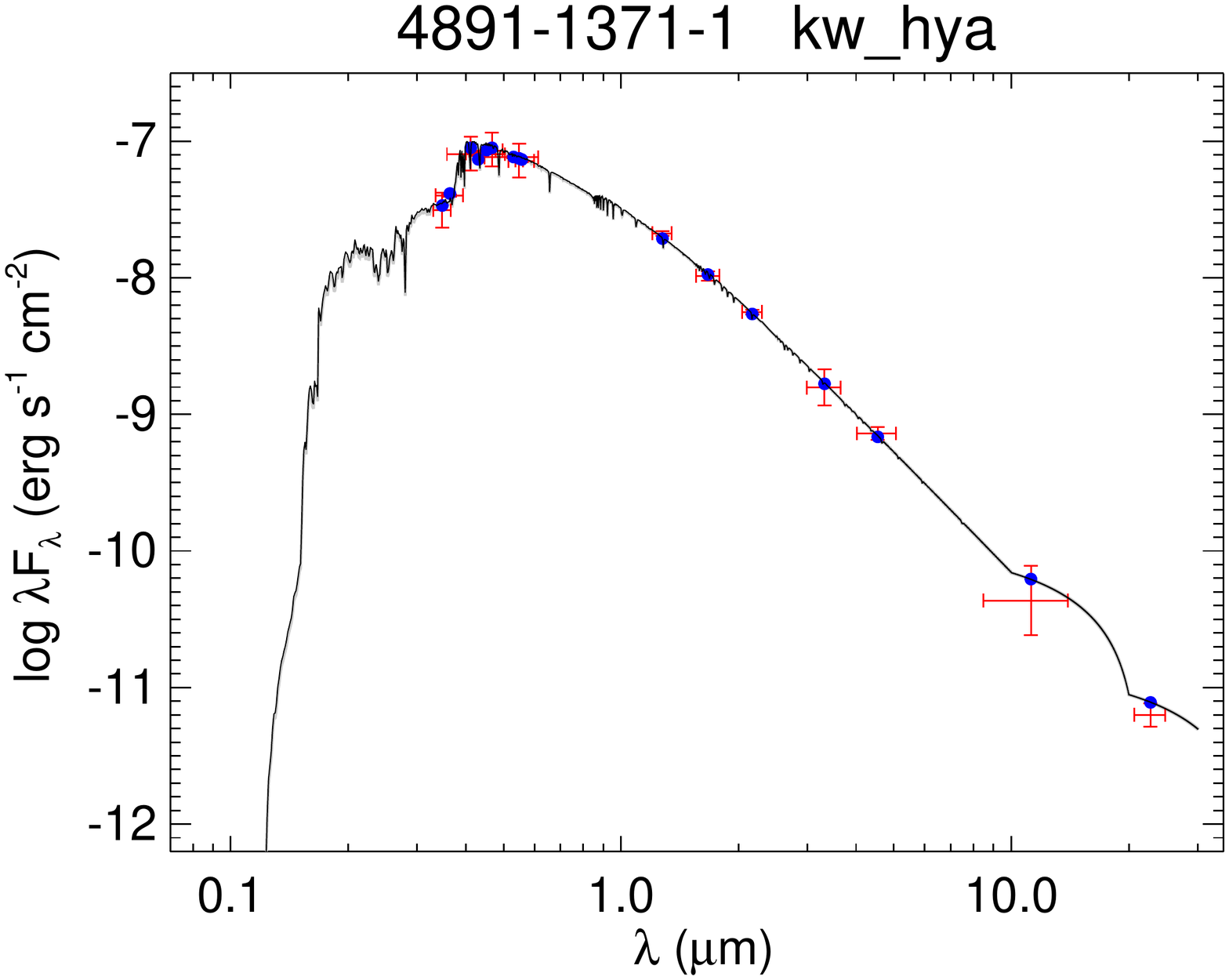}
  \includegraphics[trim=60 60 60 60,clip,width=0.49\linewidth]{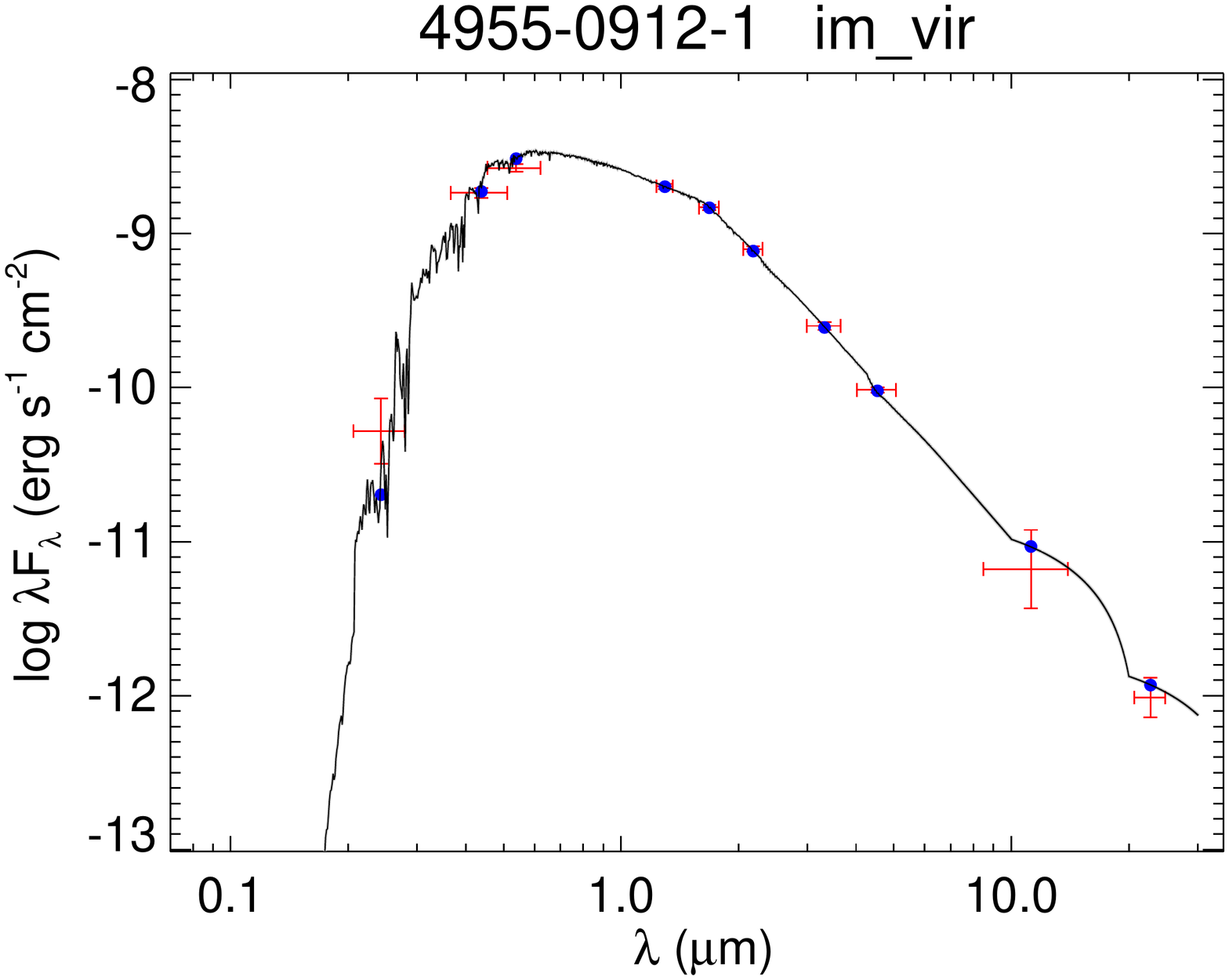}
  \caption{All labels, lines, symbols, and colors as in Figure \ref{fig:seds}.}
  \label{fig:seds_18}
\end{figure}

\begin{figure}[H]
  \centering
  \includegraphics[trim=60 60 60 60,clip,width=0.49\linewidth]{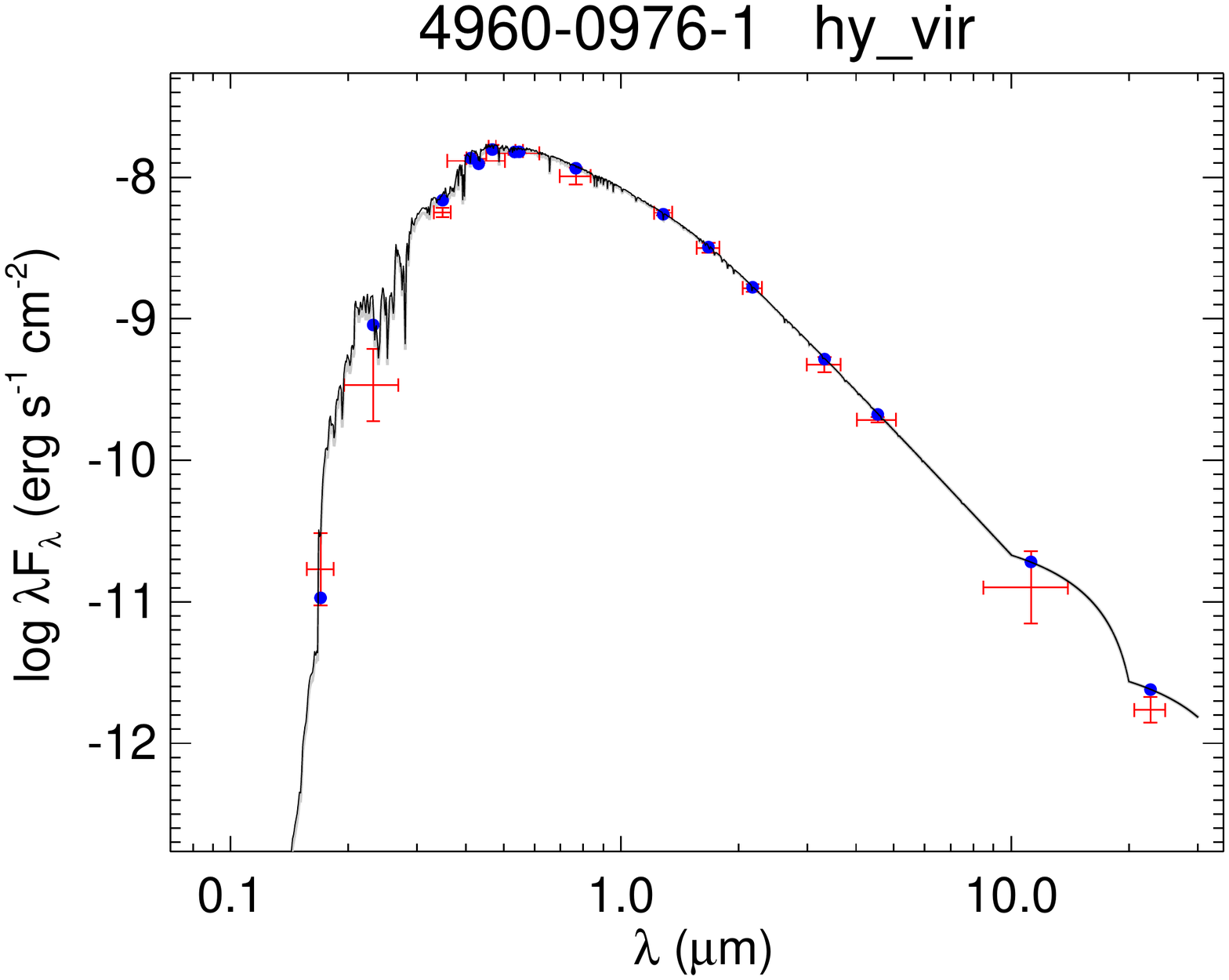}
  \includegraphics[trim=60 60 60 60,clip,width=0.49\linewidth]{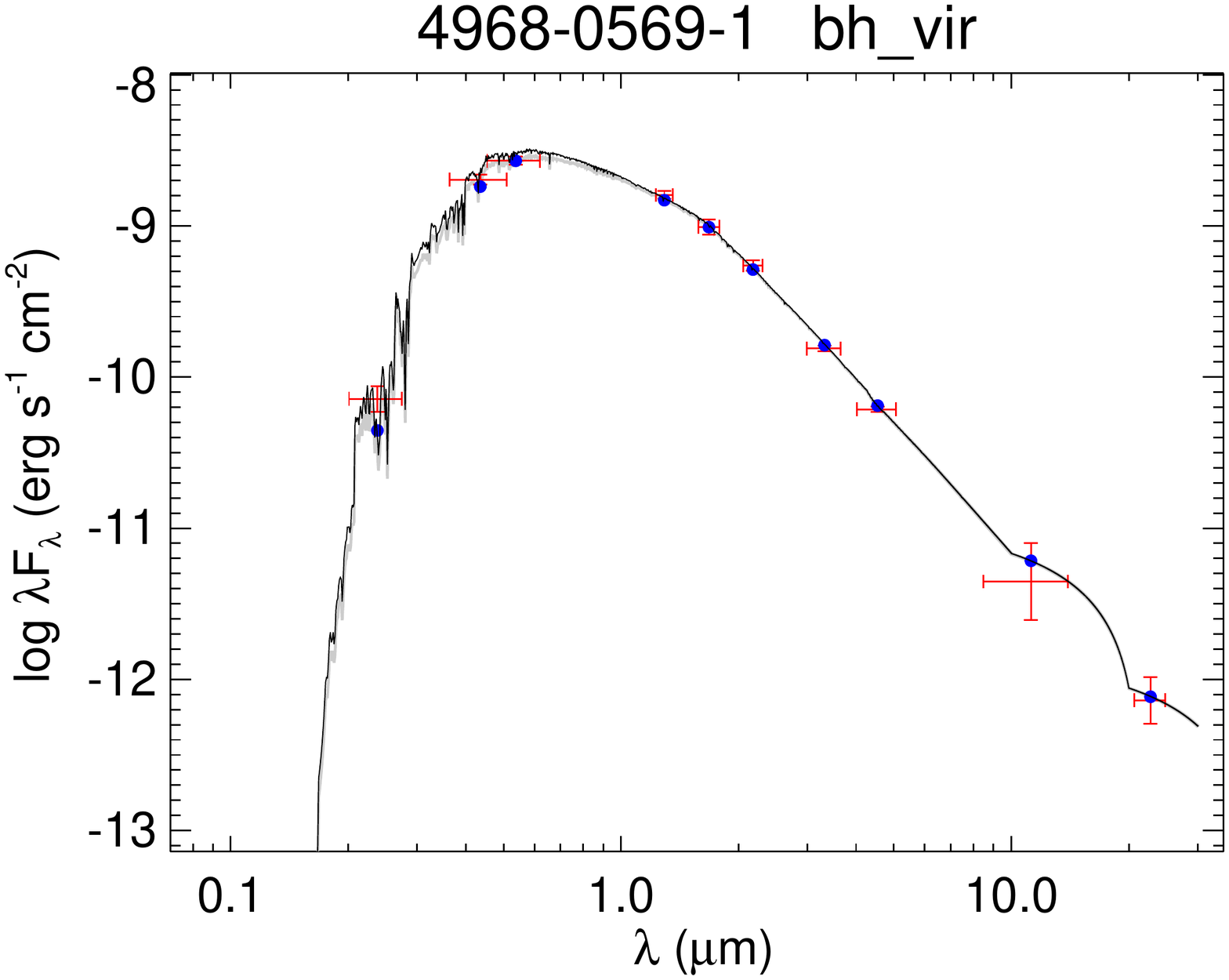}
  \includegraphics[trim=60 60 60 60,clip,width=0.49\linewidth]{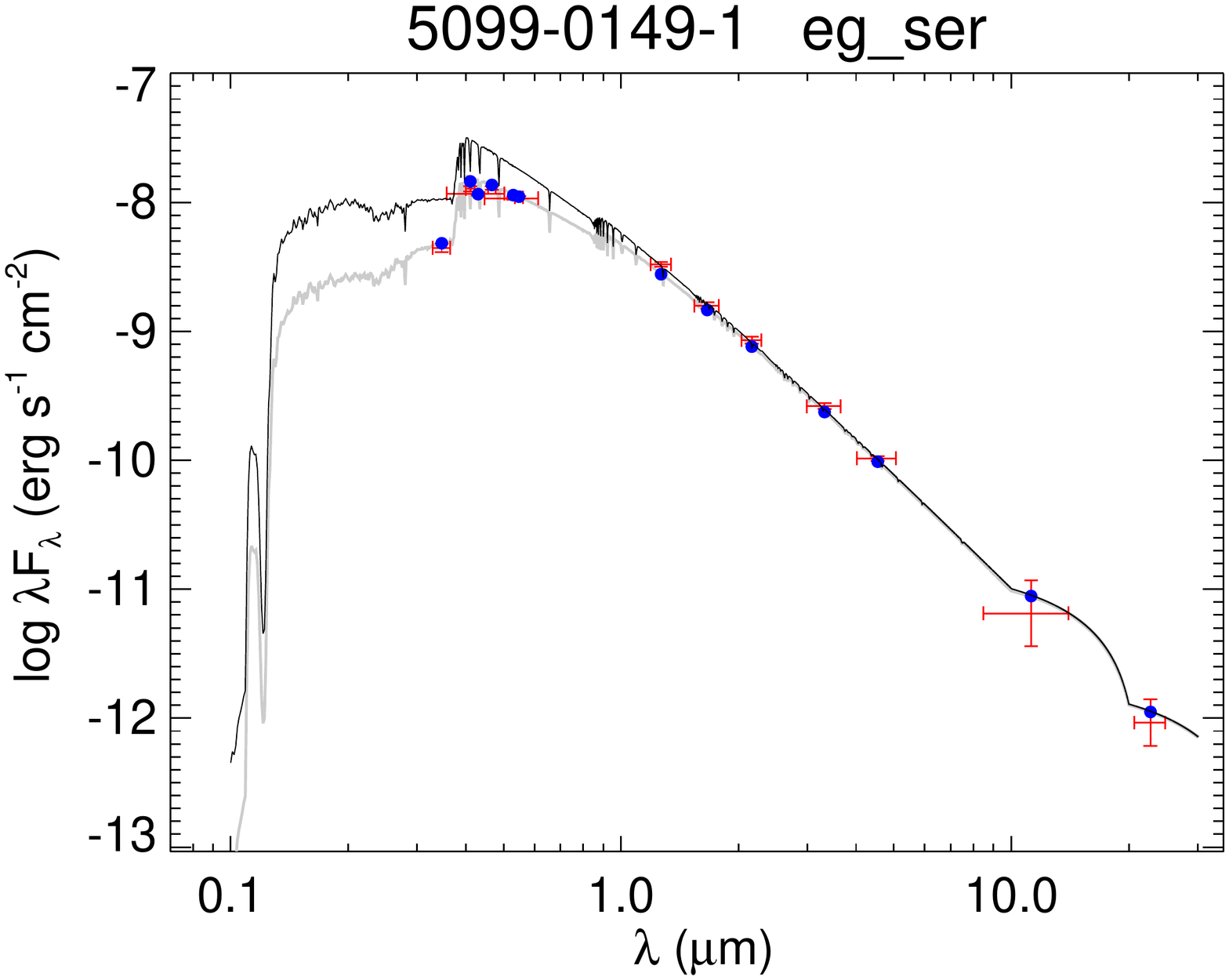}
  \includegraphics[trim=60 60 60 60,clip,width=0.49\linewidth]{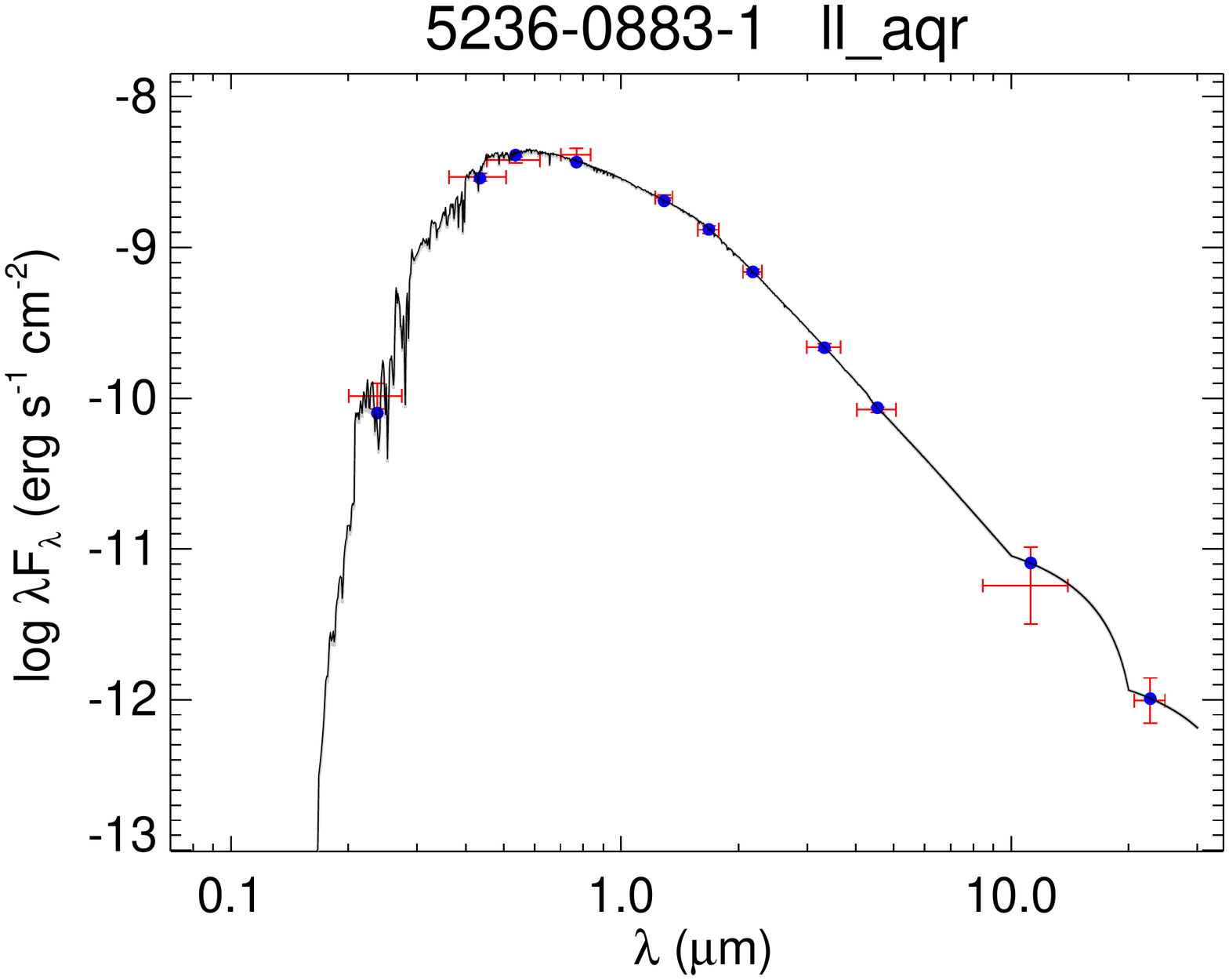}
  \includegraphics[trim=60 60 60 60,clip,width=0.49\linewidth]{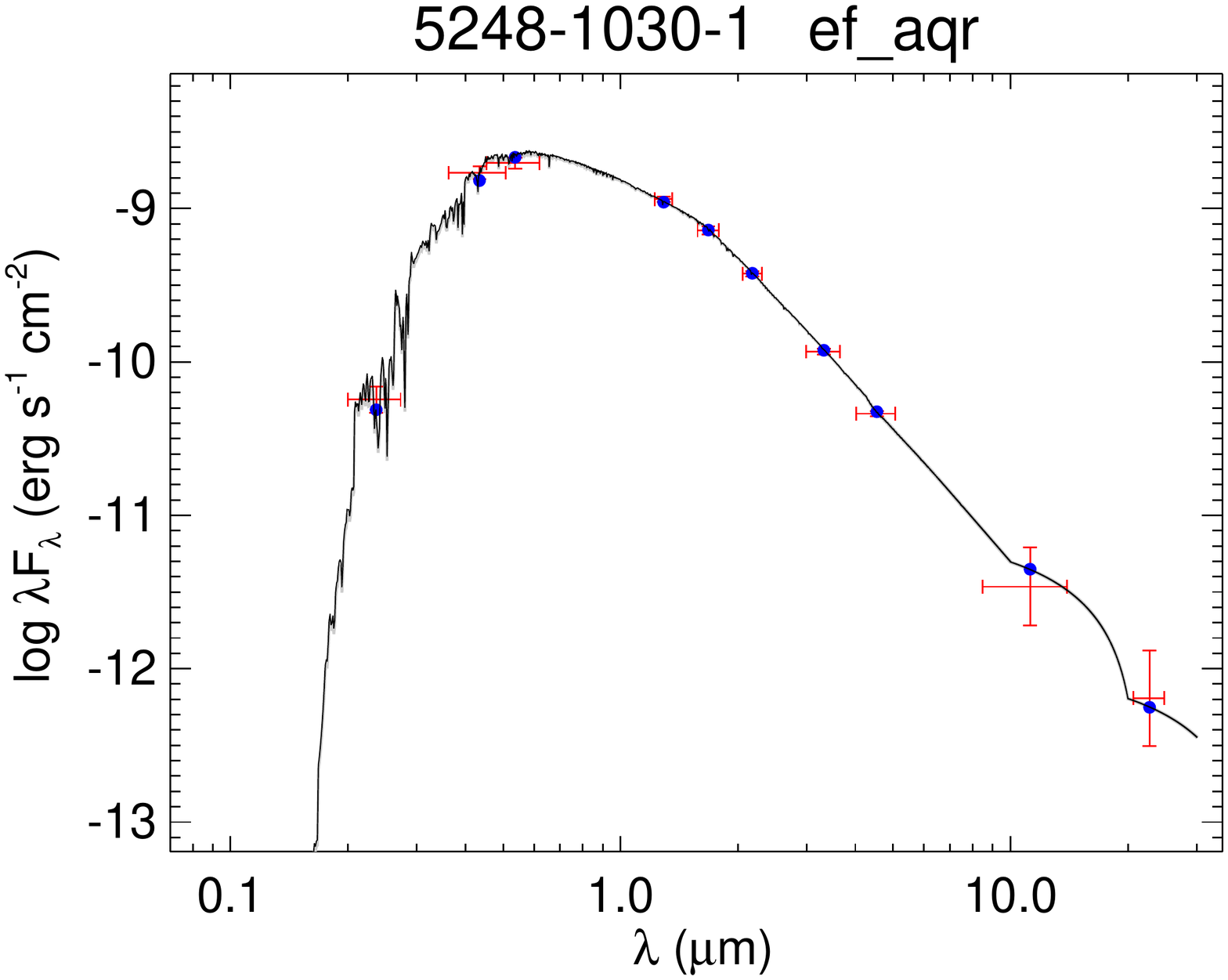}
  \includegraphics[trim=60 60 60 60,clip,width=0.49\linewidth]{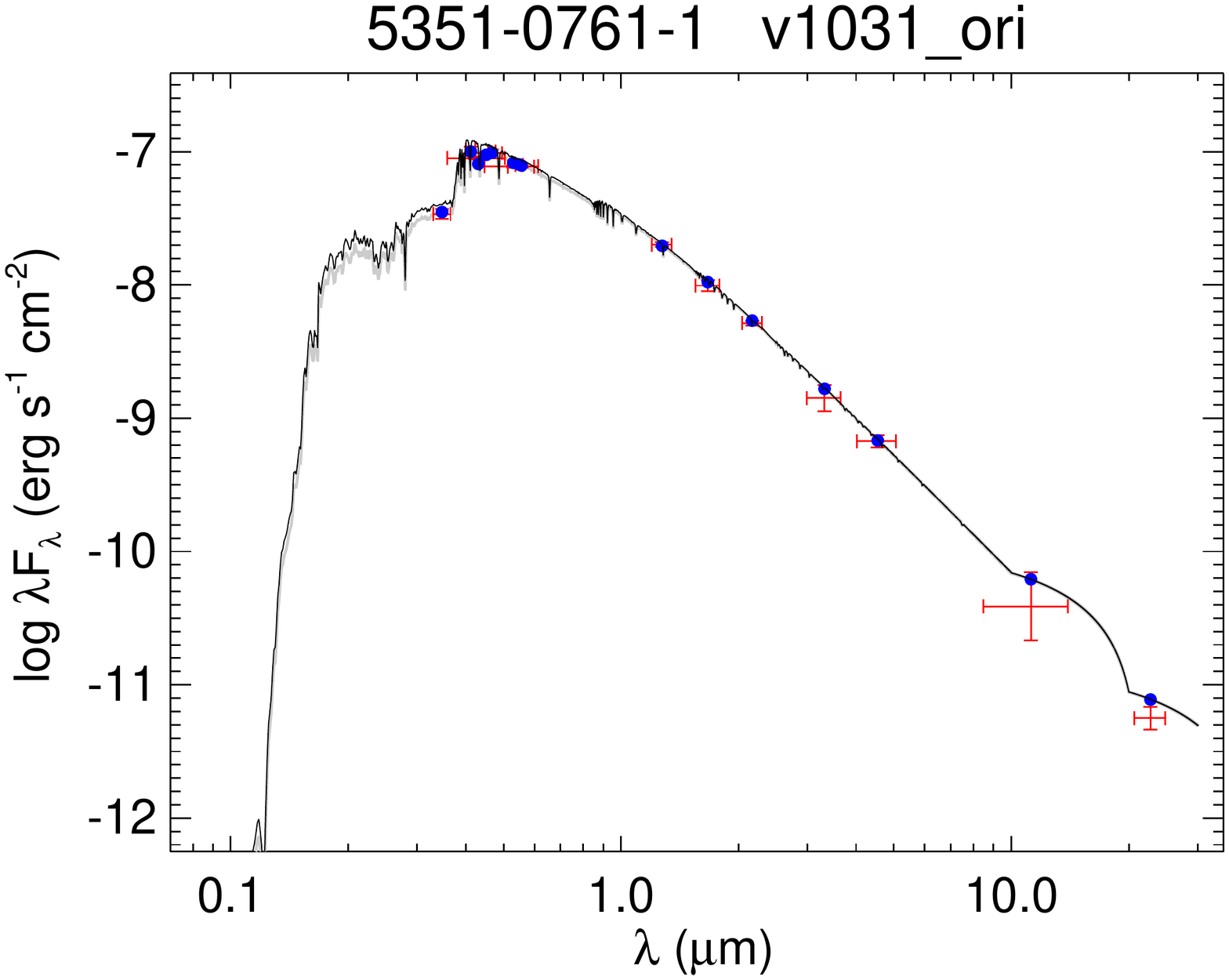}
  \caption{All labels, lines, symbols, and colors as in Figure \ref{fig:seds}.}
  \label{fig:seds_19}
\end{figure}

\begin{figure}[H]
  \centering
  \includegraphics[trim=60 60 60 60,clip,width=0.49\linewidth]{sedfigs/pv_pup.pdf}
  \includegraphics[trim=60 60 60 60,clip,width=0.49\linewidth]{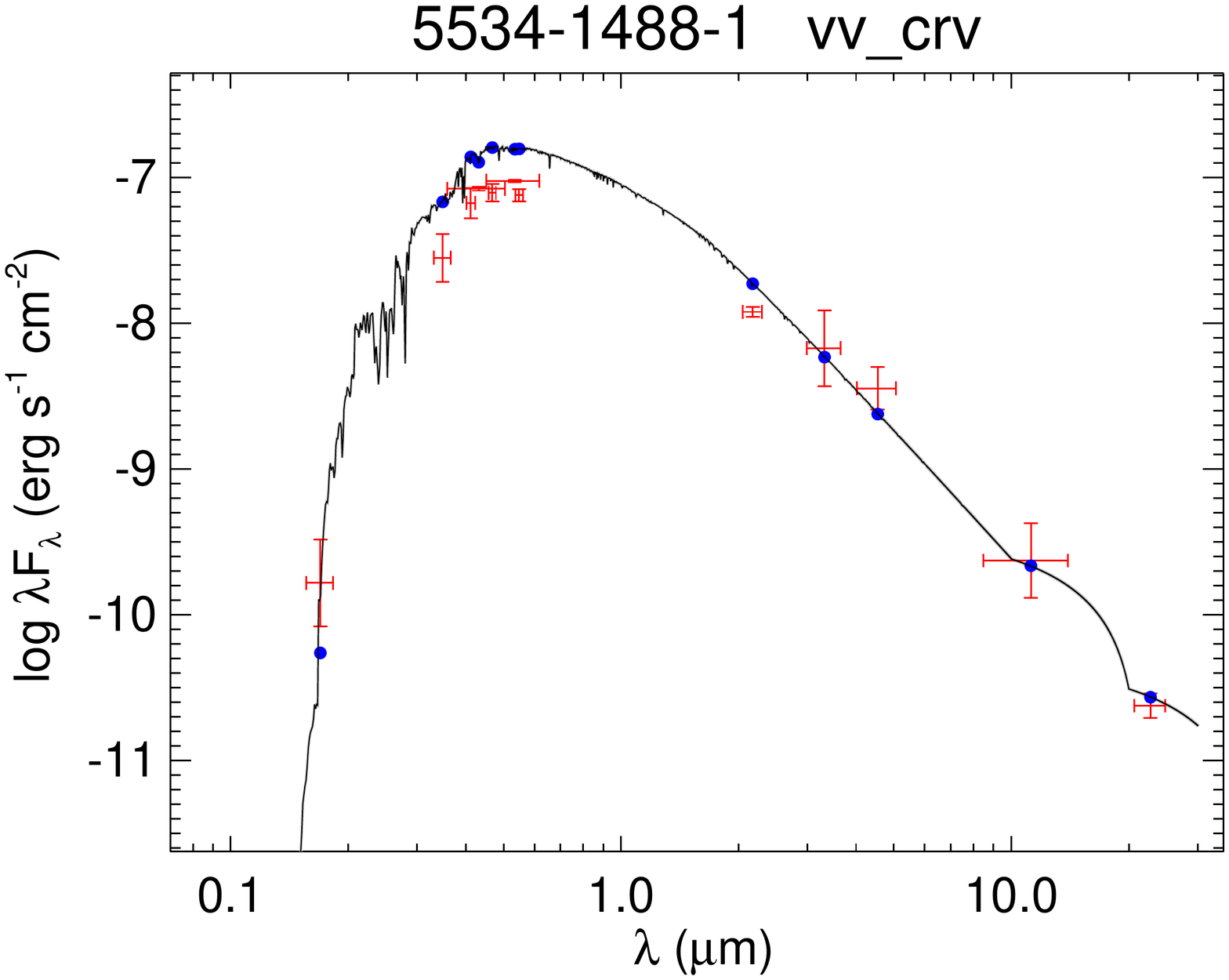}
  \includegraphics[trim=60 60 60 60,clip,width=0.49\linewidth]{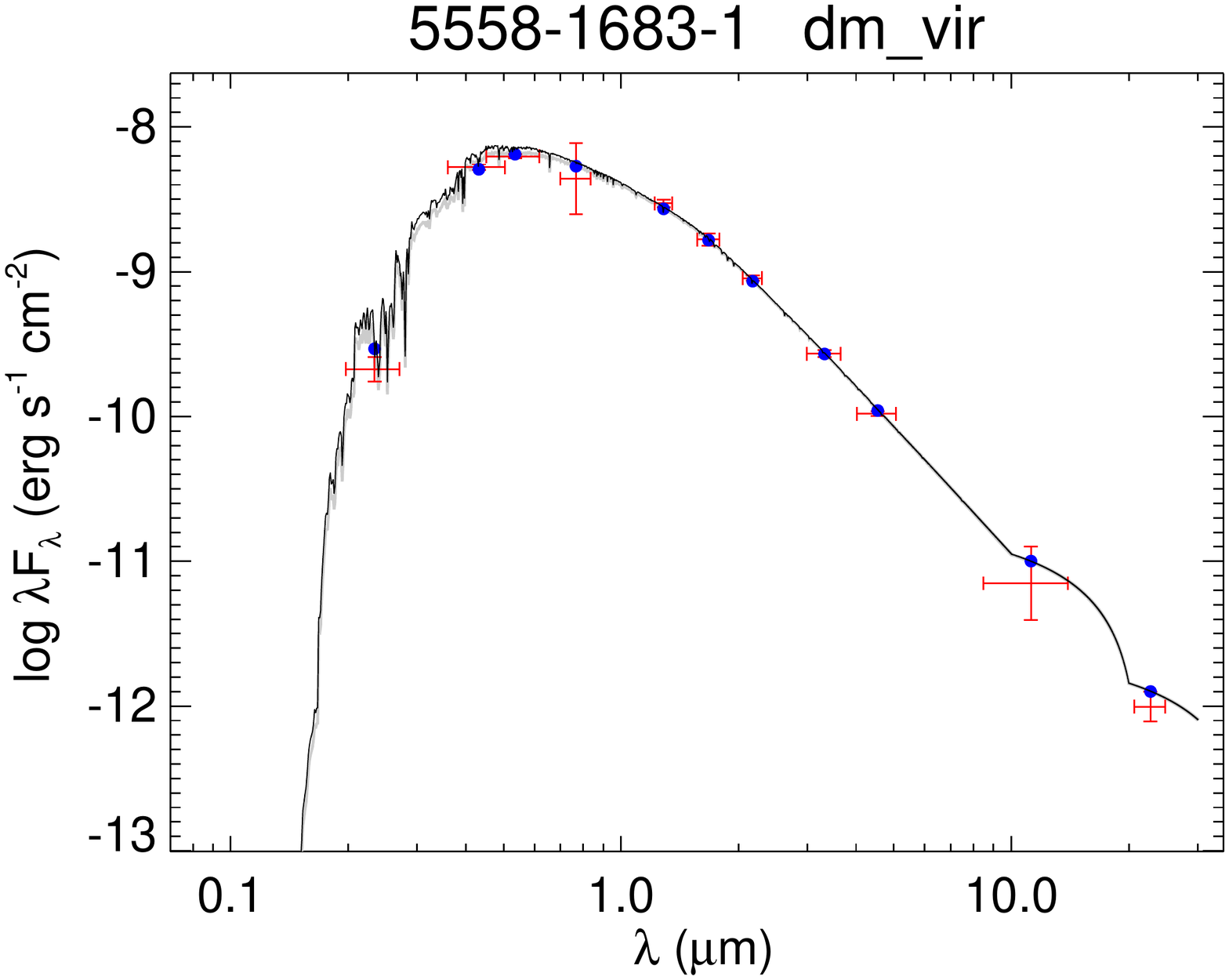}
  \includegraphics[trim=60 60 60 60,clip,width=0.49\linewidth]{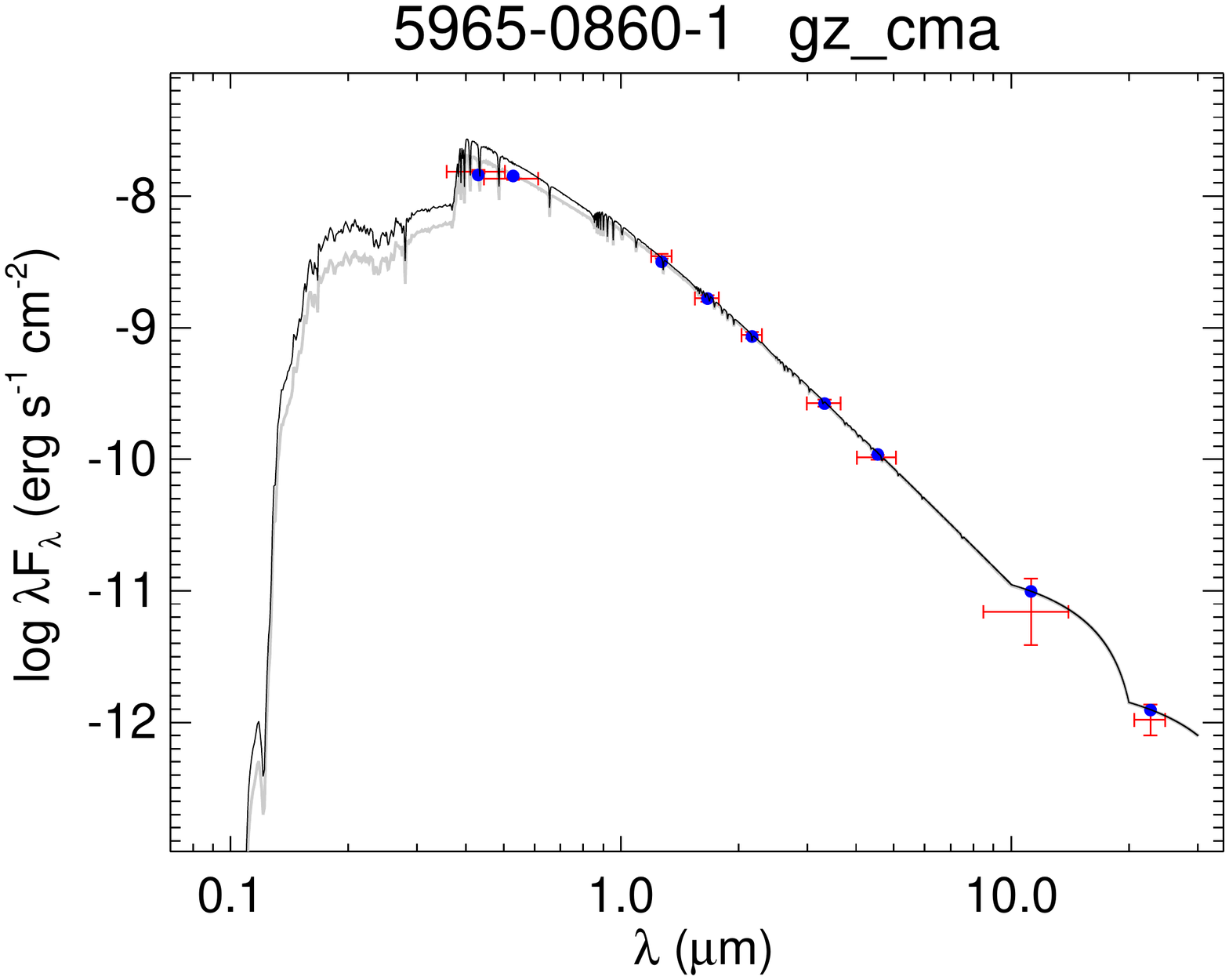}
  \includegraphics[trim=60 60 60 60,clip,width=0.49\linewidth]{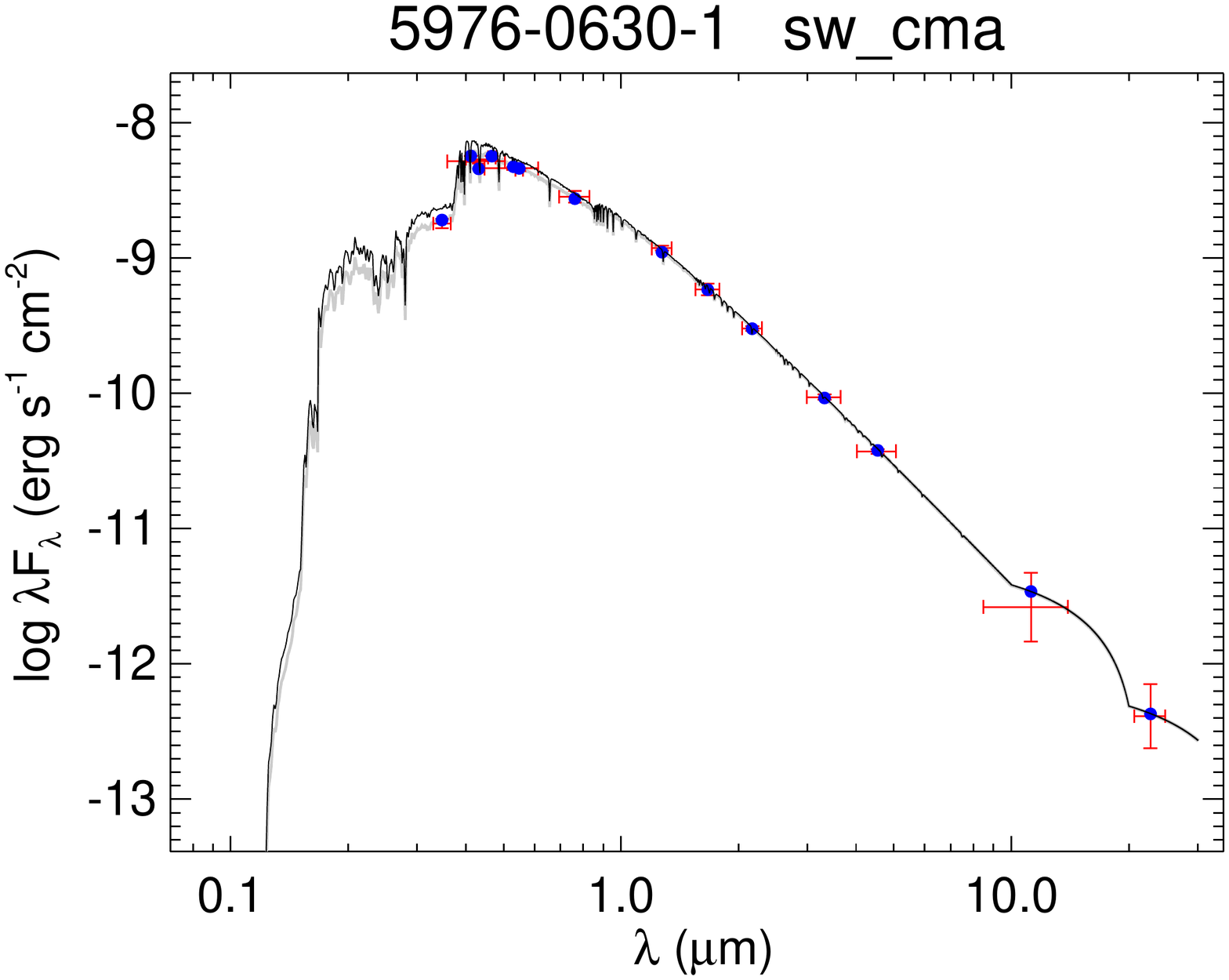}
  \includegraphics[trim=60 60 60 60,clip,width=0.49\linewidth]{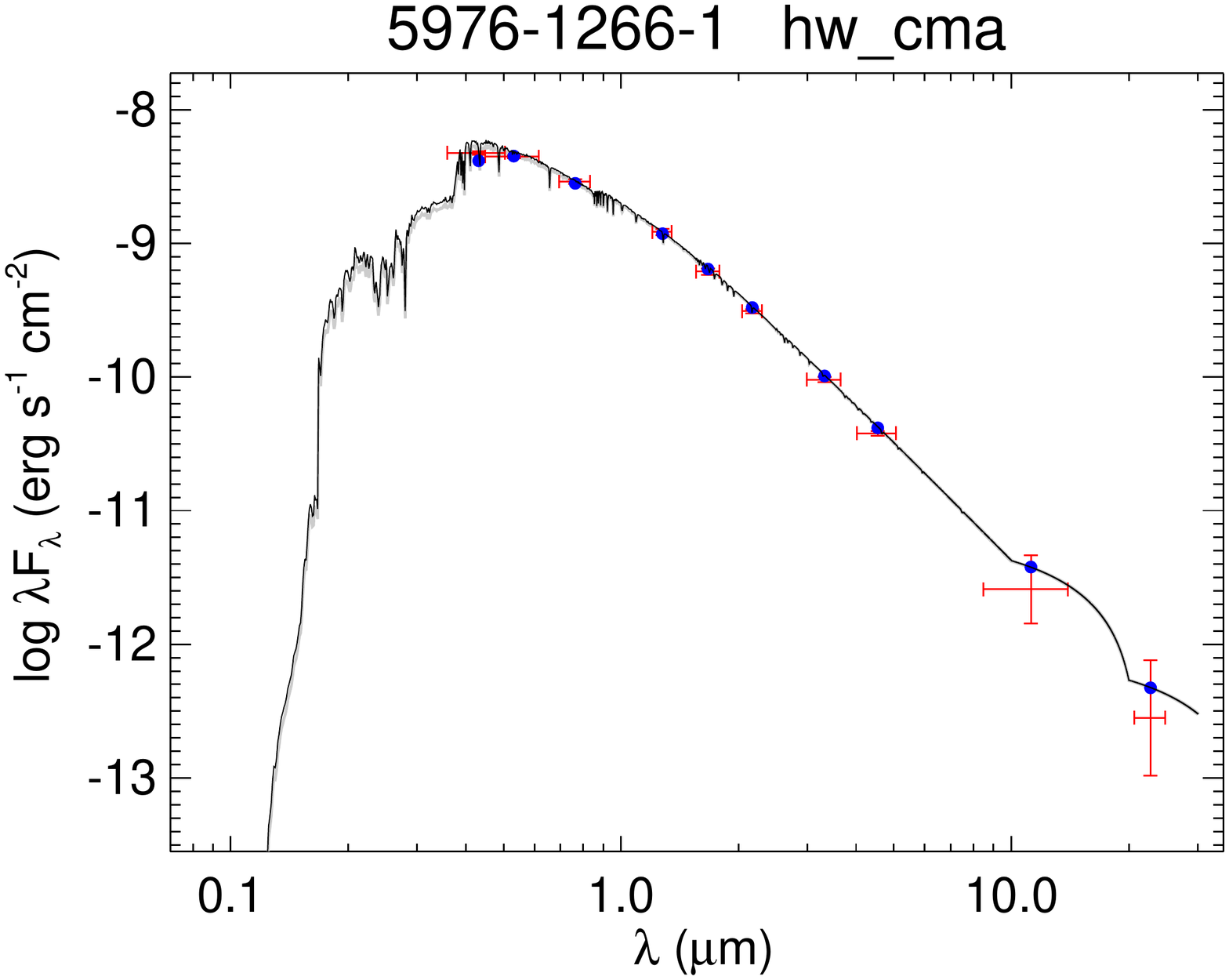}
  \caption{All labels, lines, symbols, and colors as in Figure \ref{fig:seds}.}
  \label{fig:seds_20}
\end{figure}

\begin{figure}[H]
  \centering
  \includegraphics[trim=60 60 60 60,clip,width=0.49\linewidth]{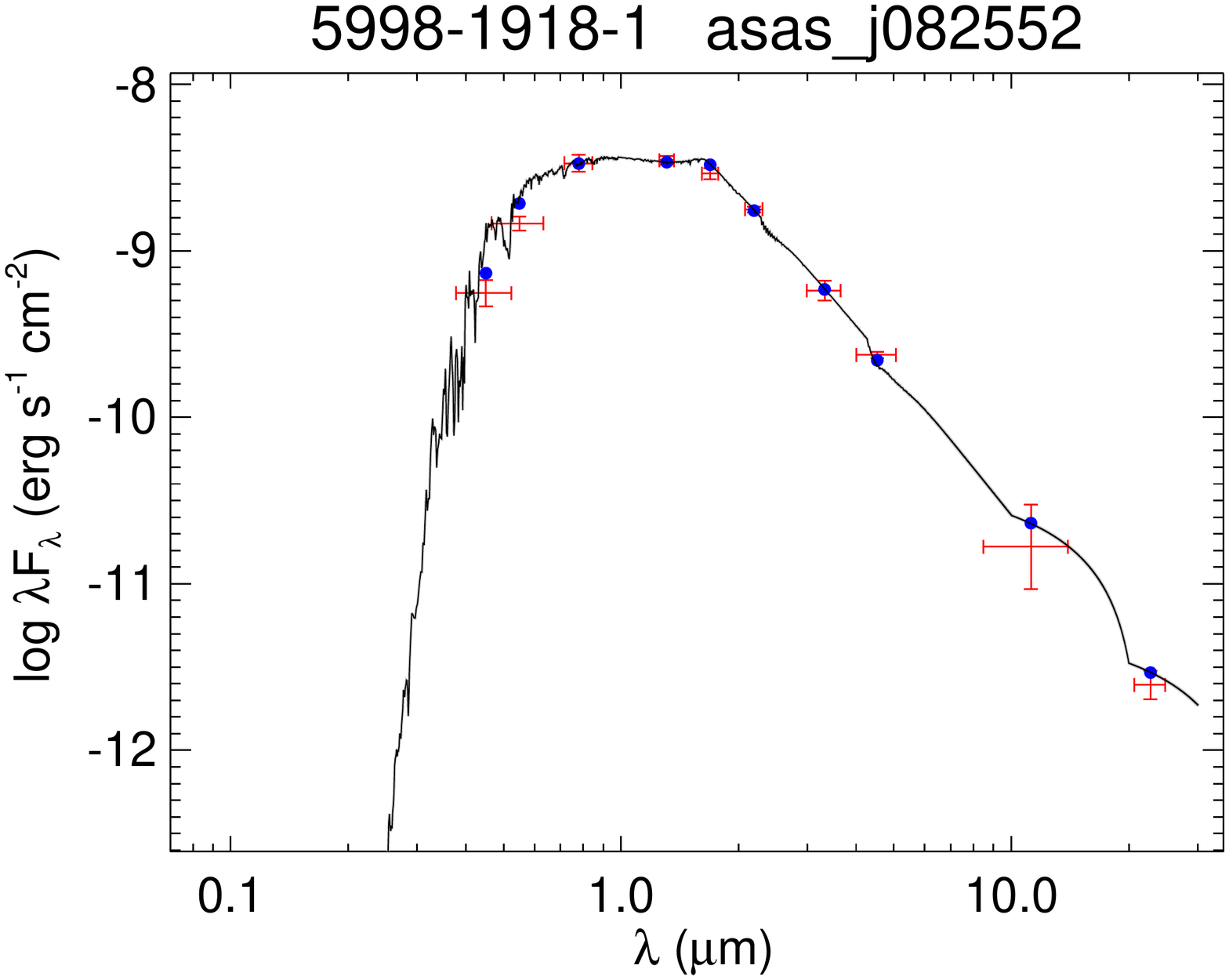}
  \includegraphics[trim=60 60 60 60,clip,width=0.49\linewidth]{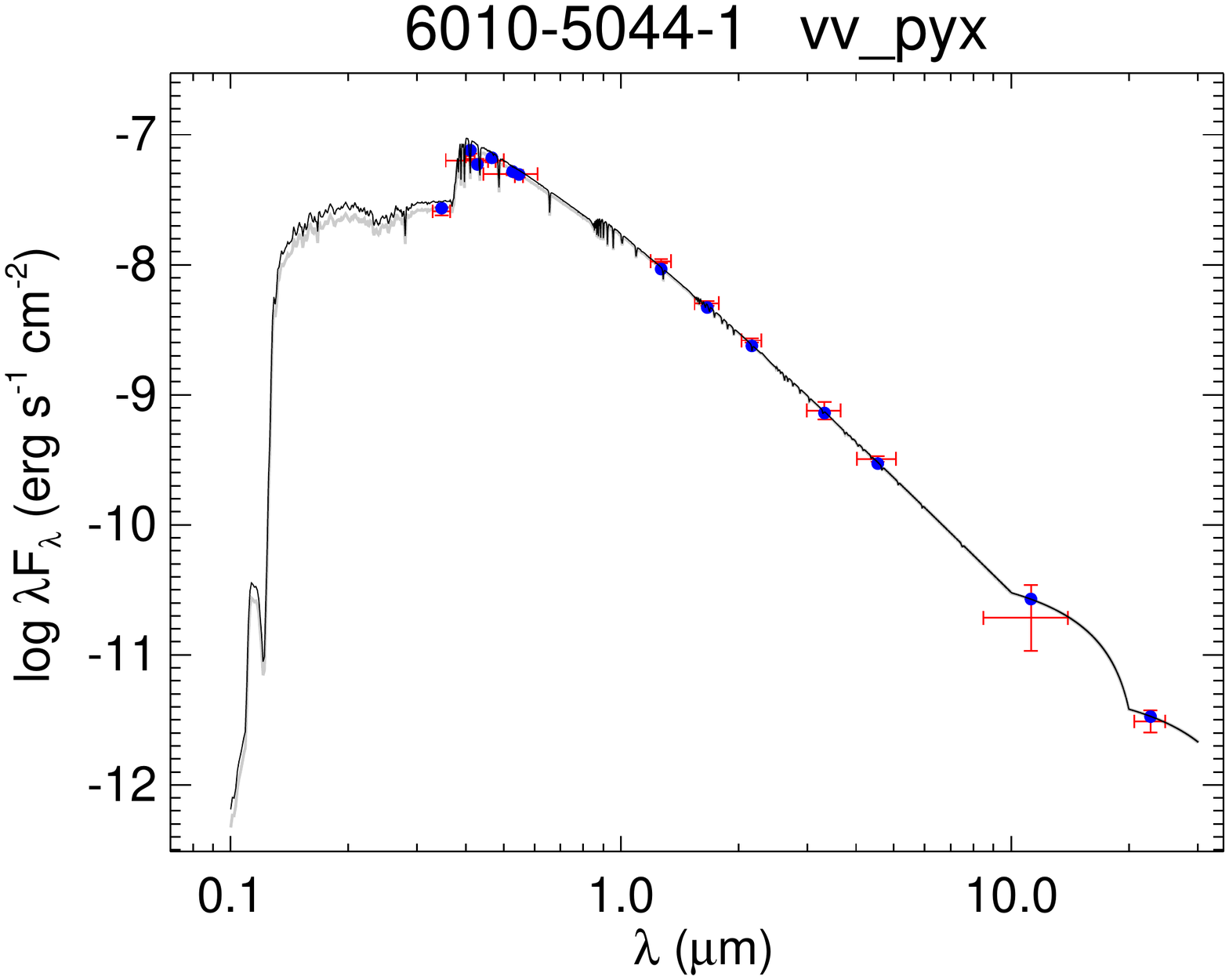}
  \includegraphics[trim=60 60 60 60,clip,width=0.49\linewidth]{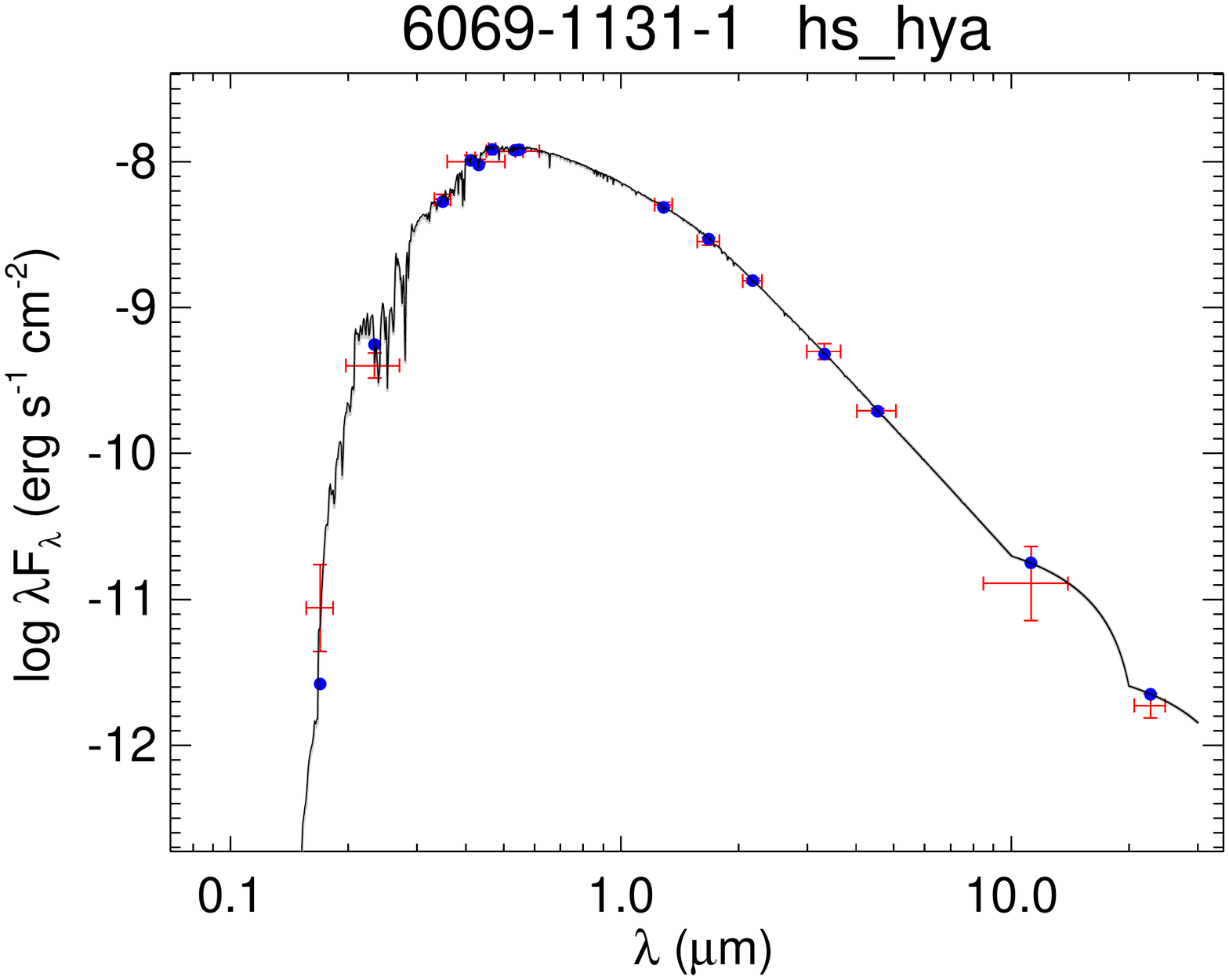}
  \includegraphics[trim=60 60 60 60,clip,width=0.49\linewidth]{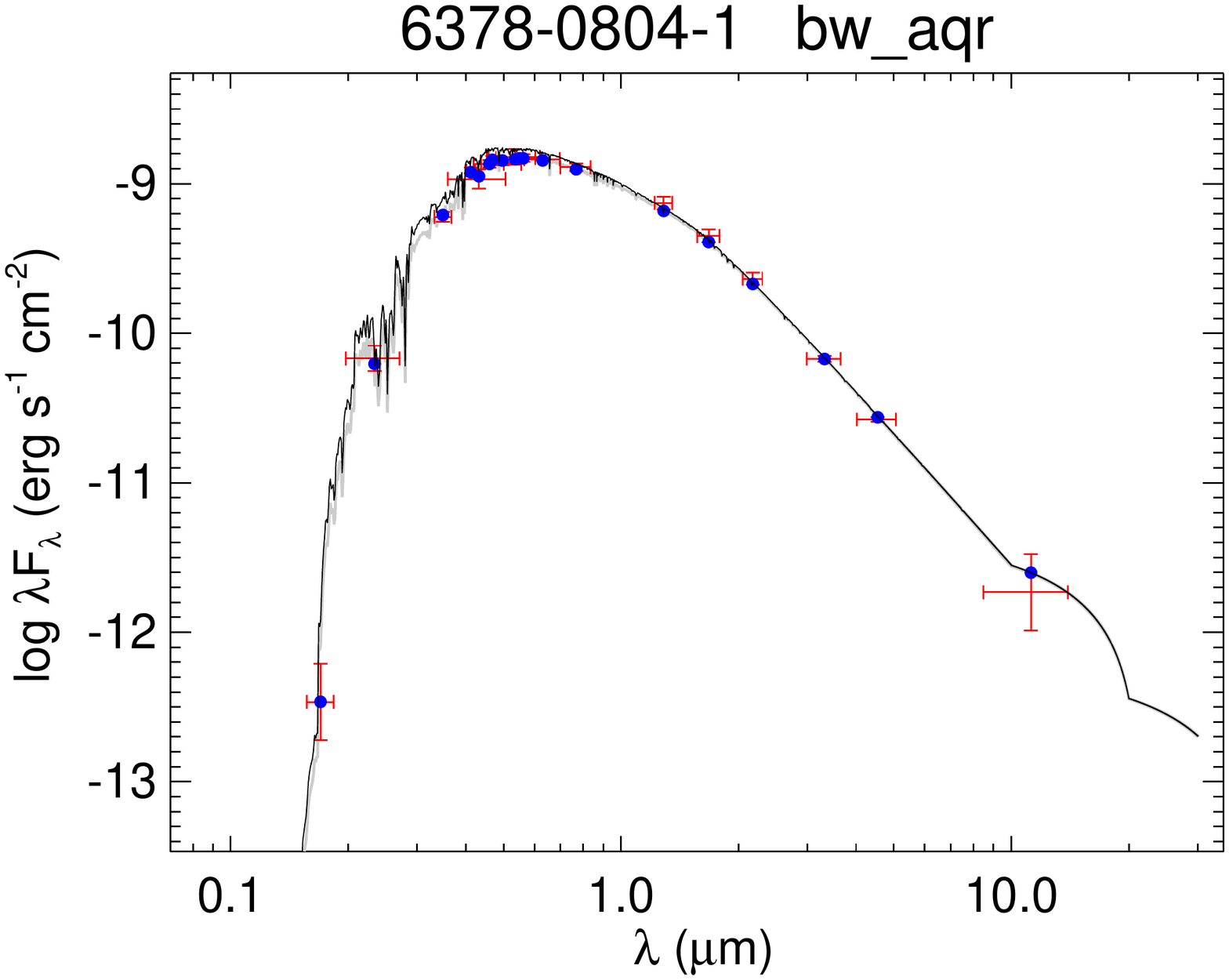}
  \includegraphics[trim=60 60 60 60,clip,width=0.49\linewidth]{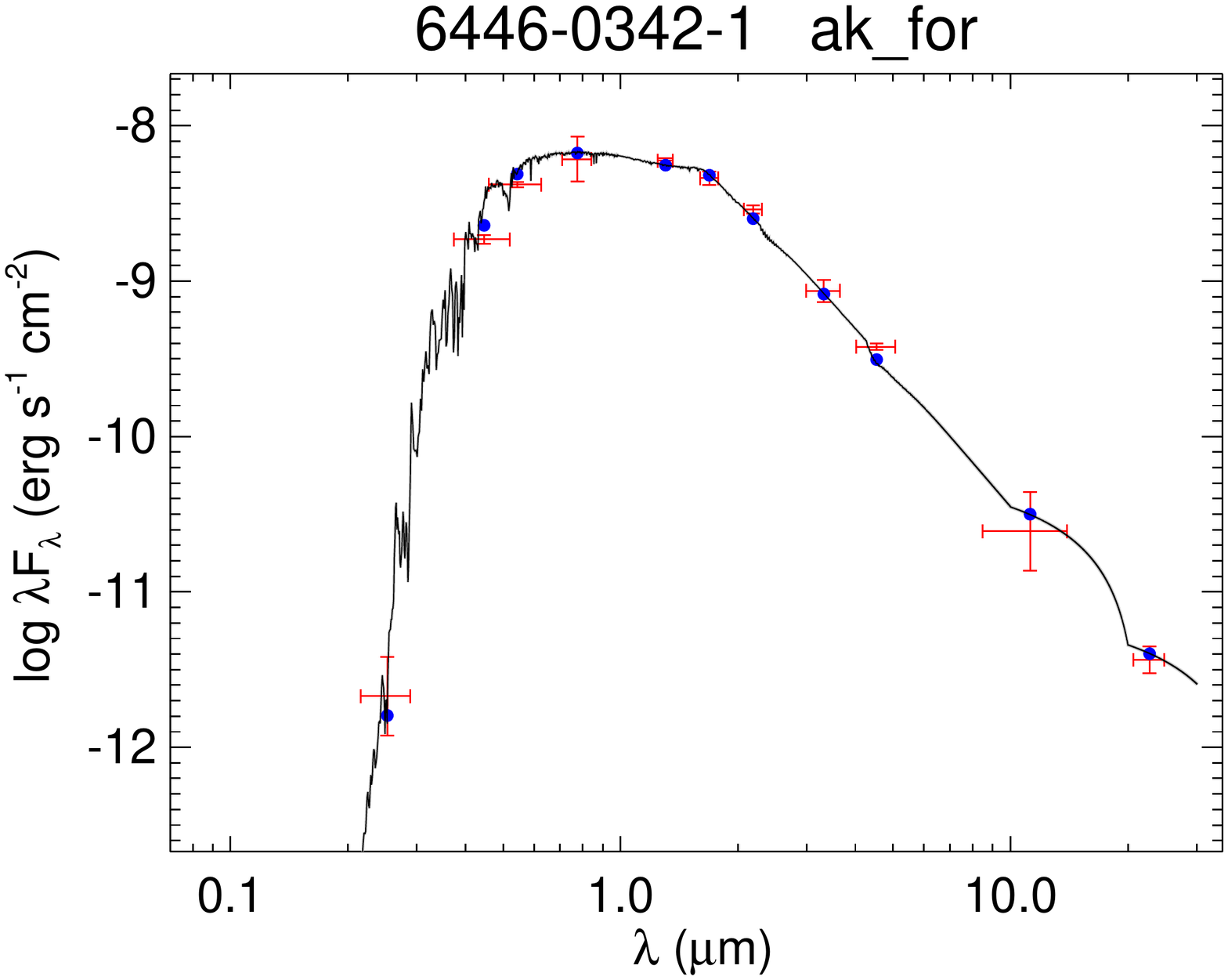}
  \includegraphics[trim=60 60 60 60,clip,width=0.49\linewidth]{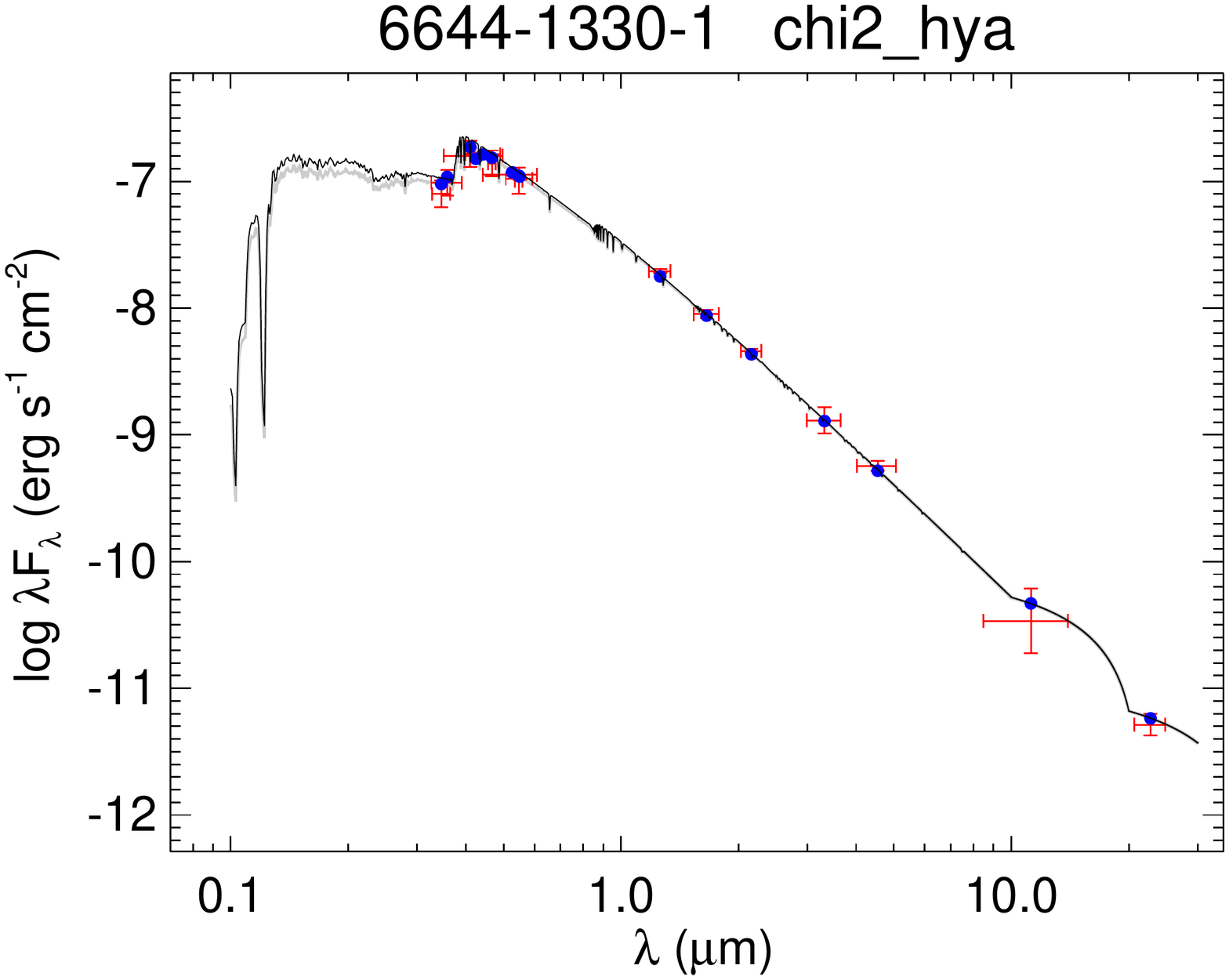}
  \caption{All labels, lines, symbols, and colors as in Figure \ref{fig:seds}.}
  \label{fig:seds_21}
\end{figure}

\begin{figure}[H]
  \centering
  \includegraphics[trim=60 60 60 60,clip,width=0.49\linewidth]{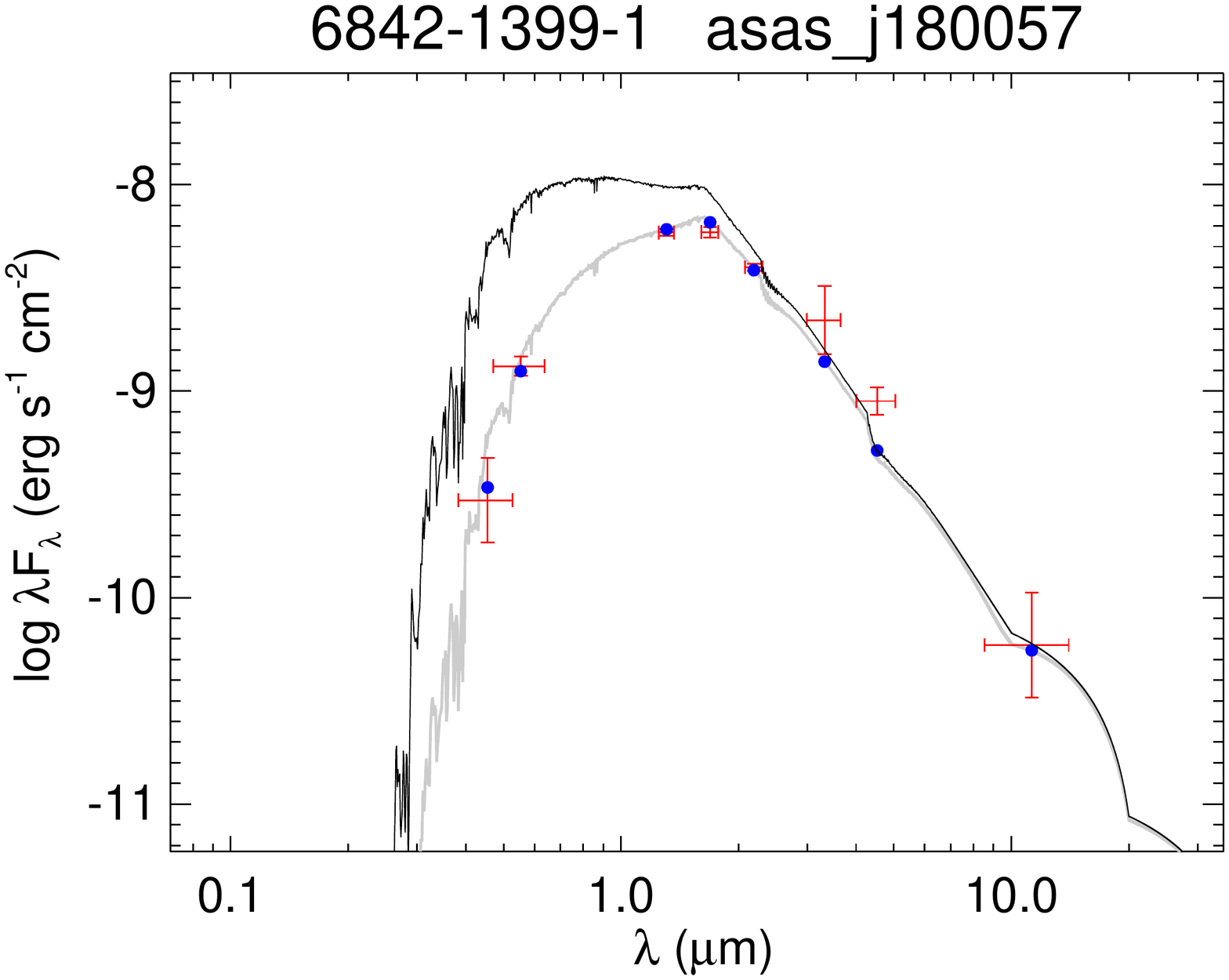}
  \includegraphics[trim=60 60 60 60,clip,width=0.49\linewidth]{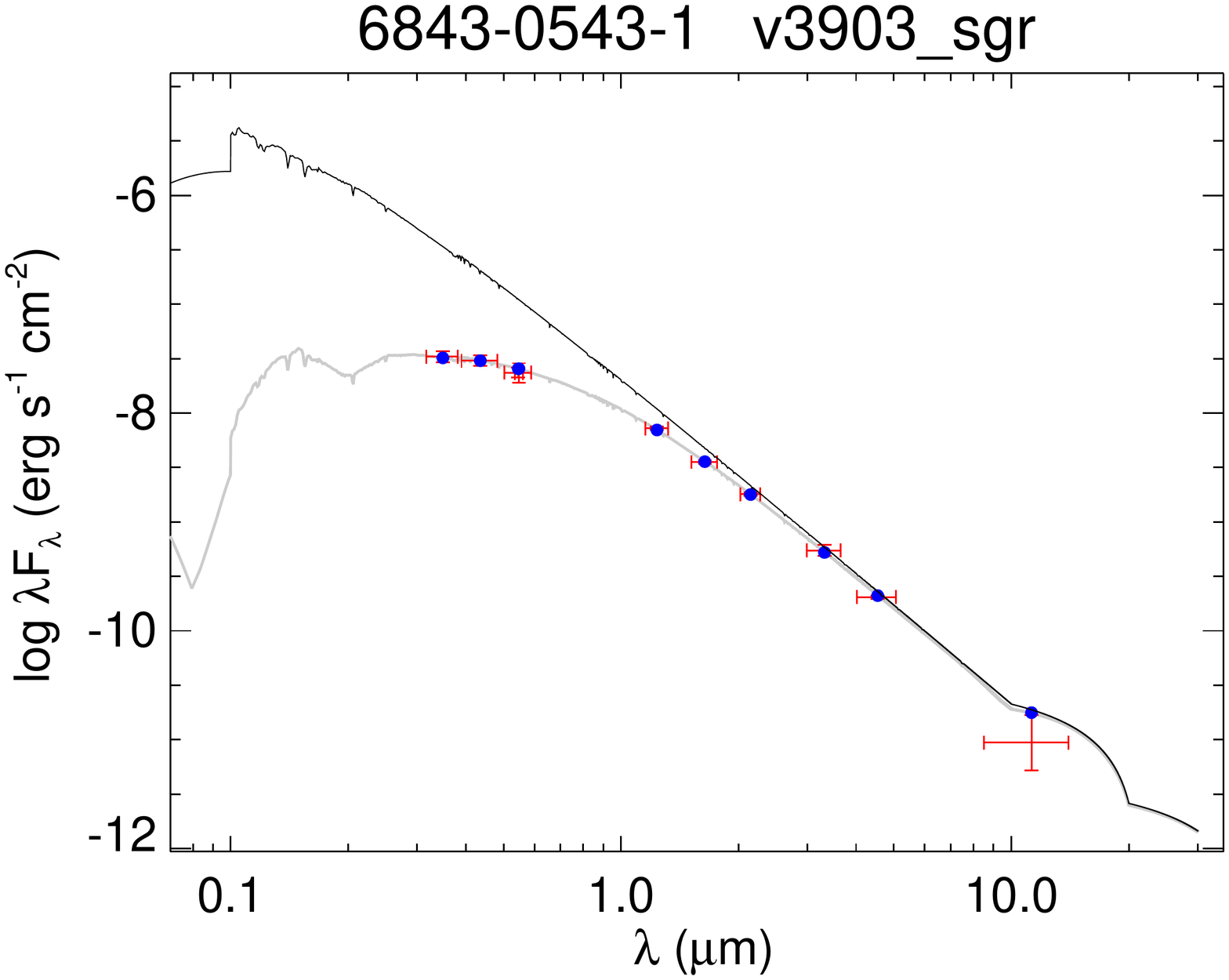}
  \includegraphics[trim=60 60 60 60,clip,width=0.49\linewidth]{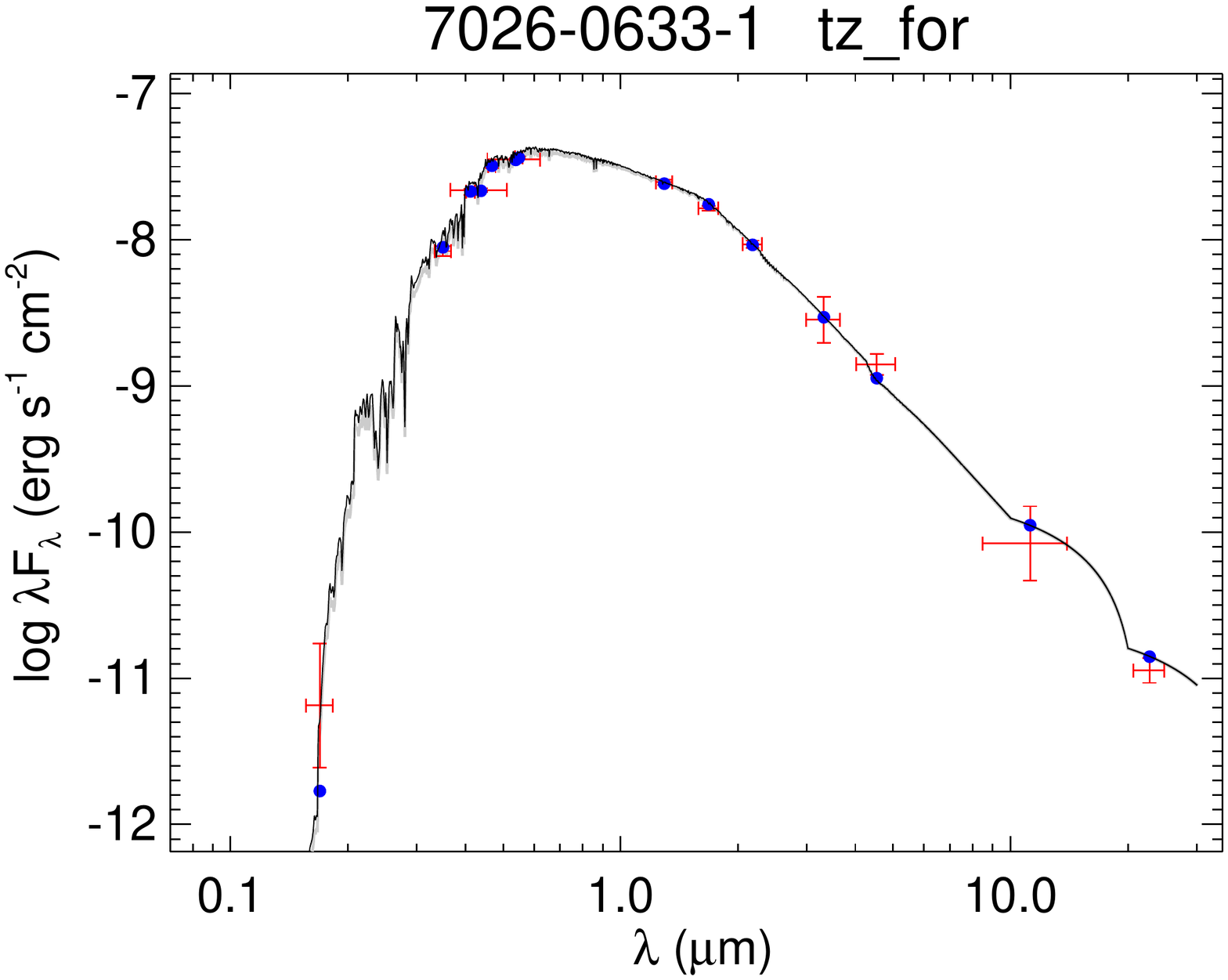}
  \includegraphics[trim=60 60 60 60,clip,width=0.49\linewidth]{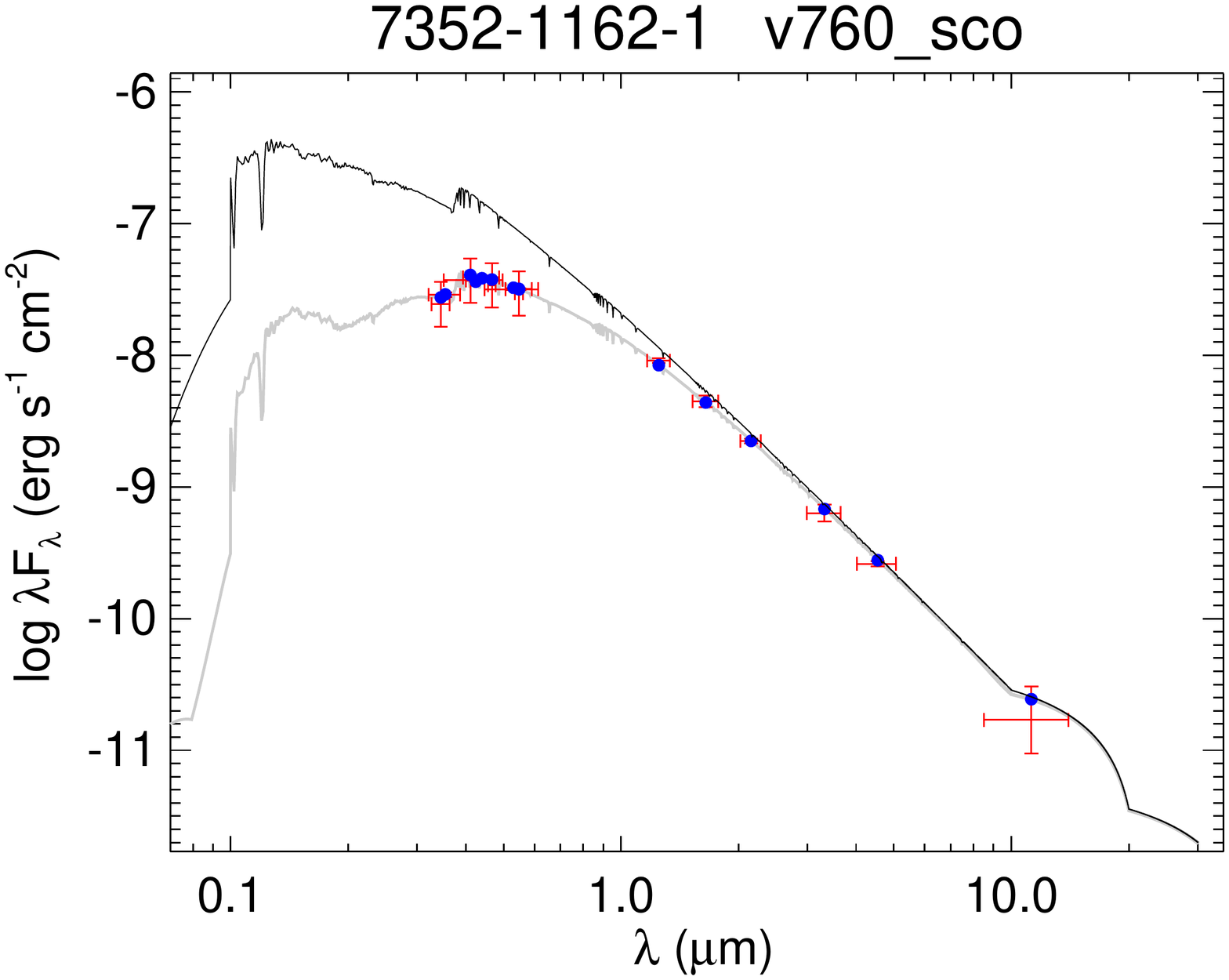}
  \includegraphics[trim=60 60 60 60,clip,width=0.49\linewidth]{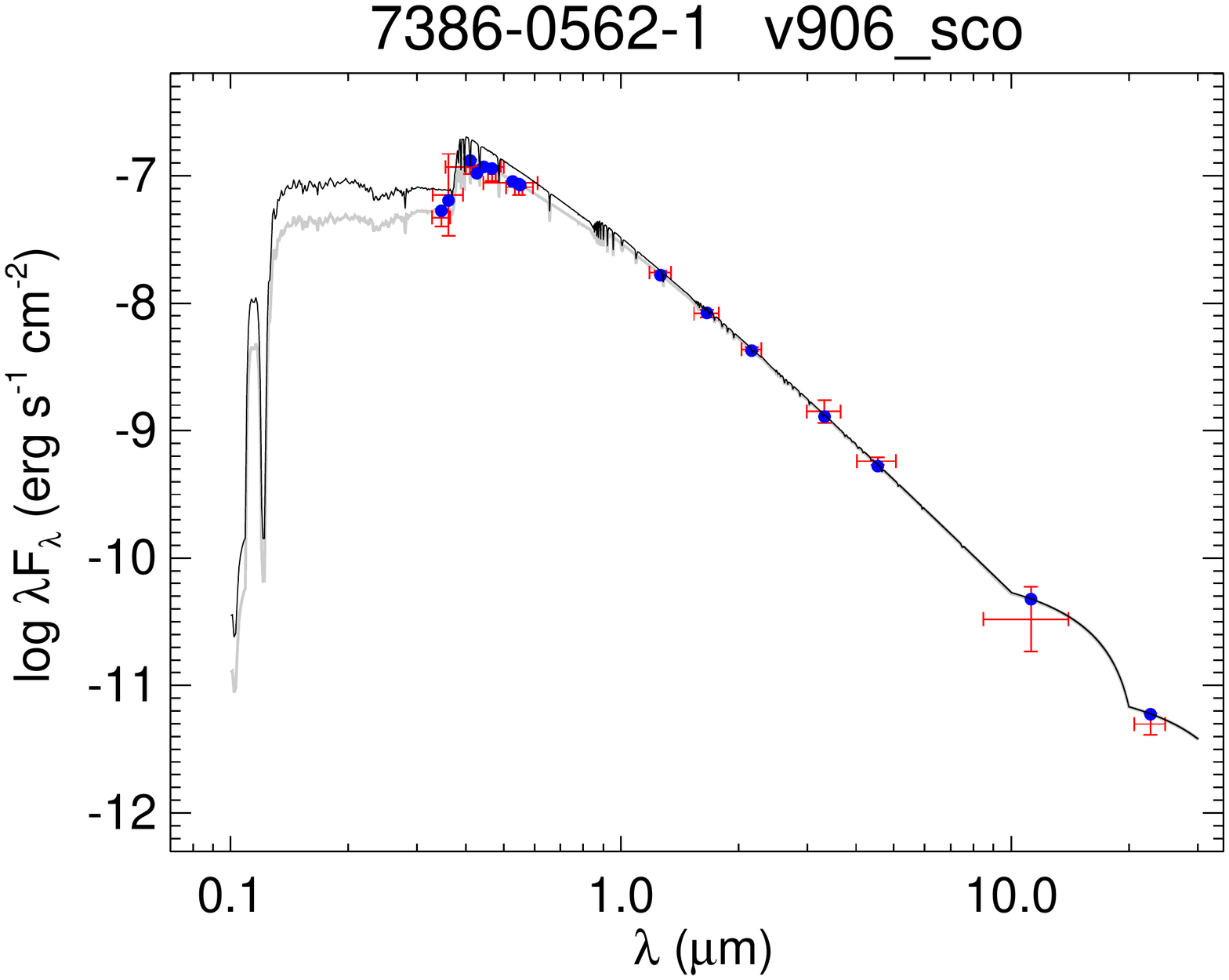}
  \includegraphics[trim=60 60 60 60,clip,width=0.49\linewidth]{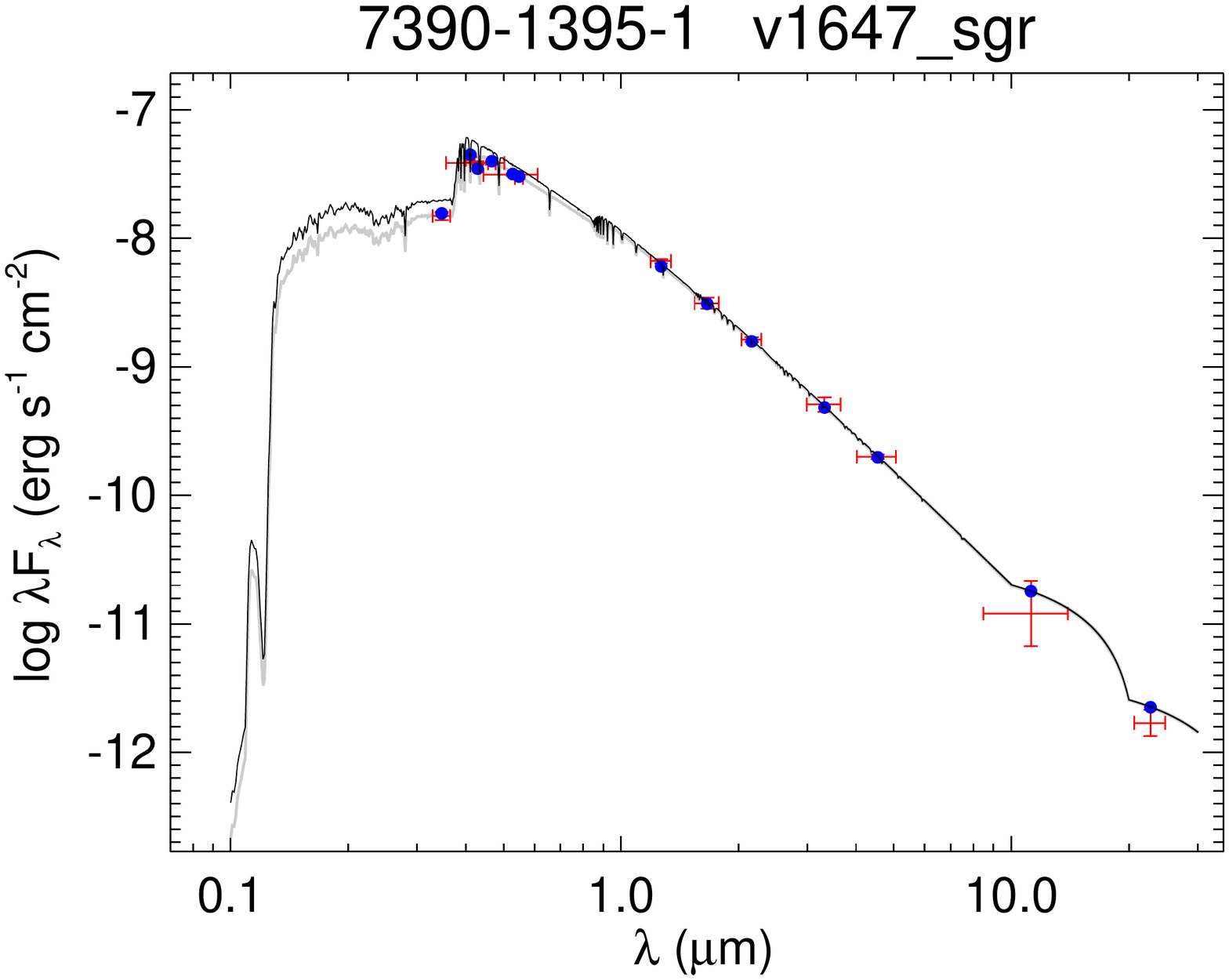}
  \caption{All labels, lines, symbols, and colors as in Figure \ref{fig:seds}.}
  \label{fig:seds_22}
\end{figure}

\begin{figure}[H]
  \centering
  \includegraphics[trim=60 60 60 60,clip,width=0.49\linewidth]{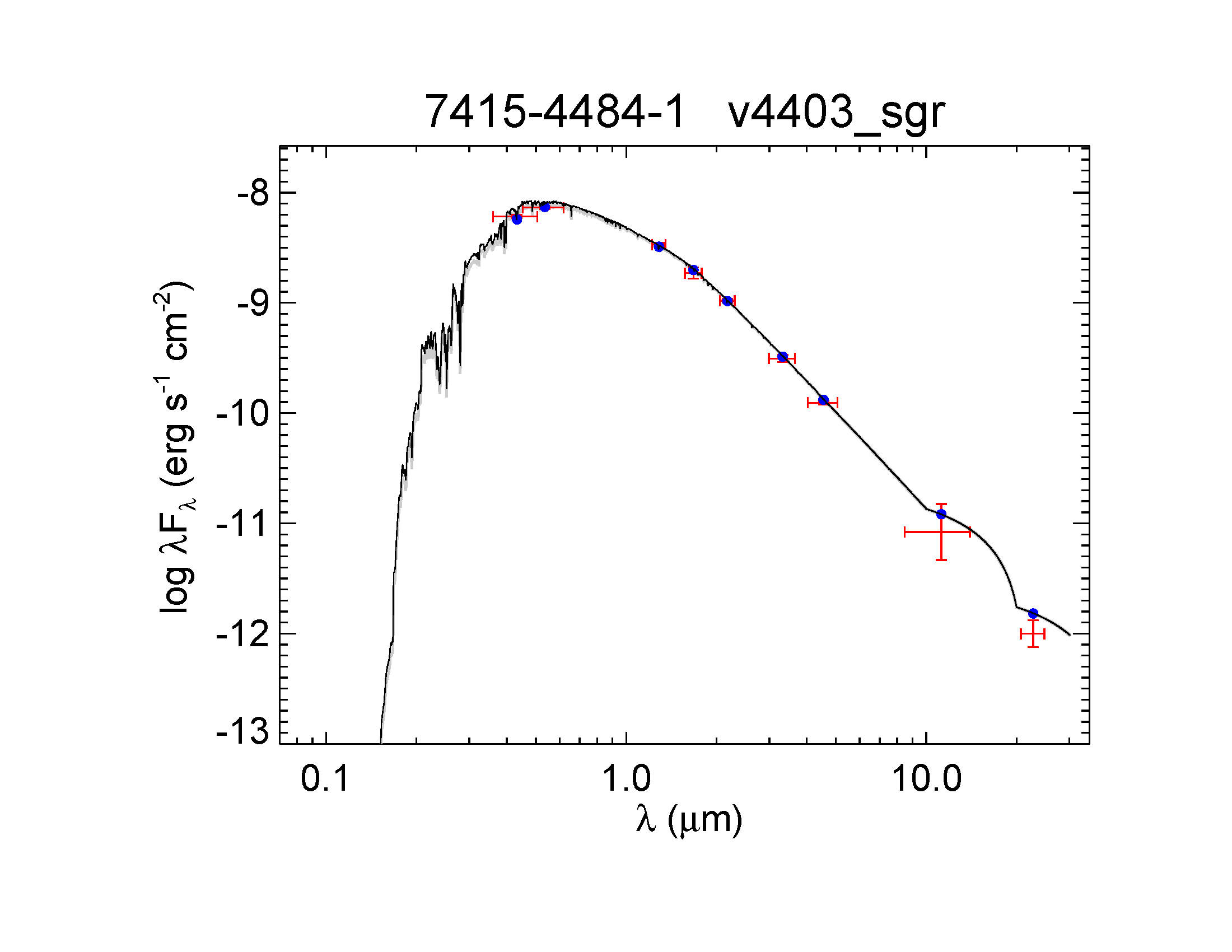}
  \includegraphics[trim=60 60 60 60,clip,width=0.49\linewidth]{sedfigs/hd_187669.png}
  \includegraphics[trim=60 60 60 60,clip,width=0.49\linewidth]{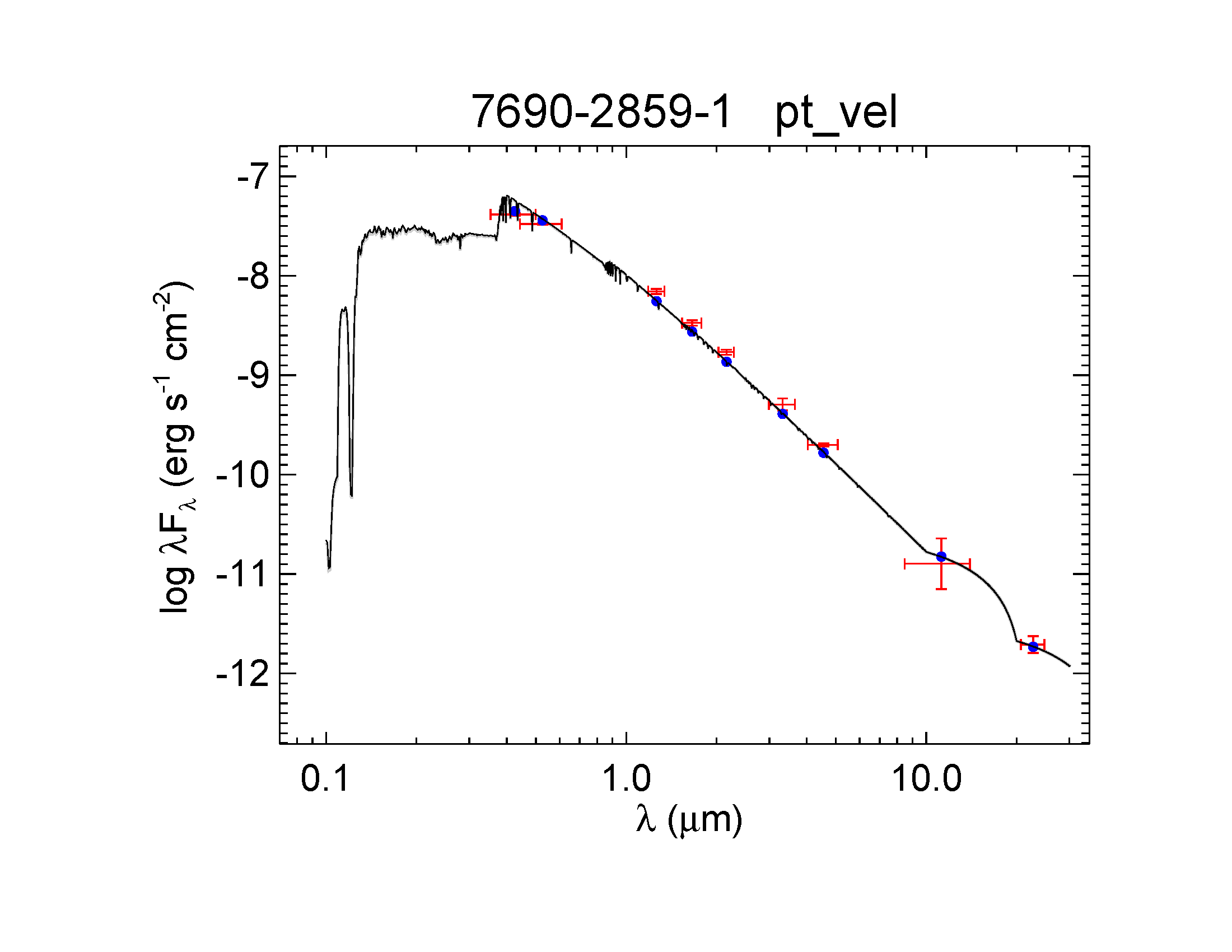}
  \includegraphics[trim=60 60 60 60,clip,width=0.49\linewidth]{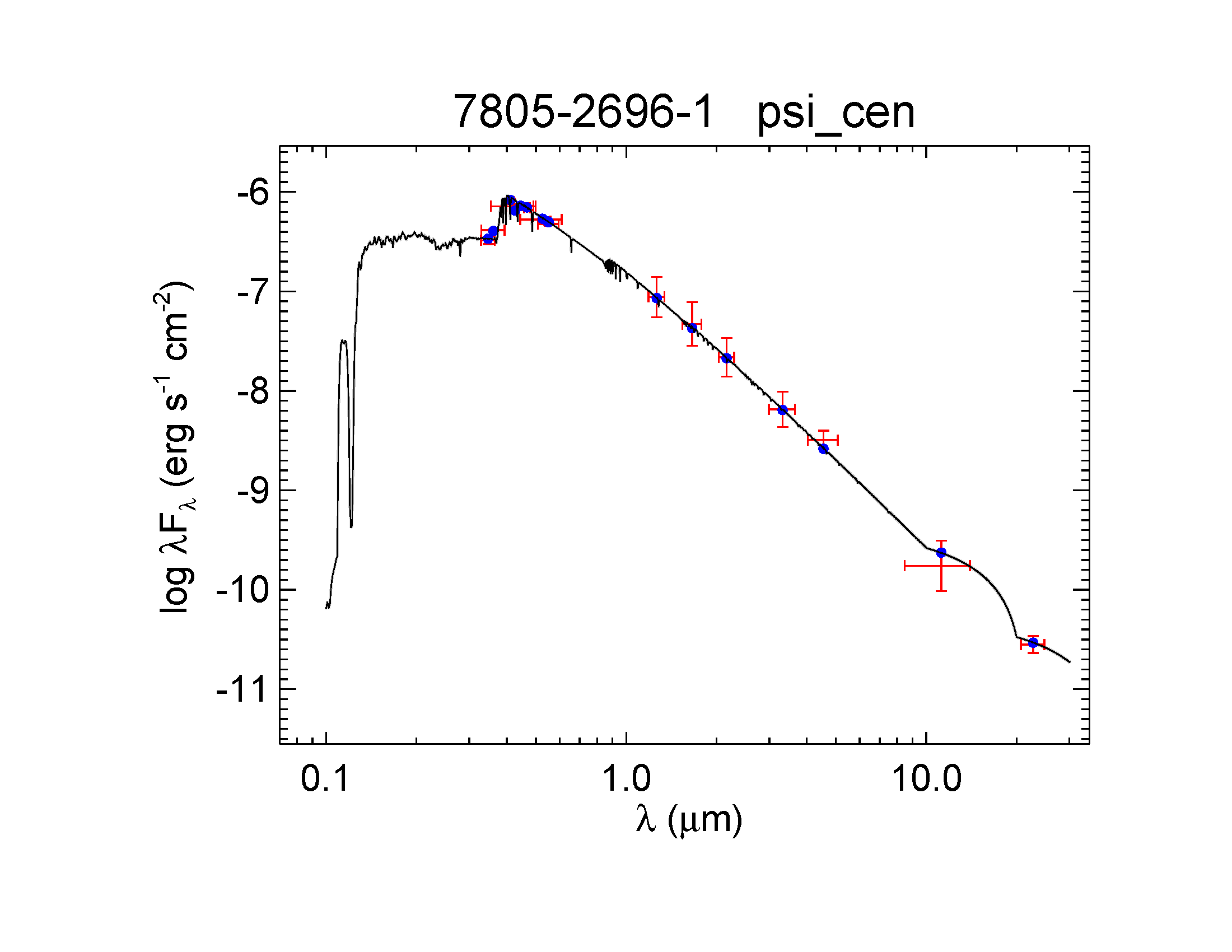}
  \includegraphics[trim=60 60 60 60,clip,width=0.49\linewidth]{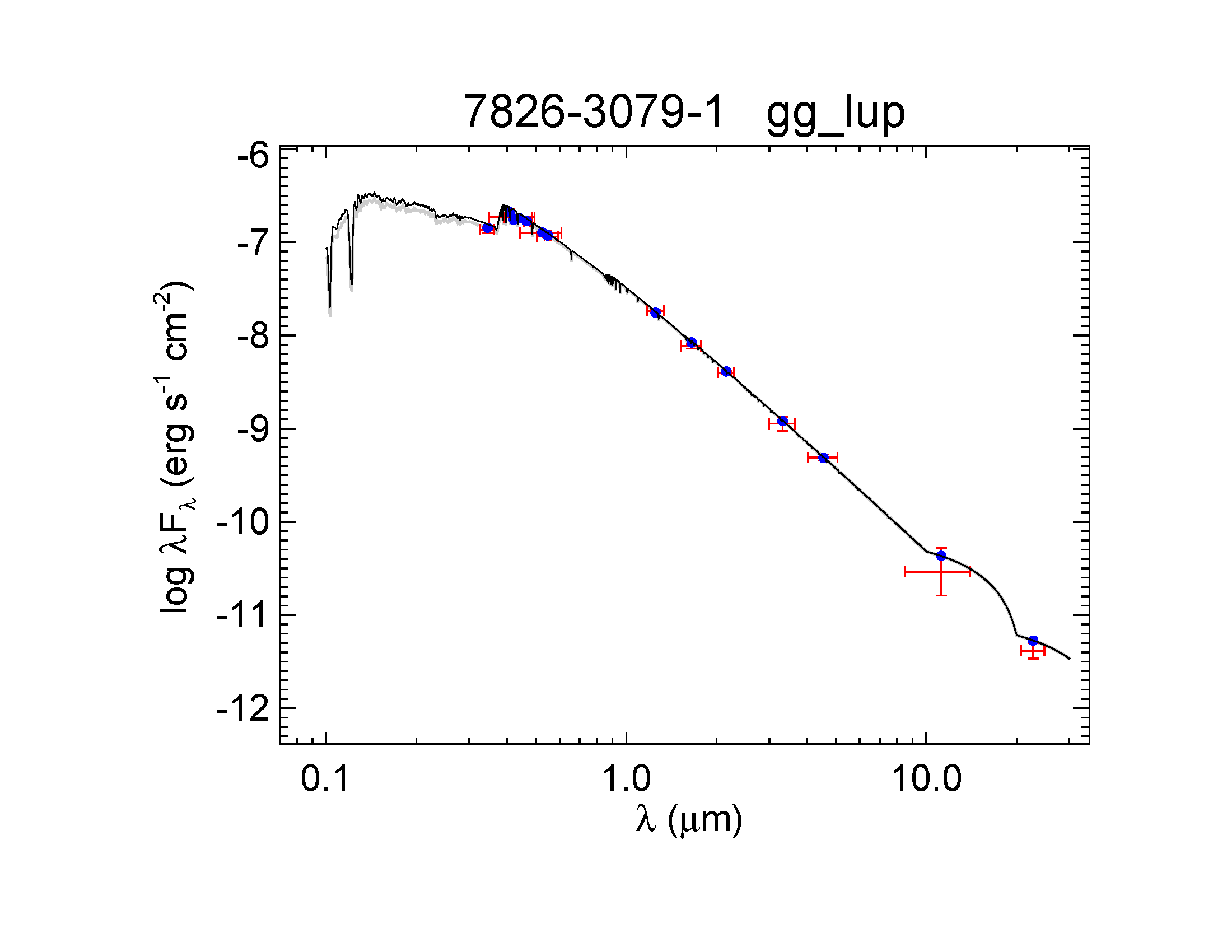}
  \includegraphics[trim=60 60 60 60,clip,width=0.49\linewidth]{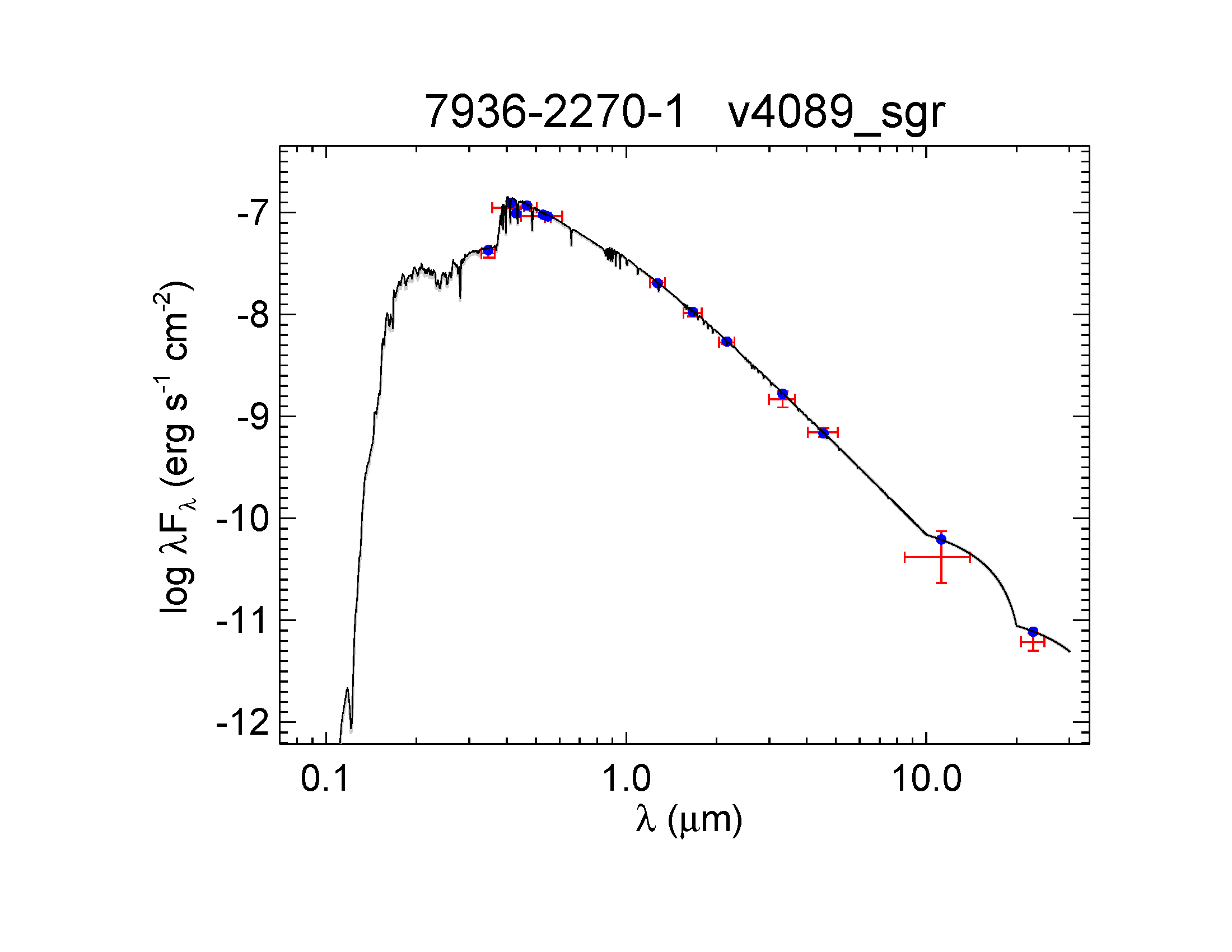}
  \caption{All labels, lines, symbols, and colors as in Figure \ref{fig:seds}.}
  \label{fig:seds_23}
\end{figure}

\begin{figure}[H]
  \centering
  \includegraphics[trim=60 60 60 60,clip,width=0.49\linewidth]{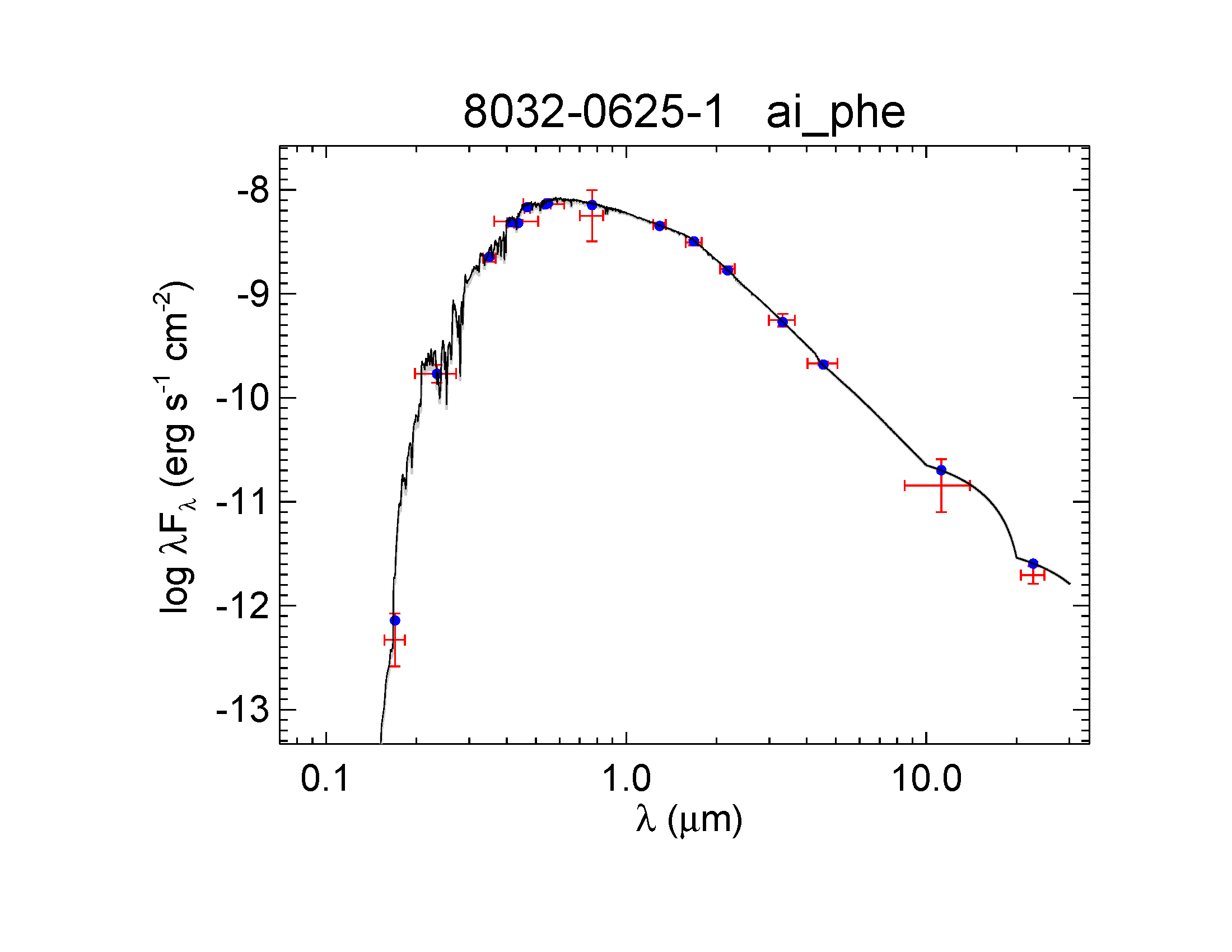}
  \includegraphics[trim=60 60 60 60,clip,width=0.49\linewidth]{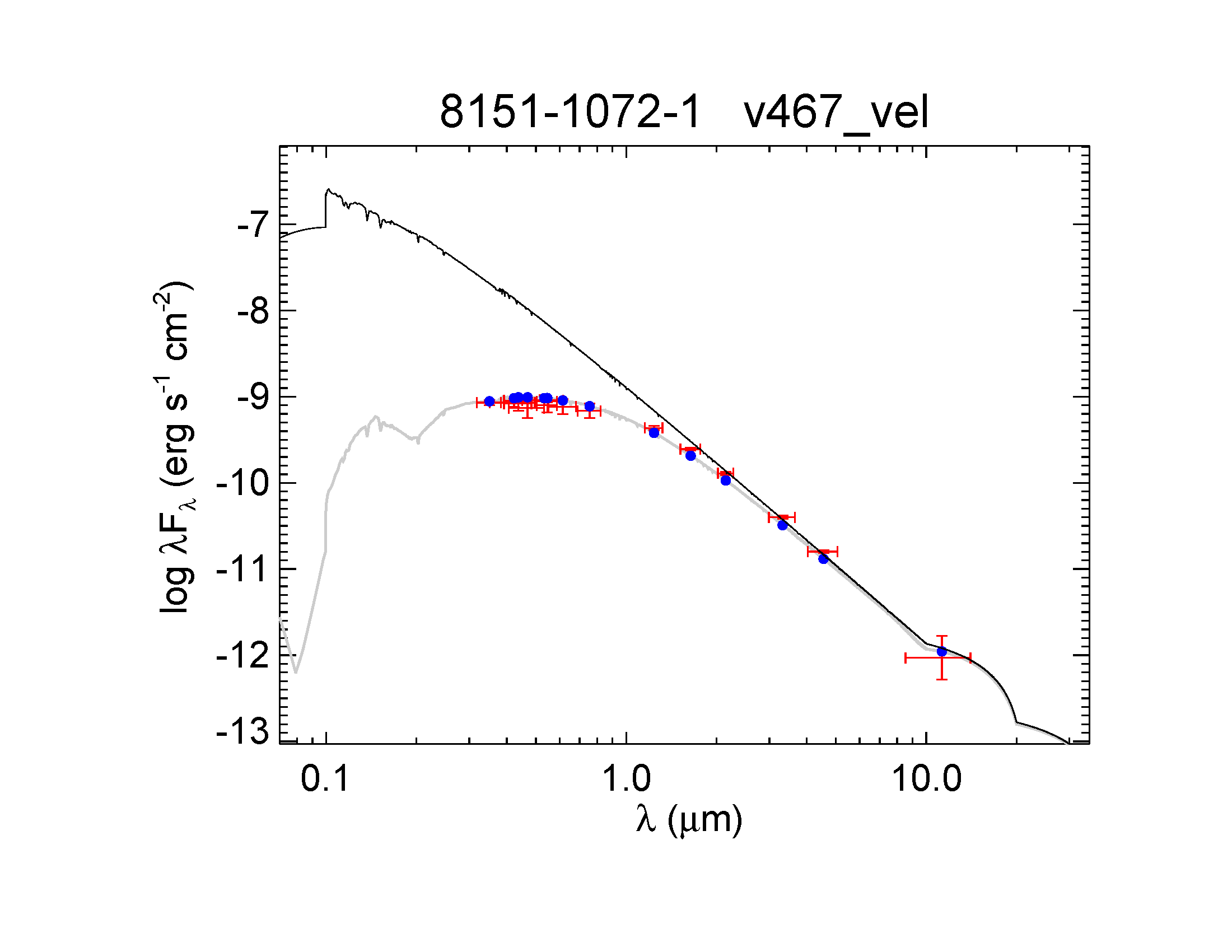}
  \includegraphics[trim=60 60 60 60,clip,width=0.49\linewidth]{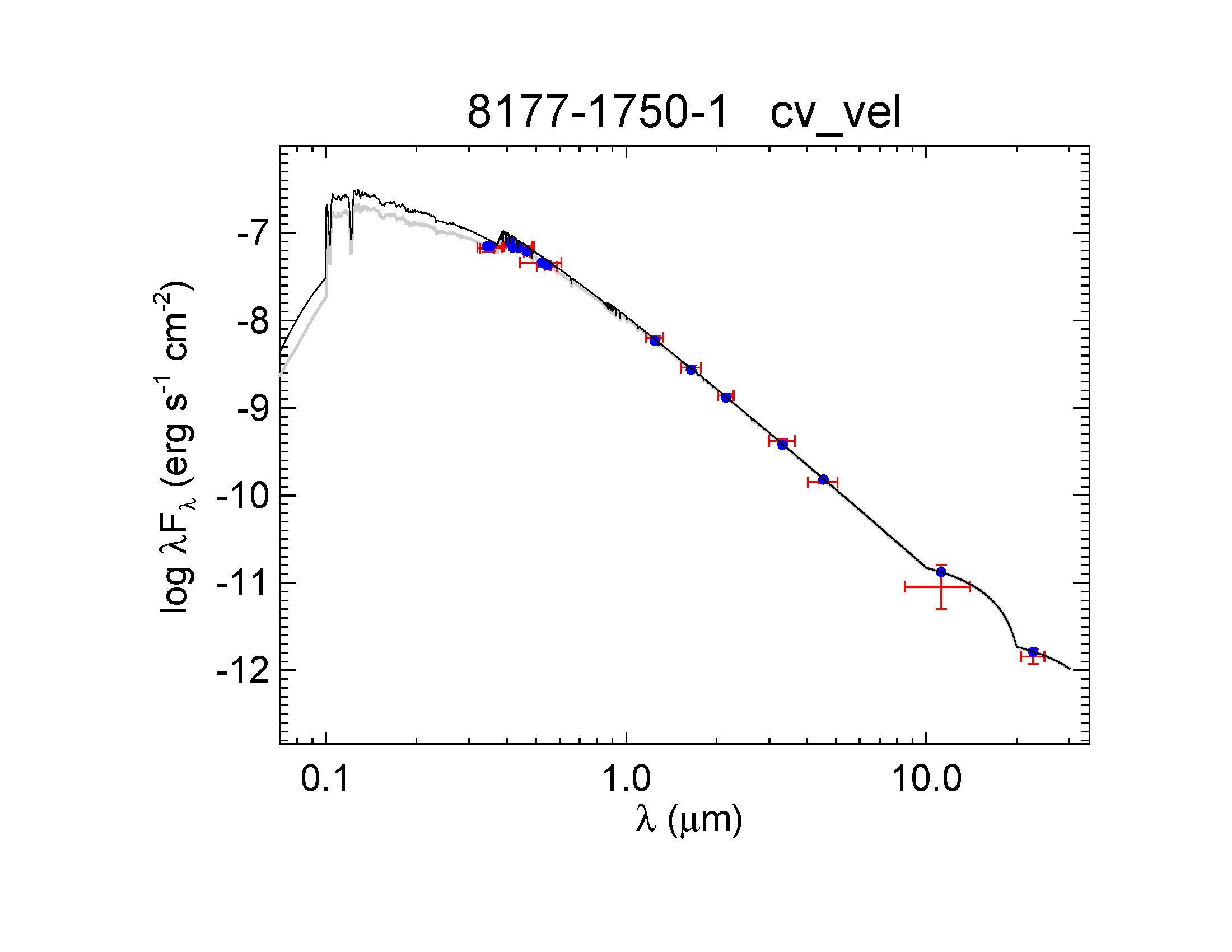}
  \includegraphics[trim=60 60 60 60,clip,width=0.49\linewidth]{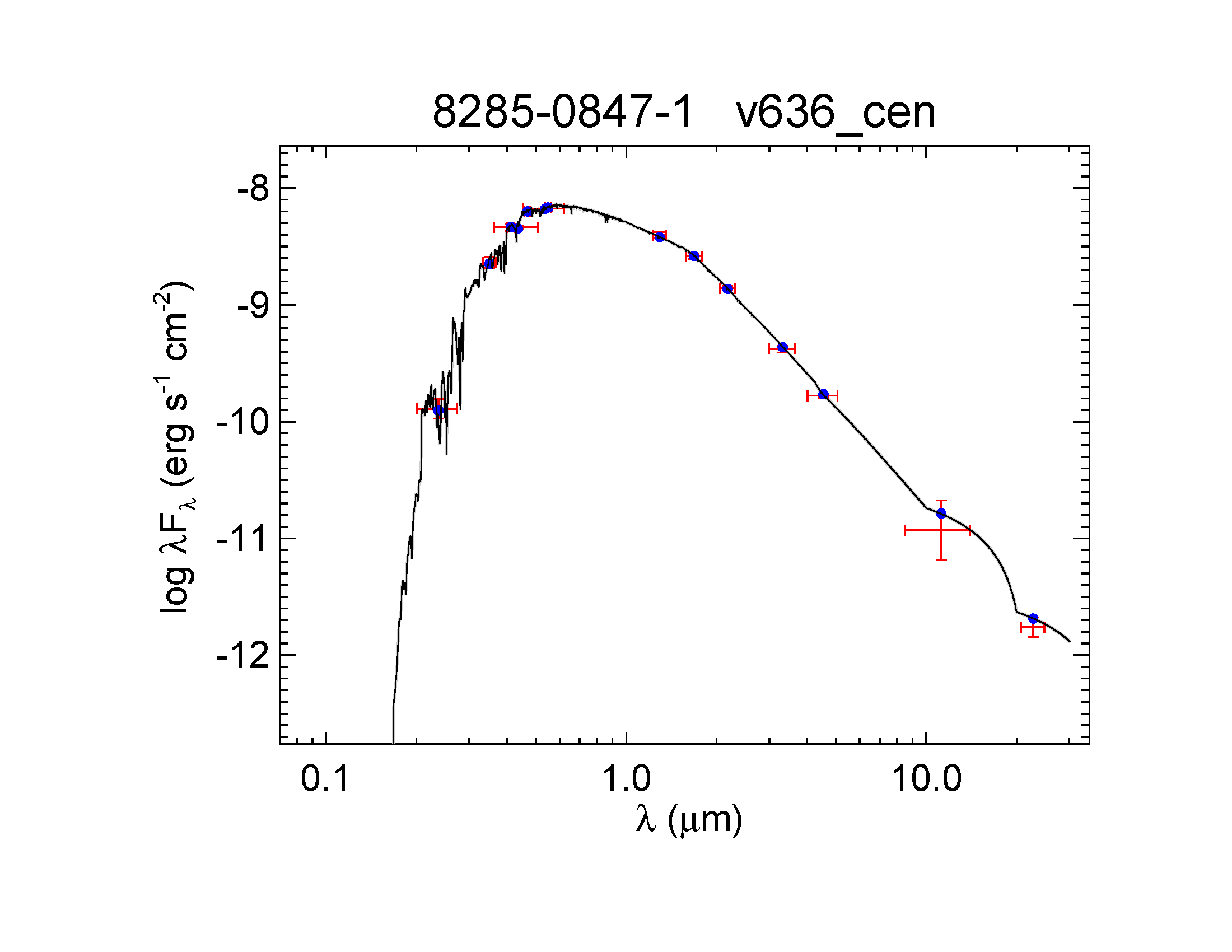}
  \includegraphics[trim=60 60 60 60,clip,width=0.49\linewidth]{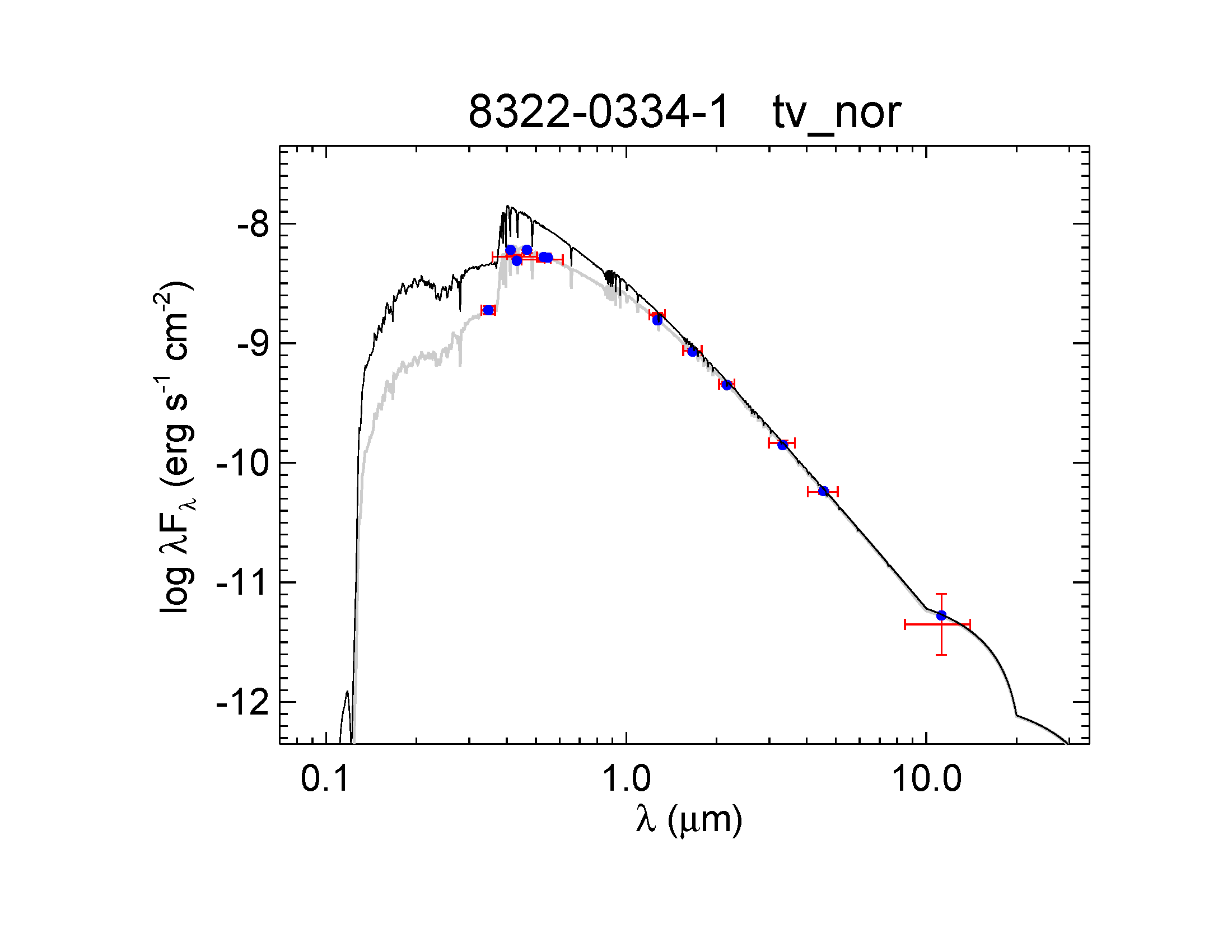}
  \includegraphics[trim=60 60 60 60,clip,width=0.49\linewidth]{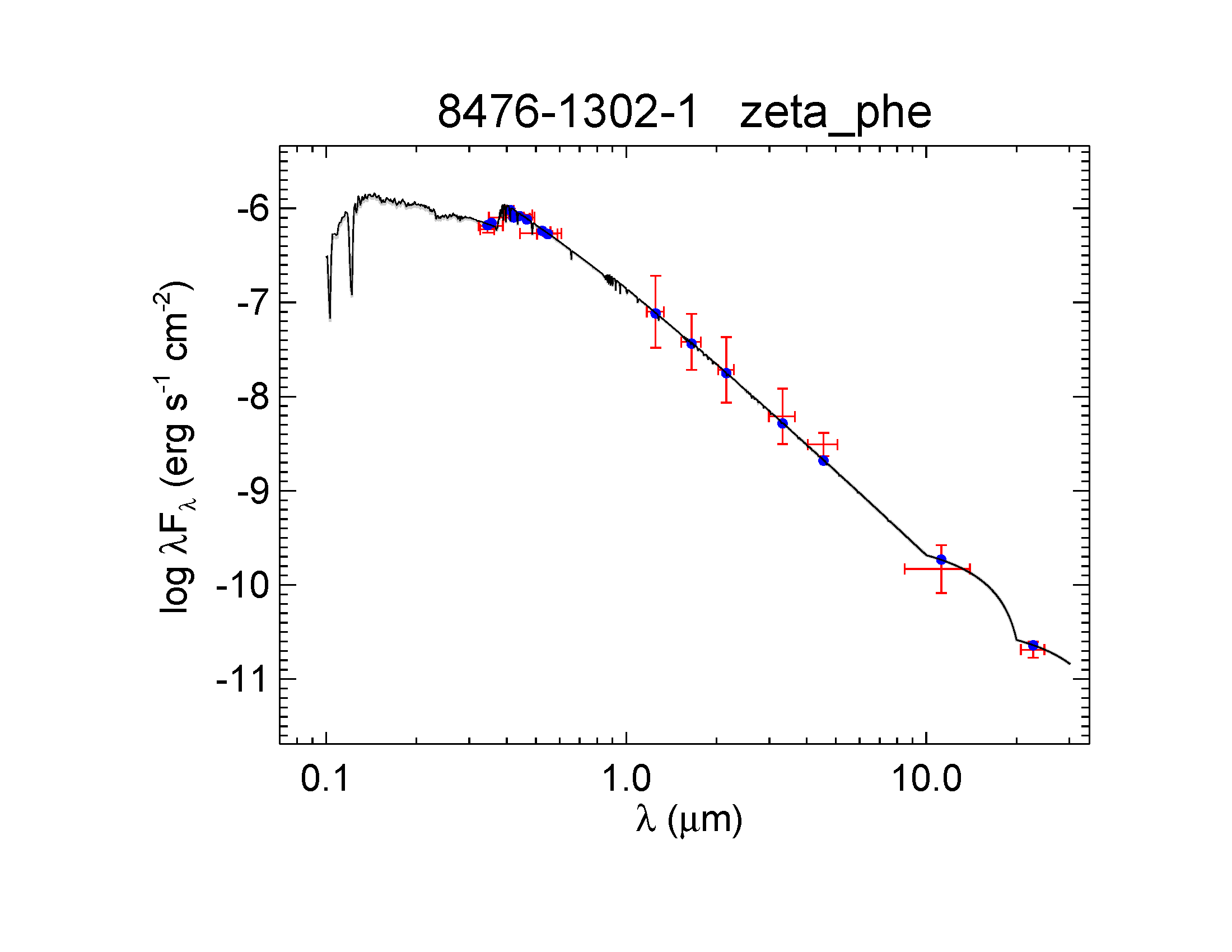}
  \caption{All labels, lines, symbols, and colors as in Figure \ref{fig:seds}.}
  \label{fig:seds_24}
\end{figure}

\begin{figure}[H]
  \centering
  \includegraphics[trim=60 60 60 60,clip,width=0.49\linewidth]{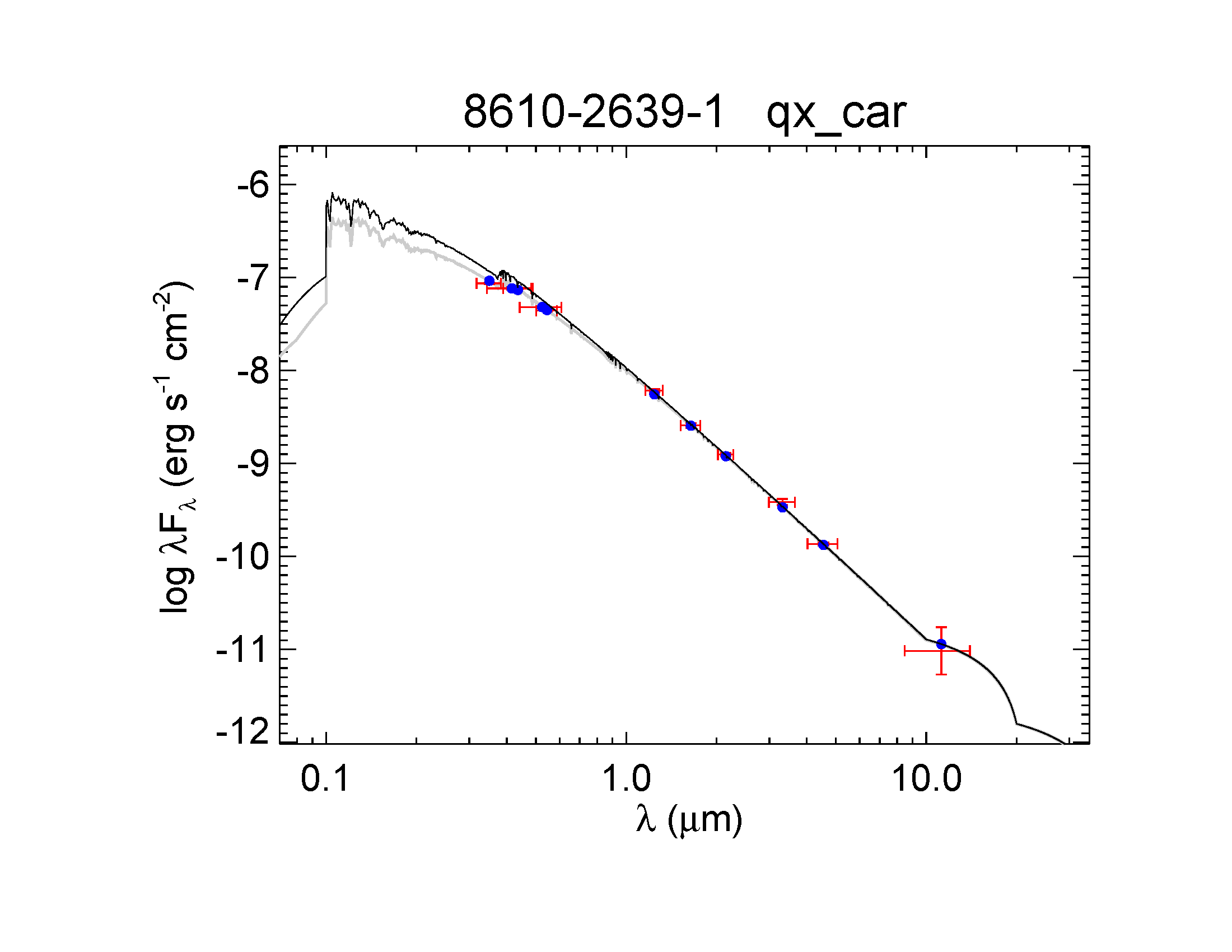}
  \includegraphics[trim=60 60 60 60,clip,width=0.49\linewidth]{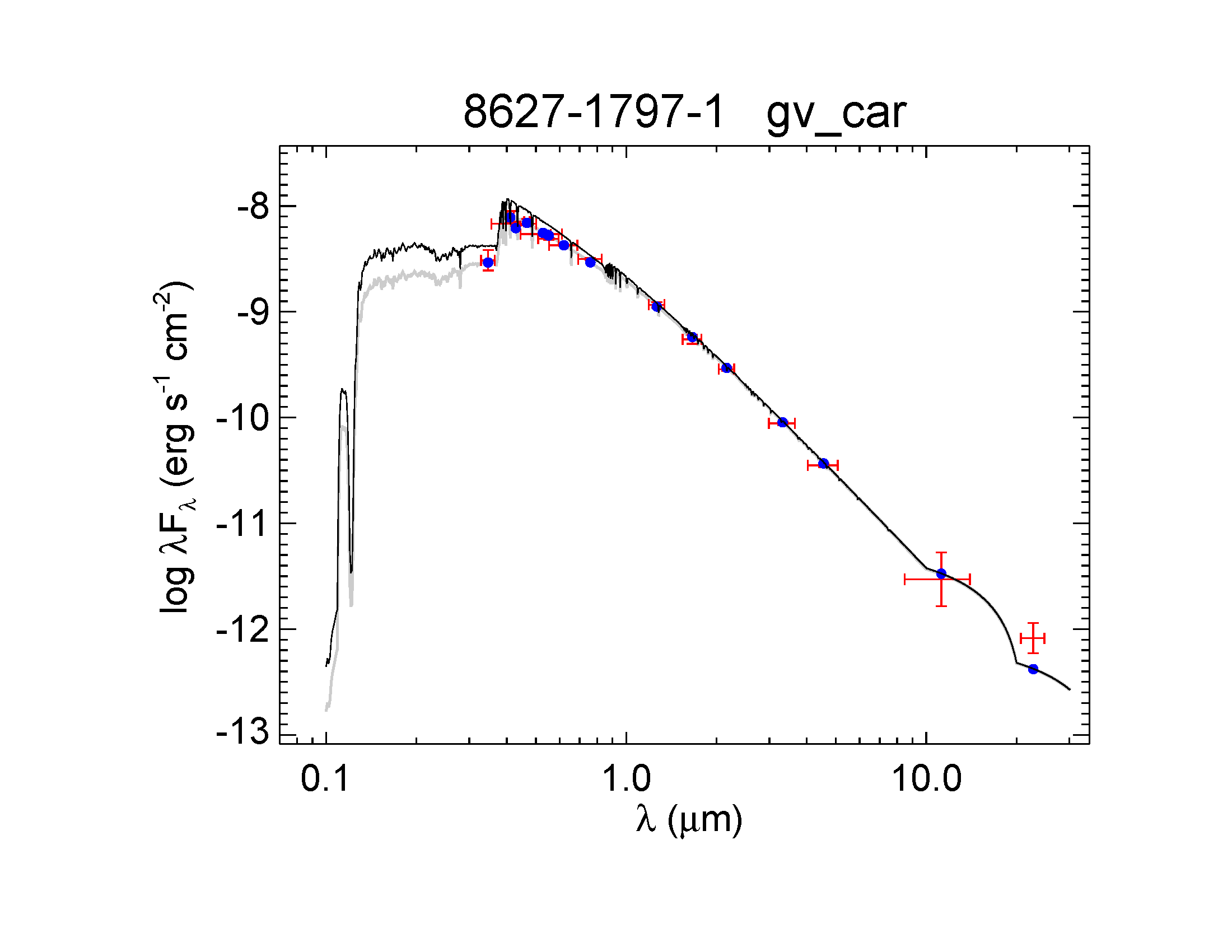}
  \includegraphics[trim=60 60 60 60,clip,width=0.49\linewidth]{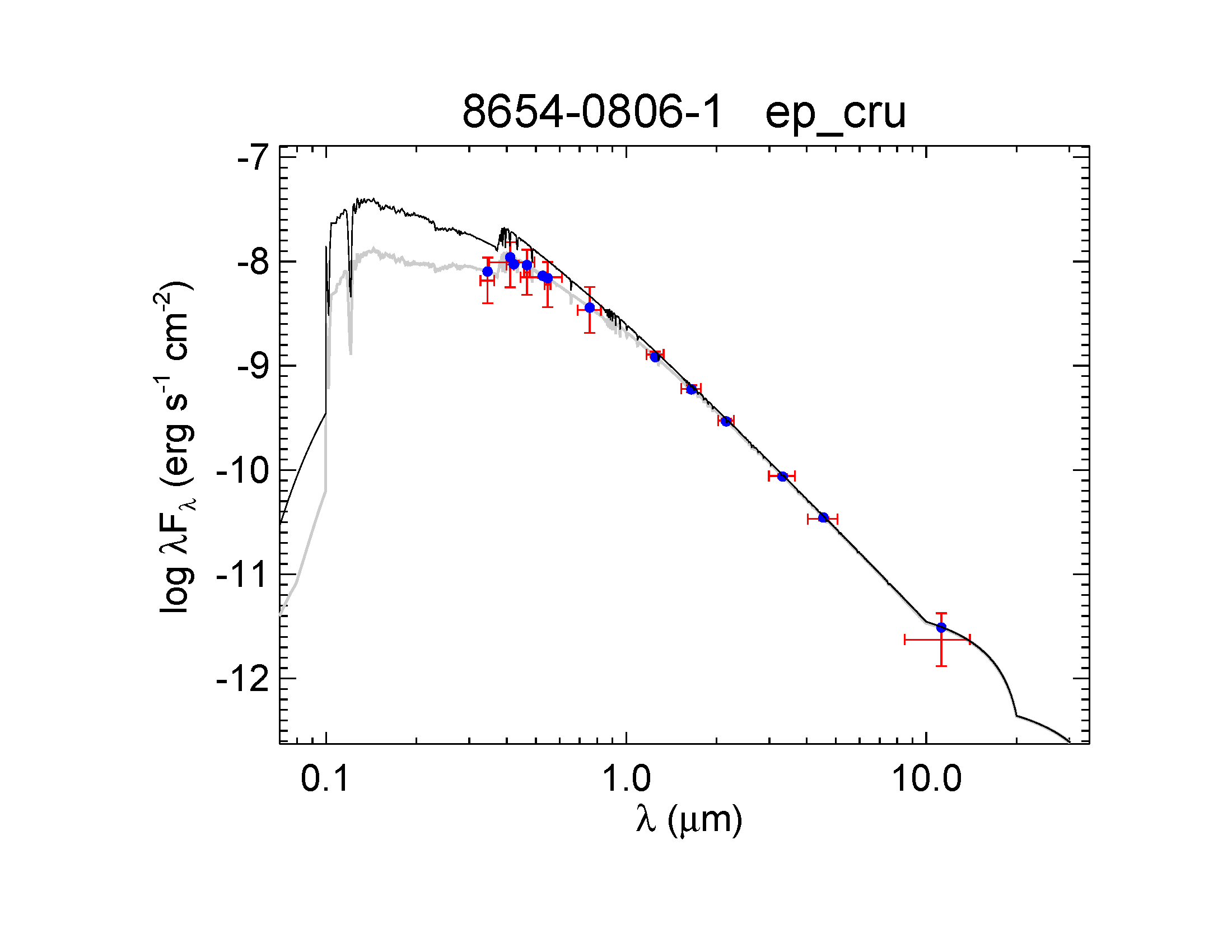}
  \includegraphics[trim=60 60 60 60,clip,width=0.49\linewidth]{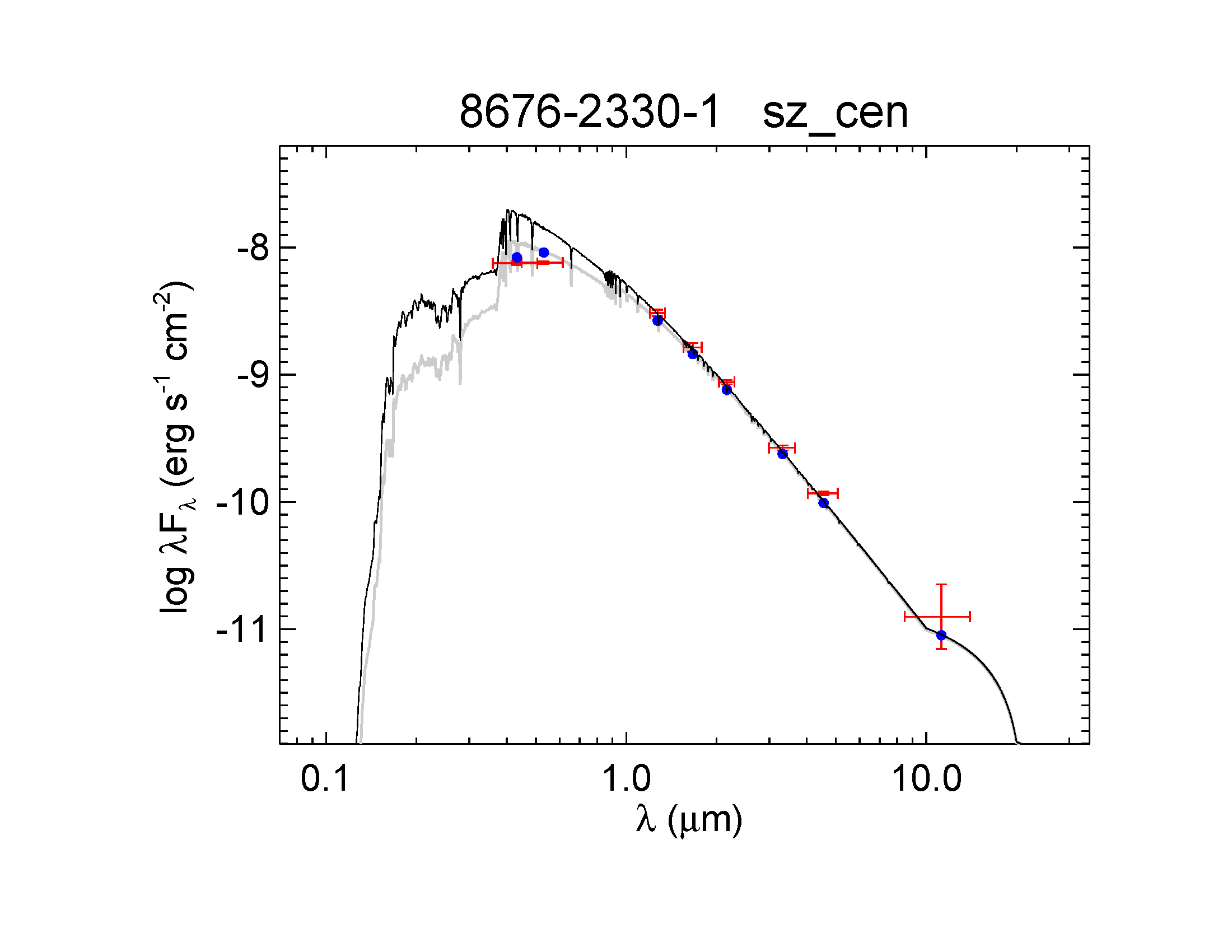}
  \includegraphics[trim=60 60 60 60,clip,width=0.49\linewidth]{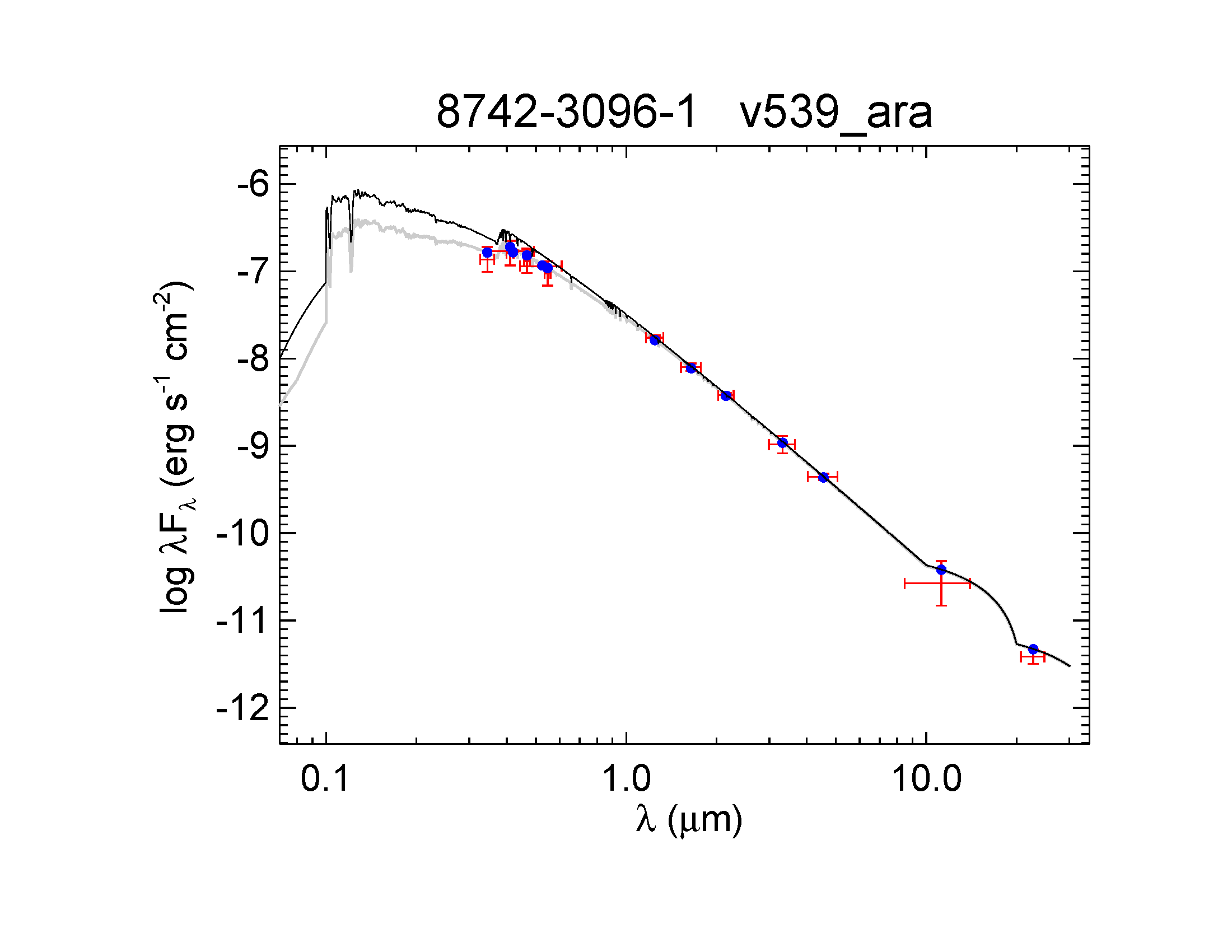}
  \includegraphics[trim=60 60 60 60,clip,width=0.49\linewidth]{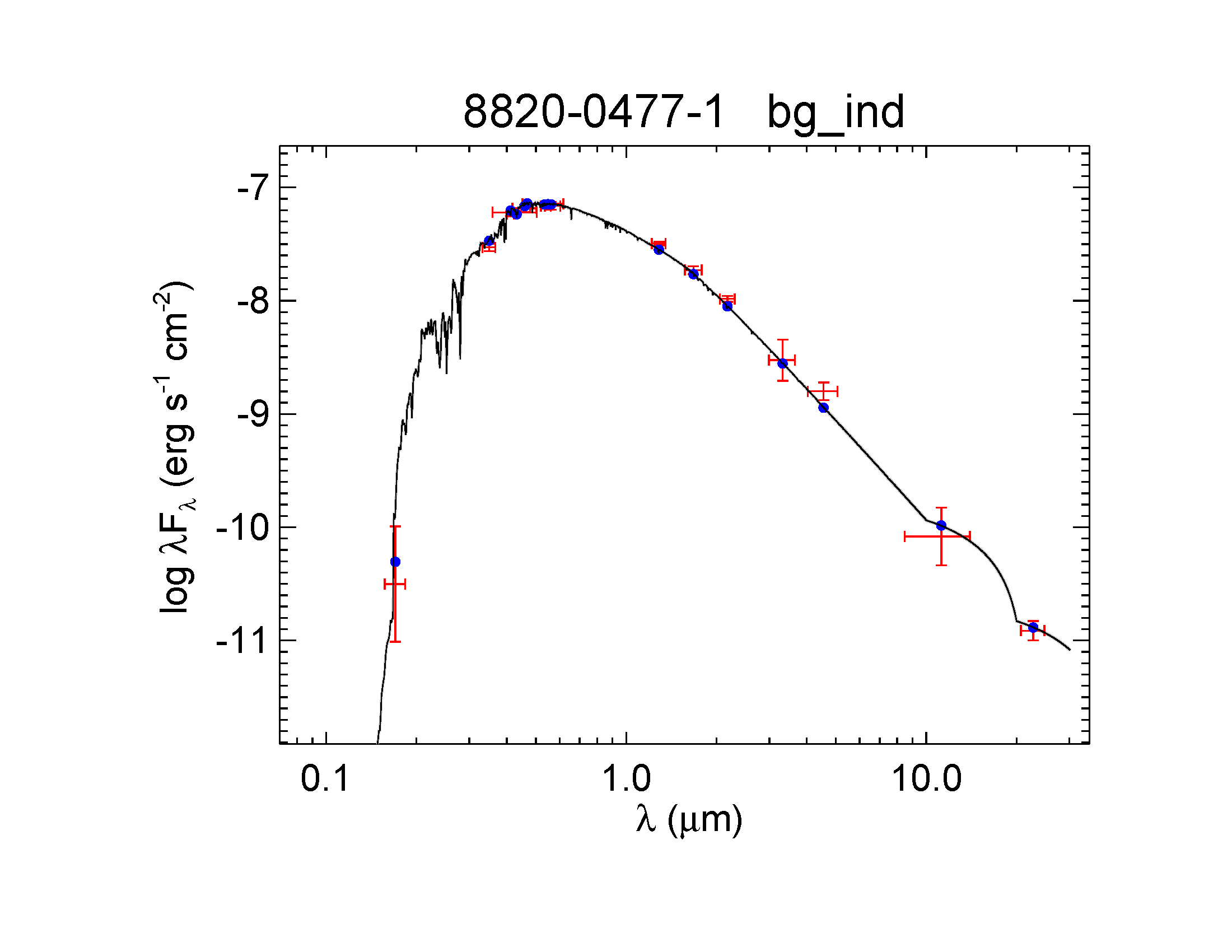}
  \caption{All labels, lines, symbols, and colors as in Figure \ref{fig:seds}.}
  \label{fig:seds_25}
\end{figure}

\begin{figure}[H]
  \centering
  \includegraphics[trim=60 60 60 60,clip,width=0.49\linewidth]{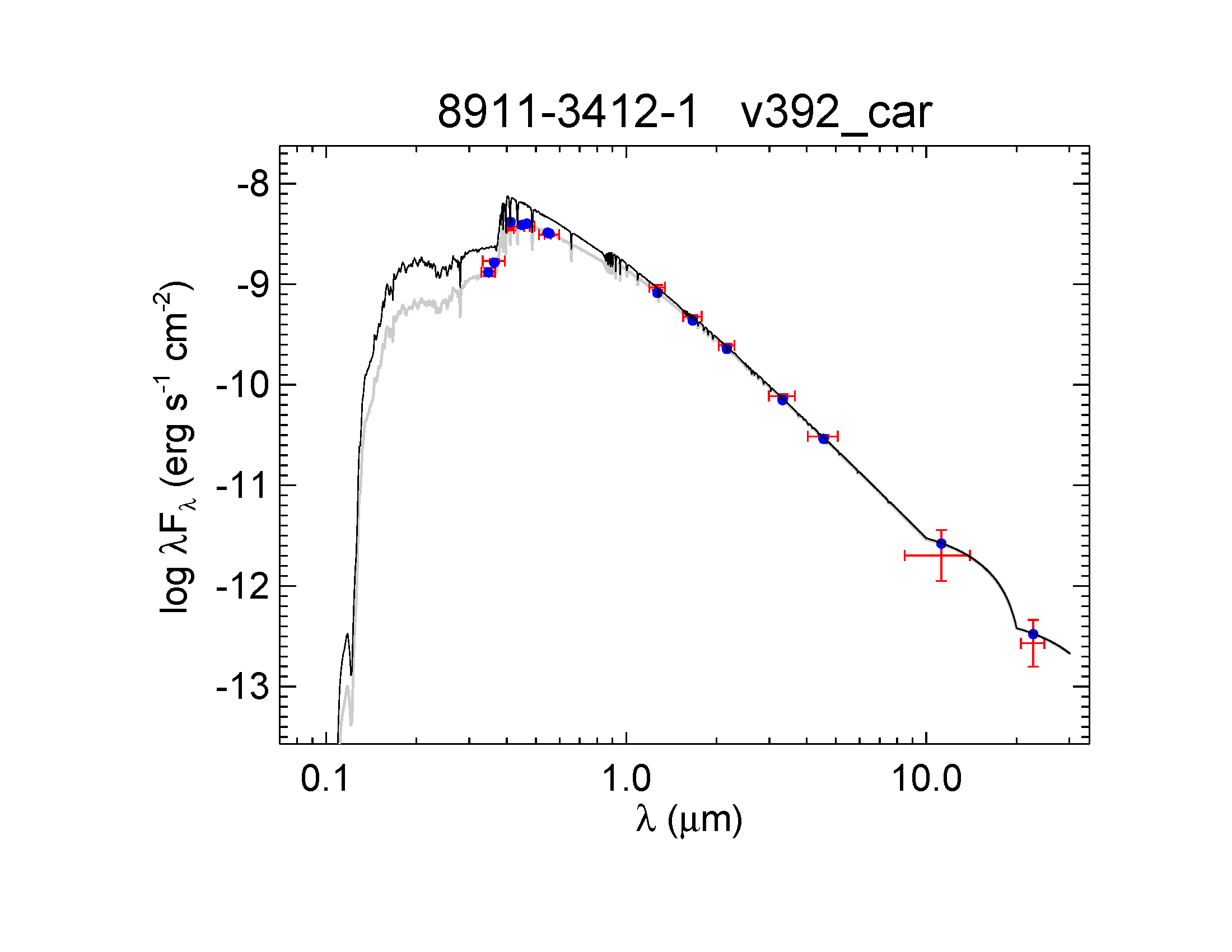}
  \includegraphics[trim=60 60 60 60,clip,width=0.49\linewidth]{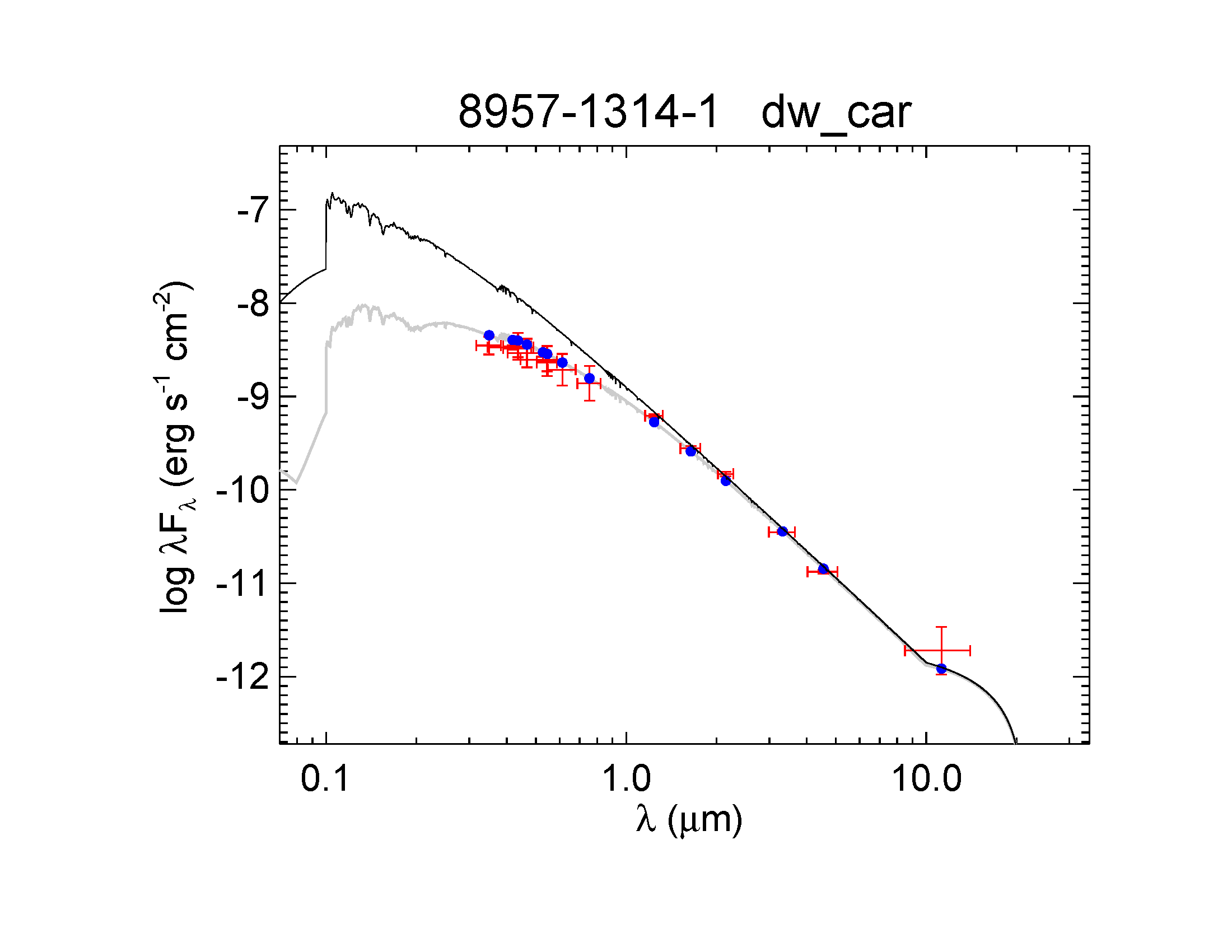}
  \includegraphics[trim=60 60 60 60,clip,width=0.49\linewidth]{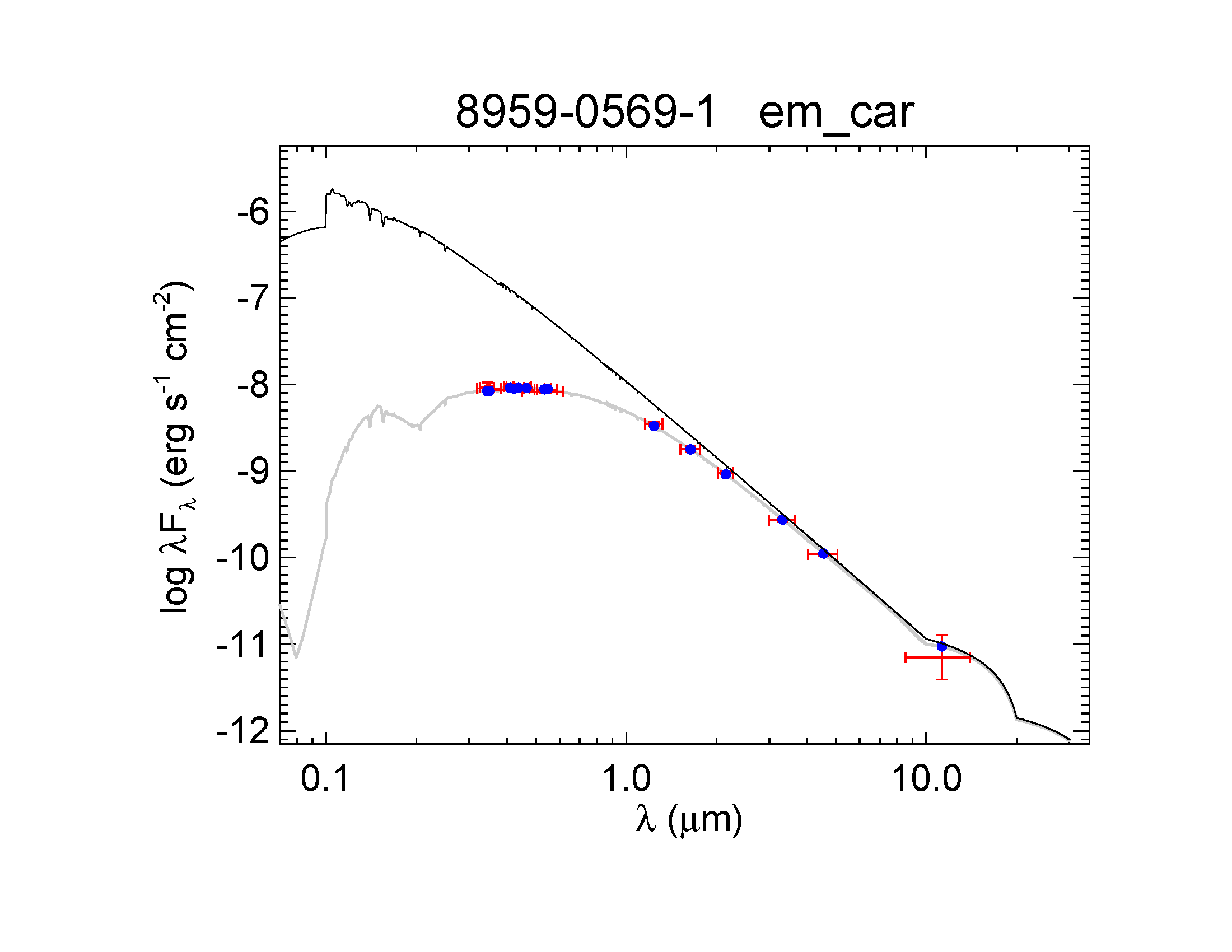}
  \includegraphics[trim=60 60 60 60,clip,width=0.49\linewidth]{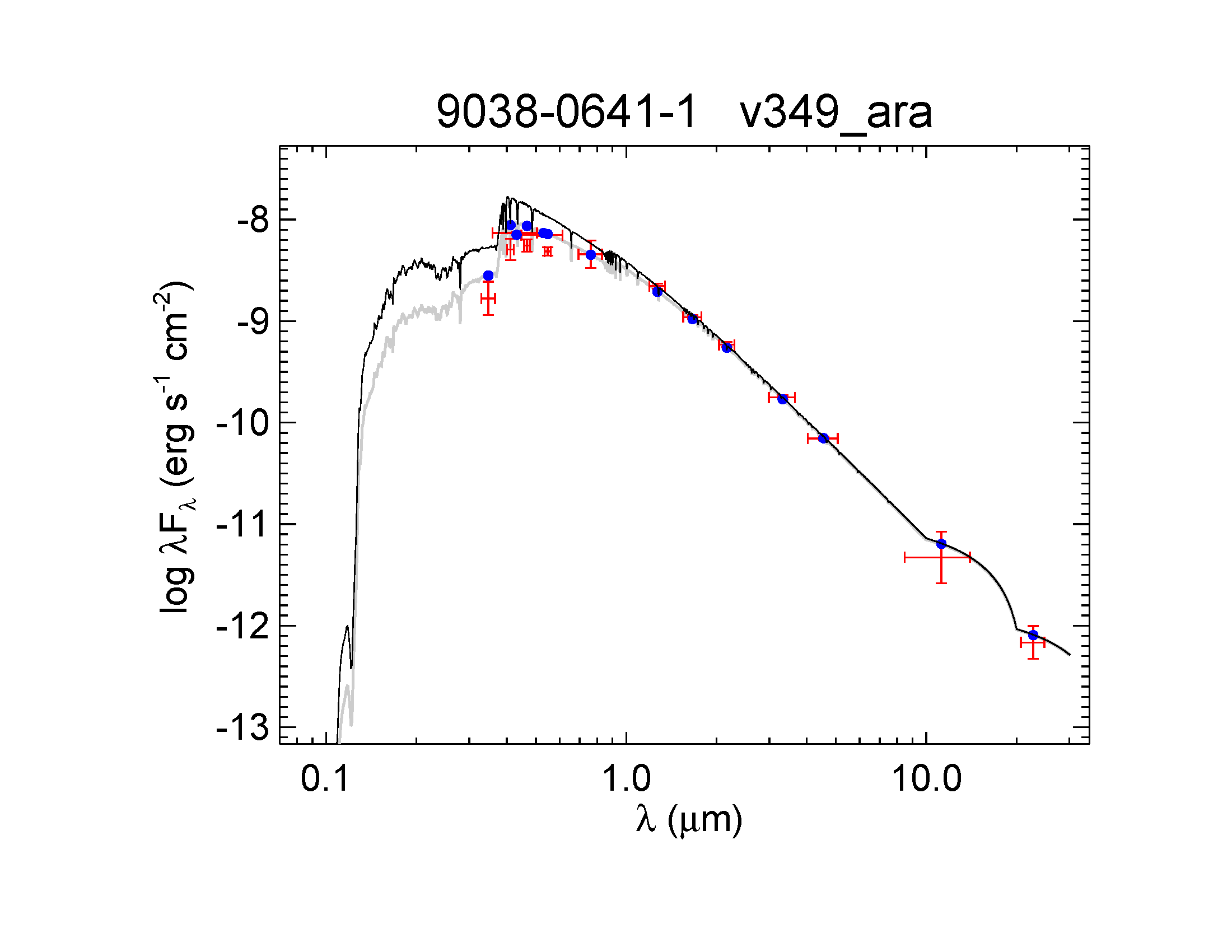}
  \includegraphics[trim=60 60 60 60,clip,width=0.49\linewidth]{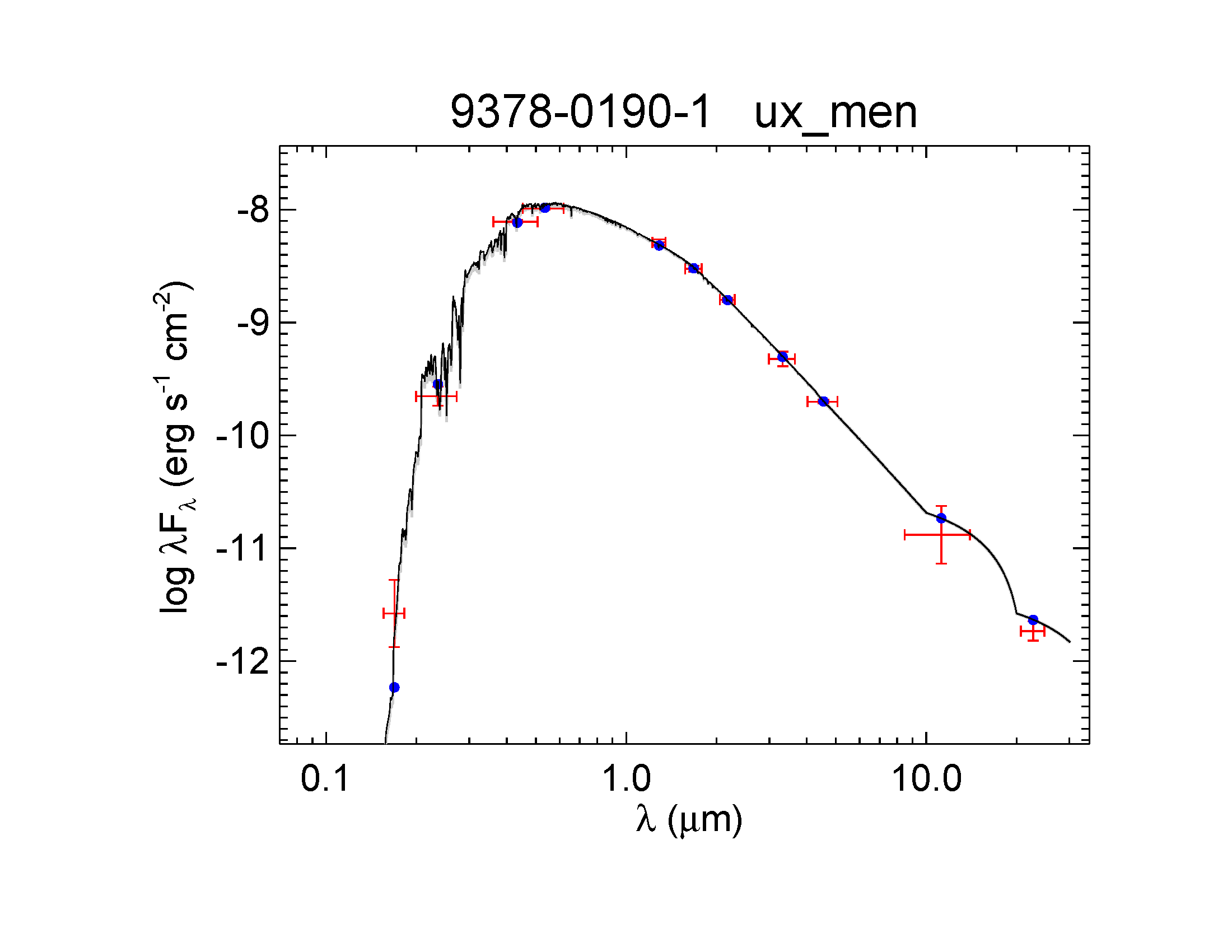}
  \includegraphics[trim=60 60 60 60,clip,width=0.49\linewidth]{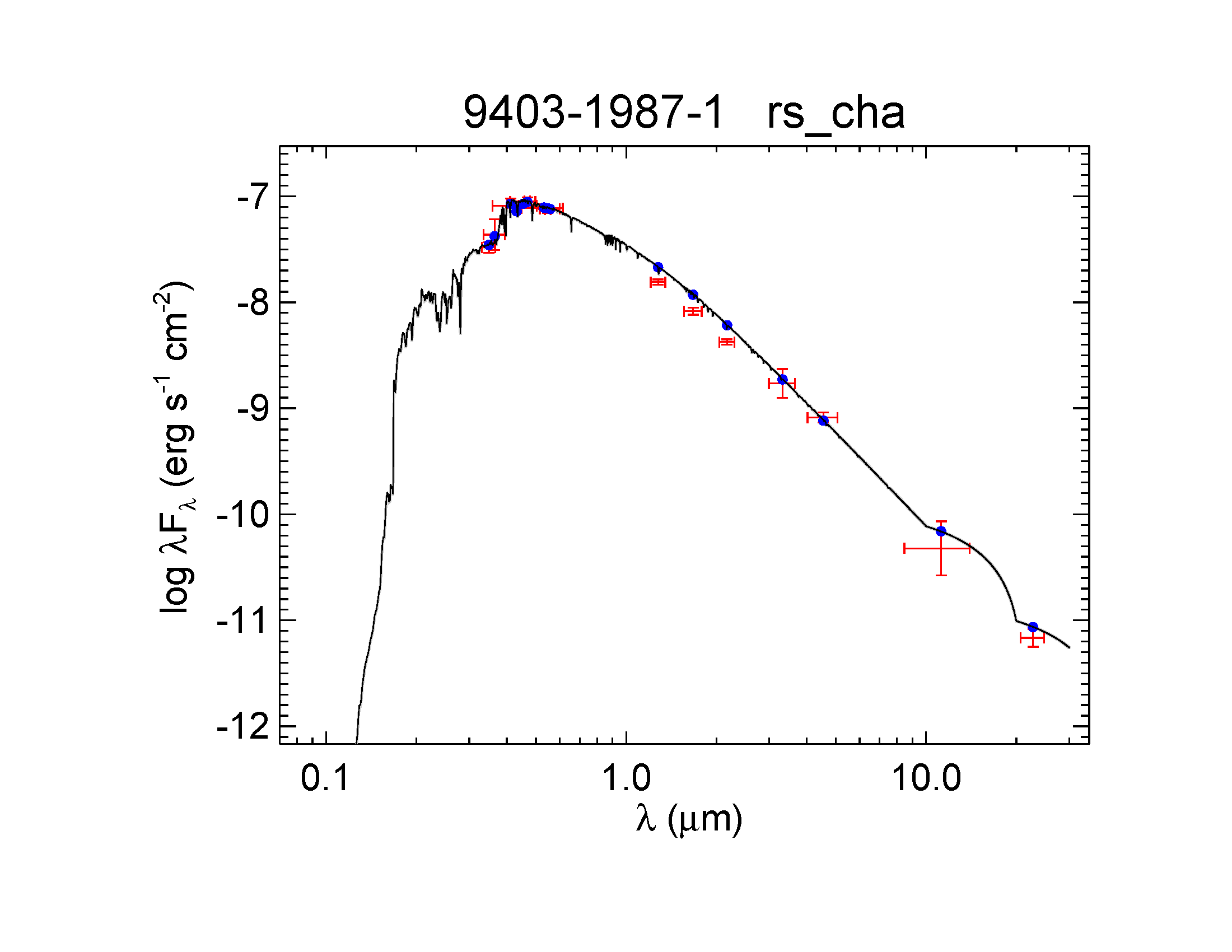}
  \caption{All labels, lines, symbols, and colors as in Figure \ref{fig:seds}.}
  \label{fig:seds_26}
\end{figure}

\begin{figure}[H]
  \centering
  \includegraphics[trim=60 60 60 60,clip,width=0.49\linewidth]{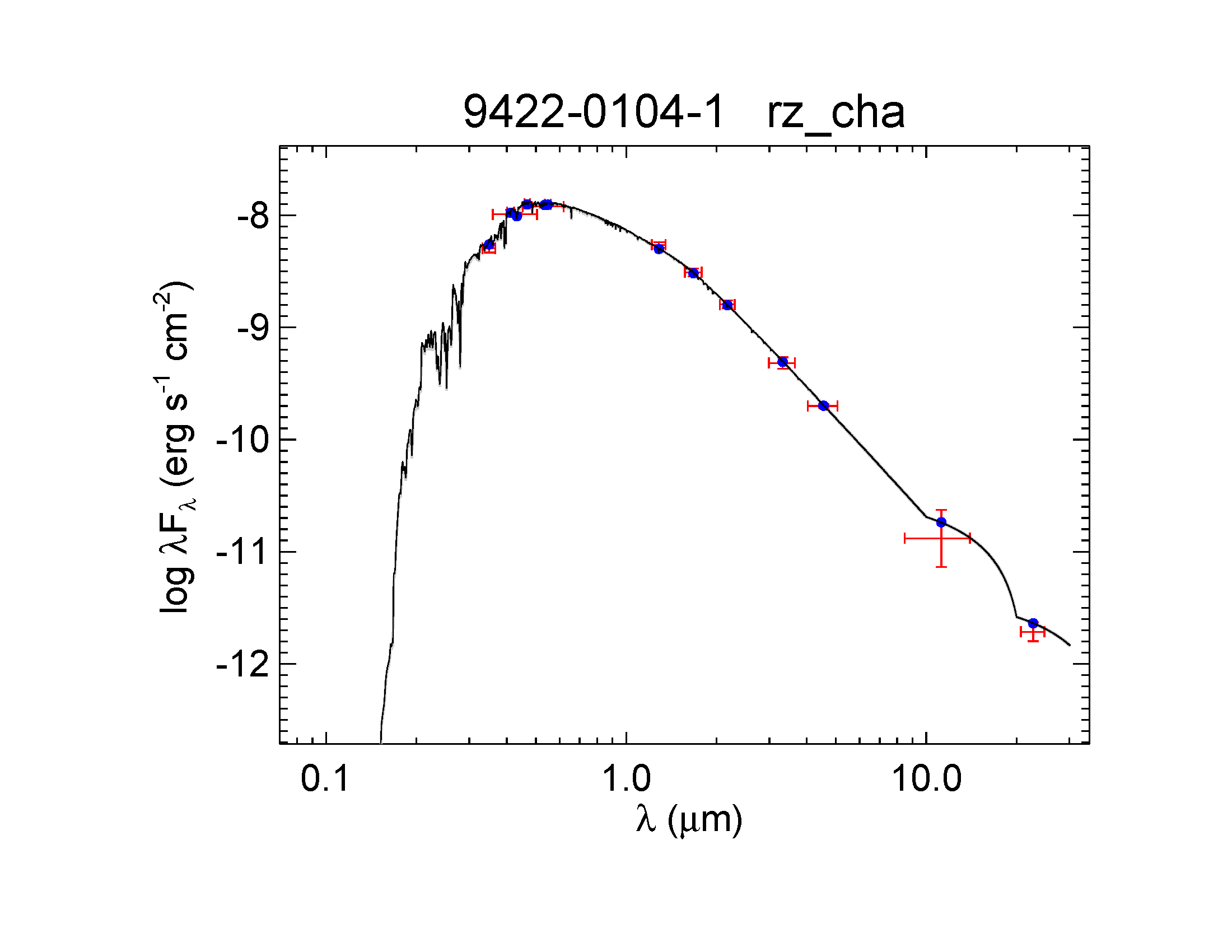}
  \includegraphics[trim=60 60 60 60,clip,width=0.49\linewidth]{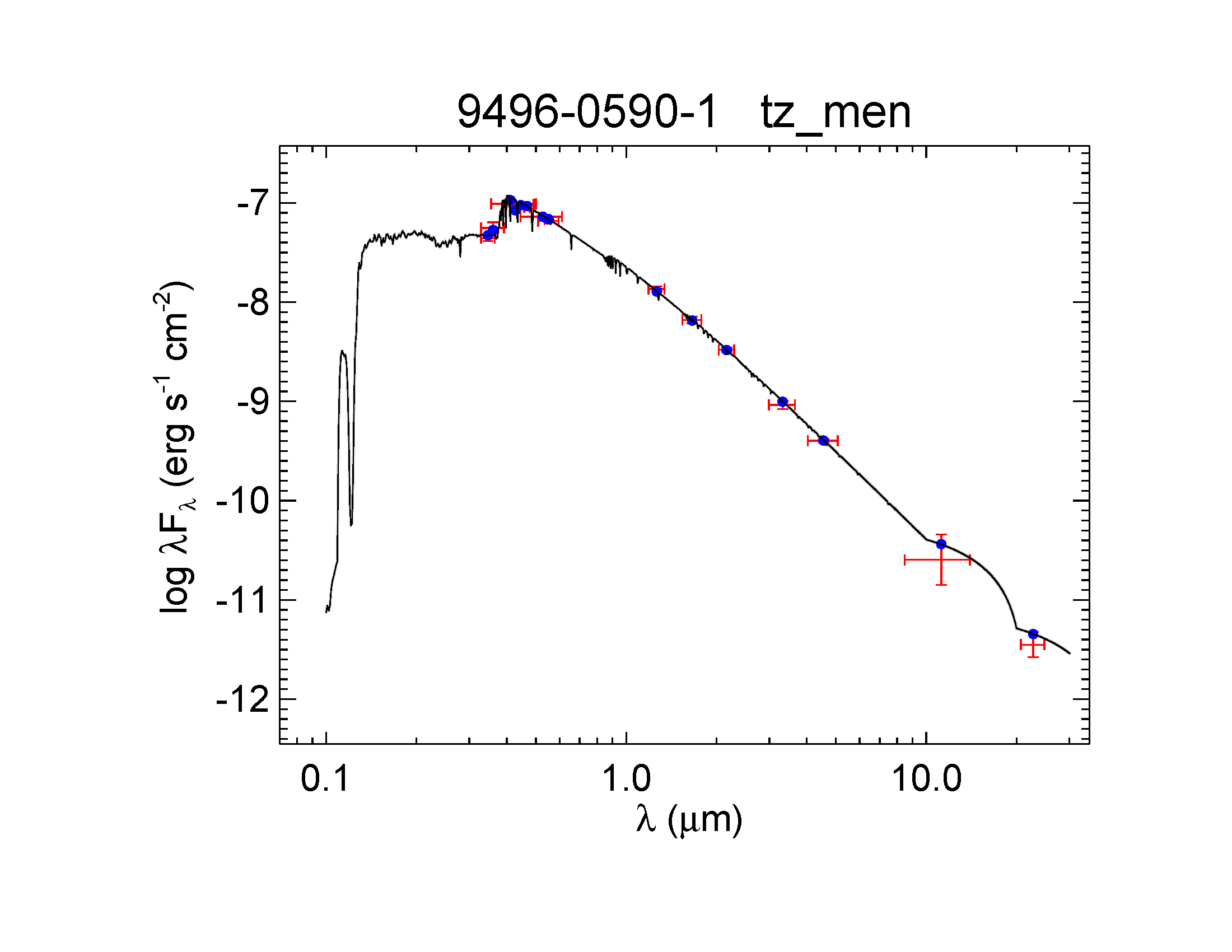}
  \caption{All labels, lines, symbols, and colors as in Figure \ref{fig:seds}.}
  \label{fig:seds_27}
\end{figure}


\begin{thebibliography}{}


\bibitem[Albrecht et al.(2013)]{Albrecht:2013} Albrecht, S., Setiawan,
J., Torres, G., Fabrycky, D.~C., \& Winn, J.~N.\ 2013, \apj, 767, 32

\bibitem[Bailer-Jones(2015)]{Bailer-Jones:2015} Bailer-Jones, C.~A.~L.\ 2015, \pasp, 127, 994

\bibitem[Bak{\i}{\c s} et al.(2008)]{Bakis:2008} Bak{\i}{\c s}, V.,
Bak{\i}{\c s}, H., Demircan, O., \& Eker, Z.\ 2008, \mnras, 384, 1657

\bibitem[Bruntt et al.(2006)]{Bruntt:2006} Bruntt, H., Southworth, J.,
Torres, G., et al.\ 2006, \aap, 456, 651

\bibitem[Burger et al.(2013)]{Burger:2013} Burger, D., Stassun, K.~G., Pepper, J., et al.\ 2013, Astronomy and Computing, 2, 40

\bibitem[{\c C}ak{\i}rl{\i} et al.(2008)]{Cakirli:2008} {\c
C}ak{\i}rl{\i}, {\"O}., Ibano{\v g}lu, C., Southworth, J., Frasca, A.,
\& Hernandez, J.\ 2008, \mnras, 389, 205

\bibitem[{\c C}ak{\i}rl{\i} et al.(2012)]{Cakirli:2012} {\c
C}ak{\i}rl{\i}, {\"O}., Dervi{\c s}o{\u g}lu, A., Sipahi, E., Ibano{\u g}lu, C.\ 2012, \na, 17, 215

\bibitem[Cardelli et al.(1989)]{Cardelli:1989} Cardelli, J.~A., Clayton, G.~C., \& Mathis, J.~S.\ 1989, \apj, 345, 245 

\bibitem[Clausen et al.(2010a)]{Clausen:2010a} Clausen, J.~V., Olsen,
E.~H., Helt, B.~E., \& Claret, A.\ 2010a, \aap, 510, A91

\bibitem[Clausen et al.(2010b)]{Clausen:2010b} Clausen, J.~V., Frandsen,
S., Bruntt, H., et al.\ 2010b, \aap, 516, A42

\bibitem[da Silva et al.(2014)]{daSilva:2014} da Silva, R., Maceroni,
C., Gandolfi, D., Lehmann, H., \& Hatzes, A.~P.\ 2014, \aap, 565, A55

\bibitem[David et al.(2016)]{David:2016} David, T.~J., Conroy, K.~E., Hillenbrand, L.~A., et al.\ 2016, \aj, 151, 112 

\bibitem[Debosscher et al.(2013)]{Debosscher:2013} Debosscher, J.,
Aerts, C., Tkachenko, A., et al.\ 2013, \aap, 556, A56

\bibitem[Dimitrov et al.(2015)]{Dimitrov:2015} Dimitrov, W.,
Kami{\'n}ski, K., Lehmann, H., et al.\ 2015, \aap, 575, A101

\bibitem[Erdem et al.(2015)]{Erdem:2015} Erdem, A., S{\"u}rgit, D.,
Engelbrecht, C.~A., \& Van Heerden, H.~P.\ 2015, \pasa, 32, e028

\bibitem[Fekel et al.(2013)]{Fekel:2013} Fekel, F.~C., Henry, G.~W.,
\& Sowell, J.~R.\ 2013, \aj, 146, 146

\bibitem[Frandsen et al.(2013)]{Frandsen:2013} Frandsen, S., Lehmann,
H., Hekker, S., et al.\ 2013, \aap, 556, A138

\bibitem[Graczyk et al.(2014)]{Graczyk:2014} Graczyk, D., Pietrzy{\'n}ski, G., Thompson, I.~B., et al.\ 2014, \apj, 780, 59 

\bibitem[Graczyk et al.(2015)]{Graczyk:2015} Graczyk, D., Maxted,
P.~F.~L., Pietrzy{\'n}ski, G., et al.\ 2015, \aap, 581, A106

\bibitem[Graczyk et al.(2016)]{Graczyk:2016} Graczyk, D., Smolec, R., Pavlovski, K., et al.\ 2016, \aap, in press (arXiv:1608.01000)

\bibitem[Groenewegen et al.(2007)]{Groenewegen:2007} Groenewegen, M.~A.~T., Decin, L., Salaris, M., \& De Cat, P.\ 2007, \aap, 463, 579 

\bibitem[Harmanec et al.(2011)]{Harmanec:2011} Harmanec, P., Bo{\v
z}i{\'c}, H., Mayer, P., et al.\ 2011, \aap, 531, A49

\bibitem[Harmanec et al.(2014)]{Harmanec:2014} Harmanec, P., Holmgren,
D.~E., Wolf, M., et al.\ 2014, \aap, 563, A120

\bibitem[Hauschildt et al.(1999)]{Hauschildt:1999} Hauschildt, P.~H., Allard, F., \& Baron, E.\ 1999, \apj, 512, 377 

\bibitem[He{\l}miniak et al.(2015a)]{Helminiak:2015a} He{\l}miniak,
K.~G., Ukita, N., Kambe, E., \& Konacki, M.\ 2015a, \apjl, 813, L25

\bibitem[He{\l}miniak \& Konacki(2011)]{Helminiak:2011} He{\l}miniak,
K.~G., \& Konacki, M.\ 2011, \aap, 526, A29

\bibitem[He{\l}miniak et al.(2014)]{Helminiak:2014} He{\l}miniak,
K.~G., Brahm, R., Ratajczak, M., et al.\ 2014, \aap, 567, A64

\bibitem[He{\l}miniak et al.(2015b)]{Helminiak:2015b} He{\l}miniak,
K.~G., Graczyk, D., Konacki, M., et al.\ 2015b, \mnras, 448, 1945

\bibitem[H{\o}g et al.(2000)]{Hog:2000} H{\o}g, E., Fabricius, C., Makarov, V.~V., et al.\ 2000, \aap, 355, L27 

\bibitem[Hubrig et al.(2012)]{Hubrig:2012} Hubrig, S., Gonz{\'a}lez,
J.~F., Ilyin, I., et al.\ 2012, \aap, 547, A90

\bibitem[Ibano{\v g}lu et al.(2008)]{Ibanoglu:2008} Ibano{\v g}lu, C.,
Evren, S., Ta{\c s}, G., et al.\ 2008, \mnras, 384, 331

\bibitem[K{\i}ran et al.(2016)]{Kiran:2016} K{\i}ran, E., Harmanec,
P., De{\u g}irmenci, {\"O}.~L., et al.\ 2016, \aap, 587, A127

\bibitem[Koo et al.(2014)]{Koo:2014} Koo, J.-R., Lee, J.~W., Lee,
B.-C., et al.\ 2014, \aj, 147, 104

\bibitem[Kurucz(2013)]{Kurucz:2013} Kurucz, R.~L.\ 2013, Astrophysics Source Code Library, ascl:1303.024 

\bibitem[Lacy et al.(2014a)]{Lacy:2014a} Lacy, C.~H.~S., Torres, G.,
Wolf, M., \& Burks, C.~L.\ 2014a, \aj, 147, 1

\bibitem[Lacy et al.(2015)]{Lacy:2015} Lacy, C.~H.~S., Torres, G.,
Fekel, F.~C., Muterspaugh, M.~W., \& Southworth, J.\ 2015, \aj, 149,
34

\bibitem[Lacy et al.(2014b)]{Lacy:2014b} Lacy, C.~H.~S., Torres, G.,
Fekel, F.~C., \& Muterspaugh, M.~W.\ 2014b, \aj, 147, 148

\bibitem[Lacy et al.(2012)]{Lacy:2012} Lacy, C.~H.~S., Torres, G.,
Fekel, F.~C., Sabby, J.~A., \& Claret, A.\ 2012, \aj, 143, 129

\bibitem[Lacy et al.(1989)]{Lacy:1989} Lacy, C.~H., Gulmen, O., Gudur,
N., \& Sezer, C.\ 1989, \aj, 97, 822

\bibitem[Lehmann et al.(2016)]{Lehmann:2016} Lehmann, H., Borkovits,
T., Rappaport, S.~A., et al.\ 2016, \apj, 819, 33

\bibitem[Maceroni et al.(2014)]{Maceroni:2014} Maceroni, C., Lehmann,
H., da Silva, R., et al.\ 2014, \aap, 563, A59

\bibitem[Maceroni et al.(2013)]{Maceroni:2013} Maceroni, C.,
Montalb{\'a}n, J., Gandolfi, D., Pavlovski, K., \& Rainer, M.\ 2013,
\aap, 552, A60

\bibitem[M{\"a}dler et al.(2016)]{Madler:2016} M{\"a}dler, T., Jofr{\'e}, P., Gilmore, G., et al.\ 2016, arXiv:1606.03015 

\bibitem[Mantegazza et al.(2010)]{Mantegazza:2010} Mantegazza, L.,
Rainer, M., \& Antonello, E.\ 2010, \aap, 512, A42

\bibitem[Matson et al.(2016)]{Matson:2016} Matson, R.~A., Gies, D.~R.,
Guo, Z., \& Orosz, J.~A.\ 2016, \aj, 151, 139

\bibitem[Maxted et al.(2015)]{Maxted:2015} Maxted, P.~F.~L., Hutcheon,
R.~J., Torres, G., et al.\ 2015, \aap, 578, A25

\bibitem[Melis et al.(2014)]{Melis:2014} Melis, C., Reid, M.~J., Mioduszewski, A.~J., Stauffer, J.~R., \& Bower, G.~C.\ 2014, Science, 345, 1029 

\bibitem[Mermilliod(2006)]{Mermilliod:2006} Mermilliod, J.~C.\ 2006, VizieR Online Data Catalog, 2168,  

\bibitem[Michalik et al.(2015)]{Michalik:2015} Michalik, D., Lindegren, L., \& Hobbs, D.\ 2015, \aap, 574, A115 

\bibitem[Michalska et al.(2013)]{Michalska:2013} Michalska, G.,
Niemczura, E., Pigulski, A., Ste{\'s}licki, M., \& Williams, A.\ 2013,
\mnras, 429, 1354

\bibitem[Milone et al.(2010)]{Milone:2010} Milone, E.~F.,
Kurpi{\'n}ska-Winiarska, M., \& Oblak, E.\ 2010, \aj, 140, 129

\bibitem[Morales et al.(2009)]{Morales:2009} Morales, J.~C., Torres,
G., Marschall, L.~A., \& Brehm, W.\ 2009, \apj, 707, 671

\bibitem[Munari et al.(2004)]{Munari:2004} Munari, U., Dallaporta, S., Siviero, A., et al.\ 2004, \aap, 418, L31 

\bibitem[Nordstrom \& Johansen(1994)]{Nordstrom:1994} Nordstrom, B.,
\& Johansen, K.~T.\ 1994, \aap, 282, 787

\bibitem[North et al.(1997)]{North:1997} North, P., Studer, M., \&
Kunzli, M.\ 1997, \aap, 324, 137

\bibitem[Paunzen(2015)]{Paunzen:2015} Paunzen, E.\ 2015, \aap, 580, A23 

\bibitem[Perryman et al.(1997)]{Perryman:1997} Perryman, M.~A.~C., Lindegren, L., Kovalevsky, J., et al.\ 1997, \aap, 323,

\bibitem[Pietrzy{\'n}ski et al.(2013)]{Pietrzynski:2013} Pietrzy{\'n}ski, G., Graczyk, D., Gieren, W., et al.\ 2013, \nat, 495, 76 

\bibitem[Pietrzy{\'n}ski et al.(2009)]{Pietrzynski:2009} Pietrzy{\'n}ski, G., Thompson, I.~B., Graczyk, D., et al.\ 2009, \apj, 697, 862 

\bibitem[Pinsonneault et al.(1998)]{Pinsonneault:1998} Pinsonneault, M.~H., Stauffer, J., Soderblom, D.~R., King, J.~R., \& Hanson, R.~B.\ 1998, \apj, 504, 170 

\bibitem[Press et al.(1992)]{Press:1992} Press, W.~H., Teukolsky, S.~A., Vetterling, W.~T., \& Flannery, B.~P.\ 1992, Cambridge: University Press, |c1992, 2nd ed.,  

\bibitem[Ratajczak et al.(2010)]{Ratajczak:2010} Ratajczak, M.,
Kwiatkowski, T., Schwarzenberg-Czerny, A., et al.\ 2010, \mnras, 402,
2424

\bibitem[Rawls et al.(2016)]{Rawls:2016} Rawls, M.~L., Gaulme, P.,
McKeever, J., et al.\ 2016, \apj, 818, 108

\bibitem[Ribas et al.(2005)]{Ribas:2005} Ribas, I., Jordi, C., Vilardell, F., et al.\ 2005, \apjl, 635, L37 

\bibitem[Ricker et al.(2015)]{Ricker:2015} Ricker, G.~R., Winn, J.~N., Vanderspek, R., et al.\ 2015, Journal of Astronomical Telescopes, Instruments, and Systems, 1, 014003 

\bibitem[Rozyczka et al.(2011)]{Rozyczka:2011} Rozyczka, M., Kaluzny,
J., Pych, W., et al.\ 2011, \mnras, 414, 2479

\bibitem[Sabby et al.(2011)]{Sabby:2011} Sabby, J.~A., Sandberg Lacy,
C.~H., Ibanoglu, C., \& Claret, A.\ 2011, \aj, 141, 195

\bibitem[Sabby \& Lacy(2003)]{Sabby:2003} Sabby, J.~A., \& Lacy,
C.~H.~S.\ 2003, \aj, 125, 1448

\bibitem[Sandberg Lacy et al.(2012a)]{SandbergLacy:2012a} Sandberg Lacy,
C.~H., Fekel, F.~C., \& Claret, A.\ 2012a, \aj, 144, 63

\bibitem[Sandberg Lacy et al.(2012b)]{SandbergLacy:2012b} Sandberg Lacy,
C.~H., Torres, G., \& Claret, A.\ 2012b, \aj, 144, 167

\bibitem[Sandberg Lacy et al.(2016)]{SandbergLacy:2016} Sandberg Lacy, C.~H.,
Fekel, F.~C., Pavlovski, K., Torres, G., \& Muterspaugh, M.~W.\ 2016,
\aj, 152, 2

\bibitem[Sandberg Lacy \& Fekel(2014)]{LacyFekel:2014} Sandberg Lacy,
C.~H., \& Fekel, F.~C.\ 2014, \aj, 148, 71

\bibitem[Sandberg Lacy \& Fekel(2011)]{SandbergLacy:2011} Sandberg
Lacy, C.~H., \& Fekel, F.~C.\ 2011, \aj, 142, 185

\bibitem[Sandberg Lacy et al.(2010)]{SandbergLacy:2010} Sandberg Lacy,
C.~H., Torres, G., Claret, A., et al.\ 2010, \aj, 139, 2347

\bibitem[Schlegel et al.(1998)]{Schlegel:1998} Schlegel, D.~J., Finkbeiner, D.~P., \& Davis, M.\ 1998, \apj, 500, 525

\bibitem[Soderblom et al.(2005)]{Soderblom:2005} Soderblom, D.~R., Nelan, E., Benedict, G.~F., et al.\ 2005, \aj, 129, 1616 

\bibitem[Soderblom et al.(1998)]{Soderblom:1998} Soderblom, D.~R., King, J.~R., Hanson, R.~B., et al.\ 1998, \apj, 504, 192 

\bibitem[Southworth et al.(2011)]{Southworth:2011} Southworth, J.,
Pavlovski, K., Tamajo, E., et al.\ 2011, \mnras, 414, 3740

\bibitem[Southworth(2013)]{Southworth:2013} Southworth, J.\ 2013,
\aap, 557, A119

\bibitem[Southworth \& Clausen(2006)]{Southworth:2006} Southworth, J.,
\& Clausen, J.~V.\ 2006, \apss, 304, 199

\bibitem[Southworth(2015)]{Southworth:2015} Southworth, J.\ 2015, Living Together: Planets, Host Stars and Binaries, 496, 164 

\bibitem[Sowell et al.(2012)]{Sowell:2012} Sowell, J.~R., Henry,
G.~W., \& Fekel, F.~C.\ 2012, \aj, 143, 5

\bibitem[Stassun et al.(2014)]{Stassun:2014} Stassun, K.~G., Feiden, G.~A., \& Torres, G.\ 2014, \nar, 60, 1 

\bibitem[Suchomska et al.(2015)]{Suchomska:2015} Suchomska, K.,
Graczyk, D., Smolec, R., et al.\ 2015, \mnras, 451, 651

\bibitem[Tamajo et al.(2012)]{Tamajo:2012} Tamajo, E., Munari, U.,
Siviero, A., Tomasella, L., \& Dallaporta, S.\ 2012, \aap, 539, A139

\bibitem[Tkachenko et al.(2014)]{Tkachenko:2014} Tkachenko, A.,
Degroote, P., Aerts, C., et al.\ 2014, \mnras, 438, 3093

\bibitem[Tomkin(2005)]{Tomkin:2005} Tomkin, J.\ 2005, The Observatory, 125, 232 

\bibitem[Tomkin \& Fekel(2006)]{Tomkin:2006} Tomkin, J., \& Fekel,
F.~C.\ 2006, \aj, 131, 2652

\bibitem[Torres et al.(2010)]{Torres:2010} Torres, G., Andersen, J., \& Gim{\'e}nez, A.\ 2010, \aapr, 18, 67 

\bibitem[Torres et al.(2014a)]{Torres:2014a} Torres, G., Vaz, L.~P.~R.,
Sandberg Lacy, C.~H., \& Claret, A.\ 2014a, \aj, 147, 36

\bibitem[Torres et al.(2009a)]{Torres:2009a} Torres, G., Sandberg Lacy,
C.~H., \& Claret, A.\ 2009a, \aj, 138, 1622

\bibitem[Torres \& Lacy(2009b)]{Torres:2009b} Torres, G., \& Lacy,
C.~H.~S.\ 2009b, \aj, 137, 507

\bibitem[Torres et al.(2014b)]{Torres:2014b} Torres, G., Sandberg Lacy,
C.~H., Pavlovski, K., et al.\ 2014b, \apj, 797, 31

\bibitem[Torres et al.(2015)]{Torres:2015} Torres, G., Sandberg Lacy,
C.~H., Pavlovski, K., Fekel, F.~C., \& Muterspaugh, M.~W.\ 2015, \aj,
150, 154

\bibitem[Torres et al.(2012)]{Torres:2012} Torres, G., Clausen, J.~V.,
Bruntt, H., et al.\ 2012, \aap, 537, A117

\bibitem[Torres et al.(2000)]{Torres:2000} Torres, G., Lacy, C.~H.~S., Claret, A., \& Sabby, J.~A.\ 2000, \aj, 120, 3226 

\bibitem[Trudel et al.(1993)]{Trudel:1993} Trudel, J.-L., Fernie,
J.~D., \& Mochnacki, S.\ 1993, \aj, 105, 2291

\bibitem[van Leeuwen(2009)]{vanLeeuwen:2009} van Leeuwen, F.\ 2009, \aap, 497, 209 

\bibitem[van Leeuwen(2007)]{vanLeeuwen:2007} van Leeuwen, F.\ 2007, \aap, 474, 653

\bibitem[Veramendi \& Gonz{\'a}lez(2015)]{Veramendi:2015} Veramendi,
M.~E., \& Gonz{\'a}lez, J.~F.\ 2015, \na, 34, 266

\bibitem[Vos et al.(2012)]{Vos:2012} Vos, J., Clausen, J.~V.,
J{\o}rgensen, U.~G., et al.\ 2012, \aap, 540, A64

\bibitem[Williams et al.(2011)]{Williams:2011} Williams, S.~J., Gies,
D.~R., Helsel, J.~W., Matson, R.~A., \& Caballero-Nieves, S.\ 2011,
\aj, 142, 5

\bibitem[Wilson et al.(2010)]{Wilson:2010} Wilson, R.\ E., Van Hammme, W., \& Terrell, D. 2010, \apj, 723, 1469

\bibitem[Zhang et al.(2009)]{Zhang:2009} Zhang, X.~B., Deng, L., \&
Lu, P.\ 2009, \aj, 138, 680

\bibitem[Zwahlen et al.(2004)]{Zwahlen:2004} Zwahlen, N., North, P., Debernardi, Y., et al.\ 2004, \aap, 425, L45 


\end{thebibliography}
\end{document}